\documentclass[12pt,twoside]{book}
\usepackage{a4}
\usepackage{epsfig}
\usepackage{ifthen}
\usepackage{amssymb}
\usepackage{lscape}

\newcommand{\ly}{Ly$\alpha$}
\newcommand{\mrm}{\mathrm}
\begin{document}

\newcommand{\twopics}[4]{
\begin{figure}
\label{#1}
\begin{minipage}{170mm}
\begin{center}
\hskip -60mm \epsfig{file=#2,width=6.0cm,angle=-90}
\epsfig{file=#3,width=6.0cm,angle=-90}\\
\end{center}
\end{minipage}
\caption{#4}
\end{figure}
}
\newcommand{\twopicsup}[6]{
\begin{figure}
\begin{minipage}{170mm}
\begin{center}
\hskip -60mm \epsfig{file=#2,width=#3,angle=0}
\epsfig{file=#4,width=#5,angle=0}\\
\end{center}
\end{minipage}
\caption{#6}
\label{#1}
\end{figure}
}
\newcommand{\singlefig}[5]{
\begin{figure}
\begin{center}
\epsfig{file=#2,width=#3,angle=#4}\\
\end{center}
\caption{#5}
\label{#1}
\end{figure}
}

\newcommand{\sevdef}[4]{
\left\{\begin{array}{r@{\quad:\quad}l}
#1 & #2 \\
#3 & #4 
\end{array}
\right.
}

\newcommand{\sevdeft}[6]{
\left\{\begin{array}{r@{\quad:\quad}l}
#1 & #2 \\
#3 & #4 \\ 
#5 & #6 \\ 
\end{array}
\right.}
\newcommand{\vect}[2]{
\left(\begin{array}{c}
#1 \\
#2 \\  
\end{array}
\right)
}
\frontmatter
\title{\Huge{The Ly$\alpha$ Emission Line as}\\ \Huge{a Cosmological Tool}} 
\author{\LARGE{Thesis for the degree of Doctor of Philosophy}\\ \\  \LARGE{Kim K. Nilsson}\\ \large{K{\o}benhavns Universitet}\\ \large{ESO -- European Southern Observatory}\\ \\ \\  \Large{Thesis supervisors:}\\ \\
\Large{Johan P.U. Fynbo}\\
\large{Dark Cosmology Centre, K{\o}benhavns Universitet, Danmark} \\ \\
\Large{Palle M{\o}ller} \\ \large{ESO -- European Southern Observatory}}
\date{2007-09-01}
\maketitle
\chapter{Abstract}
\begin{flushright}
NEC FASCES\\
NEC OPES\\
SOLA ARTIS\\
SCEPTRA PERENNANT\\
\emph{Not power, not wealth,}\\
\emph{only the reign} \\
\emph{of art and science} \\
\emph{shall persist}\\
\bf{Tycho Brahe 1546 - 1601}
\end{flushright}
This thesis deals with different aspects of a special kind of high redshift 
galaxy, namely Ly$\alpha$ emitters. Ly$\alpha$ emitters are galaxies found
through their Ly$\alpha$ emission, at redshifts larger than $z \geq 2$ where 
the emission line has been redshifted into the optical or near-infrared
regime. The thesis has two main parts; a lower redshift, observational 
part ($z \sim 3$) and a more technical/theoretical very high redshift part
($z \sim 9$).

In the first, lower redshift part I present the analysis of a narrow-band image
taken in the GOODS-S field, focused on a redshift for Ly$\alpha$ of $z = 3.15$.
The image, covering a central $\sim 7' \times 7'$ part of the GOODS-S field
revealed 25 Ly$\alpha$ emitting candidates, of which one turned out to be
a so-called Ly$\alpha$ ``blob''. Ly$\alpha$ blobs are large nebulae of 
gas emitting a large amount of light in the Ly$\alpha$ line. They can be very 
bright
($L_{Ly\alpha} \sim 10^{44}$~erg~s$^{-1}$) and have projected diameters
as large as 150 kpc. Three possible mechanisms have been proposed to explain 
this phenomenon; \emph{i)} star formation and ``super-winds'', \emph{ii)} AGN
activity and \emph{iii)} cold accretion. The blob in GOODS-S turned out to be
the first Ly$\alpha$ blob best explained by cold accretion, and was published
in a \emph{Letter to the Editor} in A\&A in June 2006 (Nilsson et al. 2006a).
The remaining 24 Ly$\alpha$ emitters were analysed together, in order to 
gain insight in the nature of galaxies selected through narrow-band imaging
for redshifted Ly$\alpha$ emission. An extensive SED fitting was performed
and the results also included the discovery of an apparent filamentary 
structure and a comparison to Lyman-Break Galaxies at similar redshifts.
The results from this analysis is accepted for publication in A\&A
in Nilsson et al. (2007).

In the second part, I discuss future, very high redshift narrow-band surveys
for Ly$\alpha$ emitters. In February 2005, the idea of buying a narrow-band 
filter for the, then half way through its construction, Visible and Infrared
Survey Telescope for Astronomy (VISTA) to search for Ly$\alpha$ emitters at 
very high redshift started. In Chapter~\ref{chapter:vista} I describe the 
science case for the filters and the process of designing and procurement of
the filters. In this chapter is also described how the initial idea for a 
narrow-band survey was developed into what has become ELVIS - Emission Line
galaxies with VISTA Survey, part of the Ultra-VISTA survey which will --
hopefully -- start observing in the Spring of 2008. The need for better
predictions of what may be observed with this type of survey was realised
early on, and in Chapter~\ref{chapter:predictions} I present a paper
in which we try to make more accurate predictions of the outcome of very
high redshift narrow-band surveys for Ly$\alpha$ emitters using two
theoretical models.

Finally, in a project unrelated to the other two parts of the thesis, I
present a search for a ``Fundamental plane'' of Ly$\alpha$ emitters in
the colour space produced by large-scale multi-wavelength surveys such
as GOODS or COSMOS. The goal of the project was to decide the most efficient
narrow-band colour selection method for detecting Ly$\alpha$ emitters at 
several different redshifts. We also wanted to optimise the selection method
to exclude interloper galaxies such as lower redshift [OII]-emitters. In
this thesis I present the results of this study.

\chapter{Acknowledgments}
I am tremendously grateful to all who have in some way helped or supported
me and my work that has lead to this thesis. I would like to thank IDA, DARK
and ESO for funding my PhD studentship and for taking me in. 
I have been helped by many people. In particular my two 
supervisors Johan Fynbo and Palle M{\o}ller have been pillars of support.
We may not always have agreed, but I have learned incredibly much from
both of you and I feel very fortunate in my supervision. I am not sure that
ELVIS would have happened if Johan had not believed in my crazy idea. Along
the same line, I should probably thank my room-mate at that winter school
in Obergurgl in February 2005 for giving me encouraging words when I woke
her up in the middle of the night to tell her about my idea. I am sorry I 
do not remember her name! I am also very grateful for help with the VISTA 
project from Will Sutherland, who has been a great help and support and who
always answered my e-mails with questions very quickly. Further input and
help with ELVIS has been thankfully received from Wolfram Freudling, Michelle
Doherty, Lisbeth Fogh-Grove, Piero Rosati, Ian Smail, Jim Emerson, Jean-Gabriel
Cuby and Gavin Dalton.

Another person that has been very important to me socially and professionally
is my fiancee Ole M{\"o}ller. I met Ole shortly after moving to Munich and
it was interest at first sight. Our personal relationship has since then
evolved into a professional dito, when we realised that Oles love for 
programming filled a void in my own skills. I am thus grateful and proud
of the fruitful joint ventures we have had so far in writing a beautiful
SED fitting code and a ``fundamental plane'' code. Ole, you have also
been an amazing support when listening to my rants during both ups and downs
in my life. Thank you.

Further, I would like to thank Alvaro Orsi, Matthew Hayes and Christian
Tapken for interesting meetings and discussions which have led to 
more or less completed projects. I would like to thank all my other co-authors
for wanting to work with me and for your assistance in producing brilliant
results! Thank you {\'A}rd{\'i}s and Jos{\'e} Mar{\'i}a for reading and
commenting on my thesis manuscript and for being good friends throughout my
PhD time. A special thanks to Klaus Meisenheimer and 
Hans-Walter Rix at 
Max-Planck-Institut f{\"u}r Astronomie in Heidelberg for believing in
me to the extent that you hired me for a post-doc seven months before I
was due to finish my PhD. I look forward to working with you. Of course I
also want to thank my thesis committee --- Sangheeta Malhotra, G{\"o}ran 
{\"O}stlin and Jens Hjorth --- for taking time to evaluate my work.

Finally, a thanks to all my friends at ESO and DARK, who made the work-days 
more pleasant, and to my mother Yvonne for loving support and my cat Matrise 
for unconditional love irrespective of my mood. To the people that have helped 
me: you are my heroes. Thank you all.

\tableofcontents
\listoffigures
\listoftables
\mainmatter
\chapter{Introduction}
\markboth{Chapter 1}{Introduction}
\section{The high redshift Universe}
\subsection{Brief history of our Universe and the redshift}
\begin{figure}[!ht]
\begin{center}
\includegraphics[width=0.70\textwidth,height=1.00\textwidth]{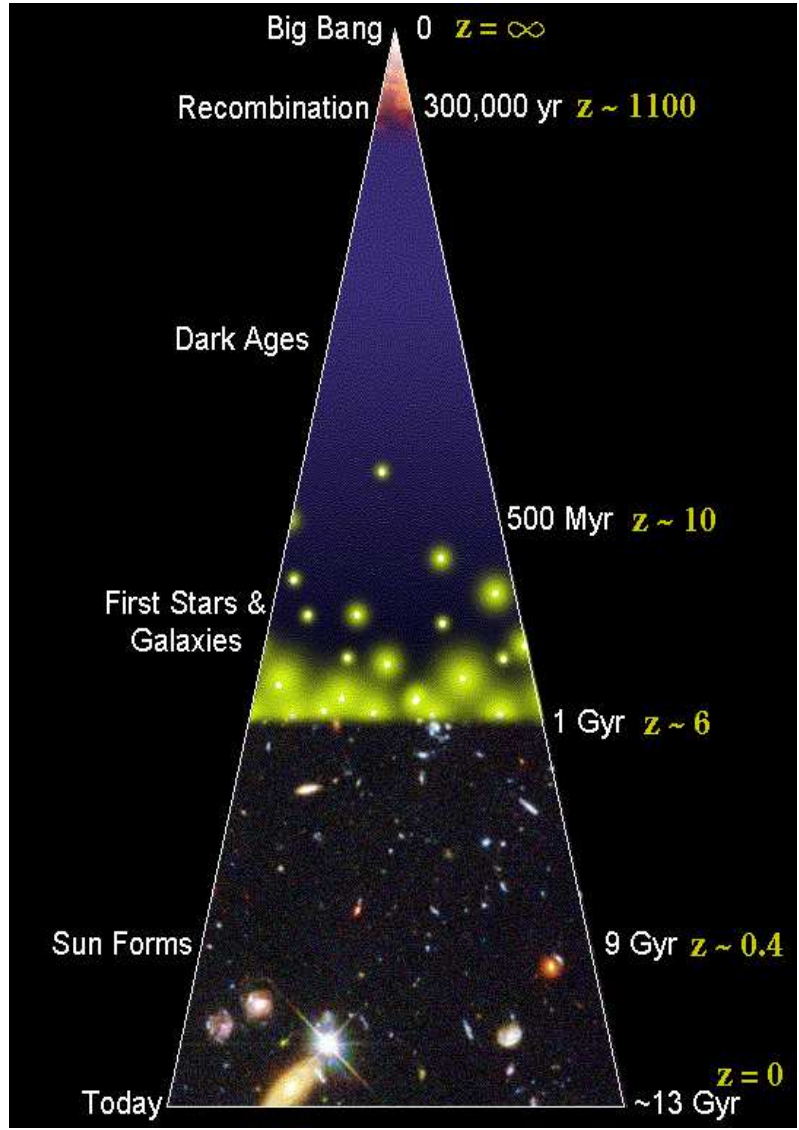}
\caption[Time-line of Universe, illustration of redshift]{
This plot illustrates the history of our Universe and marks a few note-worthy
redshifts, see text. From \emph{www.astronomy.ohio-state.edu/$\sim$pogge/TeachRes/Artwork/Cosmology/index.html}.\label{fig:redshiftill}}
\end{center}
\end{figure}

Our Universe was created in the Big Bang. In the first few fractions of a second
the Universe was extremely hot, but it expanded and cooled rapidly. After this 
initial stage of ``primeval soup'', where particles and radiation were
coupled and high energy physics governed everything, a period of cooling
began. At around $300'000$~yrs after the Big Bang, the Universe had cooled
and de-pressurised enough so that nuclei had formed and the electrons had
been captured to make neutral atoms. This is called recombination, see 
Fig.~\ref{fig:redshiftill}. At this time, the photons created in the Big Bang
and immediately thereafter were released and could begin their journey through 
the Universe. This radiation is called the Cosmic Microwave Background
(CMB) and is an almost perfect blackbody emission with a peak temperature of
2.73~K. The irregularities seen in the CMB indicate that there were slight
inhomogeneities in the matter density at recombination. These inhomogeneities
later turned into the structures we see today.

After recombination, these peaks in the matter-density distribution continued 
to contract due to gravity and after some $\sim 100$~Myrs, the first stars
were born. These were massive stars that burned their fuel quickly, exploded
and enriched their surroundings. Soon, more stars were forming and
galaxies started to take shape. Also, the first quasars may have lit up at 
this time.
These galaxies and quasars produced a large amount of UV radiation and this
radiation field ensured that the Universe would once again become ionised
and the electrons separated from the atoms. This is called the re-ionisation
and is believed to have occurred at $z \approx 6-10$, between $0.5 - 1.0$~Gyrs
after the Big Bang. The search for when exactly this happened is a hot topic
today, as our telescopes push further and further back in time.
The current record in redshift for an observed galaxy today is $z = 6.96$
(Iye et al. 2006), and observations in the near future should be able to push this
record to redshifts around $z \sim 9$, see also 
Chapter~\ref{chapter:vista} and \ref{chapter:predictions}.

\begin{figure}[!pt]
\begin{center}
\includegraphics[width=0.60\textwidth,height=0.40\textwidth]{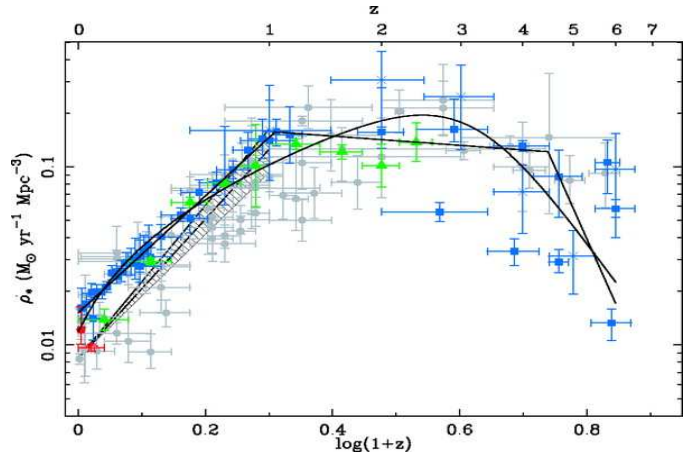}
\caption[Star formation rate density history]{
Star formation rate density history of the Universe. Data points are 
star formation rate densities from observed galaxies. Lines are best fit
extrapolations. From Hopkins \& Beacom (2006).\label{fig:sfrhistory}}
\end{center}
\vspace{-0.5cm}
\end{figure}

From redshift $\sim 6$ and on, the Universe has continued to expand, cool,
gather mass in clusters and galaxies and produce stars. The redshift at
which the star formation rate per volume peaked appears to be around redshift
$z \sim 1-2$, see Fig.~\ref{fig:sfrhistory}.
Our Sun was formed some 4~Gyrs ago, corresponding to redshift $z \approx 0.4$.

So, what is redshift? Redshift occurs as photons travel through the Universe,
which simultaneously expands. An analogy is the change in pitch when a trains 
whistle or
an ambulances sirens pass your ear. The sound waves are
compressed in front of the train, thus the pitch goes up, and elongated
behind the train, thus the decrease in pitch again. In space, the size of
the Universe changes with time according to the scale factor \emph{a(t)}.
Thus, photons travelling towards us will stretch its wavelength due to the
expanding Universe and the light will be shifted towards the red, i.e.  
redshift. Redshift is defined as:
\begin{center}
\begin{equation}\label{eq:redshift}
1 + z = \frac{\lambda_{observed}}{\lambda_{emitted}} = \frac{a(t_{z=0})}{a(t_{emitted})}
\end{equation}
\end{center}
\begin{figure}[!pb]
\begin{center}
\includegraphics[width=0.80\textwidth,height=0.50\textwidth]{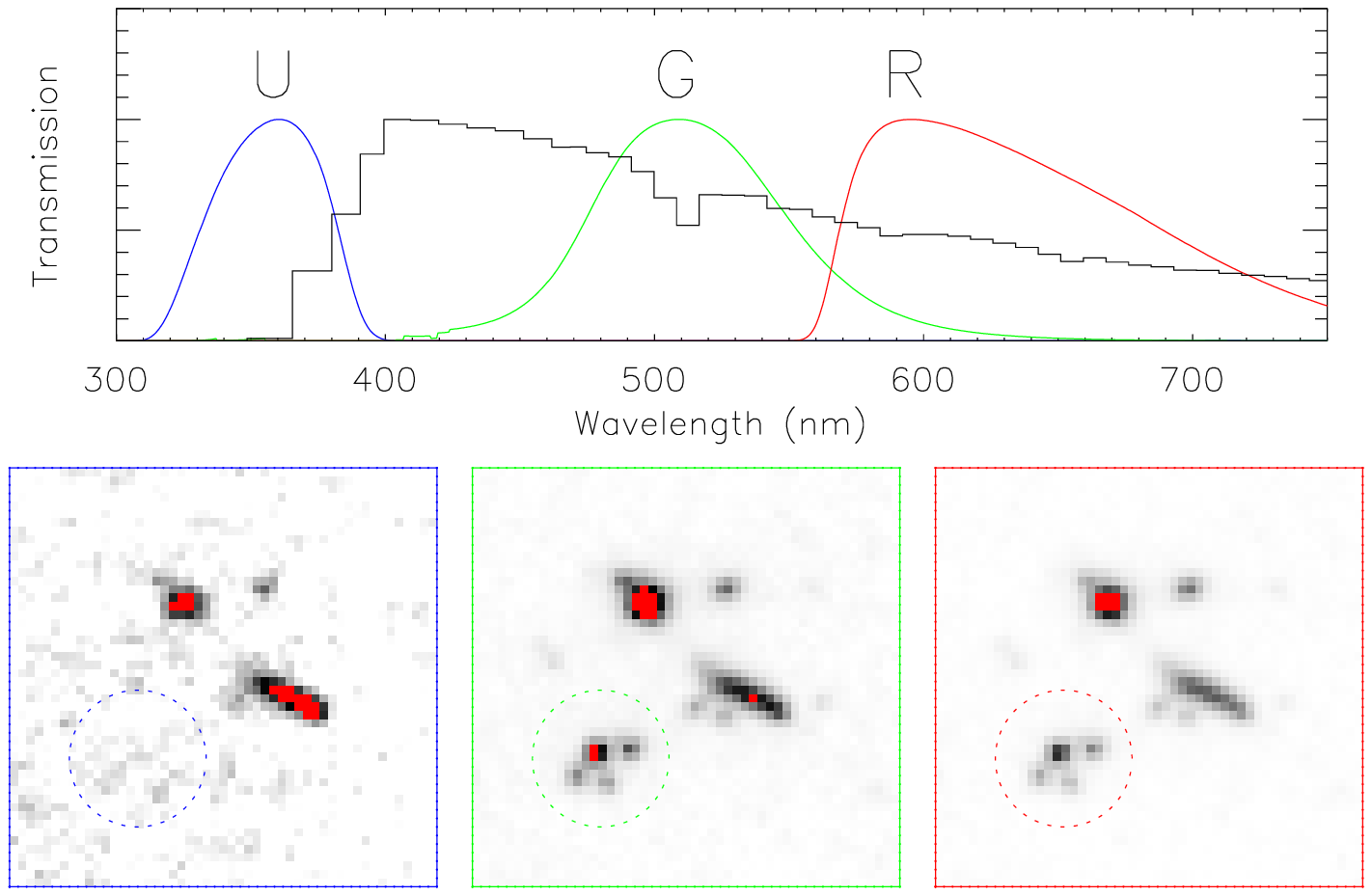}
\caption[Illustration of LBG method]{Illustration of the LBG selection method. The top panel shows the typical
shape of a galaxy spectrum, redshifted to $z \sim 3$. The Lyman
Break can be seen at approximately 400 nm. The bottom panels show how such a 
galaxy would be seen as observed through the three broad-band filters U, G and 
R that are located on the blue and the red side of the break respectively. 
The galaxy will not be observed in the filter blueward of the break, 
but is clearly seen in the red filters. Image credit: Johan Fynbo.\label{fig:lbgmethod}}
\end{center}
\end{figure}

\subsection{Methods of finding high redshift galaxies}
In our constant search for knowledge we want to understand how the stars
and galaxies were formed, and how they evolved. We can make models of
star formation, but ultimately we need observations of the young Universe.
Thus, we search for high redshift galaxies, whose light was emitted a long time
ago, corresponding to $t_{emitted}$ in Eq.~\ref{eq:redshift}. 
That means that the light we observe from a galaxy at
redshift $z \sim 6$ was emitted from a very young galaxy, when the Universe
was only $\sim 1$~Gyrs old. Looking further away in redshift means looking
further into our past. And thus finding a large, representative sample of
galaxies at different redshifts will give us insight into galaxy 
formation and evolution.

There are many methods of finding star forming galaxies in the high
redshift Universe, and each method explores different classes of
galaxies. One of the major questions in observational cosmology is 
thus what are the relations between these classes. This thesis explores
properties of high redshift galaxies selected by their Ly$\alpha$ emission 
(see below). In this introduction I discuss several different methods used to 
detect high redshift galaxies.

\subsubsection{Lyman Break Galaxies selection}
One of the most common methods in the last decade, now comprising an
impressive catalogue of thousands of galaxies, has been the
selection based on photometric redshifts gained from observations of
the ``Lyman Break'' in galaxies. The Lyman Break, located at 912 {\AA} in
the spectrum of a galaxy, represents the cut-off energy where the single
electron of a hydrogen atom is ionised. Almost all the UV light of
a galaxy is absorbed at wavelengths shorter than this, as hydrogen is 
extremely abundant in galaxies. Thus, a ``break'' in the spectrum of the 
galaxy. The Lyman Break Galaxy (LBG) selection method was pioneered
by Steidel et al. (1996; 1999; 2000; 2003), but has also been used by many 
other groups (e.g. Madau et al. 1996; Pettini et al. 2001; Bunker et al. 2004; 
Stanway et al. 2004; Ouchi et al. 2004a,b; Wadadekar et al. 2006).
An illustration of the method is found in Fig.~\ref{fig:lbgmethod}.
As can be seen, a galaxy will appear to ``drop out'' in the bluest filter, 
hence the method is sometimes also referred to as the drop-out technique. 
Spectroscopic follow-up is necessary to confirm the high redshift nature
of the galaxy.

The Lyman Break technique can be used for a wide range of redshifts, when 
different filters are used for the selection. Table~\ref{tab:lbgzs} gives
the filters and redshifts commonly used, or proposed to be used.
\begin{table}[ht!]
\begin{center}
\caption[Redshifts and filters used in the LBG method]{Redshifts and filters used in the LBG method.}\label{tab:lbgzs}
\vspace{0.5cm}
\begin{tabular}{ccccccccc}
\hline
\hline
Blue filter & Red filters & Redshift range &\\
\hline
U        & B, G, R, V         & 2.5 - 3.5   &\\
B        & G, R, V            & 3.5 - 4.5   &\\
V        & \emph{i}, \emph{z} & 4.5 - 6.0   &\\
\emph{i} & \emph{z}, J        & 6.0 - 7.5   &\\
\emph{z} & J, H, K$_s$        & 7.5 - 11.5  &\\
J        & H, K$_s$           & 11.5 - 15.5 &\\
\hline
\end{tabular}
\end{center}
\end{table}
The technique is typically spectroscopically complete to an R band magnitude of 
$R \sim 25.5$ for U-band drop-outs but the rate of confirmation falls for
higher redshift LBG candidates (Giavalisco 2002). A few studies of the 
properties of LBGs such as masses, dust content, ages etc.~have been made
so far. These studies are well summarised in Giavalisco (2002). LBGs appear
to have ages ranging from a few to several hundred Myrs or even up
to 1 Gyr. Stellar masses lie in the range 
$10^9 < \mathrm{M}_{\star} / \mathrm{M}_{\odot} < 10^{11}$. Papovich, 
Dickinson \& Ferguson (2001) find very high extinction in their sample of LBGs,
with $A_V \approx 1 - 2$, while Shapley et al. (2001) and Verma et al. (2007)
find more modest values of $A_V \approx 0.3 - 0.5$ in their samples. Thus, 
it appears that LBGs are medium mass and medium dusty galaxies, with high
star formation rates of several hundred solar masses per year (Shapley et al.
2001). Several groups have also detected clustering in
LBGs, similar to that of Ly$\alpha$ emitters (Steidel et al. 1998; Giavalisco
et al. 1998; Ouchi et al. 2004b; see also sec.~\ref{sec:introlss}).

\subsubsection{Damped Ly$\alpha$ Absorption selection}
\begin{figure}[t]
\begin{center}
\includegraphics[width=0.70\textwidth,height=0.55\textwidth]{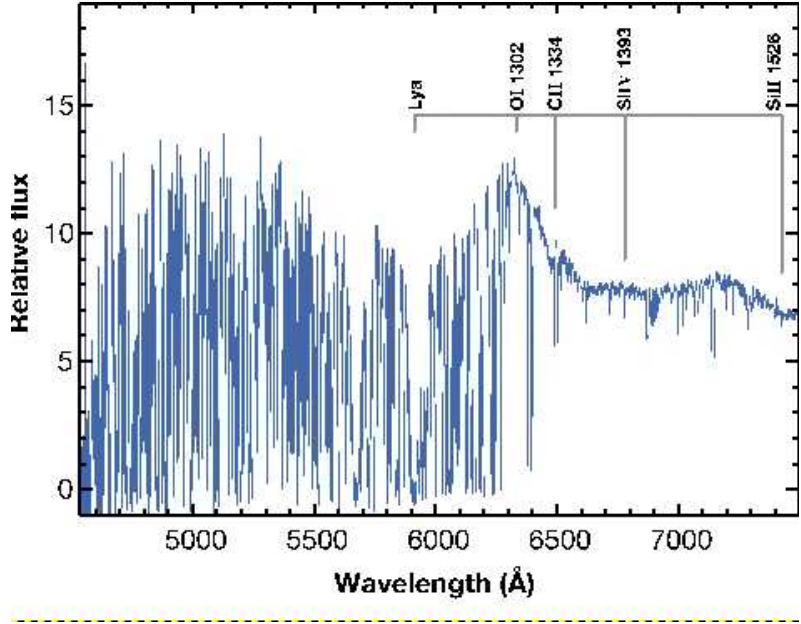}
\caption[Illustration of DLA method]{Illustration of the DLA selection method. The 
figure shows the Keck/ESI spectrum of QSO PSS0209+0517 at redshift 
$z_{QSO} = 4.17$. Two DLA systems can be observed at redshifts $z_{DLA} = 3.86$
and $z_{DLA} = 3.67$. The labelling refers to absorption lines identified
to be associated with the $z_{DLA} = 3.86$ system. From Wolfe, Gawiser \& 
Prochaska (2005).\label{fig:dlamethod}}
\end{center}
\end{figure}

When observing a quasar at high redshift, sometimes a gas cloud with a high
column density will be found along the same line of sight. This is not as uncommon
as one might think. When a cloud with an HI column density larger than
$N(H1) \geq 2\times10^{20}$~cm$^{-2}$ is observed along the sight-line of a 
quasar it is called a Damped Ly$\alpha$ Absorber (DLA) as its Ly$\alpha$
absorption line will be damped. Thus, the DLA technique is based on finding
high column density galaxies by searching for absorption lines in the spectra
of quasars. An illustration of the method can be found in Fig.~\ref{fig:dlamethod}.
The search for DLA systems began already in the mid-80s but a large sample
of DLA galaxies were not collected until ten years ago (for a review, see
Wolfe, Gawiser \& Prochaska 2005). 

A problem with this selection method is the apparent proximity of the quasar 
in the sky, making imaging follow-up difficult. However, DLA system can still
give valuables insights into properties such as the neutral gas fraction in the
Universe, chemical evolution and metallicity production across a large
redshift range and gas kinematics in galaxies. 

\subsubsection{Selection based on sub-millimeter emission}
\begin{figure}[t]
\begin{center}
\includegraphics[width=0.60\textwidth,height=0.40\textwidth]{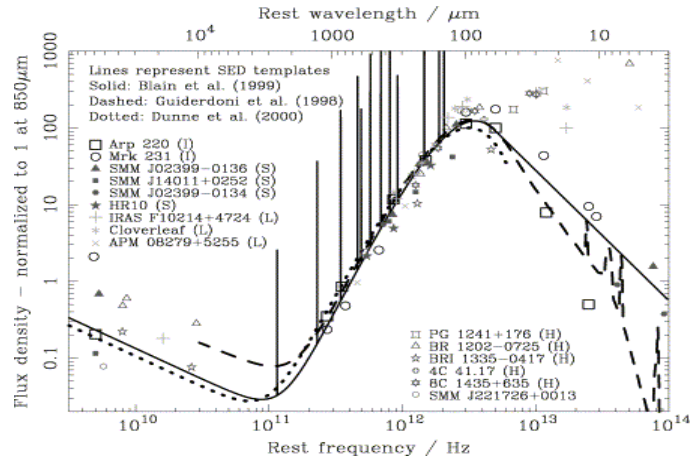}
\caption[SED of starburst galaxy]{Composite SED of a starburst galaxy. The 
peak at sub-mm wavelengths is apparent. From Blain et al. (2002).\label{fig:submmmethod}}
\end{center}
\end{figure}

The Spectral Energy Distribution (SED) of a dusty starburst galaxy will have
its peak emission at wavelengths of $\sim 0.1$~mm, see 
Fig.~\ref{fig:submmmethod}. These galaxies will,
however, not have significant emission in the optical or near-infrared parts
of the spectrum. Thus, observing the sky in the sub-mm will reveal a class of
galaxies otherwise left unstudied. Due to the difficulties in building 
detectors
for these wavelengths, it was not until the commissioning of the SCUBA
detector (Holland et al. 1999) on the James Clerk Maxwell Telescope (JCMT) in
1997 that progress was made in this field of study.
As the sample of sub-mm selected galaxies has grown to be $\sim 100$ (e.g.
Blain et al. 2002; Ivison et al. 2005), the median redshift of this sample 
appears to be
$z \sim 2$, although sub-mm emission should in theory be just as efficient
to detect galaxies between $z = 1 - 10$. 

Two, related, problems with selecting
galaxies based on sub-mm emission are that the beam sizes are still very large
(of the order 10 arcsec) and that identification of optical/infrared 
counterparts are difficult. The process of identifying a counterpart is 
generally done by first searching for a radio counterpart, since radio 
observations have smaller beam sizes and radio emissions have been shown to be 
proportional to sub-mm emission (e.g. Carilli \& Yun, 1999). Following 
identification of a radio counterpart, an optical or infrared counterpart
may be found. However, even with radio identifications, optical/infrared 
identifications can be difficult to make. The sample of galaxies with 
confirmed counterparts display a wide variety of properties including
AGN and starburst activity, as well as signs of recent mergers (Ivison et al. 
2000; Blain et al. 2002). Future
instruments such as SCUBA-II and ALMA will hopefully give us a better insight 
into the nature of sub-mm selected galaxies.

\subsubsection{Gamma Ray Burst selection}
Gamma Ray Bursts (GRBs) are violent explosions, more energetic than 
supernovae. The name derives from the fact that GRBs are found through a 
burst of gamma ray emission. GRBs are typically divided into two classes
of events, long or short duration bursts; where the long bursts are believed to
be related to relativistically beamed extreme versions of (hundreds of
times brighter than) supernovae (e.g. Hjorth et al. 2003), i.e.
the death throws of a massive star, and the short bursts are believed to occur
when two compact objects (i.e. neutron stars or black holes) collide 
(e.g. Hjorth et al. 2005). 
The method of finding high redshift galaxies using GRBs is to use the location
of the GRB to search for its host galaxy. The gamma ray event of the GRB
triggers a satellite based telescope, such as for instance the 
Swift\footnote{http://www.swift.psu.edu/} telescope, to slew to the part of the
sky where the event occurred. In general, the accuracy of the positioning from
the gamma ray telescope is poor, e.g. of the order of arcminutes, and it is 
followed up with an X-ray telescope,
as X-rays are easier to pin-point and most GRB afterglows emit strongly also in
X-rays. When the location of the GRB has been localised to a smaller error box,
optical/near-IR telescopes take over and search for the optical/near-IR afterglow 
of the GRB. Thus, the GRB can
finally be completely localised, to subarcsecond resolution, using an optical 
detection. After the GRB
has faded, this position can be observed to greater depth to search for the
host galaxy of the GRB. Host galaxies of GRBs are often blue, star-forming
galaxies (e.g. Bloom et al. 1998; Sokolov et al. 2001; Gorosabel et al. 2005;
Fruchter et al. 2006).

\subsubsection{Ly$\alpha$ selection}\label{sec:nbtech}
A very efficient method to detect high redshift galaxies is to observe the
sky with a narrow-band filter focused on the Ly$\alpha$ emission line at a
particular redshift. A narrow-band filter is a filter that allows only a 
very small range of wavelengths to pass through it and blocks all other
light. This will result in a narrow range of redshifts for Ly$\alpha$,
typically $\Delta z \approx 0.05$. In order to find the Ly$\alpha$ 
emitters, the same field is then observed with one or two broad band
filters with the same, or near the same, central wavelength as the narrow-band
filter. The broad band observations then probe the continuum of the source
and emission-line galaxies are found by comparing the narrow-band flux  
with the broad band flux. Objects with a high flux ratio of narrow-band
vs. broad band measurement are selected.

In the work that is presented in this thesis, the equivalent width of the line 
has been calculated in the selection process. Equivalent width (EW) is a
measure of the strength of an emission- (or absorption-) line and is
defined as

\begin{equation}\label{eq:ew}
EW = \frac{ F_{line} }{ f_{\lambda,cont} } 
\end{equation}

\noindent where $F_{line}$ is the flux in the emission/absorption line and
$f_{\lambda,cont}$ is the flux density in the continuum at the central
wavelength of the line. The EW is positive for emission lines and negative
for absorption lines. Object with a flat continuum, without emission or
absorption lines, should have
an EW of zero. The method used to find Ly$\alpha$ emitters in our surveys
is described in more detail in section~\ref{ch3:LEGOs}. In 
sec.~\ref{sec:lyareview}, results from past surveys for Ly$\alpha$ emitters are 
reviewed.

\section{Ly$\alpha$ in a historical perspective}
This thesis deals with high redshift galaxies found through their Ly$\alpha$ 
emission lines. The Lyman series ({\ly}, Ly{$\beta$}, Ly{$\gamma$}\ldots)
are the emission lines created when an electron falls from any higher energy 
level, back to the ground state of the simplest and most abundant 
atom in the Universe, the Hydrogen atom. The strongest line, with longest 
wavelength (1215.67 {\AA}), is the one from the 
second energy level to the first; {\ly}. The Lyman series are named after
Theodore Lyman, a physicist who discovered them during the first two decades of
the 20th century. 

In the mid 1920s, Menzel (1926) and Zanstra (1927) each separately discussed
the ionisation of emission lines in planetary nebulae, including {\ly},
and in 1945 Chandrasekhar discussed the radiative equilibrium of {\ly}. The
first observations of highly redshifted {\ly} came with the discovery of
quasars, or QSOs. It was in the 1950s that star-like objects were found
unexpectedly to have bright radio emission. In 1963 Schmidt published the first
``large redshift object'' (3C 273) and just two years later a quasar
was identified to be at redshift 2.01 through its {\ly} emission
(3C 9; Schmidt 1965). It was at the same time as these observations were
published that theoretical predictions for {\ly} emission and absorption in
the young Universe started. Gunn \& Peterson (1965) observed that
the continuum of the quasar on the blue side of the {\ly} emission line was
not absorbed and put an upper limit on the absorption by neutral hydrogen 
clouds in the line
of sight towards the quasar. Bahcall (1966) predicted the line strengths of
the Lyman lines, and finally Partridge \& Peebles (1967) published the first
article with predictions on the observability of the {\ly} from young
star forming galaxies, not QSOs, at high redshifts. Their results have a
remarkable level of agreement with the results of present day 
{\ly} surveys and sparked the field of narrow-band surveys for {\ly} emitters.

It would, however, take a bit more than two decades before the first
high redshift {\ly} emitters were to be found. Immediately following the
Partridge \& Peebles (1967) paper, some groups attempted to discover these
``primeval galaxies'' (Partridge 1974; Davis \& Wilkinson 1974; Meier 1976;
Hogan \& Rees 1979) but they were all unsuccessful. The narrow-band technique
(see sec.~\ref{sec:nbtech}) was brought into use in the late 1980s
but the first surveys were unsuccessful (e.g. Pritchet \& Hartwick 1989, 1990;
Rhee, Webb \& Katgert 1989). The first detections came in the early 1990s by
Lowenthal et al. (1991), Wolfe et al. (1992), M{\o}ller \& Warren (1993) and
Macchetto et al. (1993). From there on, narrow-band surveys became more and more
frequent and successful. Examples are the LALA survey (Rhoads et al.
2003; Dawson et al. 2003), the Building the Bridge Survey (Fynbo et al. 2003)
and the studies in the fields of radio galaxies (Venemans et al. 2007). To the
publication date of this thesis, more than 550 {\ly} emitters have
spectroscopic confirmation with redshifts ranging from $z \sim 2 - 7$, see
Fig.~\ref{fig:legozdist}.

\section{What are Ly$\alpha$ emitters?}\label{sec:lyareview}
\subsection{Objects that emit Ly$\alpha$}
A multitude of objects in the Universe emit {\ly}. To ionise the hydrogen
atom, a photon with a wavelength shorter than 912 {\AA} is needed. This kind
of UV radiation can come from several different sources. The most common sources 
are
young, massive and short-lived stars of the spectral type O and B. The more 
massive the star, the shorter it lives and the bluer its spectrum is. O and
B stars are the most massive ones known in the Universe, hence they emit
almost all of their light in the UV range of the spectrum. They are also very
luminous because they are so massive. Hence, {\ly} is often seen emitted from
regions with intense star formation, because the short lifetime of the O and
B stars make them burn out and die before they have time to move away from
their birthplace. So, O and B stars and hence star forming galaxies are often
intense sources of {\ly}.

Another type of object that often emits {\ly}, or has a {\ly} halo around it,
are quasars and active galactic nuclei (AGN, Schmidt 1965;
McCarthy 1993; Villar-Mart{\'i}n et al. 2005). These objects are believed
to consist of massive black holes accreting material from accretion disks
surrounding them. The fact that AGNs have an accretion disk is in part
crucial for the large {\ly} haloes seen around this type of object, as the
outcoming UV light is highly collimated along the axis of the disk (Weidinger
et al. 2005). 
That radio jets are observed in some AGN is also one of the primary
arguments of why there should be an accretion disk (Urry \& Padovani 1995). 
The spectrum emitted by
an AGN has three components, an X-ray power law spectrum, an infrared
bump and a strong UV continuum emission component (Collin 2001). It is this 
UV component that ionises the hydrogen, which will then emit {\ly} in two highly 
ionised cones perpendicular to the disk (Haiman \& Rees 2001; Weidinger et al. 
2005). Hence, AGN are also {\ly} emitters.

A third possibility for the production of {\ly} photons is the theory of ``cold
accretion'' (e.g. Fardal et al. 2001; Dijkstra et al. 2006a,b; Nilsson et al. 
2006a). 
The general idea is that if the Universe contains dark matter
haloes that initially have no galaxy in them but plenty of neutral hydrogen
gas, then the gas would slowly start to fall in, onto the halo because of the
gravitational potential energy. This would cause the material to heat up 
because of
the loss of potential, and this heat could theoretically be cooled off by
emitting {\ly} photons. The idea was prompted after a similar effect had
been seen in clusters of galaxies were the gas fell in, heated up and cooled
via X-ray emission (Fabian 1994). On a galaxy size scale, this emission could 
be predominantly in {\ly}.
Until recently this had never been observed. However, as can be read in
Chapter~\ref{chapter:blob}, we have published the first paper presenting a
probable observation of this phenomenon.

\subsection{Redshift distribution of Ly$\alpha$ emitters}
\begin{figure}[t]
\begin{center}
\includegraphics[width=0.70\textwidth,height=0.50\textwidth]{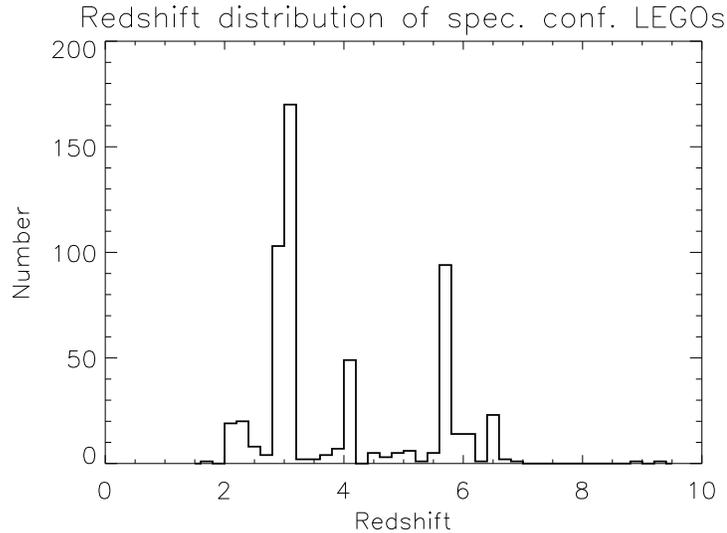}
\caption[Distribution of spectroscopically confirmed Ly$\alpha$ emitters]{Distribution of 
spectroscopically confirmed Ly$\alpha$ emitters, as of the date of submission of 
this thesis. The total number of confirmed Ly$\alpha$ emitters is 565.
\label{fig:legozdist}}
\end{center}
\end{figure}

Ly$\alpha$ can be observed with ease in the optical regime, covering 
approximately the redshift range of $2 \leq z \leq 6.5$. At even higher
redshifts, observations are possible in the near-infrared. The distribution
of spectroscopically confirmed Ly$\alpha$-emitters at the time of writing
this thesis can be found in Fig.~\ref{fig:legozdist}\footnote{I have been trying to keep an up-to-date list of all spectroscopically confirmed Ly$\alpha$ emitters. The true number may be slightly higher if I missed a publication. It will not be lower.}.
A much larger sample of candidate Ly$\alpha$-emitters exist, with 
photometric selection only. Spectroscopic follow-up generally has a 
success rate of 75~-~90\% for $z \sim 3$ surveys (Fynbo et al. 2001; 2003),
and approximately 50~-~75\% for $z = 5-7$ surveys (Taniguchi et al. 2005;
Kashikawa et al. 2006; Shimasaku et al. 2006). 

The multiply peaked distribution shown in
Fig.~\ref{fig:legozdist} can be explained in the following way. The peak at 
redshift $z \sim 3$ is most likely due to the CCD being most sensitive at
around 5000~{\AA} as well as to the abundance of [OIII] emission targeted 
narrow-band filters at most ground-based observatories. The peak at 
$z \sim 4.5$ is due to the LALA survey (Rhoads et al. 2000; Dawson et al. 2004).
At higher redshifts, $z \ge 5$, CCD detector efficiency starts to drop off
and the sky OH airglow lines gain in strength. Thus, narrow-band surveys at
these redshifts are focused on the OH airglow ``windows'' where the sky 
background is low. Such windows exist at redshift $z = 5.7, 6.5, 7.7, 8.8$\ldots
Hence the two peaks in the distribution at redshifts $z = 5.7$~and~$6.5$.
The current, undisputed, record in redshift to date is $z = 6.96$ from
Iye et al. (2006), although two tentative detections at $z \sim 9$ have been
reported in Stark et al. (2007). Three surveys have been made at $z = 8.8$ 
(Parkes, Collins \& Joseph 1994; Willis \& Courbin 2005; Cuby et al. 2007),
with upper limits as results.

\subsection{Star formation rates in Ly$\alpha$ emitters}
The star formation rate (SFR) of a galaxy denotes the amount of stellar
mass produced in that galaxy per year. The current SFR of our galaxy, 
the Milky Way, is approximately $3$~M$_{\odot}$~yr$^{-1}$ 
(Smail 2002).
When observing galaxies at high redshift, where stars or star 
forming regions are impossible to resolve, there are many
SFR indicators in use. Almost any part of the electromagnetic spectrum
can be used. Ranalli, Comastri \& Setti (2003) argue that the X-ray flux of a 
galaxy can be used as a SFR indicator based on a relationship between the X-ray
and radio/FIR fluxes. Radio and FIR fluxes have been shown to be tracers of 
star formation in Condon (1992) and Kennicutt (1998b) respectively. This is
because radio emission is dominated by synchrotron emission of accelerated
electrons in a magnetic plasma, typically from a supernovae remnant. These 
supernovae in
turn are proportional to the production rate of massive, short-lived stars,
which is proportional to the star formation rate. The FIR emission is 
dominated by thermal emission from the gas rich regions in which the stars
are formed. These regions are heated by the UV light of the newly formed
stars. The UV 
part of the spectrum ($\lambda \sim 1500$~{\AA}) has long been argued to
be a good tracer of the SFR of the galaxy due to the light from the bright,
short-lived, UV intense massive stars (Kennicutt 1998a). Further, several
emission-lines have been argued to be good tracers of star formation,
especially the H$\alpha$ emission line (Kennicutt 1983; Tresse et al. 2002)
and the [OII] emission line (Gallagher, Hunter \& Bushouse 1989; 
Hashimoto et al. 1998). Assuming a certain recombination rate between 
Ly$\alpha$ and H$\alpha$ one can derive a simple relation between
Ly$\alpha$ luminosity and star formation rate (Kennicutt 1983; Brocklehurst 
1971; Hu, Cowie \& McMahon 1998).

Star formation rates found in Ly$\alpha$ emitters are typically 
$\sim 1 - 10$~M$_{\odot}$~yr$^{-1}$ (e.g. Cowie \& Hu 1998; Hu, Cowie \&
McMahon 1998; Gronwall et al. 2007; Nilsson et al. 2007, see also 
references in Table~\ref{tab:sfrdlyalpha}). Many authors find 
discrepancies between 
Ly$\alpha$ SFR and the SFR derived from the UV continuum flux (e.g.
Ajiki et al. 2003; Taniguchi et al. 2005; Gronwall et al. 2007), with
larger UV derived SFRs than those derived from Ly$\alpha$. In Nilsson et al. 
(2007) we find
SFRs from $0.5$~to$6$~M$_{\odot}$/yr, but no discrepancy between 
continuum and emission line derived SFRs. Many studies of star formation rate
densities have been published for Ly$\alpha$. Most of these can be found in
Table~\ref{tab:sfrdlyalpha}.
\begin{table}[t!]
\begin{center}
\caption[Summary of star formation rate densities in the literature]{Summary of Ly$\alpha$ results for the cosmic star formation rate density found in the 
literature. Gronwall et al.~2007 correct for incompleteness in their sample,
Ouchi et al. (2003) and Shimasaku et al. (2006) estimates have been calculated
from the UV flux of the Ly$\alpha$ emitters. Malhotra \& Rhoads (2004) 
estimates have been derived from luminosity functions.}\label{tab:sfrdlyalpha}
\vspace{0.5cm}
\begin{tabular}{lcccccccc}
\hline
\hline
Reference & Redshift & $\rho_{SFR}$ (M$_{\odot}$~yr$^{-1}$~Mpc$^{-3}$) & \\
\hline
Palunas et al. 2004       & 2.4  & 0.0024  &\\
Madau et al. 1996         & 3.0  & 0.016   &\\
Steidel et al. 1999       & 3.0  & 0.011   &\\
Gronwall et al. 2007      & 3.1  & 0.012   &\\
Kudritzki et al. 2000     & 3.13 & 0.0064  &\\
Nilsson et al. 2007       & 3.15 & 0.013   &\\
Cowie \& Hu 1998          & 3.4  & 0.0047  &\\
Hu et al. 1998            & 3.4  & 0.006   &\\
van Breukelen et al. 2005 & 3.5  & 0.0067  &\\
Fujita et al. 2003        & 3.7  & 0.00041 &\\
Hu et al. 1998            & 4.5  & 0.01    &\\
Ouchi et al. 2003         & 4.8  & 0.0063  &\\
Ajiki et al. 2003         & 5.7  & 0.0012  &\\
Rhoads et al. 2003        & 5.7  & 0.0005  &\\
Malhotra \& Rhoads 2004   & 5.7  & 0.0018  &\\
Shimasaku et al. 2006     & 5.7  & 0.0023  &\\
Murayama et al. 2007      & 5.7  & 0.00072 &\\
Kodaira et al. 2003       & 6.5  & 0.00052 &\\
Malhotra \& Rhoads 2004   & 6.5  & 0.0036  &\\
Taniguchi et al. 2005     & 6.5  & 0.0013  &\\
\hline
\end{tabular}
\end{center}
\end{table}
The values in this table are lower than what can be seen in 
Fig.~\ref{fig:sfrhistory} by almost a factor of ten. This discrepancy
may be caused by dust extinction of the Ly$\alpha$/UV luminosity. 
Another possible explanation is that Ly$\alpha$ emitters only make up
10\% of the global star formation at high redshifts. Still, the disagreement
between measurements is large, showing that the star formation rate density
is a difficult quantity to measure. Problems with these calculations
involve dust corrections, completeness corrections, uncertain survey
volumes and not very well understood conversions between Ly$\alpha$
luminosity and star formation rate.

\subsection{SED fitting results for Ly$\alpha$ emitters}
Each galaxy has its own individual spectrum. Ideally, to extract all 
information about a galaxy, it is necessary to have the spectrum with high 
resolution.
However, this is impractical for high redshift galaxies as each galaxy would
take many hours to observe and only a small number of galaxies would be 
observed per year. The alternative is to observe the galaxy in several 
broad-, intermediate- and/or narrow-band filters which in effect
constitutes a spectrum with a very low resolution. The
spectrum of the galaxy, as observed through a number of imaging filters
is called the ``spectral energy distribution'' (SED). Once a well-sampled
SED has been obtained for a galaxy it is possible to try to find the best
fit between the SED and synthetic SEDs produced by combining populations
of stars with varying ages, metallicities, dust properties, star formation
histories, masses etc. Thus, we can gain information about the properties
of the galaxy from the best fit synthetic SED. If the fit is made in a 
statistical way, information about the probabilities of the different
properties can also be obtained. One of the main results in Nilsson et al. 
(2007) are the SED fits. Of course, to achieve a well constrained
fit to the SED, it is necessary to have a good sampling, preferably from very 
short to very long wavelengths as 
different parts of the spectrum reveal information about different 
properties of the galaxy. For instance, the UV describes the population of
massive, young stars while the infrared describes the old and dusty population.
Thus, getting galaxy properties from SED fitting is constrained to galaxy
samples in fields of the sky where large amounts of multi-wavelength data
exist.

Four papers describe SED fitting of Ly$\alpha$-emitters; Gawiser et al. 
(2006; $z = 3.1$, G06), Lai et al. (2007; $z = 5.7$, L07), Finkelstein et al. 
(2007; $z = 4.5$, F07) and Nilsson et al. (2007; $z = 3.15$, N07). The 
first and last paper examines the properties of Ly$\alpha$-emitters in 
GOODS-S, L07 uses data in GOODS-N and F07 have obtained imaging of the
LALA sample of Ly$\alpha$-emitters. All authors use the GALAXEV 
(Bruzual \& Charlot 2003) synthetic spectra models with Salpeter IMF
ranging from 0.1 - 100 M$_{\odot}$. G06, L07 and F07 use Calzetti et al. 
(1994; 1997; 2000) dust extinction laws, while N07 use Charlot \& Fall (2000)
dust models. All but N07 find the best model by creating a grid of spectra
with the different model parameters allowed and minimising the $\chi^2$
of each model, whereas N07 fit the best parameters with a Monte-Carlo 
Markov-Chain method (see more details in sec.~\ref{sec:sedfitting}). N07 and
G06 stack their sample and fit for the average parameters of the sample,
F07 stack their sample in four stacks and L07 fit for each galaxy 
individually. The results from the four papers are summarised in 
Table~\ref{tab:sedcompare}.
\begin{table}[t!]
\begin{center}
\caption[Summary of SED fitting results in the literature]{Summary of SED fitting results in the literature. Observed bands ``Ch1-4'' refer to the Spitzer 
telescope channels Ch1 (3.6~$\mu$m), Ch2 (4.5~$\mu$m), Ch3 (5.7$\mu$m) and
Ch4 (8.0~$\mu$m). F07 give no result for the fitting of the dust component.
In G06, the metallicity was set to solar metallicity, in L07 and F07 the 
metallicity was unconstrained.}\label{tab:sedcompare}
\vspace{0.5cm}
\begin{tabular}{lcccccccc}
\hline
\hline
                           & G06              & L07          & F07      & N07  \\
\hline
Redshift                   & 3.1              & 5.7          & 4.5      & 3.15            \\
Number of objects          & 40               & 3            & 76       & 23              \\
Observed bands             & UBVRIzJK         & BRViz & g'r'i'z' & UBVizJHK \\ 
                           &                  & Ch1-4 &          & Ch1-4 \\
Stellar mass (M$_{\odot}$) & $5 \times 10^8$  & $1 \times 10^{9 - 10}$ & $1 \times 10^{7 - 9}$ &  $4.7^{+4.2}_{-3.2} \times 10^8$  \\
Dust                       & A$_V \lesssim 0.1$ & E(B-V)~$\sim 0.0-0.5$ & --- & A$_V = 0.26^{+0.11}_{-0.17}$ \\
Metallicity                & Z$_{\odot}$ (set) & Unconstr. & Unconstr. & 0.005~Z$_{\odot}$ \\
Age (Myrs)                 & $\sim 500$ & $5 - 100$ & $1 - 200$ & $850^{+130}_{-420}$ \\
\hline
\end{tabular}
\end{center}
\end{table}
They are in good agreement with each other. Stellar masses of Ly$\alpha$
emitters are not very well constrained, but lie in the range 
$10^{7 - 9}$~M$_{\odot}$. Dust contents are very low. Ages are also difficult
to constrain, but lie in the range of a few to a few hundred Myrs. However, 
SED fitting is still a difficult and uncertain science and future work on this
subject will undoubtedly yield new and exciting results.

%
%
%
%

\subsection{Large scale structure results for Ly$\alpha$ emitters}\label{sec:introlss}
\begin{figure}[t]
\begin{center}
\includegraphics[width=0.70\textwidth,height=0.45\textwidth]{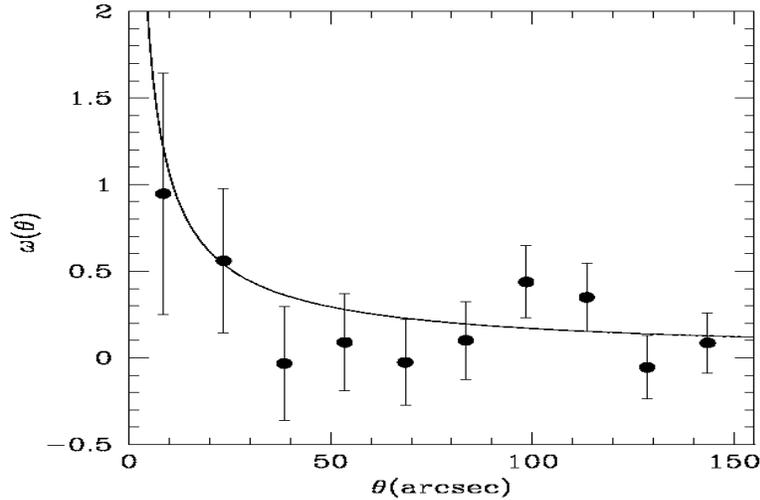}
\caption[Example of correlation function]{
Example of a correlation function. The correlation function is on the y-axis
and the angular distance in arcsec on the x-axis. This function was best fit
with an exponential form with $A_w \sim 10$ and $\beta = 0.8$. From Kova{\v c} et al. (2007).\label{fig:corrfunc}}
\end{center}
\end{figure}

The luminous matter in the Universe is homogeneously distributed, on large
scales, but on smaller scales the matter is highly clumped. Small 
perturbations in the initial mass distribution during the inflation of the 
Universe have collapsed into galaxy groups, clusters and filaments. Such
large scale structure has been observed in many surveys (Totsuji \& Kihara 
1969; Peebles 1974; Bahcall \& Soneira 1983; Hawkins et al. 2003; Eisenstein 
et al. 2005). The clustering of galaxies is generally measured by the
angular correlation function, 
$w(\theta)$, although many different suggestions have been made on how to 
calculate this
value (Hamilton 1993; Landy \& Szalay 1993; Landy, Szalay \& Broadhurst 1998).
The basic idea of the correlation function is to calculate the probability
that two galaxies will be found within two infinitesimal solid angle 
elements $\delta\Omega_{1,2}$ separated by an angle $\theta$. This probability
is
\begin{center}
\begin{equation}
P = (1 + w(\theta))\Sigma^2\delta\Omega_1\delta\Omega_2
\end{equation}
\end{center}
where $\Sigma$ is the surface density of the galaxies. Thus, the greater
the clustering at a certain angle, the greater $w(\theta)$ is. If the
sample is completely homogeneous, $w(\theta)$ is zero.
Most commonly, the correlation function is calculated by
\begin{center}
\begin{equation}\label{eq:corrfunc}
w(\theta) = \frac{DD(\Theta)-2DR(\Theta)+RR(\Theta)}{RR(\Theta)}
\end{equation}
\end{center}
In this equation, $DD(\Theta)$ is the number of pairs of observed galaxies
as a function of angle, $RR(\Theta)$ is the similar number of randomly 
generated galaxy pairs after creating a large number of random fields with
the same field geometry and sample size. $DR(\Theta)$ are pairs counted
when inserting an observed galaxy in each random frame. Correlation
functions are often assumed to follow power-laws of the form
\begin{center}
\begin{equation}
w(\theta) = A_w\theta^{-\beta}
\end{equation}
\end{center}
with amplitude $A_w$ and slope $\beta$. This is valid for projected samples
of galaxies in two dimensions. However, if the redshifts of the objects are
known it is also possible to calculate a spatial correlation function. This
will also have the form of a power law
\begin{center}
\begin{equation}
\xi = (r/r_0)^\gamma
\end{equation}
\end{center}
where $r_0$ is the so-called scale correlation length and $\gamma = \beta + 1$. 
The correlation length gives the typical distance that a type of object is
clustered on. An example of a correlation function can be seen in 
Fig.~\ref{fig:corrfunc}.

Clustering of Ly$\alpha$-emitters is a well-studied topic. Ouchi et al. 
(2003; 2005) and Shimasaku et al. (2003; 2005) present clustering results 
from the Subaru Deep Field (SDF) and Subaru-XMM Deep Field (SXDF) at redshifts 
$z = 4.8$~and~$5.7$. Ouchi et al. (2003) presents 87 candidates in SDF at
$z = 4.8$ and perform a correlation analysis on them. They find a very shallow
$\beta$ of 0.1 and decide to fix it to 0.8, similar to that discovered for
LBGs, because the Ly$\alpha$ sample is too small to determine the slope. After
fixing the slope, they find a correlation amplitude of $A_w = 29''$, 
corresponding to a correlation length of $r_0 = 3.5 \pm 0.3 h^{-1}$~Mpc.
This correlation length becomes, after correcting for sample completeness,
$r_{0,corr} = 6.2 \pm 0.5 h^{-1}$~Mpc. Shimasaku et al. (2003) and Ouchi et 
al. (2005) use a slightly different method of looking at overdensities in
surface density within certain radii when studying samples of $z = 4.8$ and
$z = 5.7$ emitters in SDF and SXDF respectively. Both papers present
overdensities with respect to field galaxies. 

Steidel et al. (1998; 2000), 
Hayashino et al. (2004) and Matsuda et al. (2005) describe the 
properties of a large cluster at $z = 3.1$ in the SSA22 field. Steidel et al.
(2000) find 77 candidates which appear, by visual inspection, to be
structured. However, the correlation analysis yielded no result. Hayashino
et al. (2004) observed the same part of the sky, at the same redshift, but with 
an area ten times larger than Steidel et al. (2000). They find 283 candidates 
which again show no clear clustering in
the correlation analysis. However, they study the surface density of objects
and find that the Ly$\alpha$ emitter candidates are four times more clustered
than what is predicted by the mass fluctuations of the CDM models. 

Several papers from Venemans et al. (2002; 2004; 2005; 2007) claim to have
detected overdensities and protoclusters of Ly$\alpha$-emitters surrounding
radio galaxies. In total their sample consists of $\sim 300$ candidates
between $2 < z < 5.2$ in 
eight fields, of which they have spectra for 139 confirmed Ly$\alpha$
emitters. They claim that at least six out of their eight fields are
overdense compared to field galaxy samples.
Kova$\check{\mathrm{c}}$ et al. (2007) have studied the clustering properties 
of $z = 4.5$ 
Ly$\alpha$-emitters in the LALA fields. They study 151 Ly$\alpha$ emitter 
candidates and find a correlation amplitude of $A_w = 6.73 \pm 1.80$
when the slope is fixed to $\beta = 0.8$. The amplitude corresponds to
a scale length of $r_0 = 3.20 \pm 0.42 h^{-1}$~Mpc, which after completeness
correction becomes $r_0 = 4.61 \pm 0.60 h^{-1}$~Mpc.

Several filamentary structures have also been discovered, including one
presented in sec.~\ref{sec:filament}. M{\o}ller \& Fynbo (2001) discovered
a 20~Mpc long filament with a cross section diameter of 1.6~Mpc at 
$z = 3.04$, consisting of 8 Ly$\alpha$-emitters. Matsuda et al. (2005) 
presented a filament in SSA22 consisting of 56 spectroscopically confirmed
Ly$\alpha$-emitters at redshift $z = 3.1$. The length of the filament is
30 Mpc and its width is 10 Mpc. In Nilsson et al. (2007; see 
sec.~\ref{sec:filament}) we find two projected, parallel filaments with a 
$4\sigma$ significance. Follow-up spectroscopy is needed to confirm the 
filamentary structure. The distance between the filaments appear to be 
$\sim 6$~Mpc and the width of the filament on average $\sim 1.5$~Mpc.

Finally, there are also studies which have shown Ly$\alpha$-emitters not to
be clustered. In particular Shimasaku et al. (2005) and Murayama et al. (2007)
find no clustering detections in their samples of 89 $z = 5.7$ emitters in 
SDF and 119 $z = 5.7$ emitters in COSMOS. Thus it is still unclear to
what extent  Ly$\alpha$ emitters trace large scale structure, and if/how this
depends on the fluxes of the emitters.

\subsection{Luminosity functions}
\begin{figure}[t]
\begin{center}
\includegraphics[width=0.65\textwidth,height=0.50\textwidth]{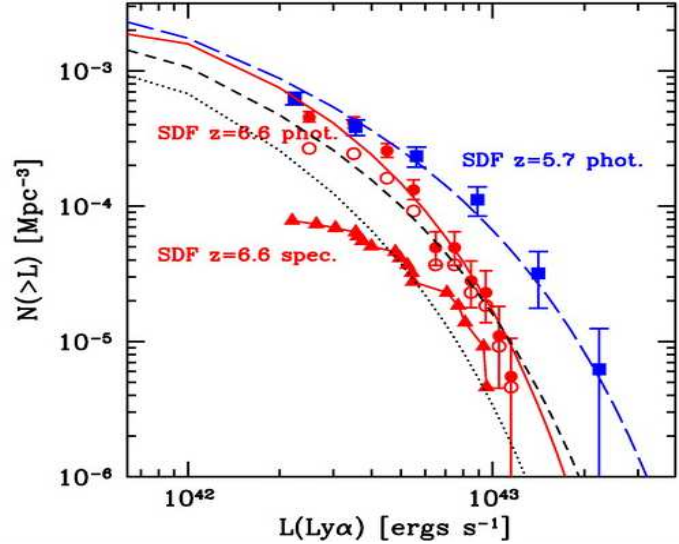}
\caption[Example of luminosity function]{
Example of a cumulative luminosity function. The \emph{x}-axis shows the 
Ly$\alpha$ luminosity of the Ly$\alpha$ emitters and the \emph{y}-axis the 
cumulative volume 
density. The red square points is the photometric sample of $z = 6.5$ 
candidates from Kashikawa et al. (2006) and the red triangles the 
spectroscopic sample from the same. The blue squares is the photometric sample 
of Shimasaku et al. (2006). The lines are the best fits made to the different 
samples. The reason that the faint end of the spectroscopic sample is
lower than the photometric is spectroscopic incompleteness. From Kashikawa et 
al. (2006).\label{fig:lumfuncex}}
\end{center}
\end{figure}

A luminosity function is a function describing the density of a certain
type of galaxies as a function of luminosity. It is always true that the
brighter galaxies are more rare than the fainter ones. Comparing luminosity
functions at different redshifts for the same type of object can 
show evolutionary changes in either density and/or luminosity of that type of 
object with time. Similarly,
comparing luminosity functions of different types of objects at the same
redshift will illuminate differences and similarities between the two 
types of objects. An example of a luminosity function can be seen in
Fig.~\ref{fig:lumfuncex}.
Often, the luminosity function is showed as a cumulative luminosity function
which means that the volume density at a certain luminosity is the density
of all objects brighter than this luminosity. Thus, given a luminosity limit,
it is easily possible to calculate the number of galaxies detected in a 
survey with a certain search volume.

Luminosity functions have been shown to be well fit by a so-called ``Schechter
function'' (Schechter 1976). 
The Schechter function has the following form:
\begin{center}
\begin{equation}
\phi(L)dL = \phi^{\star}(L/L^{\star})^{\alpha}\mathrm{exp}(-L/L^{\star})dL/L^{\star}
\end{equation}
\end{center}
where the shape of the function is determined by the three parameters $\alpha$,
$\phi^{\star}$ and $L^{\star}$. The $\alpha$ parameter determines the slope of 
the faint
end of the luminosity function, $L^{\star}$ represents the cut-off in the
bright end of the function and $\phi^{\star}$ normalises the function.

There are many difficulties in determining a luminosity function for
Ly$\alpha$ emitters. Firstly, a large sample of galaxies is needed in order
to decrease the statistical error bars on the points. Ideally, the function 
should be
made up of only spectroscopically confirmed emitters, as the sample otherwise
is contaminated by lower redshift interlopers. However, spectroscopic
follow-up is ofter confined to the bright end of the luminosity function, as 
the fainter an object is, the harder it is to detect spectroscopically.
This causes the error bars to be large in the bright end due to low number
statistics and in the faint end due to incompleteness issues. Secondly,
when detecting Ly$\alpha$ emitters by narrow-band imaging, several issues
arise due to the shape of the narrow-band filter itself. For instance, the
surveyed volume is smaller for fainter emitters, as these will only be
detected at the centre of the filter and not at the wings. Also, some
emitters may appear fainter or brighter than they are depending on how the
photometric calibration has been done and where in the filter the emitter is 
located in redshift space. For an extensive discussion on this, see
Gronwall et al. (2007). Because of these reasons, not many luminosity functions
for Ly$\alpha$ emitters have been published, and the results from the 
published ones differ significantly. 

The results from previous work is
most easily presented in a table, see Table~\ref{tab:lumfunccomp}.
In this table, the third column gives number of
candidates on which the analysis is based, separated into spectroscopic sample 
and total photometric sample. Malhotra \& Rhoads (2004) summarise several 
surveys, for more information on those surveys see that paper. The Gronwall et 
al. (2007) sample has spectroscopic follow-up, yet unpublished. The Schechter 
function parameter fits for the references with published fits can be found in
Table~\ref{tab:lfdata}, in Chapter~\ref{chapter:predictions}. As can be seen
in Table~\ref{tab:lumfunccomp}, there are not many results on this topic
yet, and many authors choose not to fit their data with Schechter functions,
or to draw any firm conclusions. It is still unclear if any evolution is 
occurring from redshift $z = 3$ to $z = 6.5$. At lower redshifts, only
one large survey has been presented so far (Gronwall et al. 2007).
At higher redshift, several results have been presented but they disagree
with each other to almost an order of magnitude in the density at the faint 
end of the luminosity function (e.g. Malhotra \& Rhoads 2004; Shimasaku et al. 
2006; Kashikawa et al. 2006, see also Fig.~\ref{fig:oldplot}). This is yet a 
subject where much improvement is to be expected in the coming decade.

\begin{landscape}
\begin{table}
\caption[Comparison of previous results for luminosity functions]{Comparison
of previous results for luminosity functions. For more information about this 
table, see text.} 
\label{tab:lumfunccomp}
\vspace{0.5cm}
\begin{tabular}{l|c|c|c|c|lcc}
\hline
\hline
Reference & Redshift & Sample size  & Area (arcmin$^2$) & Field & Conclusions  \\
          &          & (Spec./Phot.) & & &  \\
\hline
Ouchi et al. (2003) & 4.86 & ---/87 & 543 & SDF & No evolution between $z = 3.4$\\
 & & & & & and $4.8$. Bright end slope of Ly$\alpha$\\
 & & & & & continuum luminosity function\\
 & & & & & steeper than that of LBG LFs.\\
 \hline
Maier et al. (2003) & 4.8, 5.7 & 2/--- & 100 & CADIS & Fabry-Perot detections.
Small\\
 & & & & & number of detections, discuss \\
 & & & & & large scale structure influence \\ 
 & & & & & on results. Find no change in LF\\
 & & & & & between $z = 3.5$ and $5.7$.\\
 \hline
Hu et al. (2004) & 5.7 & 18/26 & 918 & SSA22 & No conclusions made \\
\hline
Malhotra \& & 5.7, 6.5 & --- & --- & --- & Compiles various
published\\
Rhoads (2004) & & & & & surveys. Find no real evolution\\
 & & & & & between $z = 4.5$ and $5.7$.  \\
 & & & & & Conclude that the Universe is \\
 & & & & & ionised at $z \sim 6$. Publish fits \\ 
 & & & & & to a Schechter function.\\
 \hline
van Breukelen & 2.5 - 4.4 & 14/--- & 1.36 & --- & IFU observations. Publish fit to\\
et al. (2005) & & & & & Schechter function. Find no\\
 & & & & & evolution between $z = 3.4$ and $5.7$.\\
 \hline
Tapken et al. & 5.7 & 8/15 & 46 & FORS & Find more Ly$\alpha$
emitters than \\
al. (2006) & & & & Deep Field & expected compared to\\
 & & & & & Malhotra \& Rhoads (2004) LF.\\
 \hline
\end{tabular}
\end{table}
\begin{table}
\caption[Comparison of previous results for luminosity functions, cont.]{Table~\ref{tab:lumfunccomp} continued.}
\vspace{0.5cm}
\begin{tabular}{l|c|c|c|c|lcc}
\hline
Shimasaku et & 5.7 & 28/89 & 725 & SDF & Publish fit for Schechter function. Find\\
al. (2006) & & & & & no change between $z = 4.5$ and $5.7$.\\ 
 \hline
Kashikawa et & 6.5 & 17/58 & 876 & SDF & Similar selection criteria as Shimasaku\\
al. (2006) & & & & & et al. (2006). Publish Schechter function fit. Large\\
 & & & & & spectroscopic follow-up rate. Find deficit in bright end\\
 & & & & & of LF by a factor of 2 compared to $z = 5.7$ but cannot\\
 & & & & & rule out cosmic variance.\\
 \hline
Gronwall et & 3.1 & ---/162 & 1008 & E-CDFS & Publish Schechter function fit.\\
al. (2007)  & & & & & Concludes that if Malhotra \& Rhoads (2004) are\\
 & & & & & correct then emitters at $z = 3.1$ are 2.5 times brighter\\
 & & & & & or more numerous than at $z = 5.7$ but if Shimasaku et al.\\
 & & & & & (2006) are correct, then the opposite result.\\
\hline
\end{tabular}
\end{table}
\end{landscape}

\subsection{Modelling of Ly$\alpha$ emitters}
Modelling Ly$\alpha$ is a difficult business. Ly$\alpha$ is a resonance line
that can be affected by many occurrences, such as e.g. dust, in- and outflows,
geometry, clumping etc. Many groups have made various models about the
Ly$\alpha$ emission line or Ly$\alpha$ emitting galaxies. Some are summarised
here.

Laursen \& Sommer-Larsen (2007) present a new three dimensional Ly$\alpha$
radiative transfer code based on smoothed particle hydrodynamics simulations.
The code is a Monte-Carlo code which propagates the Ly$\alpha$ photons through
the medium surrounding the galaxy. The code is shown to be successful in 
describing one of the observed properties of Ly$\alpha$ emitters; that
the Ly$\alpha$ emission is often more extended than the continuum emission
(M{\o}ller \& Warren 1998; Fynbo et al. 2001; 2003). A similar result is also
achieved by Furlanetto et al. (2003; 2005), who also calculate the radiative
transfer through a smoothed particle hydrodynamics simulation. Furlanetto
et al. (2005) also present luminosity functions for Ly$\alpha$ emitters at 
$z = 3$. Cantalupo et al. (2005) use the same approach to study the
luminosity of fluorescent Ly$\alpha$ sources at high redshift. Verhamme,
Schaerer \& Maselli (2006) present another radiative transfer code with which
they have studied the line profile of Ly$\alpha$. Yet another
description of the same method can be found in Tasitsiomi (2006).
The conclusion from this paper is that a $z \sim 8$ Ly$\alpha$ galaxy would
not be observable with current, ground-based observatories.

Several papers by Dijkstra et al. (2006a; 2006b; 2007a; 2007b; 2007c) also use
a radiative transfer code for Ly$\alpha$ photons, applied to a
cosmological simulation. In the two papers from 2006, the authors describe the 
effects of infall of material onto a galaxy on the Ly$\alpha$ emission. 
Dijkstra et al. (2007a) study the effect of the IGM on the emergent Ly$\alpha$
and discuss the possibility to use these galaxies to determine re-ionisation
at very high redshifts. The latter issue is also discussed in Dijkstra et al.
(2007b), in relation to recent new results regarding the luminosity function
of Ly$\alpha$ emitters at very high redshift. Finally, Dijkstra et al. (2007c)
indicate that up to 50\% of all Ly$\alpha$ emitters may harbour primeval
stars (so-called Population III stars).

After the discovery of two large Ly$\alpha$ blobs by Steidel et al. (2000),
several explanations for this phenomenon were presented in different
publications. The explanations proposed, all based on trying to find a way of
producing the energy needed to ionise the nebula, were \emph{i)} star 
formation, where the energy comes from newly formed stars,
\emph{ii)} AGN activity, where the energy comes from the accretion of
material onto a supermassive black hole, and \emph{iii)} cold accretion, where
the energy comes from loss of gravitational potential energy of cold gas
accreting onto a dark matter halo. The first
explanation, star formation, is based on the idea that a starburst galaxy will
produce a large amount of young, massive stars emitting much UV, and thus
Ly$\alpha$, radiation. These massive stars will rapidly burn out and become
supernovae, and the explosions from these supernovae will create a sort of
``superwind'' which will blow out the ionised material surrounding the galaxy.
Thus the galaxy will be enshrouded in a Ly$\alpha$ emitting cloud. The
modelling behind this scenario is described in Taniguchi et al. (2001),
Ohyama et al. (2003), Mori et al. (2004) and Wilman et al. (2005). The second
explanation, involving an AGN, involves that the AGN can be obscured from 
the line-of-sight, but still ionise the material surrounding it. This method 
is explained in Haiman \& Rees (2001) and Weidinger et al. (2004; 2005).
Finally, cold accretion, as described in e.g. Haiman, Spaans \& Quataert 
(2000), Fardal et al. (2001), Dijkstra et al. (2006a,b) and Dekel \&
Birnboim (2006), is described as infalling gas onto a dark matter halo. This
infalling gas is heated due to the loss of gravitational energy. The heat
can then be released through Ly$\alpha$ emission. This topic is further 
discussed in Chapter~\ref{chapter:blob} of this thesis.

Two papers study the theoretical aspects of filaments discovered in
Ly$\alpha$ surveys; Weidinger et al. (2002) and Monaco et al. (2005).
In Weidinger et al. (2002) it is proposed that filaments found with
relative ease through narrow-band imaging can be used as an independent measure
on cosmological parameters such as $\Omega_m$ and $\Omega_{\Lambda}$. The idea
is that, assuming that all filaments must be isotropically oriented, filaments 
will be stretched differently depending on the assumed cosmology of the 
Universe. Thus, a large
sample where the individual alignments become washed out could give an
independent measure of the cosmological parameters. In Monaco et al. (2005),
the effect of peculiar motions on the redshifts in observed filaments is
quantified. It is shown that the effect is small, but non-negligible and 
should be taken into account when drawing conclusions from filamentary
structures. The idea of determining cosmological parameters with Ly$\alpha$
emitters has been picked up in the HETDEX\footnote{www.as.utexas.edu/hetdex/}
project, described in Hill et al. (2004). HETDEX proposes to use a new
integral field unit (IFU) spectrograph called VIRUS (Visible IFU Replicable 
Ultra-cheap Spectrograph) to study several thousands of Ly$\alpha$ emitters
in the redshift range $1.8 < z < 3.8$ every night on the Hobby-Eberly 
Telescope, thus being able to study large scale structure of this type of
galaxy to extreme accuracy, and from that determine the cosmological parameters
to a great level of certainty.

Barton et al. (2004), Le Delliou et al. (2005; 2006), Thommes \& 
Meisenheimer (2005), Stark, Loeb \& Ellis (2007) and Kobayashi et al. (2007) 
make predictions for very high redshift Ly$\alpha$
surveys. Barton et al. (2004) use cosmological hydrodynamic simulations of 
galaxy formation to study the detectability of $z \geq 7$ sources in the
near future. Le Delliou et al. (2005; 2006) use a semi-analytic model based
on the $\Lambda$CDM model and show that it fits well with lower redshift
observations. Kobayashi et al. (2007) use a similar semi-analytical model
but with a different parametrisation of the Ly$\alpha$ escape fraction which
they claim to be more physical. They also claim good fits with lower redshift
results. Thommes \& Meisenheimer (2005) and Stark, Loeb \& Ellis (2007)
instead use phenomenological models, assuming that Ly$\alpha$ luminosity is
proportional to the star formation rate of a galaxy, and that the 
star formation rate is proportional to the baryonic mass of the galaxy. 
The results from these papers are luminosity functions at several
redshifts between $z = 3 - 8$. This work is followed up in 
Chapter~\ref{chapter:predictions}, where we make predictions for high redshift
narrow-band surveys.

In preparation for the many surveys for very high redshift Ly$\alpha$ emitters
coming in the next decade, several groups discuss how to draw conclusions
about the re-ionisation history of the Universe. Loeb \& Rybicki (1999) 
calculate the brightness distribution and the shape of the Ly$\alpha$
emission line in haloes surrounding high-redshift galaxies before 
re-ionisation was complete. They find that observations of Ly$\alpha$ prior 
to re-ionisation will be a good probe of the neutral IGM. Haiman \& Spaans
(1999) look at dust effects on high redshift Ly$\alpha$. They discuss how
changes in the Ly$\alpha$ luminosity function may be used to determine the
level of re-ionisation. Many authors have quantified the effect of the
Gunn-Peterson absorption (Gunn \& Peterson 1965) on the red wing of the
Ly$\alpha$ line as a function of redshift (Miralda-Escud{\'e} 1998;
Miralda-Escud{\'e} \& Rees 1998; Haiman \& Loeb 1999; Haiman 2002;
Santos 2004; Gnedin \& Prada 2004; Haiman \& Cen 2005; Tasitsiomi 2006).
This effect comes from the vast amount of neutral hydrogen in the Universe
at high redshift. Photons with wavelengths less than that of Ly$\alpha$ at its
place of origin will travel towards us, but is redshifted in the process. 
Depending on the original wavelength, it will at a certain redshift have the
wavelength of Ly$\alpha$. At high redshift, the likelihood to encounter
a neutral hydrogen cloud at this distance from the source is high,
resulting in that almost all the photons blue-ward of Ly$\alpha$ from high 
redshift galaxies are absorbed. At very high redshift, even photons in the
red wing of the Ly$\alpha$ line will be absorbed and thus the level of
absorption, i.e. the asymmetry of the line, will reveal information of the
level of re-ionised material surrounding the galaxy.

McQuinn et al. (2007) show, using a radiative transfer simulation, that 
observations of the clustering of very high redshift Ly$\alpha$ emitters may
reveal the level of re-ionisation independently of the luminosity function
and the line profile. This will be a good method to use to cross-check
with the results of other methods. 

In a very recent paper, Fernandez \& 
Komatsu (2007) use the mass-to-light ratio of Ly$\alpha$ galaxies
to make luminosity functions. The general idea of their paper is to 
use a derived mass function, and a mass-to-light ratio which is a free
parameter to calculate a luminosity function. From comparisons with observed
high redshift luminosity functions, they then argue that Ly$\alpha$ emitters are
either starburst galaxies with low escape fractions, or normal galaxies with
a higher escape fraction. The also see a hint of metallicity evolution 
between redshift $z = 5.7$~and~$6.5$, but no evidence that re-ionisation should
have ended by $z \sim 7$.

\section{This thesis}
This thesis presents work I have done regarding Ly$\alpha$ emitters. 
The different chapters describes various aspects of 
these galaxies. Both theoretical and observational, as well as low and
high redshift work is presented. The main part of the thesis is
divided into two parts. The first part concerns narrow-band 
imaging for medium redshift ($z \sim 3.15$) {\ly} emitters in the survey
field of GOODS-S. The outcome of that project was two papers (Nilsson et al.
2006a; Nilsson et al. 2007), presented in Chapter~\ref{chapter:blob} and
Chapter~\ref{chapter:goodss} respectively. In Chapter~\ref{chapter:blob} we 
described the discovery of a Ly$\alpha$ blob, which after careful analysis
turned out to be the first of its kind best explained by cold accretion 
onto a dark matter halo. In Chapter~\ref{chapter:goodss} all other results
from the original narrow-band imaging are presented, including the imaging 
and spectroscopic results we have, the photometry of the candidates and
spectra of three confirmed candidates, a tentative discovery of a 
filamentary structure, a careful SED fitting procedure and a comparison with
Lyman Break Galaxies. 

The second major project I have been heavily involved in from the start is 
a contribution to the Visible and Infrared Survey Telescope for Astronomy (VISTA). 
My original idea was to buy a set of narrow-band filters for VISTA,
to detect a sample of $z = 8.8$ Ly$\alpha$ emitters. VISTA is a new
survey telescope in the near-infrared, with a camera with a field-of-view
of 0.6~deg$^2$. It will be dedicated to Public Surveys, enabling large, 
time-demanding projects to be carried out. I have been involved in all
aspects of this project, from design and procurement of the filters, to
writing a Public Survey proposal and later a Survey Management Plan. This work,
described in Chapter~\ref{chapter:vista}, prompted a new project,
presented in Chapter~\ref{chapter:predictions}. In this chapter we 
attempt to use two different theoretical models to calculate luminosity 
functions for Ly$\alpha$ emitters at very high redshift ($z \geq 7$). We use
these results to predict numbers of galaxies detected in a number of present
and future surveys.

In the final science chapter we explore the colour space that Ly$\alpha$ emitters 
inhabit compared to that of ``normal'' galaxies or [OII]-emitters. We are
writing a paper with the aim to try to develop a method to improve the
narrow-band photometric selection technique using optical and near-infrared
colours. The results from this project
are found in Chapter~\ref{chapter:fundamental}.

In Chapter~\ref{chapter:conclusions} I summarise the work presented in this
thesis and in the final chapter, Chapter~\ref{chapter:future}, I attempt
to look forward at what may happen in this field of study in the near
future, and what my contribution could be.

\chapter{A Ly$\alpha$ blob in GOODS-S}\label{chapter:blob}
\markboth{Chapter 2}{A {\ly} blob in GOODS-S}
This paper has been published in \emph{Astronomy \& Astrophysics Letters}
in June 2006 (A\&A, 452, L23). The authors are Nilsson, K.K., Fynbo, J.P.U., 
M{\o}ller, P., Sommer-Larsen, J., \& Ledoux, C.

\section{Abstract}
We report on the discovery of a $z = 3.16$ Lyman-$\alpha$ emitting blob in the
Great Observatories Origins Deep Survey (GOODS) South field.
The discovery was made with the VLT, through
narrow-band imaging.
The blob has a total Ly$\alpha$ luminosity of
$\sim10^{43}$~erg~s$^{-1}$ and a diameter larger than 60~kpc.
The available multi-wavelength data in the GOODS field consists of
13 bands from X-rays (Chandra) to infrared (Spitzer). Unlike
other known Ly$\alpha$ blobs, this blob shows no obvious
continuum counter-parts in any of the broad-bands. In particular,
no optical counter-parts are found in deep HST/ACS imaging. For
previously published blobs, AGN (Active Galactic Nuclei) or ``superwind''
models have been found to provide the best
match to the data. We here argue that the most probable origin of the
extended Ly$\alpha$ emission from this blob is cold
accretion onto a dark matter halo.

\section{Introduction}
Narrow-band surveys for Lyman-$\alpha$ (Ly$\alpha$) emitting galaxies at high
redshift have recently revealed a number of luminous (up to $5 \cdot 10^{43}$ 
erg s$^{-1}$), very extended (from a few times ten kpc to more than 150 kpc)
Ly$\alpha$-emitting objects, so-called Ly$\alpha$ ``blobs'' (Fynbo et al. 1999;
Keel et al. 1999; Steidel et al. 2000; Francis et al.  2001;
Matsuda et al. 2004; Palunas et al.
2004; Dey et al. 2005; Villar-Martin et al. 2005).
At least three mechanisms have been suggested as energy sources for
Ly$\alpha$ blobs. These are: \emph{i)} hidden QSOs (Haiman \& Rees 2001;
Weidinger et al. 2004, 2005), \emph{ii)} star formation and superwinds from
(possibly obscured) starburst galaxies (Taniguchi et al. 2001; Ohyama et al.
2003; Mori et al. 2004; Wilman et al. 2005), and \emph{iii)} so-called cold
accretion (Haiman, Spaans \& Quataert 2000; Fardal et al. 2001; Keres et al.
2004; Maller \& Bullock 2004; Birnboim \& Dekel 2003; Sommer-Larsen 2005;
Dijkstra et al. 2006(a,b); Dekel \& Birnboim 2006).
Cooling flows are phenomena observed in galaxy clusters for more than a decade
(Fabian 1994). These are explained by gas which is cooling
much faster than the
Hubble time through X-ray emission in the centres of the clusters.
However, cooling
emission from a galaxy, or a group sized halo can be dominated by
Ly$\alpha$ emission
(e.g. Haiman, Spaans \& Quataert 2000; Dijkstra et al. 2006(a,b)).
In this \emph{Letter} we present the discovery of a
Ly$\alpha$ blob at redshift $z \approx 3.16$ located in the GOODS South
field, which we argue is the first piece of evidence for cold gas
accretion onto a dark matter halo.

Throughout this paper, we assume a cosmology with $H_0=72$ km s$^{-1}$
Mpc$^{-1}$, $\Omega _{\rm m}=0.3$ and $\Omega _\Lambda=0.7$. All magnitudes are
in the AB system.

\section{Observations and Data reduction}
\label{obs}
A 400$\times$400 arcsec$^2$ section, centred on R.A.~$= 03^h 32^m 21.8^s$,
Dec~$= -27^{\circ} 45' 52''$ (J2000), of the GOODS South field
was observed
with FORS1 on the VLT 8.2 m telescope Antu
during two visitor mode nights on December 1--3, 2002.
A total of 16 dithered exposures were obtained over the two nights
for a combined exposure time of 30 ksec, all with the narrow band
filter OIII/3000+51 and using the standard resolution collimator
(0.2$\times$0.2
arcsec$^2$ pixels). For this setup the central wavelength of the
filter is 5055 {\AA} with a FWHM of 59 {\AA}, corresponding to the
redshift range $z = 3.126$~--~$3.174$ for Ly$\alpha$.

The observing conditions were unstable during the two nights with the
seeing FWHM varying between $0.66''$ and $1.25''$ on the first night
and $1.4''$ and $3.3''$ on the second night.
The images were reduced (de-biased, and corrected for CCD pixel-to-pixel
variations using twilight flats) using standard techniques.
The individual reduced images were combined using a modified version
of our code that optimizes the Signal-to-Noise (S/N) ratio for faint,
sky-dominated sources (see M{\o}ller \& Warren 1993, for details on this
code). The modification of the code was necessitated by the highly
variable seeing. The sky background was assumed to be constant. The
FWHM of the PSF of the final combined narrow-band image is $0.8''$.

For object detection, we used the software package SExtractor
(Bertin \& Arnouts 1996).
A full description of our selection of Ly$\alpha$ emitters in the GOODS field
will be given in a subsequent paper. In this {\it Letter} we
discuss the nature of an extended, low surface brightness
blob with a centroid (of the Ly$\alpha$ emission) of
R.A.~$ = 03^h 32^m 14.6^s$  and Dec~$ = -27^{\circ} 43' 02.4''$ (J2000) detected in the combined narrow-band image.

Follow-up MOS spectroscopy was obtained in service mode using
FORS1/VLT UT2 over the time period December 2004 -- February 2005. The
total observing time was 6 hours.
We used a $1.4''$ slitlet and grism 600V resulting in a wavelength range of
4650 {\AA} to 7100 {\AA} and a spectral resolution FWHM of approximately 700.
The seeing varied between $0.77''$ and $1.2''$ during the spectroscopic
observations.

The GOODS archival data used here and their detection limits are listed in
Table~\ref{mwtab}.

\begin{table}[!ht] 
\begin{center}
\caption[Data in GOODS-S field]{Specifications of deep, multi-wavelength data available in the 
GOODS South field and the narrow-band image.
The last column gives the 3$\sigma$ limit as detected in a $2''$ radius
aperture and the narrow-band value gives
the blob flux in this aperture. }
\vspace{0.5cm}
\begin{tabular}{@{}lccccccc}
\hline
\hline
Filter/Channel & $\lambda_c$ & Filter & 3$\sigma$ limit ($2''$ aperture)\\
& & FWHM  & (erg~cm$^{-2}$~s$^{-1}$~Hz$^{-1}$)\\
\hline
X-rays (\emph{Chandra})     &  4.15 keV    &  3.85 keV    & $9.90 \cdot 10^{-34}$   \\
U (\emph{ESO 2.2-m})        &  3630 \AA    &  760 \AA     & $8.62 \cdot 10^{-31}$  \\
B (\emph{HST})              &  4297 \AA    &  1038 \AA    & $9.25 \cdot 10^{-30}$ \\
Narrow (\emph{VLT})         &  5055 \AA    &  60 \AA      & $6.68 \cdot 10^{-30}$  \\
V (\emph{HST})              &  5907 \AA    &  2342 \AA    & $4.66 \cdot 10^{-30}$  \\
i (\emph{HST})              &  7764 \AA    &  1528 \AA    & $1.50 \cdot 10^{-29}$  \\
z (\emph{HST})              &  9445 \AA    &  1230 \AA    & $3.00 \cdot 10^{-29}$ \\
J (\emph{VLT})              &  1.25 $\mu$m &  0.6 $\mu$m  & $5.31 \cdot 10^{-30}$ \\
H (\emph{VLT})              &  1.65 $\mu$m &  0.6 $\mu$m  & $1.86 \cdot 10^{-29}$ \\
Ks (\emph{VLT})             &  2.16 $\mu$m &  0.6 $\mu$m  & $1.56 \cdot 10^{-29}$ \\
Ch1 (\emph{Spitzer/IRAC})   &  3.58 $\mu$m &  0.75 $\mu$m & $2.51 \cdot 10^{-31}$  \\
Ch2 (\emph{Spitzer/IRAC})   &  4.50 $\mu$m &  1.02 $\mu$m & $6.43 \cdot 10^{-32}$ \\
Ch3 (\emph{Spitzer/IRAC})   &  5.80 $\mu$m &  1.43 $\mu$m & $5.01 \cdot 10^{-29}$ \\
Ch4 (\emph{Spitzer/IRAC})   &  8.00 $\mu$m &  2.91 $\mu$m & $4.65 \cdot 10^{-30}$ \\
\hline
\label{mwtab}
\end{tabular}
\end{center}
\end{table}

\section{Results}\label{results}

\begin{figure}[t] 
\begin{center}
\epsfig{file=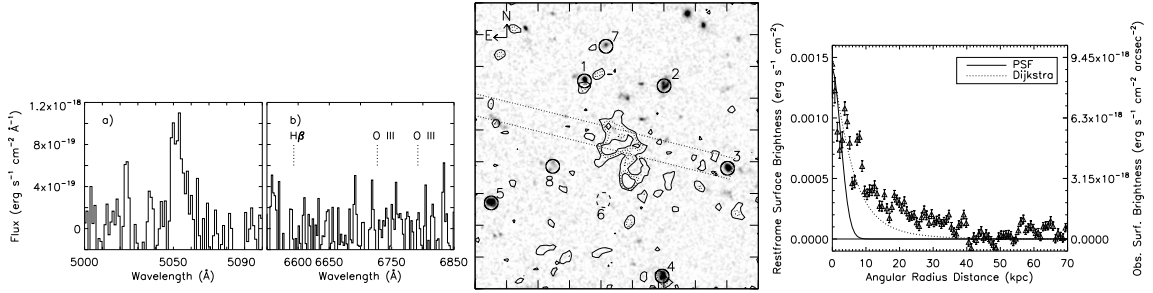,width=15.5cm,clip=}
\caption[Ly$\alpha$ blob spectrum, contour-plot and surface brightness plot]{\emph{Left}~\emph{a)} Flux calibrated spectrum of the blob emission
line. The
line has the characteristic blue side absorption, indicating high redshift.
\emph{b)} The part of the spectrum (binned with a binsize equal to half
the resolution ($1.1$~{\AA})) where H$\beta$ and [OIII] should have been observed if the emission line was [OII] at a redshift of $z \approx 0.36$.
These lines are not observed and
therefore we conclude the observed line is due to Ly$\alpha$ at $z=3.16$.
\emph{Middle} Contour-plot of narrow-band emission from the Ly$\alpha$ blob
overlaid the
HST V-band image. The narrow-band image has been continuum
subtracted by subtracting the re-binned, smoothed and scaled HST/V-band image.
Contour levels are $2 \cdot 10^{-4}$, $4 \cdot 10^{-4}$ and $6
\cdot 10^{-4}$~erg~s$^{-1}$~cm$^{-2}$ in restframe flux (corresponding
to $1.2 \cdot 10^{-18}$, $2.5 \cdot 10^{-18}$ and $3.7 \cdot 10^{-18}$ in
observed flux). The image is $18'' \times 18''$ ($18''$ corresponds to a physical size of $\sim 133$~kpc).
Numbers refer to those used in section~\ref{results}. The dotted lines indicate
the slitlet position for our follow-up spectroscopy.
\emph{Right} Plot of surface brightness as function of radius.
The flux is the sky subtracted narrow-band flux.
The PSF of the image is illustrated by the solid line, and the
dotted line is the best fit model of Dijkstra et al. 2006.
The deficit at $\sim 45$~kpc is due to the asymmetric appearance of
the blob.}
\label{surfbright}
\label{spectrum}
\label{contour}
\end{center}
\end{figure}

The spectrum of the part of the Ly$\alpha$ blob covered by the
slitlet can be seen in the left-most panel of Fig.~\ref{spectrum}.
The line has the asymmetric profile expected for a high redshift
Ly$\alpha$ emitter.
We detect no other emission lines in the spectrum.
The most likely interloper is [OII] at redshift 0.36, but no emission
is observed in the spectrum where e.g. H$\beta$ or [OIII] are expected at this
redshift, see Fig.~\ref{spectrum}. This leads us to the
conclusion that we are observing a Ly$\alpha$-emitting object at $z = 3.157$.
The observed FWHM velocity width of the emission line is $505$~km~s$^{-1}$.
The instrument FWHM of the set-up is $290$~km~s$^{-1}$, hence 
the Ly$\alpha$ intrinsic velocity width is marginally resolved. The
intrinsic width is less than $500$~km~s$^{-1}$. This is of the order or
smaller than for other published blobs,
with velocity widths of $500 - 2000$~km~s$^{-1}$ (Keel et al. 1999;
Steidel et al. 2000; Francis et al. 2001; Ohyama et al. 2003; Bower et al 2004;
Dey et al.  2005).

\begin{figure}[t] 
\begin{center}
\epsfig{file=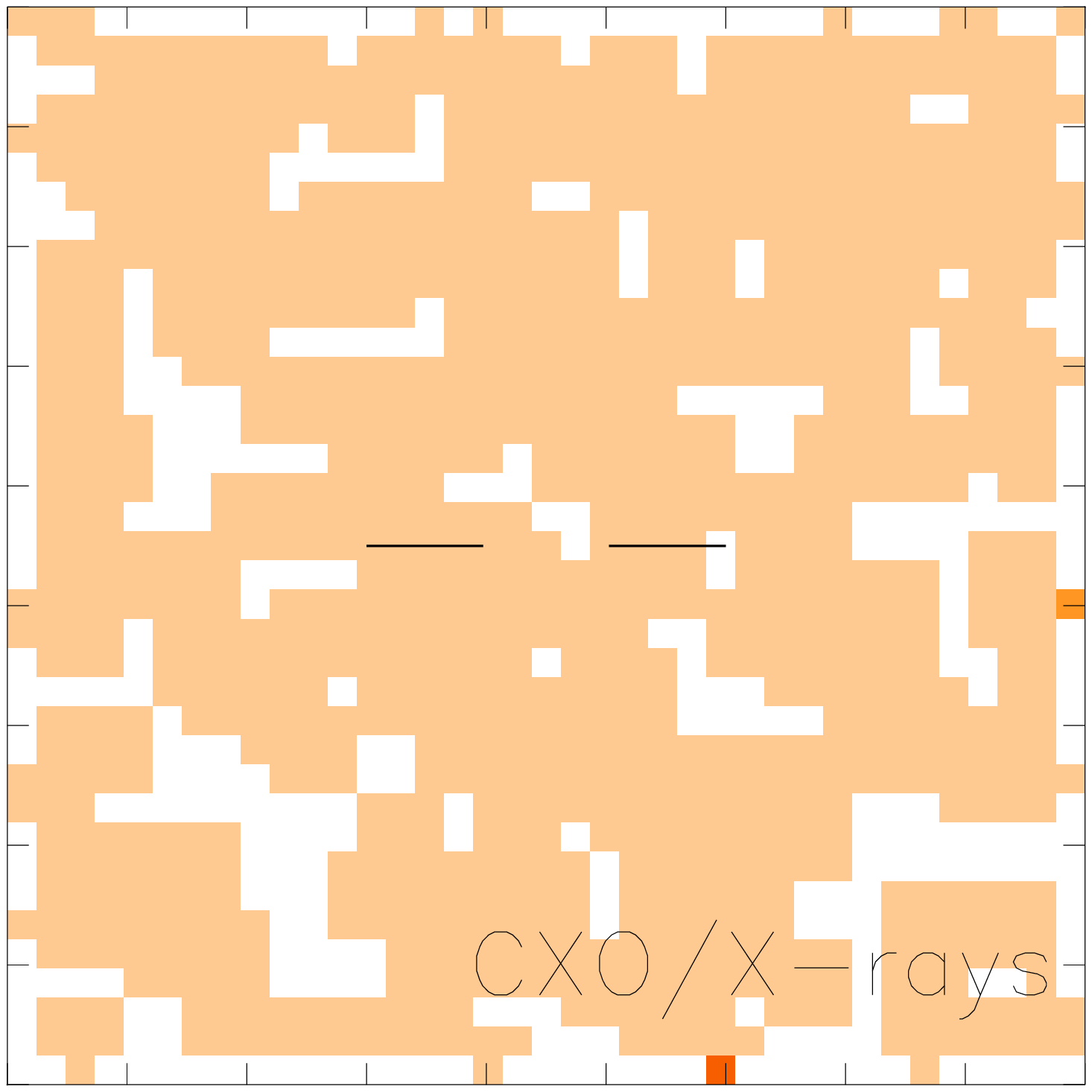,width=2.1cm,clip=}\epsfig{file=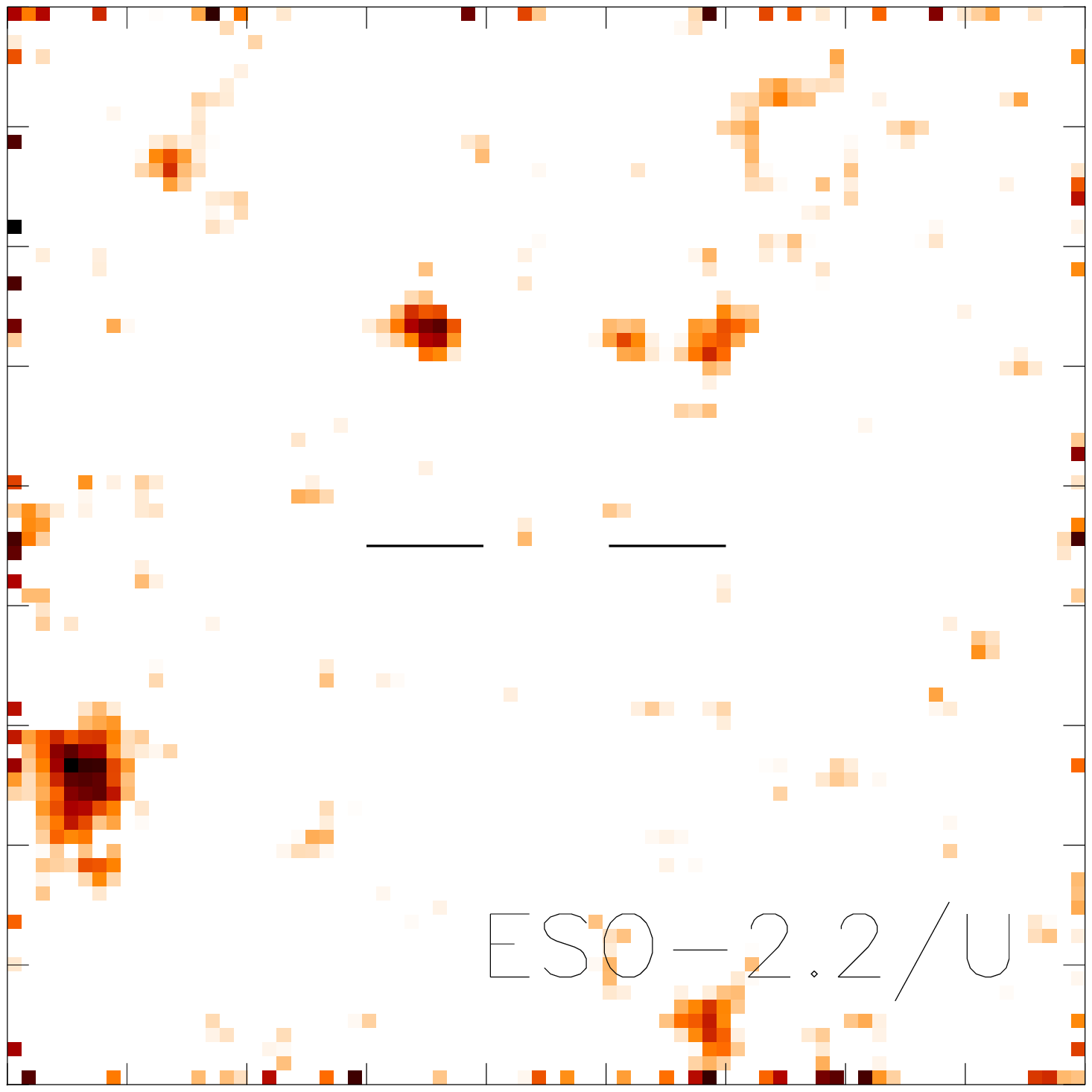,width=2.1cm,clip=}\epsfig{file=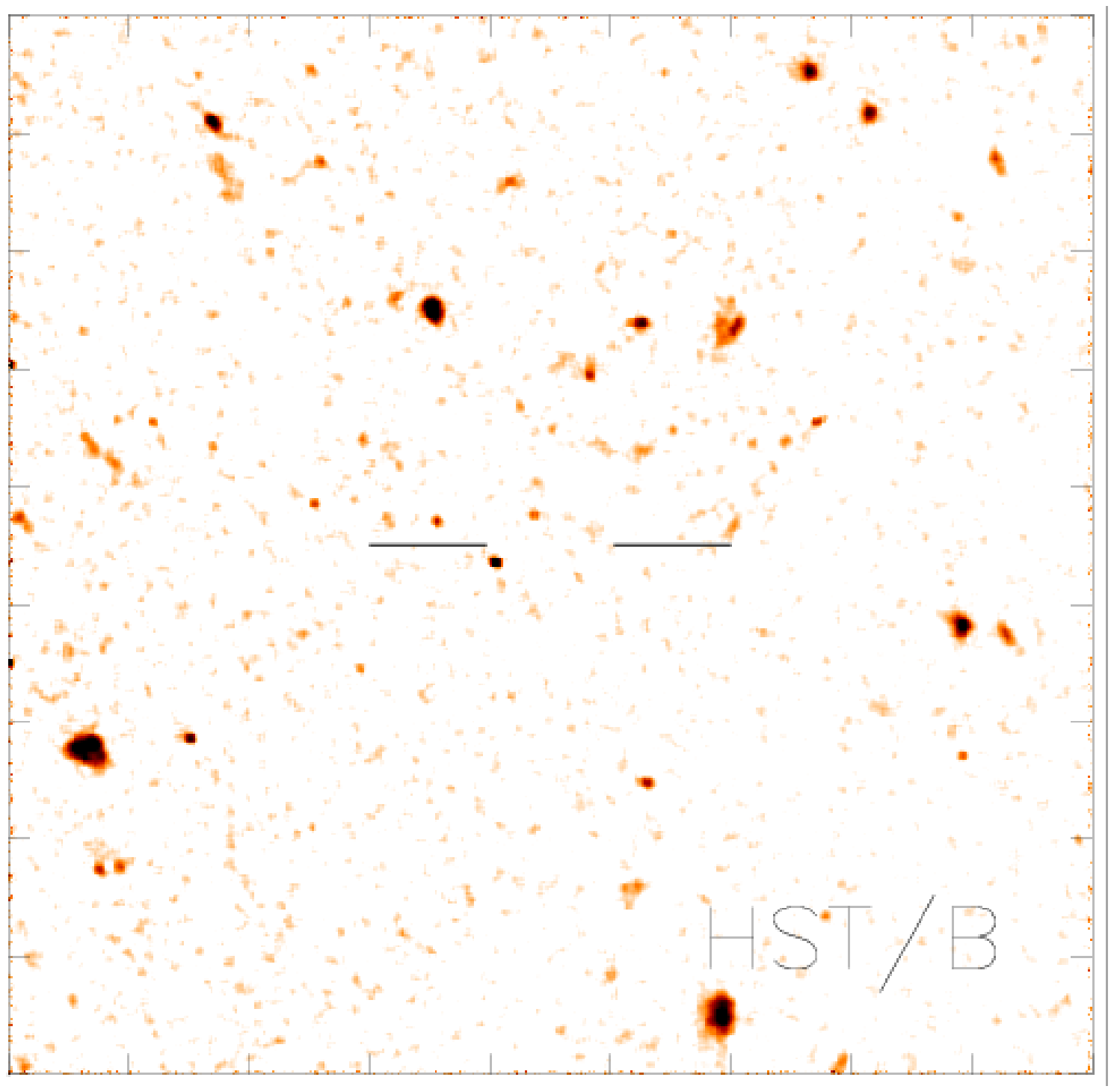,width=2.1cm,clip=}\epsfig{file=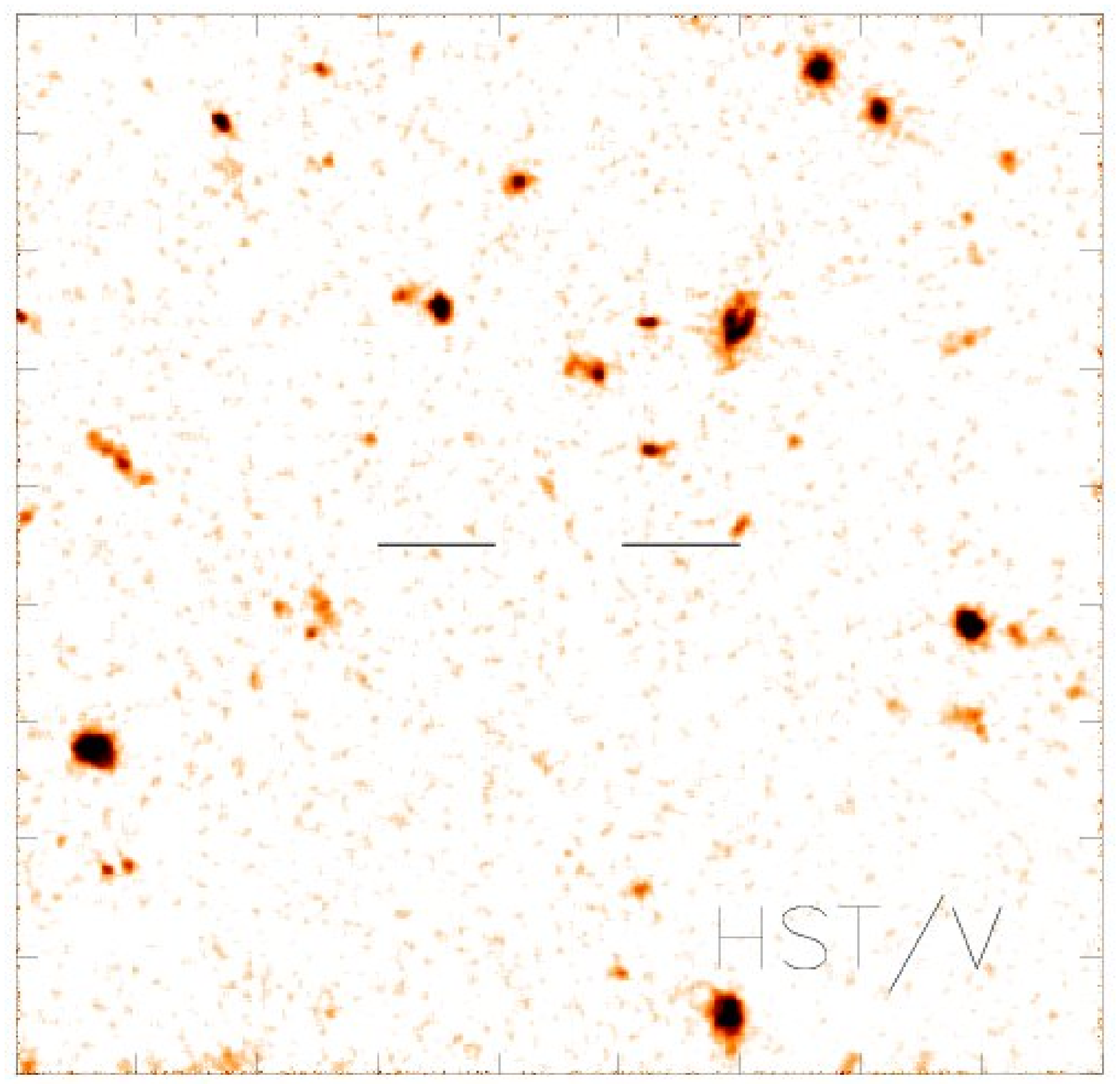,width=2.1cm,clip=}\epsfig{file=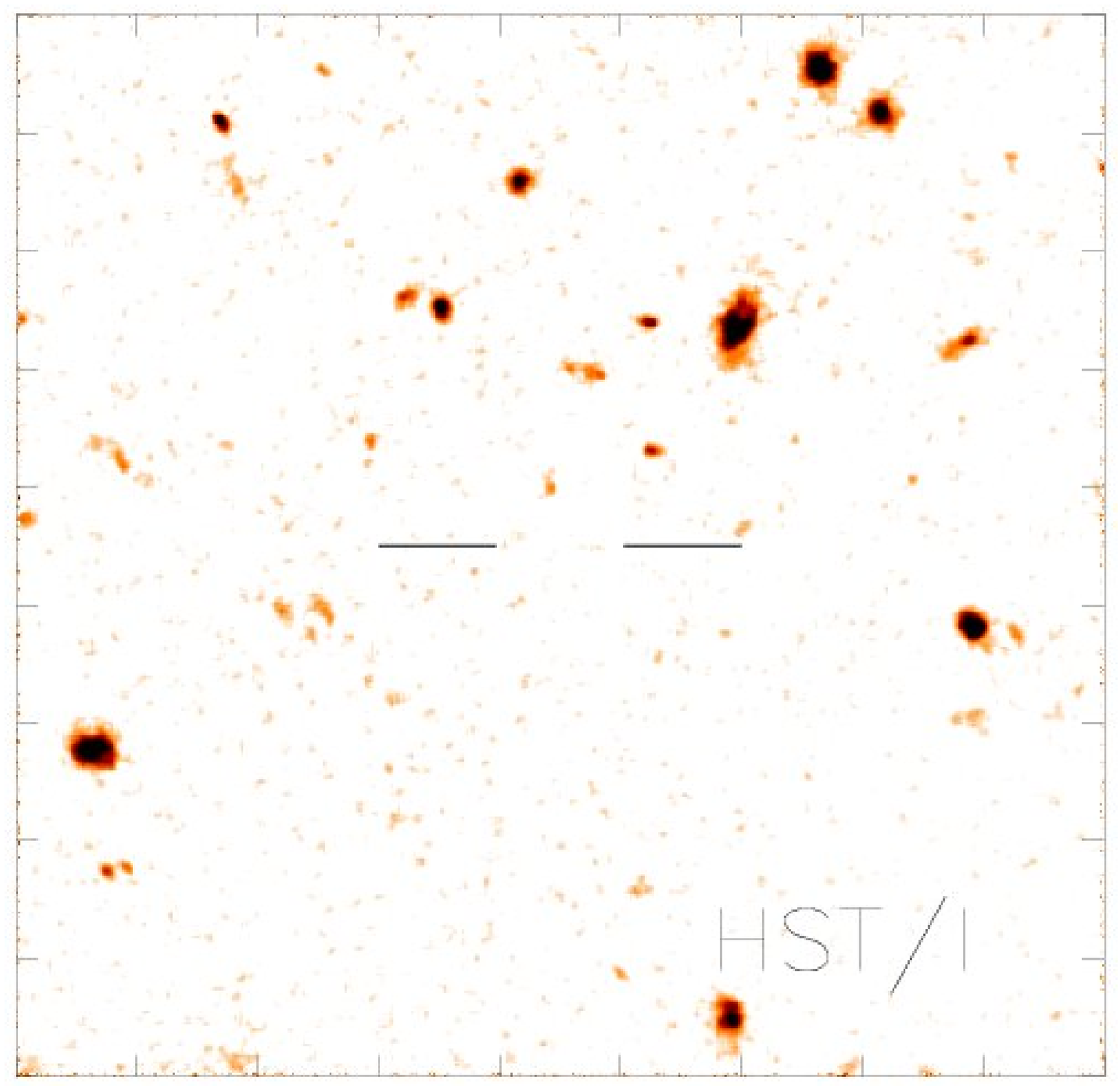,width=2.1cm,clip=}\epsfig{file=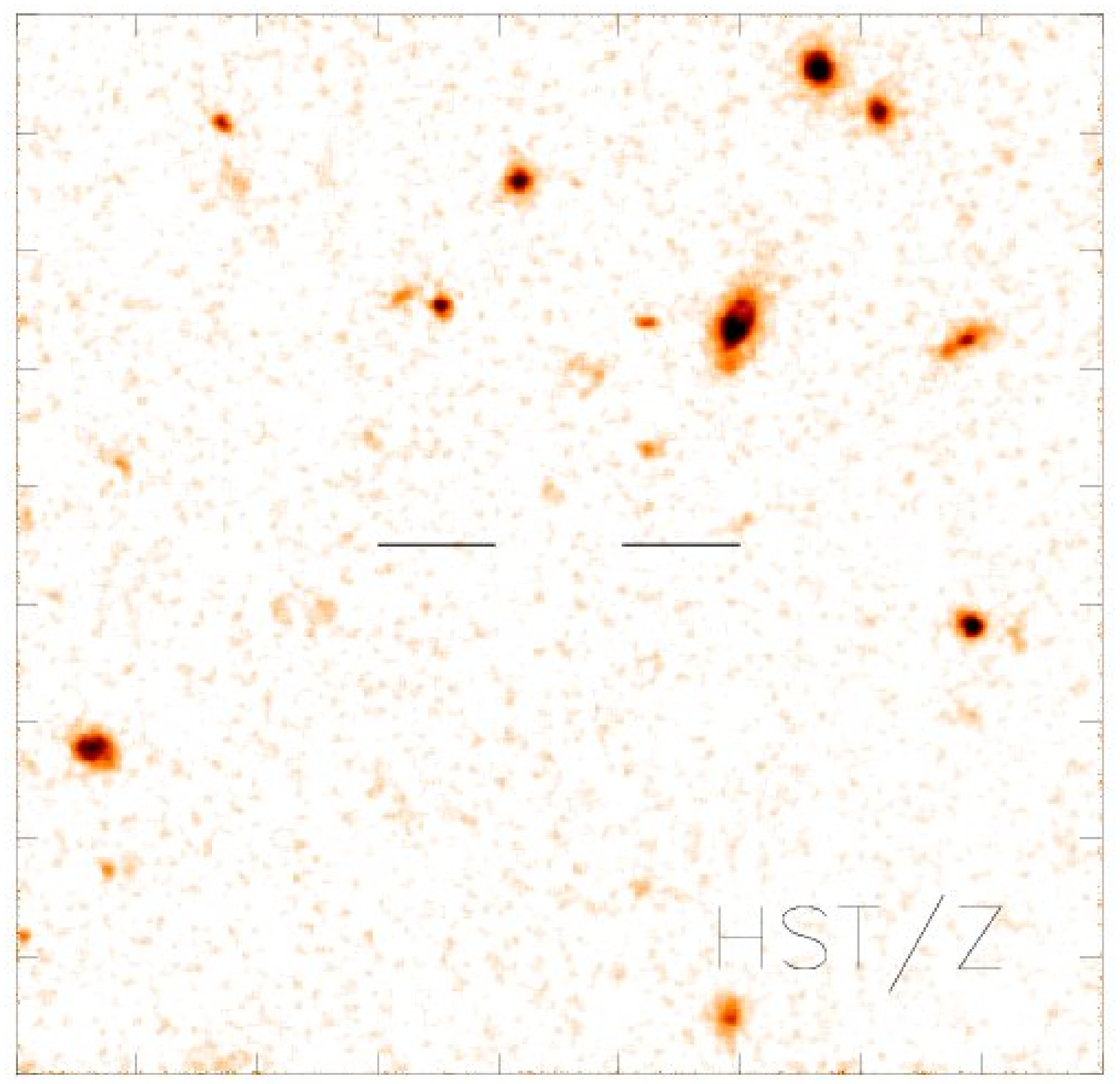,width=2.1cm,clip=}\epsfig{file=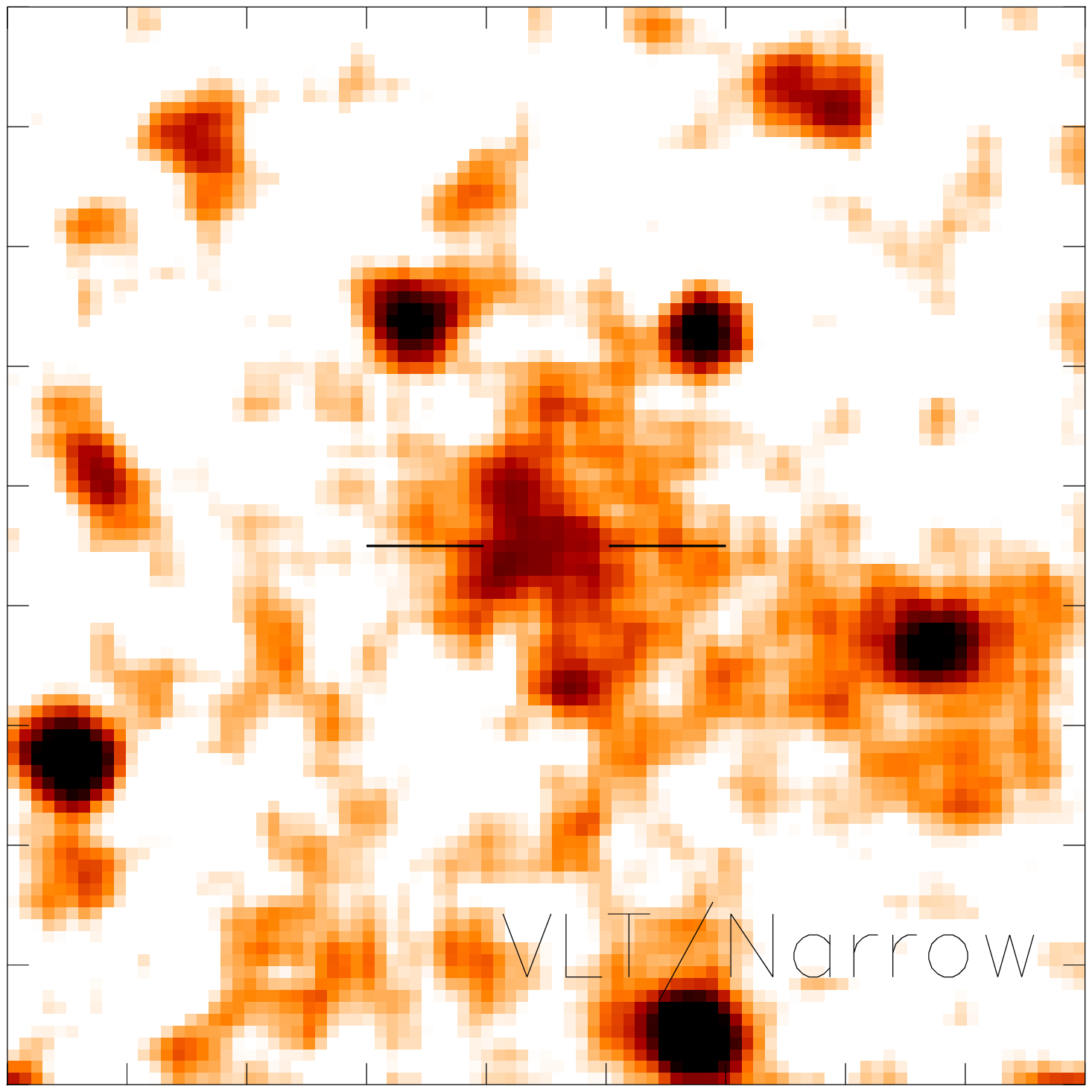,width=2.1cm,clip=}

\epsfig{file=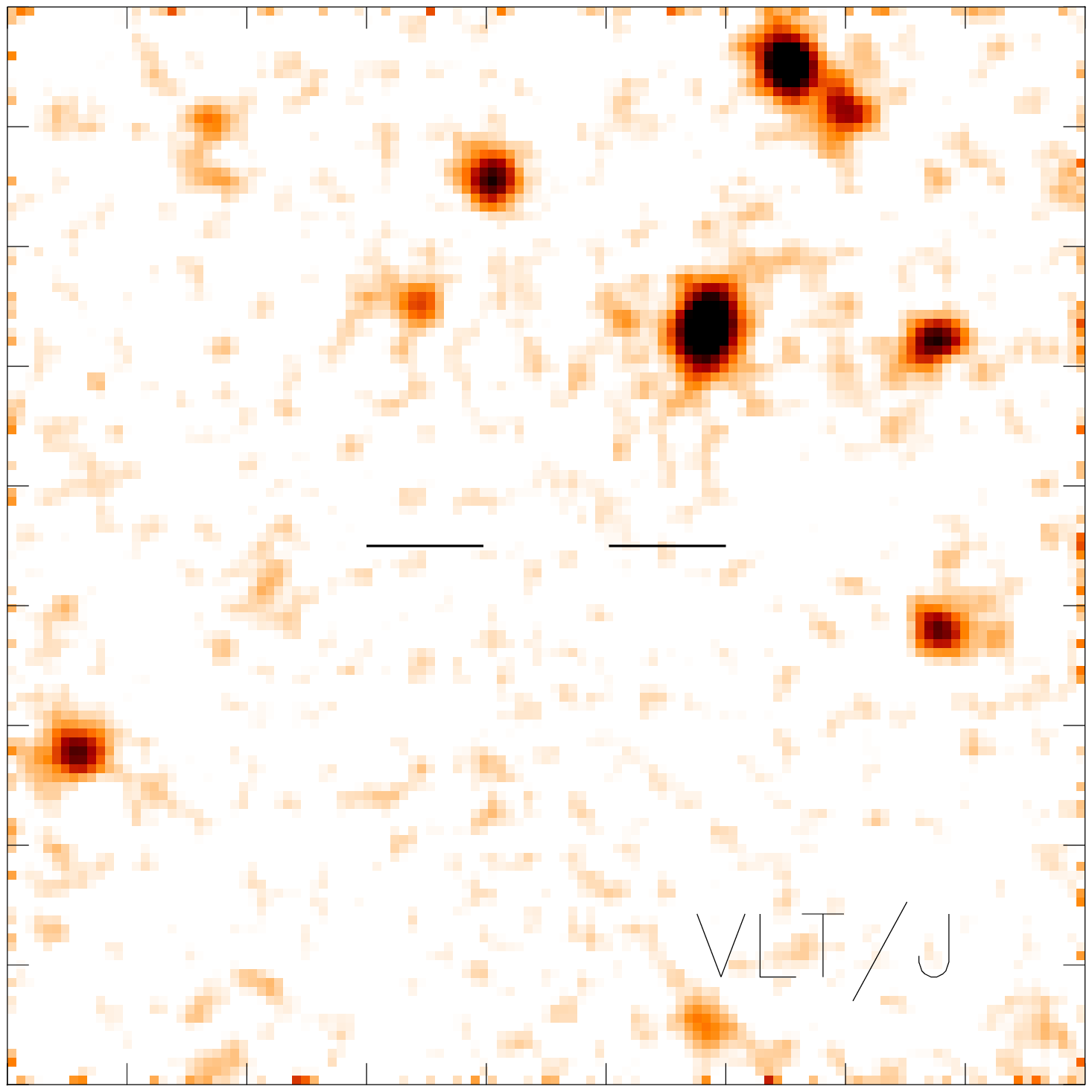,width=2.1cm,clip=}\epsfig{file=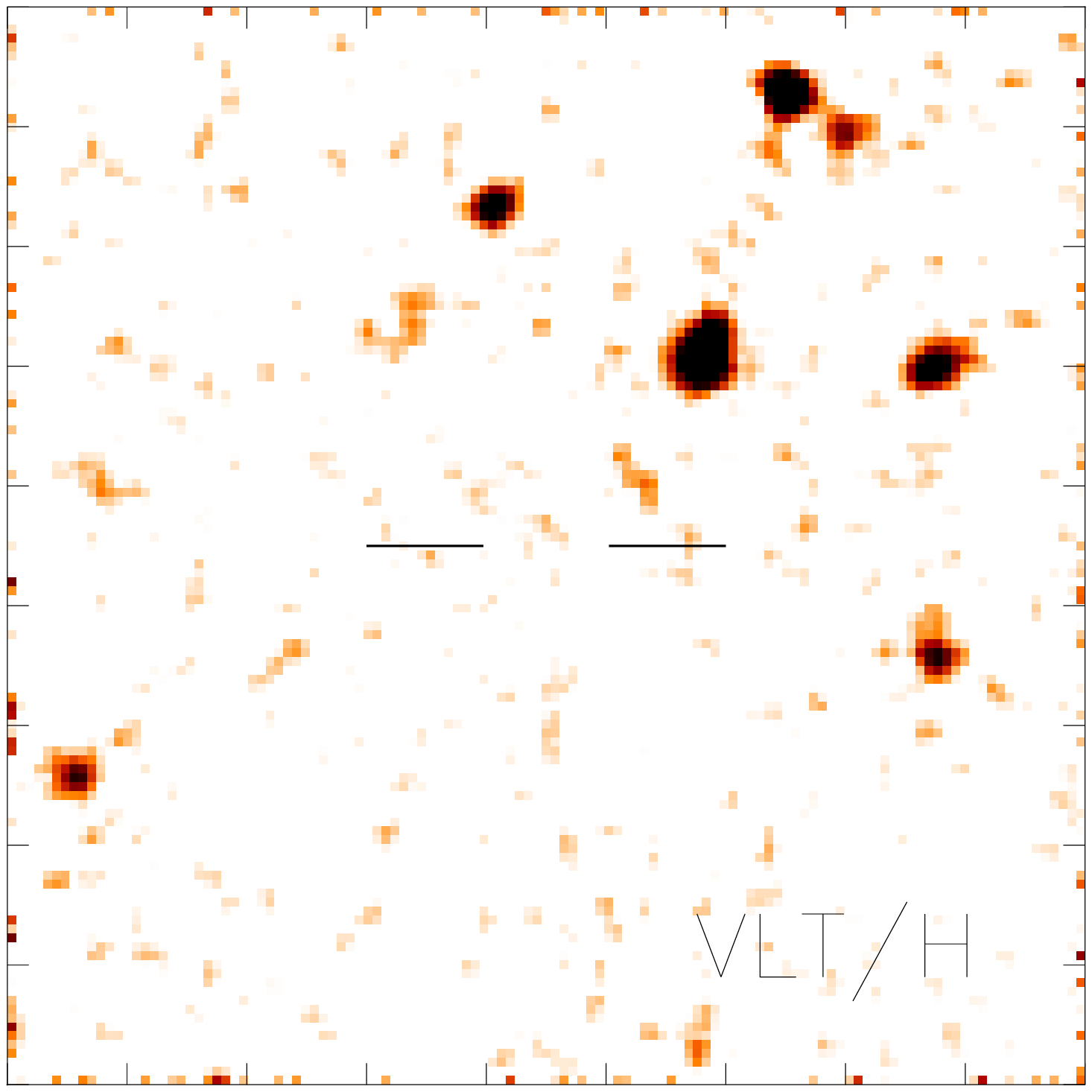,width=2.1cm,clip=}\epsfig{file=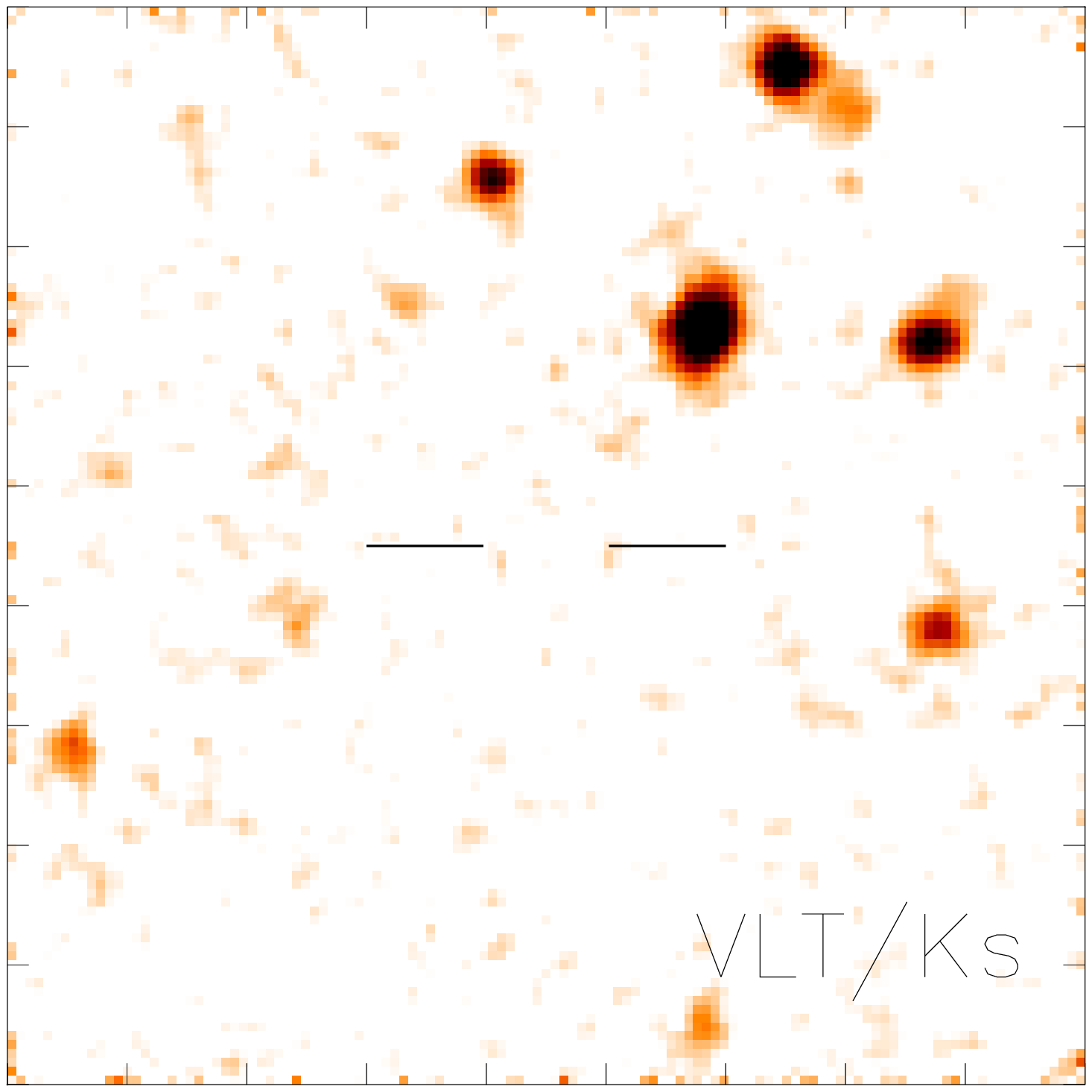,width=2.1cm,clip=}\epsfig{file=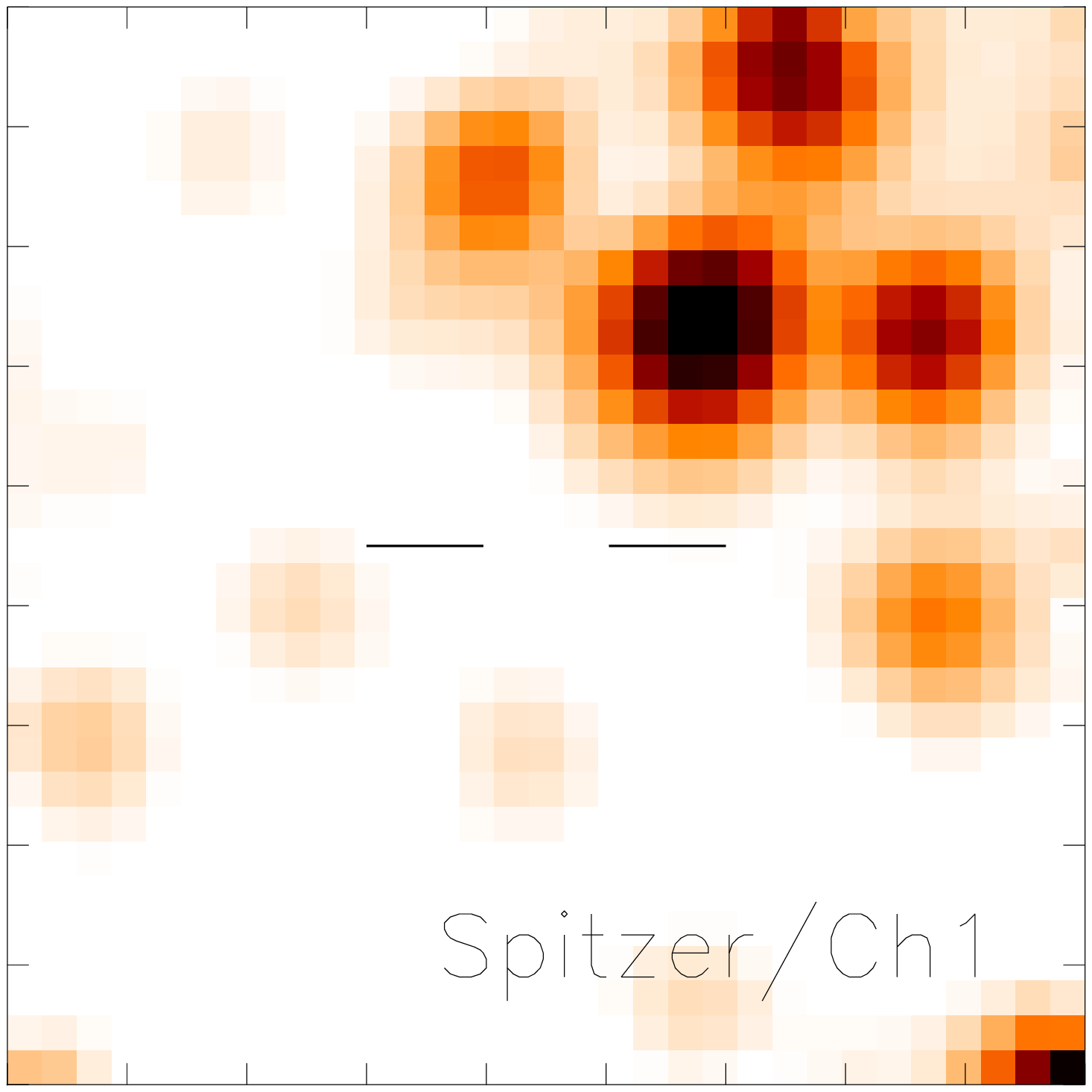,width=2.1cm,clip=}\epsfig{file=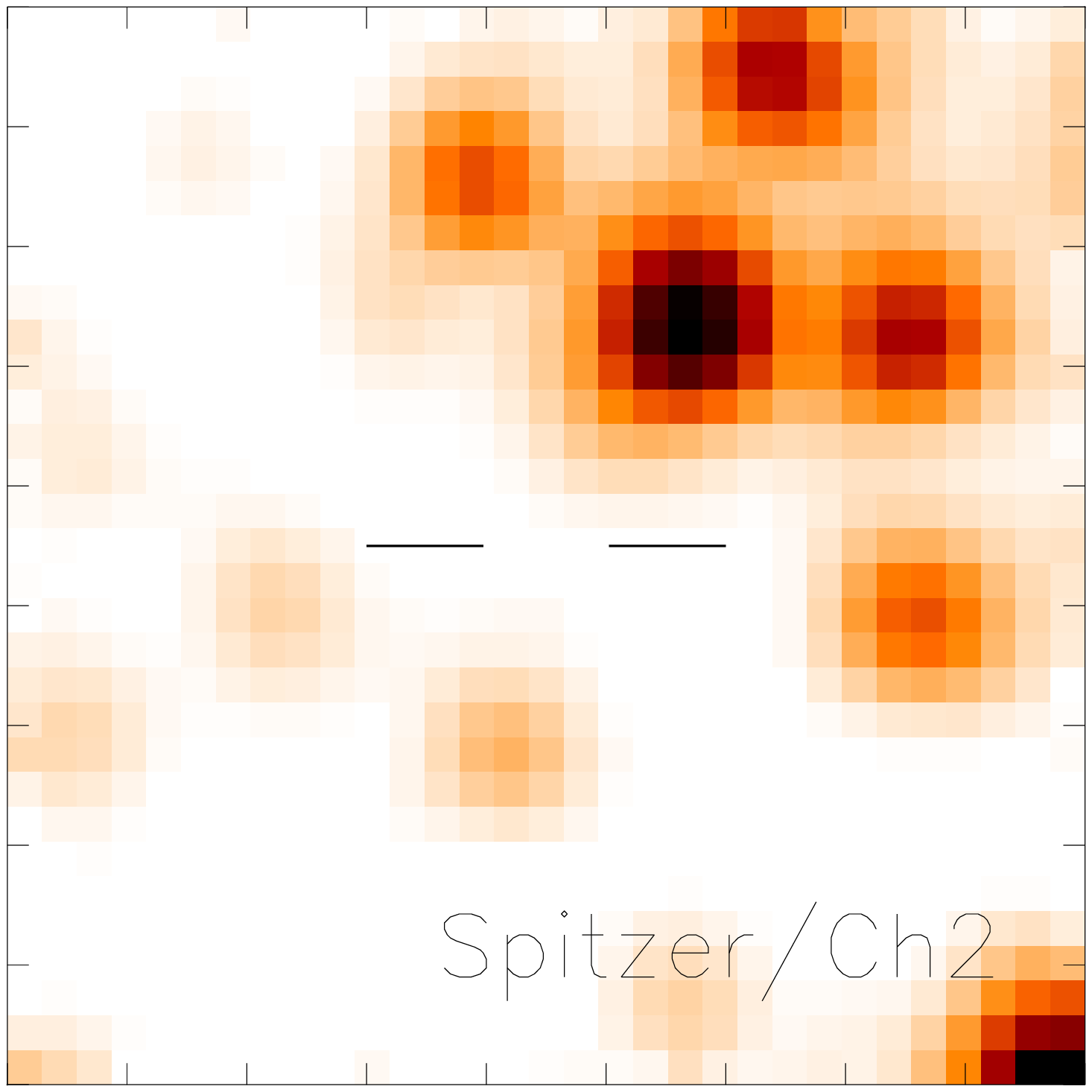,width=2.1cm,clip=}\epsfig{file=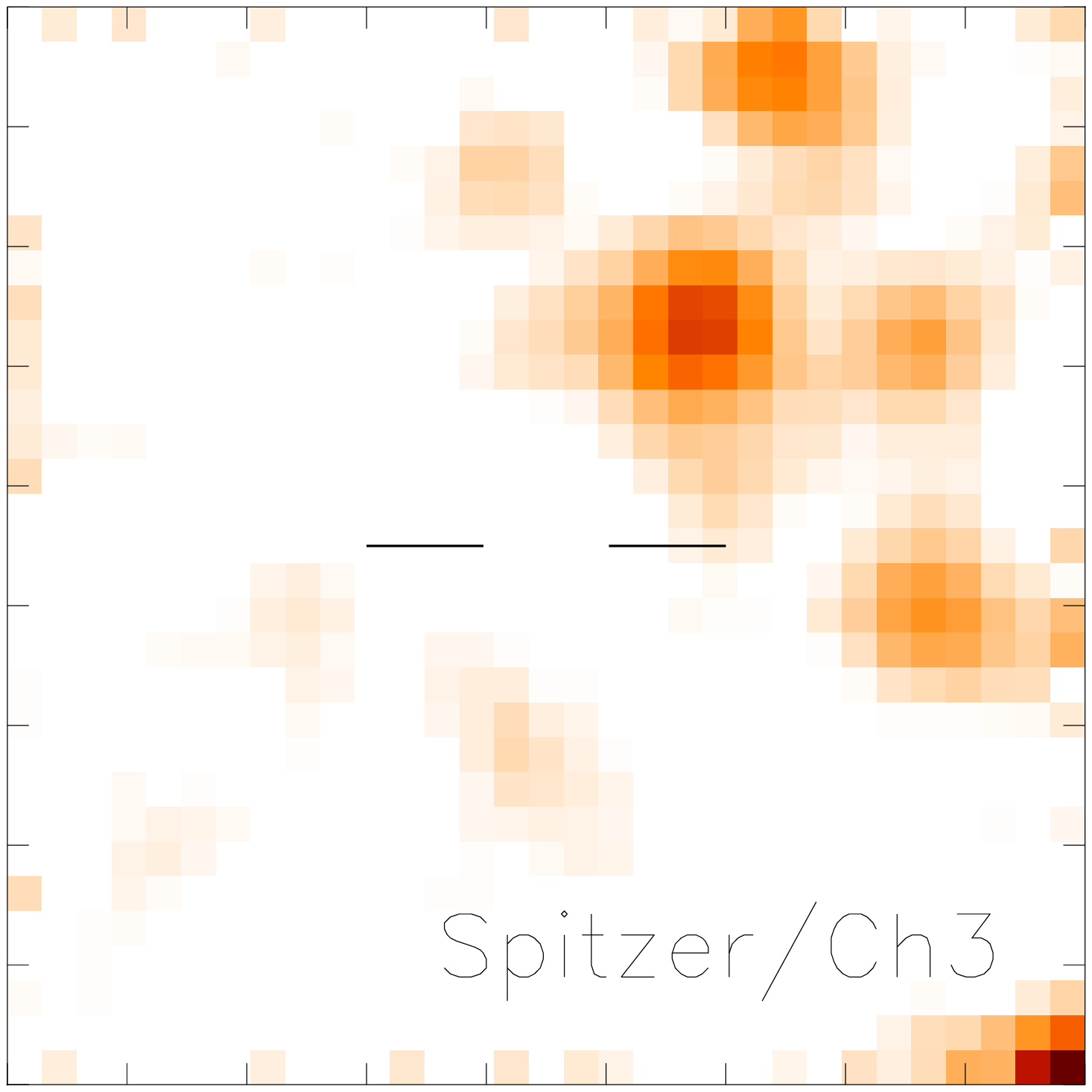,width=2.1cm,clip=}\epsfig{file=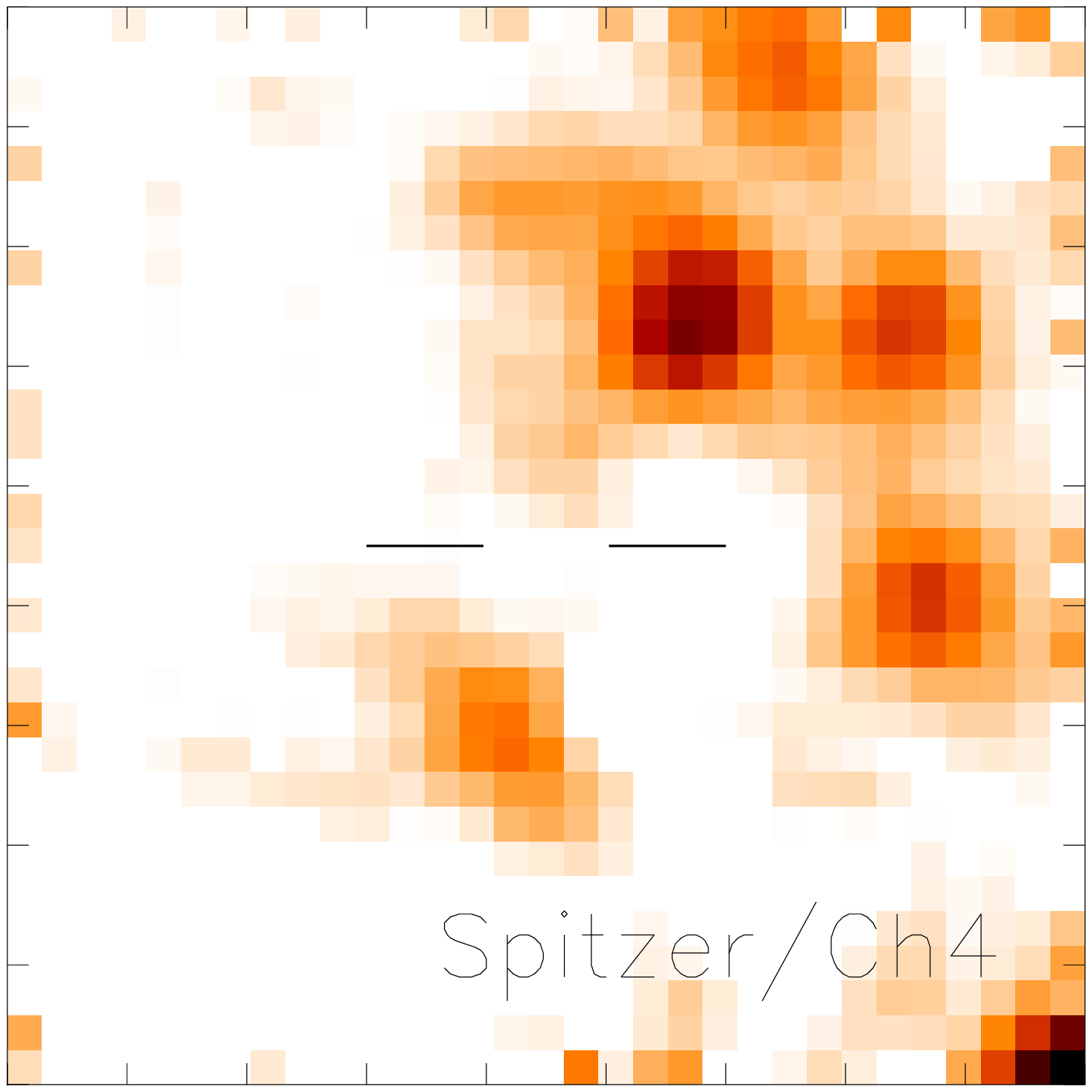,width=2.1cm,clip=}
\caption[Thumb-nail images of Ly$\alpha$ blob]{Thumbnail images of all available multi-wavelength data in the GOODS South field, centred on the Ly$\alpha$ blob. All images are $18'' \times 18''$.}
\label{thumbs}
\end{center}
\end{figure}

A contour-plot of the blob superimposed on the HST/ACS
V-band image is shown in the middle panel of Fig.~\ref{contour} and a plot of
the surface brightness of the blob is seen in the right panel of the same
figure. The full set of thumb-nail images of the blob
in all 14 bands can be found in Fig.~\ref{thumbs}. No obvious continuum
counterpart is detected in any band. The radial
size is at least 30 kpc (60 kpc diameter) with fainter emission
extending out to 40 kpc radius. This can be seen as the extension to the SW
in the contour-plot in Fig.~\ref{contour}. The significance of the
lowest contour levels is of the order of $2\sigma$ per pixel.
The total Ly$\alpha$ luminosity, in a 30 kpc radius aperture, is
L$_{\mathrm{Ly}\alpha} = (1.09 \pm 0.07) \cdot 10^{43}$~erg~s$^{-1}$.
This coincides, after correction for the smaller area sampled in the
spectrum, to the Ly$\alpha$ flux detected in the spectrum within errors. A
conservative lower limit to
the restframe equivalent width (EW) of the emission line can be calculated from
upper limits on the broad-band fluxes in the HST B and V filters in the same
aperture.
This limit is EW~$\gtrsim 220$~{\AA} in the restframe.
This is in the range of previously published Ly$\alpha$
blobs, that have a Ly$\alpha$
flux to B-band flux density range between 50~--~1500 {\AA} in the restframe
(but typically these values are derived measuring the continuum flux in a
smaller aperture than the emission line flux).

There are seven objects detected in a wide range ($\ge 8$) of energy bands,
within a $10''$~radius
surrounding the blob. 8 other objects are detected within
the V-band and one further detected in the \emph{Spitzer/IRAC}
bands. The photometric redshift of these objects was calculated using
the public \emph{HyperZ}\footnote{http://webast.ast.obs-mip.fr/hyperz/}
code by Bolzonella et al. (2000). The resulting photometric redshifts for the
eight objects with most data points ($\ge 8$)
can be found in Table~\ref{photoz}. The other eight objects detected in the
V-band have only a few detections across the spectrum and hence their
photometric redshifts are unreliable.
The redshift of object \#~3 is similar to the blob redshift and indicates
that this galaxy may be near to the blob. Object \#~6 is
an intriguing object, undetected in the deep optical and near-IR imaging
but bright in the \emph{Spitzer/IRAC} bands. Its photometric redshift is
consistent with the redshift of the blob, but with a large uncertainty.
The object is also detected in the \emph{Spitzer/MIPS} 24~$\mu$m band.
Based on the Spitzer magnitudes, and on the diagnostic colour-colour
diagram of Ivison et al. (2004), object \#~6 is best fit by a star-burst
at high redshift ($z \sim 5.5$, consistent with the photometric redshift
estimate), hence unrelated to the blob.

\begin{table} 
\begin{center}
\caption[Photometric redshifts of objects surrounding Ly$\alpha$ blob]{Photometric redshifts of objects surrounding the blob. Numbering
refers to those given in Fig.~\ref{contour}. Errors given are 1~$\sigma$.
}
\vspace{0.5cm}
\begin{tabular}{@{}lcccccc}
\hline
\hline
Obj \# & Dist. from blob & $z_{phot}$ & $\chi^2/d.o.f$ & Type & A$_V$ rest \\
& (arcsec) & & & & \\\hline
1   & 4.6 & 1.1$^{+0.40}_{-0.30}$    &  1.3   & Burst  & 0.20 \\
2   & 4.6 & 1.1$^{+0.34}_{-0.41}$    &  8.6   & Burst  & 1.20 \\
3   & 6.8 & 2.9$^{+1.41}_{-0.59}$    &  4.8   & Spiral & 1.20 \\
4   & 8.4 & 0.6$^{+1.97}_{-0.63}$    &  8.1   & Burst  & 0.40 \\
5   & 8.7 & 0.9$^{+2.46}_{-0.89}$    &  2.0   & Burst  & 0.20 \\
6   & 3.0 & 4.5$^{+4.29}_{-1.54}$    &  0.9   & Spiral & 1.20 \\
7   & 6.3 & 1.1$^{+1.24}_{-0.81}$    &  1.9   & Burst  & 1.20 \\
8   & 4.5 & 3.5$^{+1.27}_{-3.48}$    &  0.6   & Spiral & 0.00 \\
\hline
\label{photoz}
\end{tabular}
\end{center}
\end{table}

\section{Discussion}
We first consider that the Ly$\alpha$ emission of the blob may be due to
recombination of gas, photo-ionized by an AGN or a starburst galaxy.
If the blob is a ``passive'' gas cloud
illuminated and photo-ionized by a nearby AGN, then, following Cantalupo
et al. (2005), one can show that for an AGN with luminosity L$_\nu=
\mathrm{L}_{\rm LL}(\nu/\nu_{\rm LL})^{-\alpha}$, and in order to result
in a
peak blob Ly$\alpha$ surface brightness of $\Sigma_{Ly\alpha}$, the AGN has to
be located at a distance of no more than
\[
270\, \mathrm{kpc} \left(\frac{\Sigma_{Ly\alpha}}{10^{-3}\, \mathrm{erg}\,
\mathrm{s}^{-1} 
\mathrm{cm}^{-2}}\right)^{-1} 
\sqrt{\frac{L_{\rm LL}}{10^{30}\, \mathrm{erg}\, \mathrm{s}^{-1}
\mathrm{Hz}^{-1}}}
\sqrt{\frac{0.7}{\alpha}},
\]
where equality applies to the case where the blob gas is optically
thick at the Lyman limit. No AGN has been detected in the deep \emph{Chandra}
image available
within this distance. Worsley et al. (2005) argue that a significant
fraction  
of the unresolved X-ray background in hard bands consists of highly obscured
AGN. However, Worsley et al. (2005) also predict 
that the AGN responsible for this background are situated at $z \sim 0.5-1.0$.
Furthermore, Silverman et al. (2005) present a luminosity function for AGN
at higher redshift. To the detection limit of the CDFS (L$_X (z = 3.15) \approx
1.9 \cdot 10^{43}$~erg~s$^{-1}$) and with our search volume
($3 \times 3$~Mpc~$\times \Delta z = 0.05$) we expect to detect only 0.06 AGN
in our entire search volume. 
We also consider the possibility that galaxy
\#~3 can photo-ionise the blob. However, if we assume a power law for the
spectrum and extrapolating from the HST/B and HST/V detections we find that
the UV luminosity of galaxy \#~3 is not sufficient to photo-ionise the blob,
unless highly collimated towards the blob. We have no reason to believe that
this is the case.

The second possibility is that the blob Ly$\alpha$ emission
is somehow related to starburst driven, superwind outflows. A starburst would
be expected to be located within the blob to create such a
Ly$\alpha$
halo and no central continuum source has been detected. Even though a
very massive starburst can be made invisible in the
UV/optical range by dust obscuration, it should be visible in the IR, i.e. the
\emph{Spitzer/IRAC} bands.

The third option is that the Ly$\alpha$ emission is due to cold accretion of
predominantly neutral, filamentary gas onto a massive dark matter halo.
For cold accretion, the bulk of the Ly$\alpha$ emission is produced
by collisional
excitation, rather than recombination. Recently, Dijkstra et al.
2006(a,b) presented a theoretical model for Ly$\alpha$ cooling flows, along
with predictions of the emission line profile and the shape of the surface
brightness function. The S/N of our spectrum is not high enough to allow a
comparison of emission line profiles. However, the surface
brightness profile matches well the predictions for a centrally illuminated
, collapsing
cloud of Dijkstra et al. 2006(a), see Fig.~1. Further tests are needed
to determine how well their model fits.
To test whether this blob can be filamentary gas accreting
``cold'' onto a companion galaxy, we also conducted the following
experiment:
we calculated the Ly$\alpha$ surface brightness
in a 100$\times$100 kpc (projected) region for a proto-galaxy
of ``cooling'' radiation only (so all contributions from regions with young
stars were removed, as well as all emission, in general, from gas closer
than 10 kpc to any star-forming region). The calculation was based
on a cosmological simulation of the formation and evolution of an
M31-like disk galaxy (Sommer-Larsen 2005; Portinari \& Sommer-Larsen
2005).

The results at $z\sim3$ are presented in Sommer-Larsen (2005),
and get to
a surface brightness about an order of magnitude lower than the observed level.
This is interesting, and may point to a cold accretion origin
of the blob Ly$\alpha$ emission on a larger scale, such as
filamentary gas accretion onto a galaxy-group sized halo.
Another possibility is that the periods with high surface brightness
are shorter than 2.5 Myr (the resolution of the simulation). Given that in a
search volume of about
40000 co-moving Mpc$^3$, only one such blob has been detected, it is
actually comforting, that we could not reproduce the blob
characteristics, by cold accretion onto this, randomly selected, M31-like
galaxy. This has to be a rare phenomenon.

A test for the cold accretion model would be to observe the Balmer lines.
For collisionally excited hydrogen, neglecting extinction
effects, the flux in H$\alpha$ should only be about 3.5 percent of the
Ly$\alpha$ flux, whereas for recombining, photo-ionized gas this ratio is $\sim
11.5$~\% (Brocklehurst 1971). Hence, the relative H$\alpha$ luminosity is
expected to be significantly larger in the latter case. The situation is
similar for H$\beta$, and whereas the H$\alpha$ line will be very difficult to
detect from the ground, H$\beta$ should be observable.

\section{Conclusion}
We have here reported the results of an extensive multi-wavelength
investigation of a redshift $z = 3.16$ Ly$\alpha$~emitting blob discovered in
the GOODS South field. The blob has a diameter larger than 60~kpc
diameter and a total luminosity of $\mathrm{L}_{\mathrm{Ly}\alpha} \sim
10^{43}$~erg~s$^{-1}$. Deep HST imaging show no obvious optical counterpart,
and the lack of X-ray or IR emission suggest there are no AGN or dusty
starburst components associated with at least the centroid of the blob.
Two galaxies within a $10''$ radius have photometric redshifts consistent
with the redshift of the blob, but follow-up spectroscopy is needed to
establish if there is a connection.
We have run simulations of Ly$\alpha$ surface brightness arising from
cold accretion and found that such extended Ly$\alpha$ emission may be
explained by accretion
along a filament onto a galaxy group sized dark matter halo. Another
possibility is that such emission in very short lived, i.e. significantly
shorter than the 2.5 Myr resolution of our simulation. We argue that other
previously suggested origins of Ly$\alpha$ blobs (hidden AGN and
``super-winds'')
can be ruled out in this case due to the lack of detected continuum
counter-parts. Hence,
though our cold accretion simulation cannot perfectly match our data, it is the
only explanation that is plausible. Our results combined
with
the fact that previously studied blobs appear to be caused by superwinds
and/or AGN in turn implies that the energy sources for blob Ly$\alpha$
emission are diverse.

\chapter{{\ly} emitters in the GOODS-S field}\label{chapter:goodss}
\markboth{Chapter 3}{{\ly} emitters in the GOODS-S field}
This paper has been published in 
\emph{Astronomy \& Astrophysics}
in August 2007 (A\&A, 471, 71). The authors are Nilsson, K.K., M{\o}ller, P., 
M{\"o}ller, O.,
Fynbo, J.P.U., Micha{\l}owski, M.J., Watson, D., Ledoux, C., Rosati, P.,
Pedersen, K., \& Grove, L.F.

\section{Abstract}
\emph{Context} Ly$\alpha$-emitters have proven to be excellent probes of faint, star-forming
galaxies in the high redshift universe. However, although the sample of known
emitters
is increasingly growing, their nature (e.g. stellar masses, ages,
metallicities,
star-formation rates) is still poorly constrained.

\noindent \emph{Aims} We aim to study the nature of Ly$\alpha$-emitters, to find the properties
of a typical Ly$\alpha$-emitting galaxy and to compare these properties
with the properties of other galaxies at similar redshift, in
particular Lyman-break galaxies.

\noindent \emph{Methods}
We have performed narrow-band imaging at the VLT, focused on Ly$\alpha$ at
redshift
$z \approx 3.15$, in the GOODS-S field. We have identified a sample of
Ly$\alpha$-emitting candidates, and we have studied their Spectral Energy
Distributions (SEDs).

\noindent \emph{Results}
We find that the emitters are best fit by an SED with low metallicity
($Z/Z_{\odot} = 0.005$) , low
dust extinction (A$_V \approx 0.32$) and medium stellar masses
of approximately $10^9$~M$_{\odot}$. The age is not very well constrained.
One object out of 24 appears to be a high redshift Ly$\alpha$-emitting dusty
starburst galaxy. We find filamentary
structure as traced by the Ly$\alpha$-emitters at the
4$\sigma$ level. The rest-frame UV SED of these galaxies
is very similar to that of Lyman Break Galaxies (LBGs) and comply
with the selection criteria for $U$-band drop-outs, except
they are intrinsically fainter than the current limit for LBGs.

\noindent \emph{Conclusion}
Ly$\alpha$-emitters are excellent probes of galaxies in the distant universe,
and represent a class of star-forming, dust and AGN free, medium mass objects.

\section{Introduction}
The possibility to use the Ly$\alpha$ emission line to study galaxies in early
stages of their formation was outlined already by Partridge \& Peebles (1967)
nearly 40 years ago, but early surveys (see Pritchet 1994 for a review) failed
to produce anything other than upper limits. The unexpected faintness of the
objects caused it to take almost three decades before the narrow-band technique
was successfully used to identify the first high redshift Ly$\alpha$ emitting
galaxies that were not dominated by Active Galactic Nuclei (e.g.,
Lowenthal et al. 1991; M\o ller \& Warren 1993; Hu \& McMahon 1996; Petitjean et
al. 1996; Francis et al. 1996; Cowie \& Hu 1998). It is only recently, with
the advent of 8 m
class telescopes and sensitive detectors, that larger samples of Ly$\alpha$
selected objects have been reported (e.g. Steidel et al. 2000; Malhotra \&
Rhoads 2002; Fynbo et al. 2003; Ouchi et al. 2003; Hayashino et al.
2004; Venemans et al. 2005).

Already during the early studies, two interesting suggestions were raised.
First, it
was found that there is a tendency for Ly$\alpha$ selected objects to ``line
up'' as strings in redshift space, and that they therefore may be excellent
tracers of filaments at high redshifts (Warren \& M\o ller 1996; Ouchi et al.
2004a,b; Matsuda et al. 2005). Secondly, they were found to
have very faint broad band magnitudes and therefore could be good
tools in detecting faint, high redshift galaxies (Fynbo et al.
2001; Fujita et al. 2003; Venemans et al. 2005; Gawiser et al. 2006).

The first of those suggestions was explored theoretically via modeling of
structure formation including assignment of Ly$\alpha$ emission to the models
(Furlanetto et al. 2003; Monaco et al. 2005), and has been confirmed
observationally (M{\o}ller \& Fynbo 2001; Hayashino et al. 2004), who also
proposed to use such
structures for a new cosmological test to measure $\Omega_{\Lambda}$ by looking
at the ``length-to-radius'' ratio of filaments observed from the side or
end-on. This proposed test was subsequently explored in detail by
Weidinger et al. (2002).

The second suggestion has gained significant interest because there are now
several additional, but independent, ways of identifying high redshift but
optically faint galaxies, e.g. Damped Ly$\alpha$ Absorbers (DLA) galaxies
(Wolfe et al. 1986; M\o ller et al. 2002; Wolfe, Gawiser \& Prochaska 2005),
Gamma-Ray Burst (GRB) host
galaxies (Fruchter et al. 2006), sub-mm galaxies (Chapman et al. 2004).
Most
galaxies from such searches are however too faint to be identified in current
ground-based optical flux limited samples. A significant project
(the Building the Bridge Survey, BBS) aimed at
addressing those issues is currently underway at the ESO Very Large Telescope
(VLT) (Fynbo et al. 2001; Fynbo et al. 2003).  The very faintness of the
objects, however, renders it difficult to make any detailed comparisons. While
some DLA galaxies have been imaged with HST (Warren et al. 2001; M{\o}ller et
al. 2002) and likewise GRB hosts (Jaunsen et al. 2003; Fynbo
et al. 2005), and sub-mm galaxies (Smail et al. 2004; Pope et al. 2005; Schmitt
et al. 2006), only a very small subset of the Ly$\alpha$ selected galaxies have
been imaged with HST (Pascarelle et al. 1996; Venemans
et al. 2005; Overzier et al. 2006) and furthermore, such images are mostly too
shallow for a detailed study.

The GOODS-S (Giavalisco et al. 2004) provides a unique opportunity to obtain a
deep, high resolution, multi-band data set of a complete and unbiased sample of
Ly$\alpha$-emitters (or LEGOs for Ly$\alpha$ Emitting Galaxy-building Objects;
M{\o}ller~\&~Fynbo (2001)).
We have therefore started a program to collect a complete, unbiased sample of
LEGOs in the GOODS-S. This allows a detailed study of the global properties, e.g.
photometry and morphology of LEGOs, as well as SED fits including photometry
from the large, available multi-wavelength data-set.

This paper is organised as follows; in section 2 we present the
imaging observations, data reductions and candidate selection process.
In section 3 we present spectroscopic observations of three candidates
as well as the results from these observations. Sections 4, 5 and 6 contain the
discussion of various aspects of the LEGO candidate sample; first the basic
characteristics of the LEGO sample, then the SED fitting and finally a
comparison to Lyman-Break Galaxies. The conclusion is presented in section 7.

Throughout this paper, we assume a cosmology with $H_0=72$
km s$^{-1}$ Mpc$^{-1}$ (Freedman et al. 2001), $\Omega _{\rm m}=0.3$ and
$\Omega _\Lambda=0.7$.
Magnitudes are given in the AB system. We survey a co-moving volume of
$\approx 3300$~Mpc$^3$.

\section{Imaging}
\subsection{Narrow band observations and data reduction}
\label{ch3:obs}
\begin{figure}[!t] 
\begin{center}
\epsfig{file=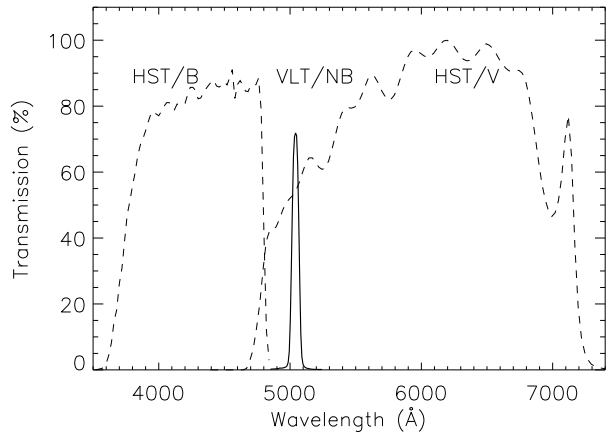,width=9.0cm}
\caption[Transmission of selection filters]{Transmission of selection filters. The VLT FORS1 narrow-band
filter is drawn with a solid line. Dashed lines show the HST $B$ and $V$ filters.}
\label{ch3:filters}
\end{center}
\end{figure}

A 400$\times$400 arcsec$^2$ section of the GOODS-S field centred on 
R.A.~$ = 03^h 32^m 21.9^s$  and Dec~$ = -27^{\circ} 45' 50.7''$ (J2000) 
was observed with FORS1 on the VLT 8.2 m telescope Antu
during two visitor mode nights on December 1-3, 2002. The log of
observations is given in Table~1. 
A total of 16 dithered exposures were obtained over the two nights
for a combined exposure time of 30000 seconds, all with the narrow band
filter OIII/3000+51 and using the standard resolution collimator (0.2$\times$0.2
arcsec$^2$ pixels). For this setup the central wavelength of the
filter is 505.3 nm with a FWHM of 5.9 nm which corresponds to the
redshift range $z = 3.131-3.180$ for Ly$\alpha$.
The transmission curve of the filter is shown in Fig.~\ref{ch3:filters}.
The four spectrophotometric standards Feige110, LDS749B, LTT3864,
and LTT3218 were observed on the same nights.

\begin{figure}[!t] 
\begin{center}
\epsfig{file=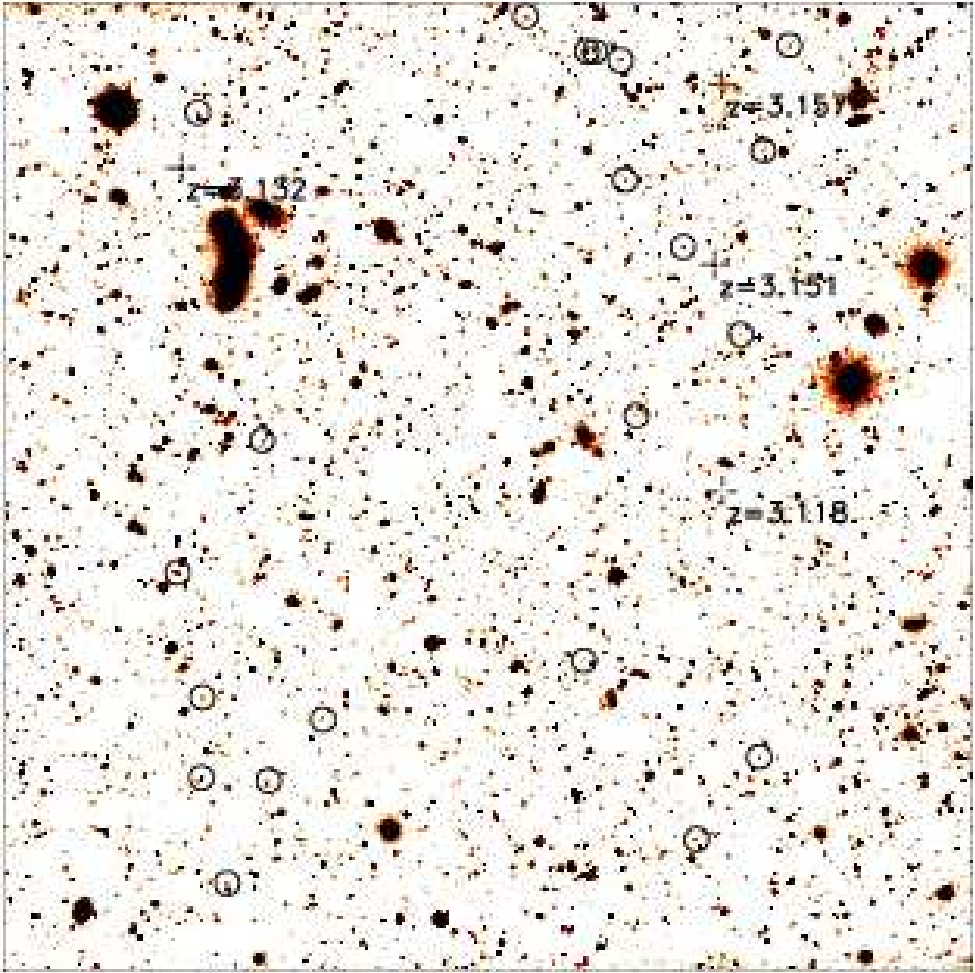,width=13.0cm,angle=90}
\caption[The VLT/narrow-band image in GOODS-S]{The VLT/narrow-band image of the 400$\times$400 arcsec$^2$ field with
the positions of selected LEGO candidates (see Sect.~\ref{ch3:LEGOs}) shown with 
circles. Spectroscopically confirmed candidates are marked with crosses and
their redshifts are indicated. North is
up and East is left. The right, uppermost spectroscopically confirmed candidate
is the Ly$\alpha$ blob (Nilsson et al. 2006a).}
\label{ch3:field}
\end{center}
\end{figure}

The observing conditions were unstable during the two nights with the
seeing FWHM, as measured on the images, varying between $0.66''$ and
$1.25''$ on the first night
and  between $1.4''$ and $3.3''$ on the second night.
The images were reduced (de-biased and corrected for CCD pixel-to-pixel
variations) using standard techniques.
The individual reduced images were combined using a modified version
of our code that optimizes the Signal-to-Noise (S/N) ratio for faint,
sky-dominated sources (see M\o ller \& Warren, 1993, for details on this
code). The 5~$\sigma$ detection limit of the combined narrow-band image
as measured in circular apertures with radius twice the
full width half maximum of point sources, i.e. with radius $1.6''$, is
mag(AB)$ = 26.1$. The combined narrow-band image is shown in Fig.~\ref{ch3:field}.

\begin{table}[!t] 
\begin{center}
\caption[Log of imaging observations with FORS1]{Log of imaging observations with FORS1. }
\vspace{0.5cm}
\begin{tabular}{@{}lcccccc}
\hline
\hline
date & total exp. & seeing range & \\
\hline
01-02.12.2002   &   5.54 hours   &   $0.66''$-$1.25''$  &  \\
02-03.12.2002   &   2.78 hours   &   $1.43''$-$3.30''$    &  \\
\hline
\label{ch3:journal}
\end{tabular}
\end{center}
\end{table}

\subsection{Selection of LEGOs in the fields}
\label{ch3:LEGOs}

For the selection of LEGO candidates we used the narrow-band image as detection
image and the HST/ACS $B$- (F435W) and $V$-band (F606W) images as
selection images. The HST data is part of the public data in GOODS-S
(Giavalisco et al. 2004).  This selection set-up, with a broad-band filter on
either side of the narrow-band filter, appears to be one of the most efficient
configurations for selection of emission-line objects (Hayes \& {\"O}stlin
2006; Hayes priv. communication). The HST images were re-binned to the pixel
size of the narrow-band image. Due to the smaller field-of-view of the HST/ACS
images, the narrow-band image was cut into six sub-images to match the size of
the HST images.

Our selection method consists of three consecutive steps. First,
using the software package SExtractor (Bertin \& Arnouts 1996), we select
all objects
identified in the narrow-band image. The narrow-band image is
scanned with a detection threshold equal to the background sky-noise
and requiring a minimum area of 5 connected pixels above this
threshold. Centred on each candidate object we then extract
photometry from the narrow, $V$, and $B$-band images using identical
circular apertures of $2''$ diameter.
Note that through this process we make no attempt to identify
broad band counterparts to the narrow band objects. The broad band
photometry is extracted in apertures defined solely from the centroid
positions in the narrow band image. There is always a small but finite
possibility that an unrelated for- or background object could fall
inside the $2''$ aperture which would complicate the search for emission
line objects. This complication is minimised by not re-centering the
aperture on objects in the broad band images.

The second step is to accept only candidates that are
detected at S/N$>5$ in the narrow band circular aperture and are found at
least 20 pixels from the edge of each image. This leaves us with a
catalogue of 2616 narrow band objects within the resulting
$385 \times 400$~arcsec$^2$ field.  The term ``narrow band objects'' is used
here only to
underline that while they are $5 \sigma$ detections in the narrow band
image, they may or may not have been detected in the broad band images.


In the third step we select the subset of our catalogue comprising of
potential emission line objects. Via interpolation of the flux levels
in the B and V bands it is easy to calculate the continuum flux at the
central wavelength of the narrow band filter, and from there to obtain the
equivalent width (EW) of a potential line and its associated propagated
statistical error. Note that we here
calculate the ``EW of the aperture'', which means that if there is only
one object in the aperture, then we find the EW of that object. If
there are additional unrelated neighbours inside the aperture, then
the calculated EW will be smaller than the actual EW. Our listed EWs
(see Table~\ref{ch3:selecttable})
are in that sense conservative lower limits.
For the present work we are interested only in those
with positive values of EW (emission line objects), and only the
subset of those where the significance of the line is high enough to
provide a high probability that it is reliable, i.e. providing a high
efficiency of spectroscopic follow-up work. For the current field we
already have a few
spectroscopic confirmations (see Sec.~\ref{ch3:specsec}) and
we conservatively chose to cut at the EW significance level (2.9$\sigma$)
where all confirmed LEGOs are included. This corresponds to a
formal probability of 99.6\% for confirmation of each object and we
find 106 such objects. Of these, three are associated with
a Ly$\alpha$-blob that we found in this field (Nilsson et al., 2006a).
Ly$\alpha$-blobs (e.g. Fynbo et al. 1999; Steidel et al. 2000;
Matsuda et al. 2004) are large luminous nebulae, with sizes up to 150 kpc,
emitting solely Ly$\alpha$ emission. The Ly$\alpha$ luminosity can reach
$10^{43}$~erg~s$^{-1}$.
79 of the 106 objects with EW excess are ruled out after visual inspection due to 
stellar artifacts, saturated
sources or for lying on parts of the image where the HST sky background is
poorly constrained.
Thus, we are finally left with a sample of 24 LEGO candidates. These candidates
are presented in Table~\ref{ch3:selecttable}. The formal probability of
99.6\% for confirmation of the emission line results in 0.10 spurious detections
in our sample of 24 objects. A colour-colour plot of the
narrow-band sources can be found in Fig.~\ref{ch3:colcol}.

\begin{table}[!t]
\caption[First selection]{Data on first selection. Coordinates are in J2000. Equivalent
widths are calculated from the ``first selection'', i.e. with 2$''$ radius
apertures, centred on the narrow-band source centroid, see section~\ref{ch3:LEGOs}. S/N is the signal-to-noise of the equivalent width.}
\label{ch3:selecttable}
\vspace{0.5cm}
\centering
\begin{tabular}{lcccccccc}
\hline\hline
LEGO & EW$_{\mathrm{obs}}$ & $\sigma_{\mathrm{EW}}$ & S/N  & R.A. & Dec\\
GOODS-S & (\AA) & (\AA) &  & &\\
\hline
1  & 896 & 57 & 16  & 03:32:14.83 & -27:44:17.5  \\
2  & 59  & 15 & 3.8 & 03:32:17.62 & -27:43:42.3 \\
3  & 172 & 24 & 7.1 & 03:32:18.56 & -27:42:48.4 \\
4  & 901 & 48 & 19  & 03:32:31.46 & -27:43:37.2 \\
5  & 184 & 30 & 6.0 & 03:32:30.02 & -27:48:37.1 \\
6  & 517 & 23 & 22  & 03:32:30.82 & -27:47:52.8 \\
7  & 153 & 32 & 4.7 & 03:32:13.40 & -27:47:43.9 \\
8  & 85  & 20 & 4.3 & 03:32:30.79 & -27:47:19.2 \\
9  & 100 & 24 & 4.2 & 03:32:12.49 & -27:42:45.8 \\
10 & 52  & 16 & 3.3 & 03:32:13.30 & -27:43:29.9 \\
11 & 154 & 33 & 4.6 & 03:32:27.03 & -27:47:28.3 \\
12 & 151 & 33 & 4.6 & 03:32:13.99 & -27:44:47.0 \\
13 & 53 & 18 & 2.9  & 03:32:14.58 & -27:45:52.5 \\
14 & 37  & 4  & 8.3 & 03:32:18.82 & -27:42:48.3 \\
15 & 88  & 11 & 7.9 & 03:32:31.56 & -27:46:26.9 \\
16 & 34 & 7.5 & 4.4 & 03:32:20.72 & -27:42:33.8 \\
17 & 202 & 46 & 4.4 & 03:32:28.93 & -27:45:31.5 \\
18 & 269 & 55 & 4.9 & 03:32:17.26 & -27:45:21.0 \\
19 & 47  & 12 & 3.9 & 03:32:15.80 & -27:44:10.3 \\
20 & 58 & 15 & 3.9  & 03:32:30.96 & -27:43:14.2 \\
21 & 53 & 16 & 3.4  & 03:32:28.73 & -27:47:54.1 \\
22 & 58 & 15 & 4.0  & 03:32:17.77 & -27:42:52.1 \\
23 & 362 & 57 & 6.3 & 03:32:15.37 & -27:48:18.5 \\
24 & 136 & 24 & 5.7 & 03:32:18.89 & -27:47:03.6 \\
\hline
\end{tabular}
\end{table}

\subsection{Continuum counterparts and final photometry}\label{ch3:finalphot}
\begin{figure}[!t]
\begin{center}
\epsfig{file=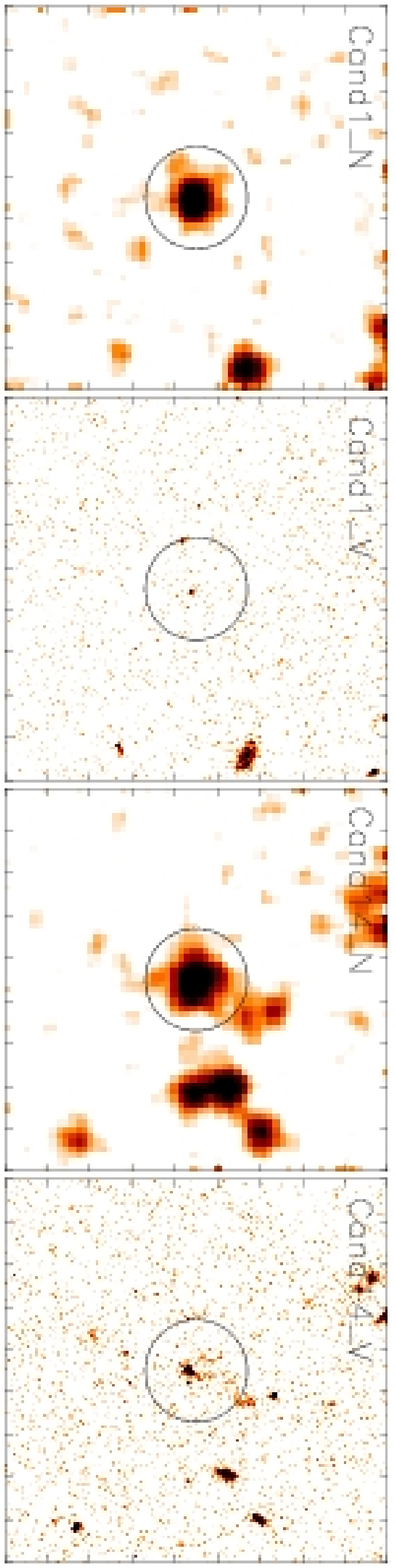,width=4.0cm,angle=90}
\caption[Illustration of broad-band counterpart selection problem]{
In this figure, we illustrate the difficulty in determining which broad-band
counterpart was associated with the narrow-band source. Images are $12''$
across and are centred on the narrow-band source. The circles mark the 
apertures used in the initial photometry, see section~\ref{ch3:LEGOs}. The two 
left panels show a simple case with
only one possible counterpart. The two right panels show a more complex case.
In this case, all objects within the circle were assumed to be counterparts.
In both cases a presumably unrelated object is seen at the edge of the
aperture.}
\label{ch3:psfill}
\end{center}
\end{figure}

Following the selection process outlined above, we examined the narrow-band and 
broad-band images to find continuum counterparts to the narrow-band objects.
The process of finding counterparts to the narrow-band objects is complicated 
by the very different PSF characteristics of the narrow- and broad-band images.
This is illustrated in Fig.~\ref{ch3:psfill}.
For some of the candidates, no obvious counterpart was detected, but rather
several counterpart candidates were found with small offsets. To determine
what continuum objects were associated with the narrow-band source, two
co-authors separately inspected the images visually. If only one
counterpart was in the
vicinity of the centroid of the narrow-band source, this object was identified
as
sole counterpart. If several sources were detected, their magnitudes were
measured and the statistical probability that they would appear in the area
surrounding the narrow-band centroid was evaluated from number counts of
galaxy searches. We then separately determined
which counterparts we considered credible counterparts, and the lists were
compared. Only counterparts assigned by both authors were accepted as
counterparts. This yielded 2 LEGO candidates without counterparts, 14
candidates with single counterparts and 8 LEGOs with two or more counterparts.
Aperture photometry was then performed on all
candidates in the narrow--band and their selected counterparts in the
HST broad band images. We used aperture radii of two times the
FWHM of each image. Aperture corrections were calculated for each image for
point sources (see Table~\ref{ch3:mwtab}). For candidates where multiple
components were assigned, small apertures were placed either \emph{i)}
centred on each counterpart separately if the counterparts are further apart
than two times the radius of the aperture, \emph{ii)} centred on a
coordinate half way between the counterparts if the distance between
counterparts is less than one times the radius of the aperture or
\emph{iii)} apertures slightly shifted from the central coordinate of the
counterparts to ensure that different apertures do not overlap if the distance
between counterparts is less than two, but more than one, times the aperture
radius. 
The magnitudes, EWs and star formation rates (SFRs) for all candidates can be
found in Table~\ref{ch3:photometry}, see also Sec.~\ref{ch3:DiscLEGO}. 
Multiple candidates are marked with a star. To investigate how correct our
method is for measuring fluxes of multiple objects, we also measured the
photometry using larger apertures; for the six multiple
counterparts with distances between the counterparts less than two times the
radius of the aperture (LEGO~GOODS-S\_\#~9, 10, 12, 14, 19 and 22), we
applied new apertures with radii half of the distance between counterparts
plus the original aperture radius. For five candidate counterparts, the
difference in flux measured was within 1.5$\sigma$ of the previously measured
value. Hence we conclude that our original measurements are correct. The
remaining candidate, LEGO~GOODS-S\_\#~14, has a very complicated morphology,
see also Fig.~\ref{ch3:psfill}.
For this candidate, the flux increased to M$_B = 25.49 \pm 0.11$,
M$_V = 24.47 \pm 0.04$, M$_I = 24.36 \pm 0.13$ and M$_{z'} = 24.49 \pm 0.13$.
This reduces the observed EW to $91 \pm 8$~{\AA}.

In summary we first performed simple circular aperture photometry in
order to select candidates based on our conservative emission-line
definition (Table~\ref{ch3:selecttable}). For the selected candidates, we
then searched for
continuum counterparts in the high resolution HST images and carried
out detailed final photometry where such counterparts were found. This
final photometry provides the relevant magnitudes that describe the
objects and is reproduced in Table~\ref{ch3:photometry}.

\begin{landscape}
\begin{table}
\caption[Final photometry of GOODS-S LEGOs]{\small{Final photometry of candidates in narrow- and broad-bands
(Sec.~\ref{ch3:finalphot}) and observed Ly$\alpha$ EWs. Magnitudes are
calculated using two times full width half maximum apertures, including aperture
corrections, centred on each identified counterpart. Errors
are $1 \sigma$, upper limits are $3 \sigma$. Equivalent widths are
calculated again from the magnitudes printed in this table. Star formation
rates are from the
Ly$\alpha$ fluxes. Emitters marked in bold are spectroscopically confirmed, see
Sec.~\ref{ch3:specsec}. Candidates marked with a star have multiple counterparts.
For these candidates, the total magnitude is given here.} }
\label{ch3:photometry}
\vspace{0.5cm}
\begin{tabular}{c|c|c|c|c|c|c|c}
\hline
\hline
\small{ID} & \small{M$_{narrow}$} & \small{M$_B$} & \small{M$_V$} & \small{M$_i$} & \small{M$_{z'}$} & \small{EW$_{\mathrm{obs,Ly}\alpha}$ ({\AA})} & \small{SFR$_{\mathrm{Ly}{\alpha}}$ } \\
 & & & & & & & \small{(M$_{\odot}$/yr)} \\
\hline
\small{\bf{1}} &\small{$\bf{24.35 \pm 0.07}$} & \small{$\bf{27.87 \pm 0.17}$} &\small{ $\bf{27.08 \pm 0.07}$} & \small{$\bf{27.33 \pm 0.28}$} & \small{$\bf{27.07 \pm 0.22}$} & \small{$\bf{1006 \pm 71}$} & \small{$\bf{3.94 \pm 0.24}$}  \\
\small{2} & \small{$25.36 \pm 0.17$} & \small{$28.02 \pm 0.35$} &\small{ $27.29 \pm 0.10$} & \small{$27.39 \pm 0.26$} & \small{$ > 27.52$      } &\small{ $434 \pm 82$} & \small{$1.55 \pm 0.22$ }  \\
\small{3} & \small{$24.97 \pm 0.06$} & \small{$ > 28.08      $} & \small{$28.05 \pm 0.38$} & \small{$ > 27.67      $} & \small{$ > 27.24$      } &\small{ $912 \pm 52$} & \small{$2.22 \pm 0.11$ } \\
\small{\bf{4}} & \small{$\bf{24.19 \pm 0.02}$} & \small{$\bf{27.63 \pm 0.02}$} & \small{$\bf{26.91 \pm 0.11}$} & \small{$\bf{27.25 \pm 0.30}$} & \small{$\bf{27.07 \pm 0.32}$} &\small{ $\bf{956 \pm 15}$} & \small{$\bf{4.55 \pm 0.07}$}  \\
\small{5} & \small{$25.29 \pm 0.03$} & \small{$ > 28.32      $} & \small{$28.45 \pm 0.40$} & \small{$ > 28.68      $} & \small{$ > 26.89      $} & \small{$881 \pm 27$ }& \small{$1.65 \pm 0.04$} \\
\small{6} & \small{$24.03 \pm 0.04$} & \small{$ > 28.23    $  } & \small{$27.12 \pm 0.15$} & \small{$ > 27.52      $} & \small{$ > 27.31$      } & \small{$1639 \pm 64$} & \small{$5.28 \pm 0.19$} \\
\small{7} & \small{$25.45 \pm 0.21$} & \small{$ > 28.05    $  } & \small{$27.47 \pm 0.09$} & \small{$27.34 \pm 0.28$} & \small{$ > 27.26$      } & \small{$436 \pm 103$} & \small{$1.42 \pm 0.24$ }\\
\small{8} & \small{$25.23 \pm 0.14$} & \small{$27.89 \pm 0.13$} & \small{$27.62 \pm 0.21$} & \small{$27.39 \pm 0.37$} & \small{$ > 27.37$      } & \small{$532 \pm 81$} & \small{$1.74 \pm 0.21$} \\
\small{9$^{\star}$} & \small{$24.75 \pm 0.08$} & \small{$26.46 \pm 0.09$} & \small{$25.98 \pm 0.06$} & \small{$25.64 \pm 0.13$} & \small{$26.03 \pm 0.15$} & \small{$169 \pm 16$} & \small{$2.73 \pm 0.18$} \\
\small{10$^{\star}$} & \small{$25.02 \pm 0.12$} & \small{$26.84 \pm 0.09$} & \small{$26.21 \pm 0.09$} & \small{$26.24 \pm 0.08$} & \small{$26.50 \pm 0.20$} & \small{$179 \pm 27$} & \small{$2.13 \pm 0.22$} \\
\small{11} & \small{$25.28 \pm 0.16$} & \small{$ > 28.27     $ } & \small{$ > 28.36      $} & \small{$ > 27.86      $} & \small{$ > 27.60$      } & \small{$ > 834    $} & \small{$1.66 \pm 0.22$} \\
\small{12$^{\star}$} & \small{$25.97 \pm 0.24$} & \small{$ > 27.76     $} & \small{$26.11 \pm 0.09$} & \small{$27.43 \pm 0.29$} & \small{$ > 27.46       $} & \small{$76 \pm 34$} & \small{$0.89 \pm 0.18$} \\
\small{\bf{13}} & \small{$\bf{25.63 \pm 0.23}$} & \small{$\bf{27.24 \pm 0.15}$} & \small{$\bf{27.13 \pm 0.12}$} & \small{$\bf{27.24 \pm 0.28}$} & \small{$\bf{ > 27.08}$} & \small{$\bf{178 \pm 57}$} & \small{$\bf{1.21 \pm 0.23}$} \\
\small{14$^{\star}$} & \small{$23.97 \pm 0.05$} & \small{$25.64 \pm 0.05$} & \small{$24.58 \pm 0.03$} & \small{$24.39 \pm 0.04$} & \small{$24.64 \pm 0.06$} & \small{$110 \pm 8$} & \small{$5.58 \pm 0.26$} \\
\small{15$^{\star}$} & \small{$24.22 \pm 0.05$} & \small{$26.59 \pm 0.09$} & \small{$25.73 \pm 0.04$} & \small{$25.25 \pm 0.04$} & \small{$25.22 \pm 0.06$} & \small{$296 \pm 18$} & \small{$4.42 \pm 0.21$} \\
\small{16} & \small{$24.92 \pm 0.11$} & \small{$ > 27.72      $} & \small{$26.87 \pm 0.12$} & \small{$26.12 \pm 0.08$} & \small{$25.46 \pm 0.08$} & \small{$473 \pm 59$} & \small{$2.33 \pm 0.23$} \\
\small{17$^{\star}$} & \small{$25.09 \pm 0.08$} & \small{$26.94 \pm 0.16$} & \small{$25.86 \pm 0.05$} & \small{$25.62 \pm 0.08$} & \small{$25.33 \pm 0.08$} & \small{$138 \pm 15$} & \small{$1.98 \pm 0.14$} \\
\small{18} & \small{$25.37 \pm 0.16$} & \small{$27.39 \pm 0.12$} & \small{$26.11 \pm 0.04$} & \small{$25.73 \pm 0.04$} & \small{$25.84 \pm 0.10$} & \small{$149 \pm 32$} & \small{$1.54 \pm 0.21$} \\
\small{19$^{\star}$} & \small{$25.43 \pm 0.07$} & \small{$27.02 \pm 0.06$} & \small{$26.01 \pm 0.03$} & \small{$25.80 \pm 0.08$} & \small{$26.16 \pm 0.07$} & \small{$102 \pm 11$} & \small{$1.45 \pm 0.10$} \\
\small{20} & \small{$25.99 \pm 0.15$} & \small{$27.08 \pm 0.14$} & \small{$27.04 \pm 0.09$} & \small{$26.23 \pm 0.11$} & \small{$26.02 \pm 0.13$} & \small{$90 \pm  23$} & \small{$0.87 \pm 0.11$} \\
\small{21} & \small{$25.48 \pm 0.06$} & \small{$ > 27.88      $} & \small{$ > 27.86    $ }  & \small{$ > 27.44      $} & \small{$ > 27.35      $} & \small{$ > 443    $} & \small{$1.38 \pm 0.07$} \\
\small{22$^{\star}$} & \small{$25.05 \pm 0.15$} & \small{$26.80 \pm 0.12$} & \small{$25.65 \pm 0.04$}  & \small{$25.65 \pm 0.08$} & \small{$25.60 \pm 0.08$} & \small{$115 \pm 26$} & \small{$2.07 \pm 0.27$} \\
\small{23} & \small{$25.30 \pm 0.15$} & \small{$ > 27.78      $} & \small{$25.99 \pm 0.05$} & \small{$25.47 \pm 0.04$} & \small{$25.43 \pm 0.07$} & \small{$174 \pm 35$} & \small{$1.64 \pm 0.21$} \\
\small{24} & \small{$25.46 \pm 0.12$} & \small{$27.12 \pm 0.10$} & \small{$26.51 \pm 0.06$} & \small{$26.02 \pm 0.07$} & \small{$26.04 \pm 0.12$} & \small{$146 \pm 25$} & \small{$1.41 \pm 0.15$} \\
\hline
\end{tabular}
\end{table}
\end{landscape}

\section{Spectroscopy}\label{ch3:specsec}
\subsection{Observations and reductions}
Follow-up Multi-Object Spectroscopy (MOS) was obtained in service mode with
FORS1/VLT UT2 over the time period December 2004 -- February 2005. The total
observing time of 5.5 hours was granted to confirm the redshift of the
Ly$\alpha$-blob found in this field, as published in Nilsson et al.~(2006a).
In addition, we had the opportunity to add three of our compact emitters on 
the mask.
The mask preparation was done using the \emph{FORS Instrumental Mask Simulator}.
Stars were placed on the remaining slits for calibration purposes. The
MOS slitlets had a width of $1.4''$ and the combination of grism 600V and order
sorting filter GG435 was used. The grism covers the wavelength range
4650 {\AA} to 7100 {\AA} with a resolving power of approximately 700.
The seeing varied between $0.77''$ -- $1.2''$.

The bias subtraction, flat-fielding and wavelength calibration was performed 
using the FORS1 pipeline. The individual, 2--dimensional, reduced science
spectra were combined using a $\sigma$--clipping for rejection of cosmic ray
hits. The sky was subsequently subtracted with MIDAS by averaging the values
of all pixels on either side of the spectrum and expanding this value to the
size of the frame. One--dimensional spectra were extracted by summing the
column values over the spectra. The spectra were then flux calibrated with
three stars that were on our slits, by measuring the stellar fluxes in the
narrow-band image in $2''$ diameter apertures and comparing to the integrated
flux in the narrow-band from the 1--d spectra. Because LEGOs are often extended
in Ly$\alpha$ (M{\o}ller \& Warren 1998; Fynbo et al. 2001) they are likely
to have higher Ly$\alpha$ fluxes than those listed in Table~\ref{ch3:spectable}.

\subsection{Results of first spectroscopic follow-up}
\begin{figure}[!t]
\begin{center}
\epsfig{file=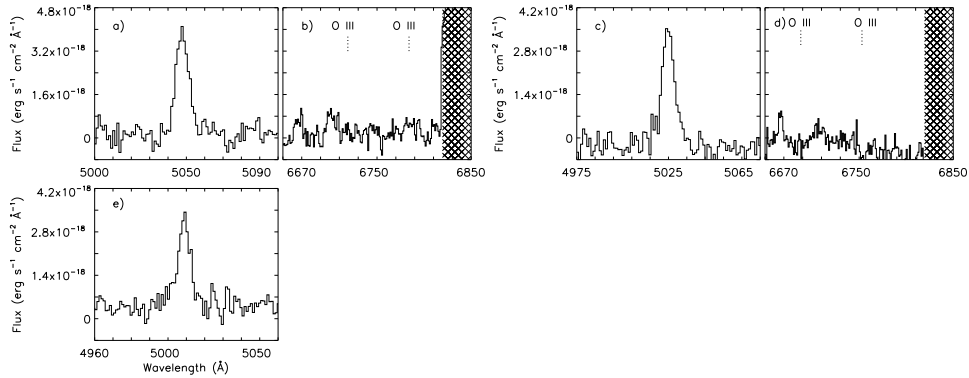,width=13.0cm}
\caption[Spectra of confirmed GOODS-S LEGOs]{Spectra of confirmed LEGOs in GOODS-S. \emph{a)} LEGO\_GOODS-S\#1
emission line, \emph{b)} LEGO\_GOODS-S\#1 expected position of [OIII] lines if
the detected emission line is [OII], \emph{c)} LEGO\_GOODS-S\#4
emission line, \emph{d)} LEGO\_GOODS-S\#4 expected position of [OIII] lines if
the detected emission line is [OII],\emph{e)} LEGO\_GOODS-S\#13
emission line. Our spectrum did not cover the position of [OIII] if the
emission line is [OII] for LEGO\_GOODS-S\#13. Hatched areas
mark the positions of bright sky lines.}
\label{ch3:specandim}
\end{center}
\end{figure}

The spectra of our three confirmed high-redshift Ly$\alpha$-emitters can be
found in Fig.~\ref{ch3:specandim}. Neither candidate show any other emission lines
in their spectra. To consider if the emission line could be [OII], we study
line ratios compared to other lines that should be observed in such a case.
In Fynbo et al. (2001), the expected line
ratios of H$\beta$, [OII], [OIII] and NeII for [OII]-emitters were presented.
For the
spectroscopic sample, the 3$\sigma$ upper limits of the ratio
log(F$_{[OIII]}$/F$_{[OII]}$) are given in Table~\ref{ch3:spectable} if the
emission lines is [OII].
The limit will be the same for all other lines, as none are detected.
As in Fynbo et al. (2001), we conclude that these are highly unlikely values for
[OII]-emitters, and that the detected emission line is redshifted Ly$\alpha$.
Details from the spectroscopic follow-up is presented in Table~\ref{ch3:spectable}.

\begin{table}[!t]
\caption[Data on spectroscopically confirmed LEGOs]{Names, Ly$\alpha$-fluxes, redshifts and line ratios for
spectroscopically confirmed LEGOs. The line ratios refer to the upper limit
to the [OIII] line, if the emission line is [OII].}
\vspace{0.5cm}
\label{ch3:spectable}
\centering 
\begin{tabular}{cccccc}
\hline\hline
LEGO\_GOODS-S\# & F$_{\mathrm{Ly}\alpha}$ & $z$ & log(F$_{[OIII]}$/F$_{[OII]}$)\\
& (erg s$^{-1}$ cm$^{-2}$) & &\\
\hline
1 & $3.31 \times 10^{-17}$ & 3.151 & $< -0.62$ \\
4 & $2.79 \times 10^{-17}$ & 3.132 & $< -0.54$ \\
13 & $2.84 \times 10^{-17}$ & 3.118 & $< -0.55$ \\
\hline
\end{tabular}
\end{table}

In the following sections we analyse the entire sample of confirmed LEGOs and
LEGO candidates together. We expect the contamination of low redshift emitters
to be small. Our previous surveys have had spectroscopic success rates of
75~-~90~\% (Fynbo et al. 2001; Fynbo et al. 2003). Hence, we expect that more
than 18 of our 24 candidates are true Ly$\alpha$-emitters.

\section{Basic characteristics of LEGOs}
\subsection{SFR, surface density and sizes}\label{ch3:DiscLEGO}
For our final sample of LEGO candidates, we calculate the star formation rate
(SFR) as derived from Kennicutt (1983) by:
\begin{equation}
\mathrm{SFR} = \frac{L_{\mathrm{H}\alpha}}{1.12 \times 10^{42}} \mathrm{M}_{\odot} \mathrm{yr}^{-1}
\end{equation}
where the H$\alpha$ luminosity, $L_{\mathrm{H}\alpha}$, is obtained with the
conversion between Ly$\alpha$ and H$\alpha$ luminosities of Brocklehurst (1971)
of L(Ly$\alpha$) = $8.7 \times$~L(H$\alpha$).
The SFR values of the LEGO candidates can be found in Table~\ref{ch3:photometry}.
The mean SFR, as derived from the Ly$\alpha$-emission for all candidates is
$1.8$~M$_{\odot}/\mrm{yr}$. The total SFR is
$43$~M$_{\odot}/\mrm{yr}$, yielding a star formation rate
density $\rho_{\mrm{SFR}}$ of $0.013$~M$_{\odot}/\mrm{yr}/\mrm{Mpc}^3$.
This value is in very good agreement with other results for high redshift
galaxies at this redshift of
e.g. Madau et al. (1996; 0.016~M$_{\odot}/\mrm{yr}/\mrm{Mpc}^3$), Steidel et
al. (1999; 0.05~M$_{\odot}/\mrm{yr}/\mrm{Mpc}^3$) and Cowie \& Hu
(1998; 0.01~M$_{\odot}/\mrm{yr}/\mrm{Mpc}^3$). The results from Steidel et al.
(1999) has been obtained from integrating the extrapolated luminosity
function down to a luminosity of $0.1$~$L_{\star}$, and the results are also
corrected for dust by multiplying by a factor of 4.7. Hence, their results
uncorrected for dust is
0.011~M$_{\odot}/\mrm{yr}/\mrm{Mpc}^3$. There is very good agreement
between the dust uncorrected measurements of the SFR density at $z \sim 3$.

We find a surface density of LEGOs at redshift $z = 3.15, \delta z = 0.05$ in
the GOODS-S field of 0.53 objects
per arcmin$^2$. We can compare this with all $V$ band sources in the
GOODS-S field by extracting all sources with $V$ band magnitudes between
25 -- 28 in the available online catalog. We find $23885$ such sources in the
entire GOODS-S area, covering approximately $160$~arcmin$^2$, corresponding to
a surface density of $149$~arcmin$^{-2}$. If we assume a homogeneous density of
LEGOs between redshift $3.0 - 3.5$, then the surface density of LEGOs, scaled
with our candidate sample, will be $\approx 5.3$~arcmin$^{-2}$.
Thus, approximately 4~\% of all $V$-detected sources with a magnitude of
$V = 25 - 28$ were selected as Ly$\alpha$-emitters in the redshift
range $z = 3.0 - 3.5$.

\begin{figure}[!t]
\begin{center}
\epsfig{file=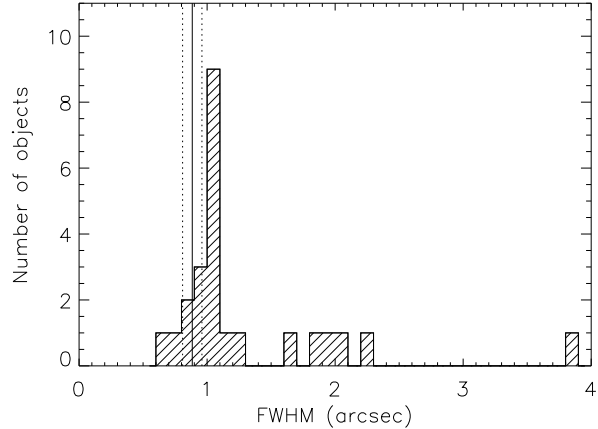,width=9.0cm,clip=}
\caption[Sizes of GOODS-S LEGOs]{Histogram plot over the size of our candidates in the narrow-band
image. The bin size is 0.1 arcsec, the solid line represents the PSF of the
image and the dotted lines the 1$\sigma$ error on the PSF. The PSF was determined
from 28 objects with SExtractor keyword CLASS\_STAR greater than 0.9 and fluxes
in the range of our LEGO candidates. The object in the
highest FWHM bin is the GOODS-S blob (Nilsson et al., 2006a).}
\label{ch3:extplot}
\end{center}
\end{figure}

We measured the sizes of our candidates in the narrow-band images using the
FLUX\_RADIUS option in
SExtractor. This gives the half width half maximum of each source. Object
LEGO\_GOODS-S\#~11 was excluded because it was blended with another, unrelated
object. The histogram of the sizes, compared to a point source, is presented
in Fig.~\ref{ch3:extplot}.
Most objects appear to be barely resolved. However, there is a
tail of larger objects extending towards the GOODS-S blob.
In a future paper, we will present a complete morphological study of the
candidate sample.

\subsection{Filamentary structure}\label{sec:filament}
Previous studies of Ly$\alpha$ emitters have reported the identification
of filamentary structures (e.g. M{\o}ller \& Fynbo 2001; Hayashino et al. 2004;
Matsuda et al. 2005). The volume we survey here
is approximately 3 times as long along the line-of-sight as it is wide which
means that several 
filaments could be crossing the volume at different redshifts and with
different position angles. If that was the case, they would likely
blend together to wash out individual structures. Bearing this in
mind we can still ask the question of whether the objects on our candidate
list are randomly distributed across the field, or if they appear to
be systematically aligned.

\begin{figure}[!t] 
\begin{center}
\epsfig{file=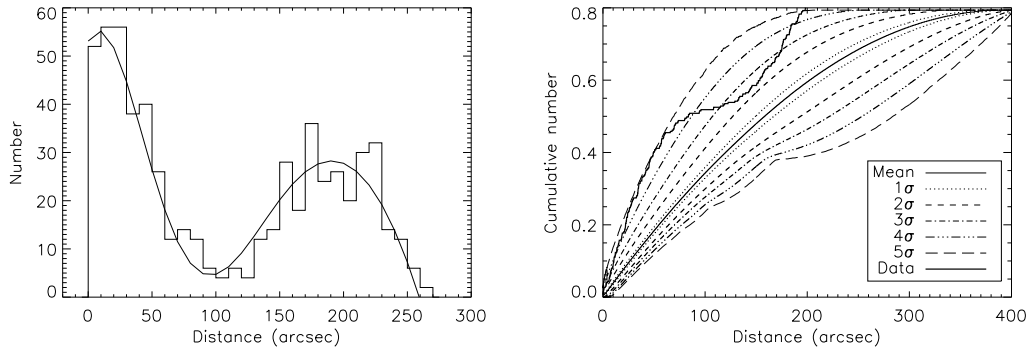,width=14.0cm,clip=}
\caption[Filamentary structure of GOODS-S LEGOs]{\emph{Left} Histogram distribution of distances between candidate
objects in the x-direction, after re-alignment by $2.1^{\circ}$. Histogram is
binned over 50 pixels. Solid curve is a double gaussian fit to the data. \emph{Right} Kolmogorov-Smirnov test of the distribution. Lines mark the simulated 
mean, and the 1 -- 5$\sigma$ contours. Thick line represents our data.}
\label{ch3:fil}
\end{center} 
\end{figure}

Inspecting Fig.~\ref{ch3:field} we note that the candidates do in fact appear to be
aligned in two filamentary structures approximately
along the y-axis of the CCD (oriented N-S). To investigate whether this is a
significant
effect, we have calculated the distances between all objects and plotted a
histogram of the projected distances between all candidates in Fig.~\ref{ch3:fil}.
The histogram shows a definite division in two components, one
describing the typical width of a filament, and the other the projected
distance between the filaments.
To optimise the search, we calculated the angle of rotation that minimised the
variance of the x-coordinates around a mean. For the two filaments, these angles
are $1.75^{\circ}$ and $2.52^{\circ}$ respectively. For the analysis, we rotate
the filament by the average, $2.13^{\circ}$. We then generated $10^7$
uniform random sets of coordinates, with the same number of objects,
and repeated the same analysis of calculating distances between pairs.
To establish how reliable our observed distribution is, we perform a
Kolmogorov-Smirnov test (e.g. Peacock 1983) on the
simulated distributions. The test showed that the likelihood of the alignment
being random is less than $2.5 \times 10^{-4}$, hence a near 4 sigma detection,
see Fig.~\ref{ch3:fil}. We fitted a double gaussian function to the histogram plot.
This
gives the typical width of the filaments as the FWHM of the first peak, and the
distance between them as the mean distance to the second peak. We find that the
typical width is $\approx 250$~pixels, corresponding to 370 kpc at redshift
3.15, in good agreement with the findings of M{\o}ller \& Fynbo (2001) who find
a spectroscopically confirmed filament with 400 kpc radius from a set of
Ly$\alpha$-emitters at $z = 3.04$. The distance between
the first and second peak is $\approx 950$~pixels, corresponding to 1.4 Mpc at
this redshift. To further verify this filamentary structure would require more
spectroscopic data, that would also enable a 3D-plot of the filaments in space.  
\section{SED fitting}\label{sec:sedfitting}
The imaging available in the \emph{GOODS-S} field is extensive. The
data, in 14 publicly available broad-bands, used here is presented in
Table~\ref{ch3:mwtab}.
\begin{landscape}
\begin{table}[!t]
\begin{center}
\caption[Data in GOODS-S]{\small{Deep, multi-wavelength data available in the GOODS-S field.
The fifth column refers to the 3$\sigma$ detection limit in
the sky in a $2 \times$~FWHM diameter aperture. The last column gives the
3$\sigma$ detection limit as measured in a $2''$ radius aperture.} }
\vspace{0.5cm}
\begin{tabular}{@{}lcccccccc}
\hline\hline
\small{Filter/Channel} & \small{$\lambda_c$} & \small{FWHM} & \small{Aperture} & \small{Aperture} & \small{3$\sigma$ limit} & \small{3$\sigma$ limit ($2''$ aperture)}\\
& & & \small{radius} & \small{correction} & \small{($2 \times$~FWHM aperture)} & \\
& & & \small{(arcsec)} & &\small{ ($\mathrm{erg} \cdot \mathrm{cm}^{-2} \cdot \mathrm{s}^{-1} \cdot \mathrm{Hz}^{-1}$)} &\small{ ($\mathrm{erg} \cdot \mathrm{cm}^{-2} \cdot \mathrm{s}^{-1} \cdot \mathrm{Hz}^{-1}$)}\\
\hline
\small{X-rays (\emph{Chandra})     }  & \small{ 4.15 keV   } &  \small{3.85 keV   } & \small{2.25} & \small{ ---} & \small{$9.90 \cdot 10^{-34}$} & \small{$9.90 \cdot 10^{-34}$}   \\
\small{$U$ (\emph{ESO 2.2-m})      }  & \small{ 3630 \AA   } &  \small{760 \AA    } & \small{3.00} & \small{ ---} & \small{$1.10 \cdot 10^{-30}$} & \small{$8.62 \cdot 10^{-31}$}    \\
\small{$B$ (F435W, \emph{HST})     }  & \small{ 4297 \AA   } &  \small{1038 \AA   } & \small{0.12} & \small{1.20} & \small{$7.09 \cdot 10^{-32}$} & \small{$9.25 \cdot 10^{-30}$}  \\
\small{$V$ (F606W, \emph{HST})     }  & \small{ 5907 \AA   } &  \small{2342 \AA   } & \small{0.12} & \small{1.18} & \small{$4.02 \cdot 10^{-32}$} & \small{$4.66 \cdot 10^{-30}$}   \\
\small{\emph{i} (F814W, \emph{HST})}  & \small{ 7764 \AA   } &  \small{1528 \AA   } & \small{0.12} & \small{1.25} & \small{$1.25 \cdot 10^{-31}$} & \small{$1.50 \cdot 10^{-29}$}   \\
\small{$z'$ (F850LP, \emph{HST})   }  & \small{ 9445 \AA   } &  \small{1230 \AA   } & \small{0.12} & \small{1.34} & \small{$1.88 \cdot 10^{-31}$} & \small{$3.00 \cdot 10^{-29}$}  \\
\small{$J$ (\emph{VLT})       }       & \small{ 1.25 $\mu$m} &  \small{0.6 $\mu$m } & \small{0.60} & \small{1.22} & \small{$1.78 \cdot 10^{-30}$} & \small{$5.31 \cdot 10^{-30}$} \\
\small{$H$ (\emph{VLT})       }       & \small{ 1.65 $\mu$m} &  \small{0.6 $\mu$m } & \small{0.60} & \small{1.21} & \small{$4.11 \cdot 10^{-30}$} & \small{$1.86 \cdot 10^{-29}$}  \\
\small{$Ks$ (\emph{VLT})      }       & \small{ 2.16 $\mu$m} &  \small{0.6 $\mu$m } & \small{0.60} & \small{1.37} & \small{$4.06 \cdot 10^{-30}$} & \small{$1.56 \cdot 10^{-29}$} \\
\small{$Ch1$ (\emph{Spitzer}) }       & \small{ 3.58 $\mu$m} &  \small{0.75 $\mu$m} & \small{1.30} & \small{2.41} & \small{$1.14 \cdot 10^{-31}$} & \small{$2.36 \cdot 10^{-30}$}  \\
\small{$Ch2$ (\emph{Spitzer}) }       & \small{ 4.50 $\mu$m} &  \small{1.02 $\mu$m} & \small{1.80} & \small{1.98} & \small{$2.07 \cdot 10^{-32}$} & \small{$2.07 \cdot 10^{-30}$} \\
\small{$Ch3$ (\emph{Spitzer}) }       & \small{ 5.80 $\mu$m} &  \small{1.43 $\mu$m} & \small{1.80} & \small{1.60} & \small{$7.50 \cdot 10^{-29}$} & \small{$7.50 \cdot 10^{-30}$} \\
\small{$Ch4$ (\emph{Spitzer}) }       & \small{ 8.00 $\mu$m} &  \small{2.91 $\mu$m} & \small{2.10} & \small{1.85} & \small{$6.87 \cdot 10^{-30}$} & \small{$6.87 \cdot 10^{-30}$} \\
\small{$MIPS$ (\emph{Spitzer})}       & \small{ 24.0 $\mu$m} &  \small{4.70 $\mu$m} & \small{6.00} & \small{ ---} & \small{$1.23 \cdot 10^{-28}$} & \small{$2.12 \cdot 10^{-29}$}  \\
\hline
\label{ch3:mwtab}
\end{tabular}
\end{center}
\end{table}
\end{landscape}
With this data-set, we wish to perform an SED fitting, in order to constrain
properties such as stellar mass M$_\ast$, dust content A$_V$,
metallicity and age of the LEGOs. Only one of our
candidates (LEGO\_GOODS-S\#16) is detected in bands other than the HST bands. This  
object is especially interesting as its SED is extremely red. It is excluded
from the SED fitting, and is discussed in Section~\ref{ch3:number19}. For the
rest of the sample, the LEGOs are only detected in the HST bands and hence
we choose to stack the entire sample of 23 candidates. We can then draw
conclusions on the general properties of this type of object. After stacking,
we get a faint detection in the $K_s$ band. The stacked magnitudes are
given in Table~\ref{ch3:tabsed}. The lack of X-ray, MIPS 24$\mu$m and radio
detections (no counterparts to any of our candidates to a 3$\sigma$ limit
of 24~$\mu$Jy, Kellermann et al. in preparation)
indicates that the AGN fraction among these objects is low.

\begin{table}[!t]
\begin{center}
\caption[Stacked magnitudes of LEGOs]{Stacked magnitudes for the GOODS-S Ly$\alpha$-emitters. Errors
in the HST magnitudes were set to a conservative value of 0.08. Upper limits
are 3$\sigma$.}
\vspace{0.5cm}
\begin{tabular}{@{}lccc}
\hline
\hline
Band & Centr. Wavelength (\AA) & Obs. Magnitude & Mag. Error \\\hline
U & 3710 & $> 25.95$ & --- \\
B & 4297 & 27.57 & 0.08\\
N & 5055 & 24.96 & 0.08\\
V & 5907 & 26.74 & 0.08\\
\emph{i} & 7764 & 26.52 & 0.08\\
\emph{z} & 9445 & 26.56 & 0.08\\
J & 12500 & $> 26.28$ & --- \\
H & 16500 & $> 25.55$ & --- \\
Ks & 21500 & 25.26 & 0.29\\
Ch1 & 35800 & $> 23.06$ & --- \\
Ch2 & 45200 & $> 23.62$ & --- \\
Ch3 & 57200 & $> 24.37$ & --- \\
Ch4 & 79000 & $> 23.84$ & --- \\
\hline
\end{tabular}
\end{center}
\label{ch3:tabsed}
\end{table}

\subsection{Fitting method}
We used the GALAXEV code (Bruzual \& Charlot, 2003) to simulate composite
stellar populations, in order to fit the stacked SED of the LEGOs. The fitting
was performed according to a Monte Carlo Markov Chain method (see e.g. Gilks et
al. 1995 for an introduction).
In outline, the method works as follows; an initial set of parameter values is
chosen according to a uniform, random and logarithmic distribution within the allowed
parameter space.
\begin{table}[!t]
\begin{center}
\caption[Parameter space sampled during SED fitting]{Parameter space sampled during the SED fitting. Metallicity is allowed
to have three different values ($Z/Z_{\odot} = 0.005$,~$0.2$~or~$1.0$). The
dust components are the two components of the Charlot \& Fall (2000) dust
model used by GALAXEV.
We fit the SED with a constant star forming model, where the star formation
rate in solar masses per year is given by ``SF-rate''.
}
\vspace{0.5cm}
\begin{tabular}{@{}lcccccccccc}
\hline
\hline
Parameter & Min. value & Max. value \\\hline
Metallicity ($Z/Z_{\odot}$) & 0.005 & 1.0 \\
Dust-$\tau$ & 0 & 4\\
Dust-$\mu$ & 0 & 1\\
SF-rate & 0.01 & 100\\
Age (Gyrs) & 0.001 & 1.5\\
\hline
\end{tabular}
\label{ch3:tabsedpars}
\end{center}
\end{table}
A summary of the parameter space is given in Table\,\ref{ch3:tabsedpars}. Given
the set of parameters, a corresponding $\chi^2$ value is calculated by
running the GALAXEV code, creating a
high-resolution spectrum with 6900 wavelength points from 91~{\AA}~to
160~$\mu$m. To obtain the magnitudes in each band,
we apply the transmission curves for the filters of
the various observed wavebands; $U$, $B$, $V$, $i$, $z'$, $J$, $H$, $K_s$
and the four Spitzer bands, $Ch1 - Ch4$,
In this analysis, we exclude the Spitzer $MIPS$ band as it is very
difficult to stack images in this band because of the source confusion. At this
redshift, the $MIPS$ band is also contaminated by a PAH emission feature.
We then
normalise the output spectrum so that the magnitude in the model $z'$-band equals
that of the observed $z'$-band. The $\chi^2$ is then calculated
by comparing the magnitudes in all the other bands.
We incorporate points with upper limits in the following way;
if the predicted flux lies below the upper limit, no value is added to the
total $\chi^2$, if the predicted flux lies above the
observed one, a $\chi^2$ is added, assuming the error on the upper limit is 0.1 in 
magnitude. Hence, models with flux above the upper limit may be acceptable if the
flux in this band is only slightly above the limit. In most cases though, the
model will be rejected due to high $\chi^2$.

Once the $\chi^2$ has been calculated for a particular model, a new
random set of parameters is chosen, by adding a ``step vector''.
The step size in each parameter is chosen
randomly in a logarithmic interval between 1\% and 100\% of the total size of
the parameter space. This step vector is equally likely to be positive or negative.
The choice of logarithmic step sizes is a natural choice if no assumptions
about the scale
of change that affect the solution are to be made and ensures a fast
convergence. When a step has been calculated and the new parameters have been calculated, the new model
is accepted into the chain with a probability
proportional to the exponential difference between the old and new $\chi^2$.
If the new
model is accepted, it is added as the next step in the chain, i.e. the parameters
of that particular model are printed in the output file. If the model is not
accepted,
a copy of the old parameters is added as the next step. This procedure is
repeated until a chain with 30000 elements has been created. The
independence of individual steps in the walk ensures
that the resulting chain is Markovian in character, i.e. that the resulting
chain after many iterations is a representation of the full probability
distribution function in the chosen parameter space. The output
file can then be used to study the distribution in each parameter, and to
determine the mean and the confidence levels within each parameter. It can also
be used to study dependencies between parameters, such as e.g. in Fig.~\ref{ch3:contours}.

\subsection{Results from SED fitting}
\begin{figure}[ht]
\begin{center}
\epsfig{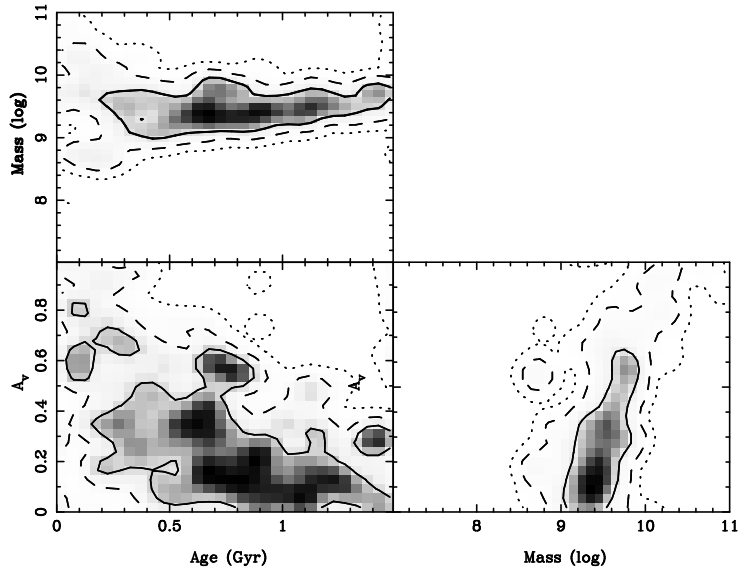}
\caption[Results of SED fitting]{Contour-plots of the mass, age
and dust parameters in our SED fitting. Contours indicate 1, 2 and 3$\sigma$
levels.  }
\label{ch3:contours}
\end{center}
\end{figure}

\begin{figure}[!t] 
\begin{center}
\epsfig{file=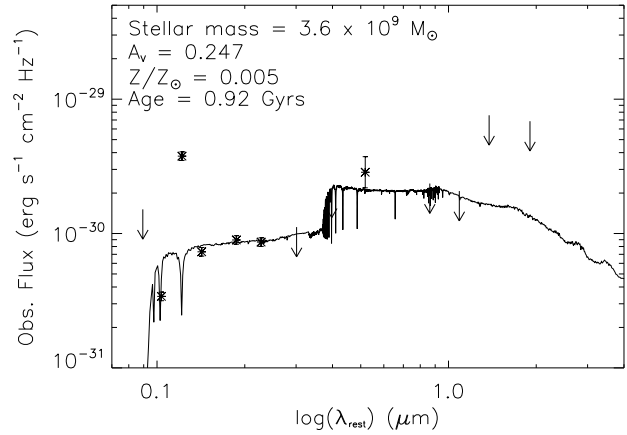,width=9.0cm,clip=}
\caption[Average spectrum of best fit SEDs]{The average spectrum of 10 models with $\chi^2\sim1$. The spectrum is
calculated by 1) creating the 10 spectra with the particular parameters of the
models with GALAXEV, 2) averaging the 10 spectra together, weighted with the $\chi^2$,
so that the total flux density is $F_{\nu}=\left[ \sum_i F^i_{\nu}\times (1 / \chi^2_i) \right]/\sum_i (1/\chi^2_i)$.
The parameters of the model spectrum are as indicated on the plot. They are
the weighted average parameters of the 10 models used, weighted in the same way as the
spectra. Data points from stacked SED.
Upper limits are represented with arrows. The point well off the SED is the
narrow-band magnitude. }
\label{ch3:plotsed}
\end{center}
\end{figure}

The parameters we wish to fit are stellar mass, dust content,
star formation rate, metallicity and age.
Redshift is set to
be the central Ly$\alpha$ redshift of the narrow-band filter, i.e. $z = 3.15$.
We use the Salpeter initial mass function (IMF) from 0.1 to 100~M$_{\odot}$
and the extinction law of Charlot \& Fall (2000). We also
incorporate the effect of the Ly$\alpha$ forest according to the model of
Madau (1995). We use constant star formation histories.
The metallicity was allowed
to be either $Z/Z_{\odot} = 0.005$,~$0.2$~or~$1.0$. Of the 30000 runs, 73\%
were
with the lowest metallicity, 21\% with the medium metallicity and 6\% had solar
metallicity. Hence the best fit models have a very low metallicity.
For the rest of the analysis, we choose to only look at the models which
have the lowest metallicity, as these
models seem to be preferred. For these models, the best fit parameters are
M$_\ast = 4.7^{+4.2}_{-3.2} \times 10^8$~M$_{\odot}$, 
age~$ = 0.85^{+0.13}_{-0.42}$~Gyrs,
A$_V = 0.26^{+0.11}_{-0.17}$, and star formation rate
SFR~$ = 0.66^{+0.51}_{-0.31}$~M$_{\odot}$~yr$^{-1}$, where the errors are $1 \sigma$. Contour-plots
of the three parameters mass, age and dust are shown in Fig.~\ref{ch3:contours}.
The degeneracies between the different parameters can be seen. In
Fig.~\ref{ch3:plotsed}, the weighted GALAXEV spectrum of a subset of 10 models
with $\chi^2 \sim 1$ is shown with the measured SED
overplotted. The $U$-band and Spitzer/IRAC data are too shallow to be useful
in constraining the data, and the narrow-band data-point is plotted only for
reference.

\subsection{Object LEGO\_GOODS-S\#16}\label{ch3:number19}
\begin{figure}[!t] 
\begin{center}
\epsfig{file=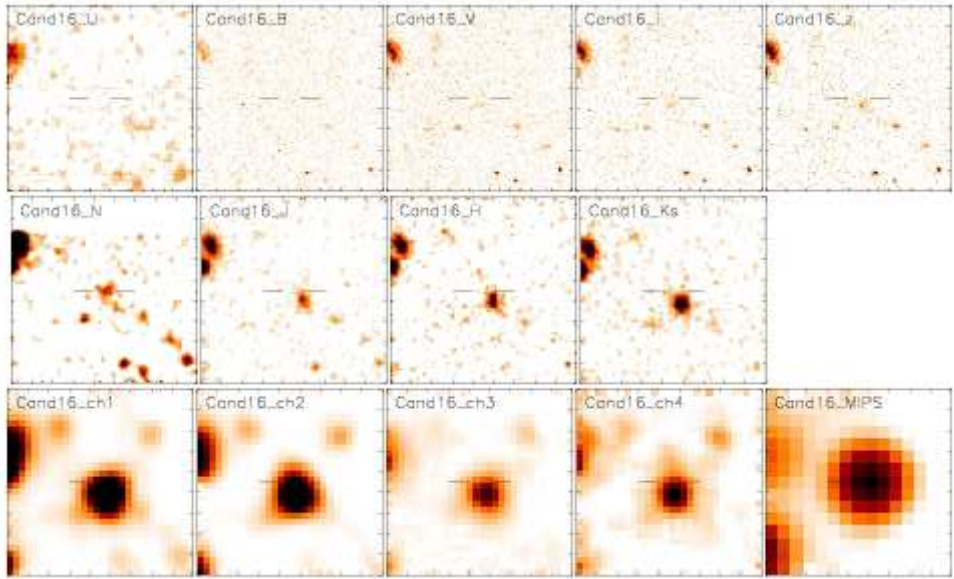,width=8.0cm,clip=,angle=90}
\caption[Thumb-nail images of LEGO\_GOODS-S\#16]{Thumb-nail images, $18''$ across, of LEGO\_GOODS-S\#16.}
\label{ch3:thumbs19}
\end{center}
\end{figure}

One candidate emission line object, LEGO\_GOODS-S\#16, was detected in all
available GOODS-S bands, except the $U$ and $B$ bands, and in X-rays. The fluxes of
the object can be found in Table~\ref{ch3:sed19}, and the thumb-nail images seen in Fig.~\ref{ch3:thumbs19}.
As there is significant excess emission in our narrow-band filter, we expect
the source to be
either an [OII]-emitter at $z = 0.36$ or a Ly$\alpha$-emitter at $z = 3.15$.
\begin{table}[!t]
\begin{center}
\caption[SED of object LEGO\_GOODS-S\#16]{SED of object LEGO\_GOODS-S\#16. Upper limit is 3$\sigma$. 
Continuum sources are offset from the narrow-band source by approximately
$0.8$~arcseconds, corresponding to 5.9 kpc at $z = 3.15$.}
\vspace{0.5cm}
\begin{tabular}{@{}lccc}
\hline
\hline
Band & Centr. Wavelength ($\mu$m) & Obs. flux ($\mu$Jy) & Flux Error ($\mu$Jy) \\\hline
B         &  0.430  &$> 0.03$  &   ---    \\NB        &  0.506  &    0.39  &   0.042 \\
V         &  0.591  &    0.06  &   0.008 \\
\emph{i}  &  0.776  &    0.13  &   0.010 \\
\emph{z}  &  0.945  &    0.24  &   0.018 \\
J         &  1.25   &    3.47  &   0.157 \\
H         &  1.65   &    7.40  &   0.420 \\
Ks        &  2.15   &   12.51  &   0.565 \\
Ch1       &  3.58   &   24.87  &   0.188 \\
Ch2       &  4.50   &   30.86  &   0.200 \\
Ch3       &  5.80   &   31.85  &   0.710 \\
Ch4       &  8.00   &   21.31  &   0.990 \\
MIPS      & 24.00   &  271.11  &   7.001 \\
\hline
\end{tabular}
\label{ch3:sed19}
\end{center}
\end{table}
This object was first fit with the same type of stellar SED as the other
sample, but
with the redshift set to be either $z = 0.36$ if the narrow-band emission
is [OII] or $z = 3.15$ if the emission is Ly$\alpha$. This fitting yielded
no good fit, with $\chi^2 \gtrsim 500$.

\begin{figure}[!t]
\begin{center}
\epsfig{file=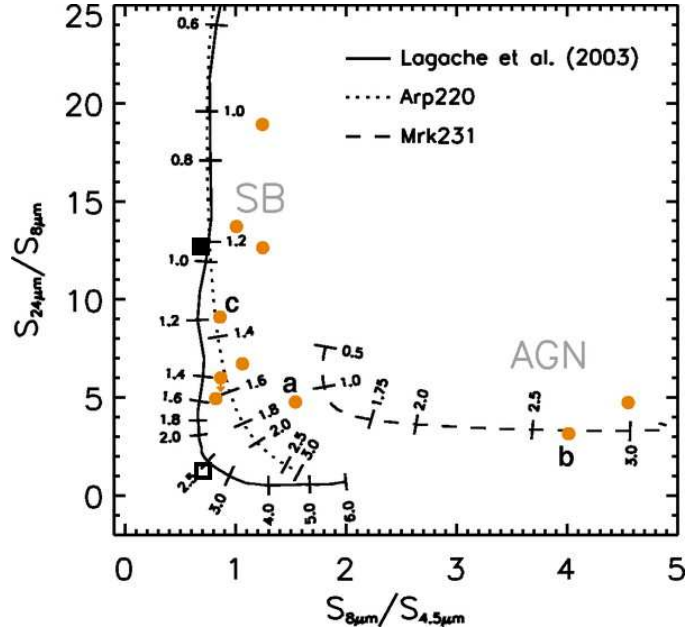,width=9.0cm,clip=}
\caption[Selection diagram of AGN/starburst galaxies using Spitzer data]{Diagram of Ivison et al. (2004). The x- and y-axes show the colours
in the Spitzer bands and the solid lines mark the locations of AGN and starburst
galaxies (SB) depending on redshift. The redshifts are marked along the lines.
The orange dots mark the location of a set of sub-mm galaxies presented in Ivison et al
(2004). The solid square marks the location of the colours of LEGO\_GOODS-S\#16,
indicating a lower redshift starburst galaxy. The open square marks the colours
of this galaxy  if the MIPS 24~$\mu$m flux is decreased by a factor of 10 (see text).
This point is indicative of a redshift $z \sim 3$ starburst galaxy. }
\label{ch3:ivison}
\end{center}
\end{figure}

Two types of objects could show MIR colours similar to those of this galaxy;
\emph{i)}
obscured AGNs (e.g. Lacy et al. 2004; Stern et al. 2005b) and \emph{ii)}
ULIRG/dusty starburst galaxies (e.g. Ivison et al. 2000; Klaas et al. 2001).
The infrared colours can be plotted in the diagnostic colour-colour
diagram of
Ivison et al.~(2004), see Fig.~\ref{ch3:ivison}.
In this diagram, Ivison et al.~(2004) plot
the colours of Arp 220, Mrk 231 and a theoretical starburst spectrum as
observed at different redshifts.
The comparison with the colours of LEGO\_GOODS-S\#16 shows the object
to be more likely a
low-redshift starburst galaxy. However, Ivison et al. (2004) do not take PAH
emission
into account. The most important PAH lines are at 3.3, 6.2, 7.7, 8.7, and 11.2 $\mu$m
(corresponding to 13.7, 25.7, 32.0, 36.1, 46.5 at $z = 3.15$). Especially the second
line falls on top of the MIPS 24~$\mu$m band. This would explain the extreme rise in
flux in this band. In Fig.~\ref{ch3:grasilmodels}, we see that this emission could easily
add a factor of ten to the flux in this band, hence the 24~$\mu$m/8~$\mu$m colour of
LEGO\_GOODS-S\#16 can be adjusted downwards by a factor of ten.
The extra line emission explains why a $z \sim 3$ star-burst galaxy would appear to be at
lower redshifts.
To understand if this is a low or high redshift object, we have also attempted to get a
photometric redshift estimate for this galaxy using the \emph{HyperZ} code
(Bolzonella et al. 2000), both with and without including the Spitzer data
points. The results were similar for both cases. When we do not include the
Spitzer data,
the best fit is redshift $z = 1.7$ with a $\chi^2$ of approximately 8. For the
redshift $z = 0.4$, the fit then has a $\chi^2 \approx 126$ and for the
Ly$\alpha$ redshift of $z = 3.15$, the $\chi^2 \approx  35$. When the Spitzer
points are included, the higher redshift is even more favoured. Hence, it seems
unlikely that this is a lower redshift source.

We wish to distinguish whether LEGO\_GOODS-S\#16 is an obscured AGN or a starburst 
galaxy. Several papers have presented infrared colours for obscured
(and unobscured) AGN (Johansson et al. 2004; Lacy et al. 2004; Stern et al.
2005b; Alonso-Herrero et al. 2006) and especially two papers publish selection
criteria for obscured AGNs (Lacy et al. 2004; Stern et al. 2005b).
The colours of LEGO\_GOODS-S\#16 are inconsistent with those selection
criteria and we therefore rule out an AGN nature of this
galaxy. This conclusion is further supported by the non-detection in
X-rays. In order to study if the SED of
the galaxy could be fitted
by a starburst spectrum, we tried to fit the SED with a GRASIL
(Silva et al. 1998) model of a starburst galaxy. GRASIL is a
spectral stellar synthesis code, which takes into account the dust
obscuration of starlight in both molecular clouds and the diffuse medium.
Hence it is perfectly suited for the investigation of starburst
galaxies. We could fit the photometric data of LEGO\_GOODS-S\#16 by a
relatively old burst ($\sim1$ Gyr) at $z = 3.15$ with a significant amount
of dust.
The results are shown in Fig.~\ref{ch3:grasilmodels}.
As can be seen in the Figure, these models reproduce the trends in
the observed SED relatively well.
This appears to be a redshift $z = 3.15$ dusty starburst galaxy,
with a region where the dust amount is smaller and Ly$\alpha$ emission can
escape, offset from the central parts of the galaxy. It would be of great
interest to get sub-mm imaging of this object in order to constrain the SED
better.

\section{Comparison to Lyman-Break Galaxies}
\begin{figure}[!ht]
\begin{center}
\epsfig{file=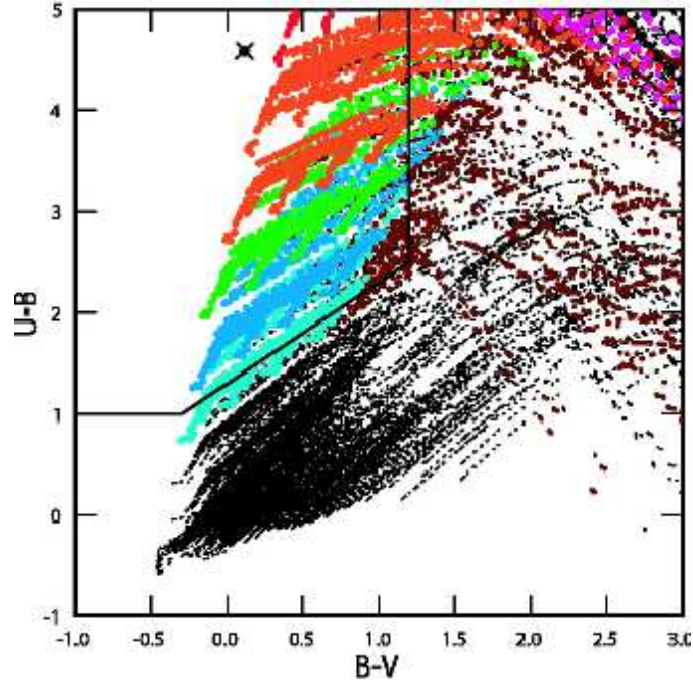,width=9.0cm,clip=}
\caption[Selection diagram for Lyman Break galaxies at redshift $z \sim 3$]{Colour-colour plot from Wadadekar et al. (2006) showing simulated
galaxy colours, see that paper for details. The solid line marks the area (upper
left corner) where redshift $z \approx 3$ LBG reside. The large star in the
upper left corner marks the colours of our best-fit synthetic model, well within
the selection boundaries for high-redshift LBGs.}
\label{ch3:lbgcol2}
\end{center}
\end{figure}

\begin{figure}[!pt]
\begin{center}
\epsfig{file=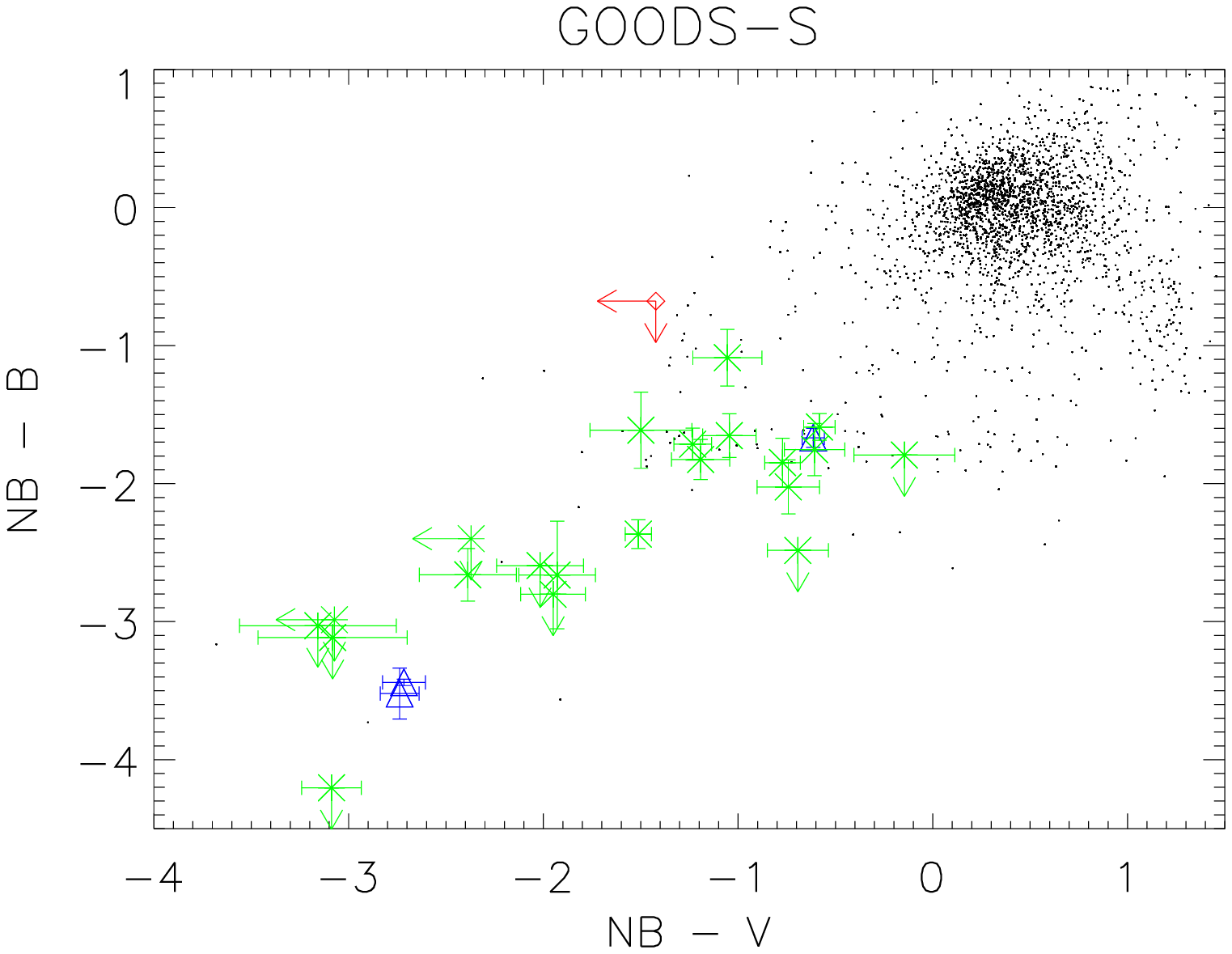,width=9.0cm}
\caption[Colour-colour plot]{
Colour-colour plot. Dots mark the whole sample of 2616 objects selected with
SExtractor. Points with errors mark the selected emission-line objects.
Stars (in magenta) mark
the candidate sample, the triangles (in blue) the spectroscopically confirmed
LEGOs and the open diamond (in red) the blob (Nilsson et al., 2006a). Dots
detected in the same region of the plot as
the selected sample, but that are not selected, consist of objects that were
discarded in the visual inspection.}
\label{ch3:colcol}
\end{center}
\end{figure}

\begin{figure}[!pb] 
\begin{center}
\epsfig{file=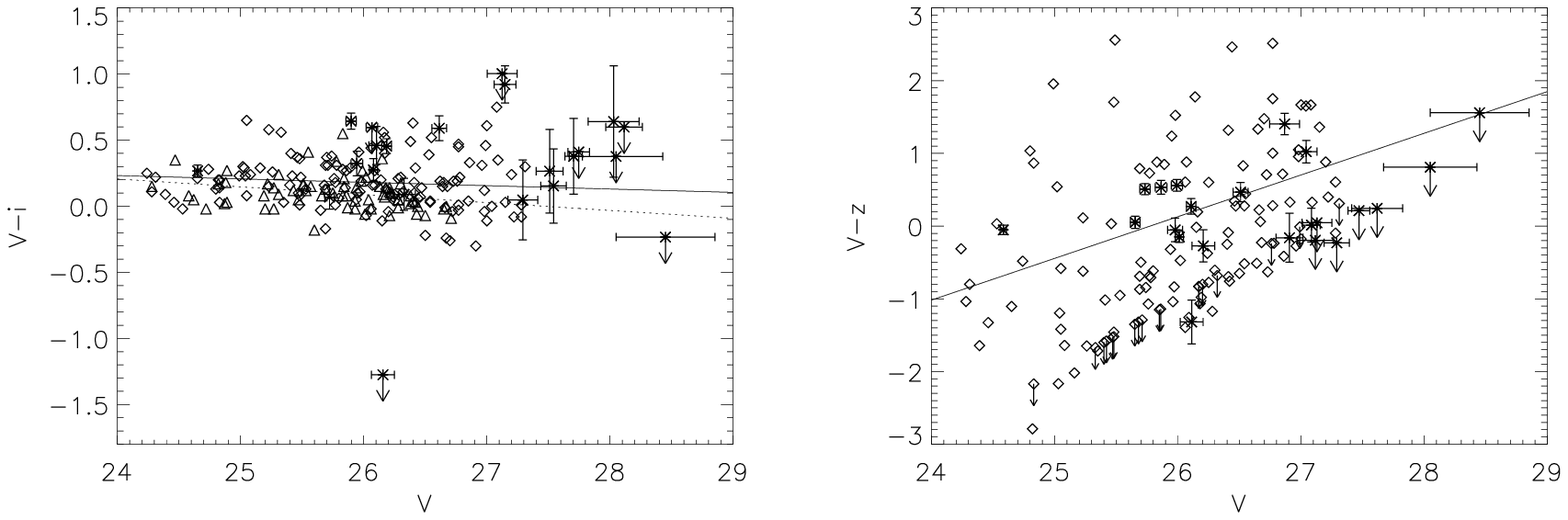,width=13.0cm,clip=}
\caption[Restframe UV colours of GOODS-S LEGOs]{Colours of our candidates (stars with error bars) compared to colours
of the sample of
faint LBGs of Wadadekar et al. (2006; diamonds) and Madau et al.
(1996; triangles). Lines represent the best fit
to the LBG data, the solid line the fit to the sample by Wadadekar et al. (2006)
and the dotted line the fit to the Madau et al. (1996) sample.
\emph{Left} $V$ minus $i$ colours, \emph{Right}
$V$ minus $z'$ colours. Small arrows indicate upper limits for the
Wadadekar et al. (2006) sample.}
\label{ch3:lbgcol}
\end{center}
\end{figure}

\begin{figure}[!t] 
\begin{center}
\epsfig{file=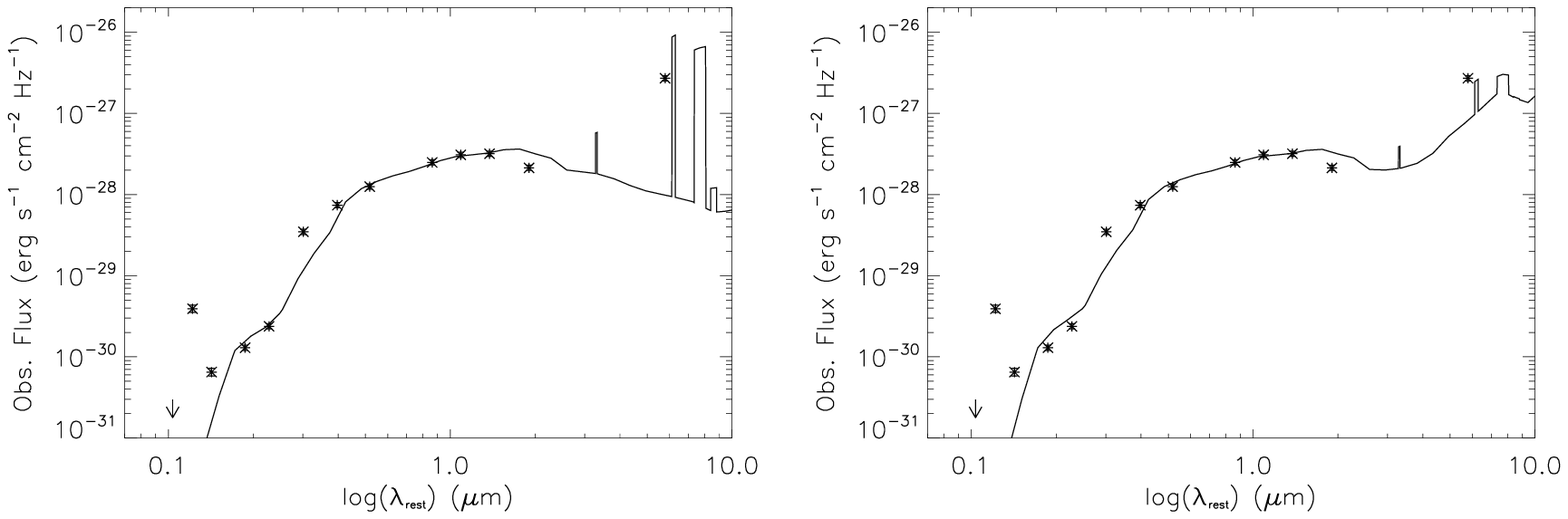,width=14.0cm,clip=}
\caption[GRASIL model fits to the SED of LEGO\_GOODS-S\#16]{Two preliminary GRASIL model fits to the SED of LEGO\_GOODS-S\#16,
both with $\chi^2 \sim 50$. Points with error bars (of the same size as the
point) are our data, with errors
being purely statistical. The point off the curve at lower wavelengths is the
narrow-band detection.}
\label{ch3:grasilmodels}
\end{center}
\end{figure}

We wish to compare our sample of LEGOs to a sample of faint Lyman Break Galaxies
(LBGs) in order to determine the similarities and differences of the two
populations of high-redshift galaxies. First, we want to know
if our LEGOs would be detected as LBGs and so we apply the LBG selection
criteria for $U$-band drop-outs of Wadadekar et al.
(2006; $U-B > 1.0$, $U-B > B-V + 1.3$ and
$B-V < 1.2$) as well as the criteria of Madau et al. (1996; $U-B > 1.3$,
$U-B > B-i + 1.2$ and $B-i < 1.5$) to our sample. However, our $U$-band data,
and in the case of the faintest candidates also the HST data, is
too shallow to get a useful measurement on the $U-B$ colour. Instead, we take the
best fit spectrum from the SED fitting (see Fig.~\ref{ch3:plotsed}) and convolve
it with the $U$ (F300W), $B$ and $V$ filter sensitivities and calculate the
colours. For this model spectrum, these colours become $U-B = 4.51$,
$B-V = 0.24$ and $B-i = 0.69$ which well satisfy the selection criteria for
$U$-band drop-outs, see Fig.~\ref{ch3:lbgcol2}. However, many of our LEGOs are
very faint and the stacked
$B$ magnitude is fainter than the lower limit of the selection of Madau et al.
(1996), and about half of our sample are fainter than the $V$ cut-off in the
sample of Wadadekar et al. (2006), see Fig.~\ref{ch3:lbgcol}.

Secondly, we wish to compare the observed optical colours (restframe UV colours)
of our LEGOs to the LBGs in order to establish if our LEGO candidates
have the same UV continuum colours as LBGs on the red side of the Lyman break.
In Fig.~\ref{ch3:lbgcol}, we plot the colours
of the two samples of faint LBGs published by Wadadekar et al. (2006) and
Madau et al. (1996) against the colours of our candidates.
All samples are drawn from survey data-sets such as GOODS-S and HDF-N, hence
there is no bias in photometry. In the plot, we see that the LEGO candidates
are drawn from a fainter sub-sample of the high-redshift galaxy population.
However, for the brighter candidates among our sample, the LEGOs appear to have
UV colours similar to LBG galaxies.

\section{Conclusion}

We have performed deep narrow-band imaging of part of the GOODS-S field.
The image revealed a set of 24 LEGO candidates, at a redshift of
$z \approx 3.15$. Of these, three candidates have been observed
spectroscopically and are confirmed. The spatial distribution of the candidates
appear
to be in a filamentary structure, with a 4$\sigma$ confidence, however to
confirm this and to plot the filament in 3D-space, we would need spectroscopic
redshifts. We have studied the entire candidate sample in all bands available
from X-rays to infrared in the GOODS-S data-set. From the SED fitting we
conclude that the LEGOs on average have low metallicity ($Z/Z_{\odot} = 0.005$),
have stellar masses in the range of $1 - 5 \times 10^9$~M$_{\odot}$
and low dust extinction (A$_V \sim 0.3$).
The candidates have ages in the range of 100~--~900~Myrs . We also
find one galaxy, LEGO\_GOODS-S\#16, which is best fit by a dusty starburst
galaxy at $z = 3.15$
with Ly$\alpha$-emission escaping from an area slightly offset from the central
core.

The comparison
to a sample of $U$-band drop-out galaxies in the GOODS-S field show that the
colours of LEGOs are consistent with the selection criteria for $U$-band
drop-outs except they are too
faint to be detected from their continuum flux. They also have colours similar
to those of LBGs at redshift $z \approx 3$.
In agreement with previous results (e.g. Gawiser et al. 2006), we conclude
that Ly$\alpha$-emitters at redshift $z \sim 3.1$ are dust- and AGN-free,
star-forming galaxies with small to medium masses.

\chapter{{\ly} emitters with VISTA}\label{chapter:vista}
\markboth{Chapter 4}{{\ly} emitters with VISTA}
\section{Introduction}
In this chapter, I describe a project which started in February 2005. The
idea was to make use of the large field-of-view of the camera of a future 
telescope called VISTA (Visible and Infrared Survey Telescope for Astronomy), 
described further in the next section, to find very high redshift 
($z \sim 9$) Ly$\alpha$
emitters through narrow-band imaging in the near-infrared. The cost of 
such filters made this a rather
expensive project and funding for the filters very generously came from
two sources; IDA -- Instrumentcenter for Dansk Astrofysik (DK) and the Dark
Cosmology Centre (DK). In the spring of 2006, we extended the science case to
involve all emission-lines (H$\alpha$, [OIII], H$\beta$ and [OII]) and
eventually submitted a proposal for a Public Survey with VISTA, entitled
``ELVIS -- Emission Line galaxies with VISTA Survey''. This proposal was
well received by the ESO Public Survey Panel, but was asked to merge with
a number of other surveys, with the aim to observe small, but deep, areas of the
sky with broad-band filters. Thus, ELVIS was merged into the Ultra-VISTA 
survey. I have been involved in all steps of this
project, from the original idea to ordering/inspecting the filters, and from 
proposal writing to
the accepted Public Survey. The following sections include a short
introduction to VISTA, a section on the design and procuration of the filters
and a description of the ELVIS Survey. This project also highlighted the need
for better estimates of number densities of very high redshift 
Ly$\alpha$ emitters.
The results of such a project are presented in Chapter~\ref{chapter:predictions}.

\section{VISTA -- Visible and Infrared Survey Telescope for Astronomy}
\begin{figure}[t]
\begin{center}
\includegraphics[width=0.75\textwidth,height=0.70\textwidth]{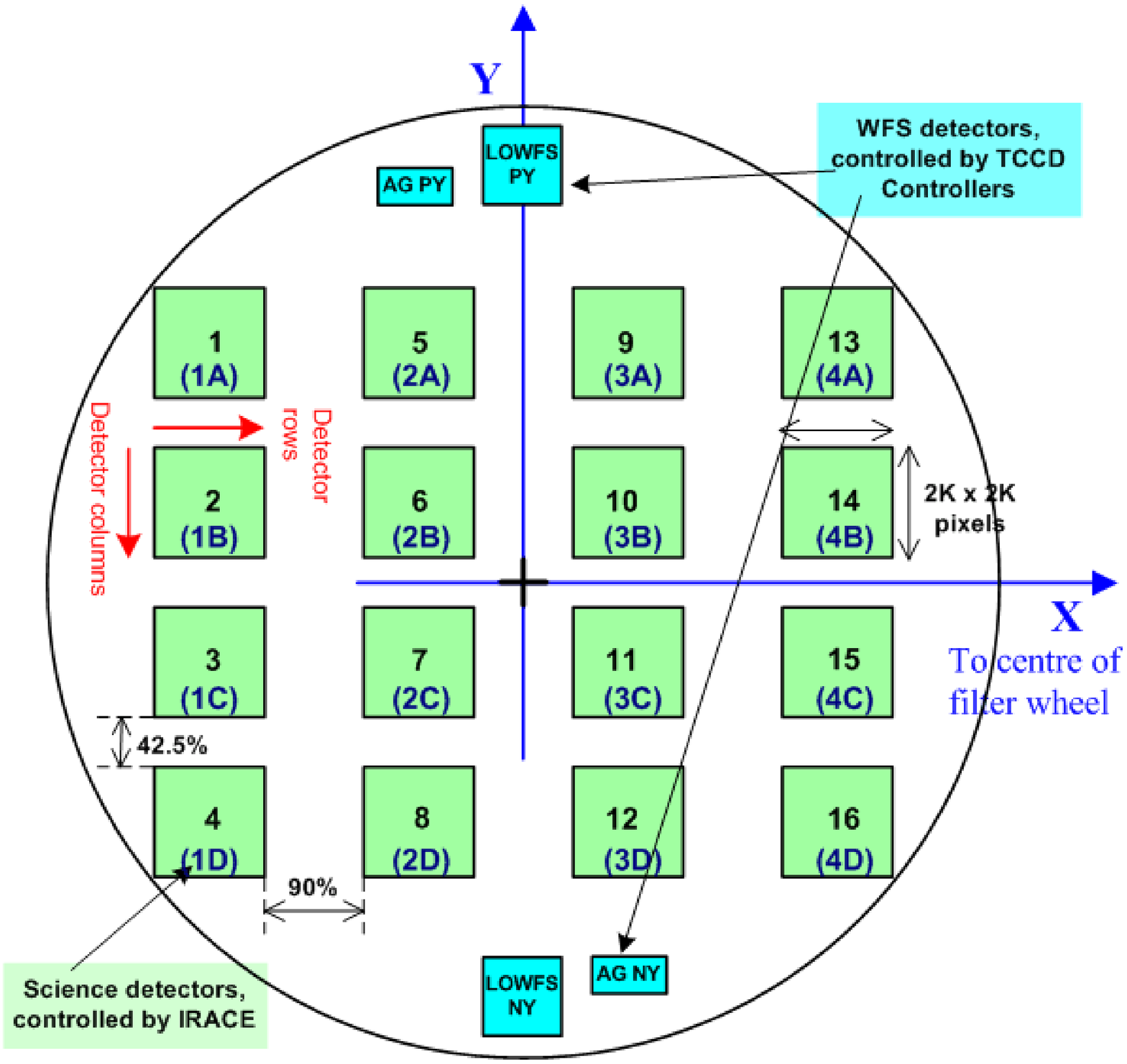}
\caption[VISTA IR camera array]{VISTA IR camera array consisting of 16 detectors. The single
``paw-print'' or exposure covers a field-of-view of 0.6 sq.degrees.
The array has large gaps between the detectors and hence needs to be
moved around to fill in a full mosaic. From \emph{www.vista.ac.uk}.\label{fig:vistaarray}}
\end{center}
\end{figure}

\begin{figure}[!pt]
\begin{center}
\includegraphics[width=0.45\textwidth,height=0.30\textwidth]{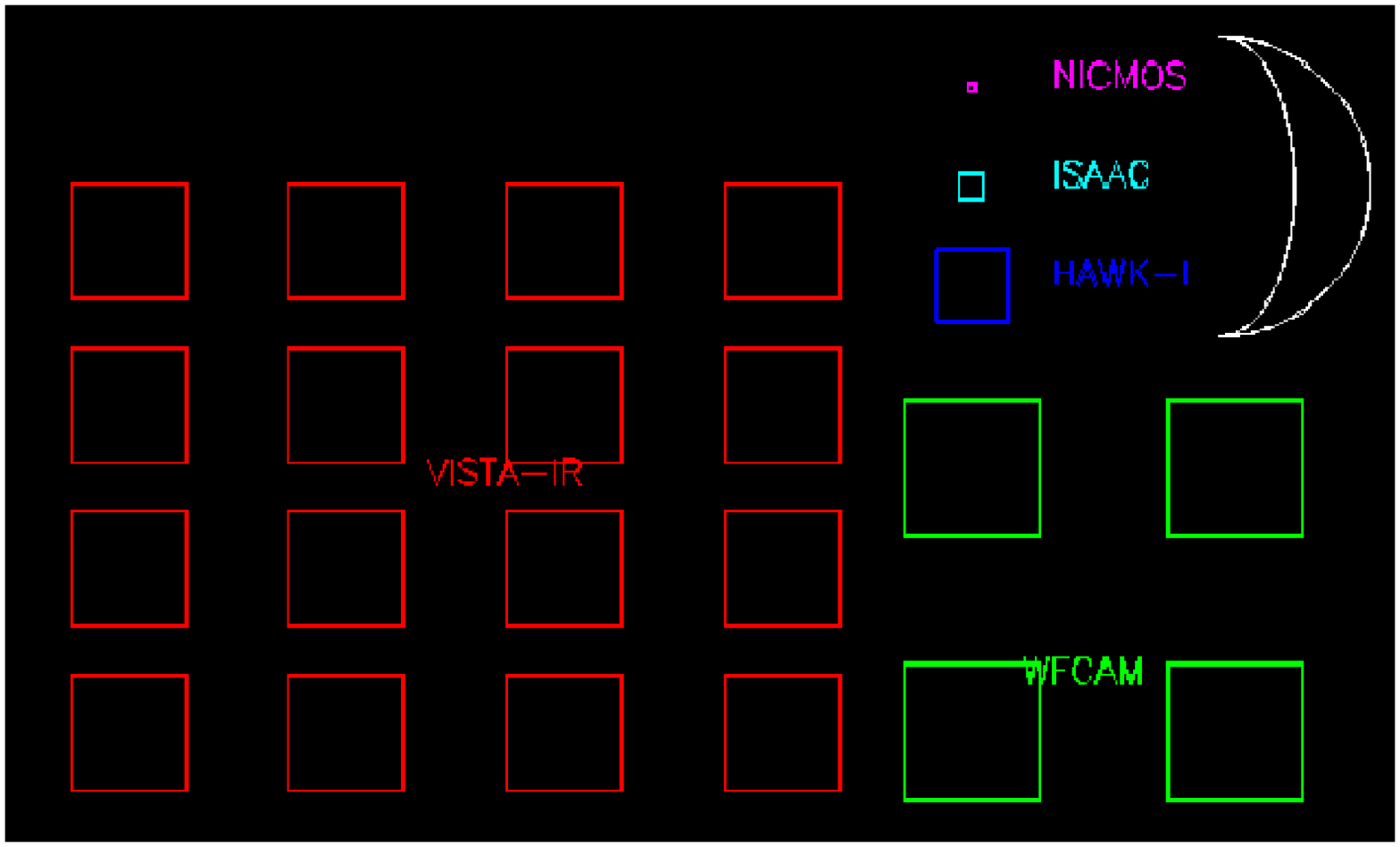}\includegraphics[width=0.45\textwidth,height=0.30\textwidth]{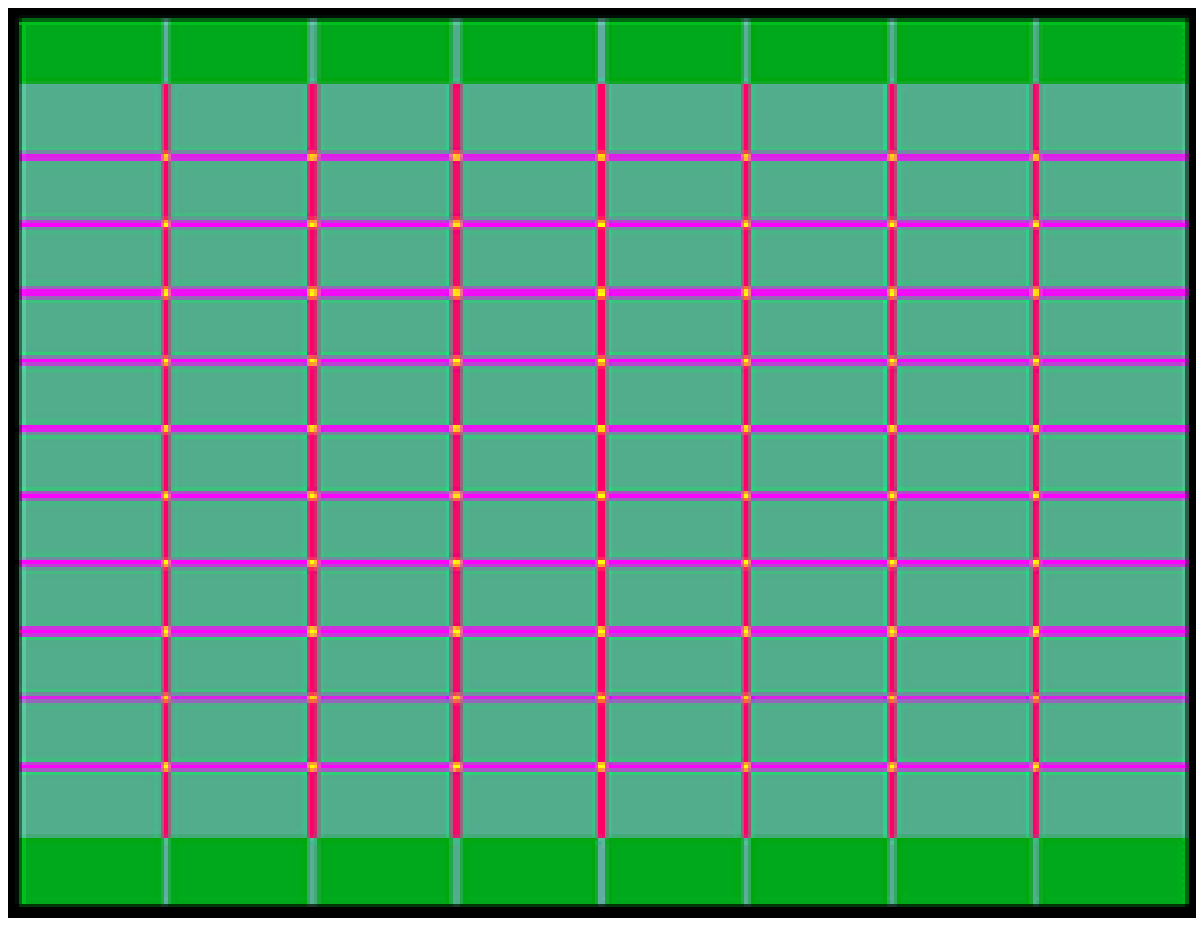}
\caption[VISTA Field-of-view comparison and coverage map]{\emph{Left} Field-of-view comparison between VISTA and other current 
or near future IR arrays as well as the size of the moon. \emph{Right}
Coverage map of the VISTA IR array after off-setting the array in six
different positions to get a uniform coverage of the sky. The colour scheme
shows effective exposure time, where dark green is one time the
single paw-print exposure time, light green is two times, magenta is three times,
red is four times and yellow is six times the single paw-print exposure time.
The average, almost homogeneous, coverage is two times the single
paw-print exposure time. Both images from \emph{www.vista.ac.uk}.\label{fig:vistafovcomp}}
\end{center}
\end{figure}

New advances in technology will always prompt new surveys of the night-time
sky. The first large area sky survey in the near infrared, the 
Two Micron Sky Survey (TMSS; Neugebauer \& Leighton 1969), was carried 
out in the 1960's and the next large area survey did not happen until
the late 1990's with the 2MASS Survey (Skrutskie et al. 2006) covering 
the entire sky in the three main bands J, H and K$_s$. A survey called
UKIDSS was started in 2005 (Lawrence et al. 2007). It has several parts, both
wide/shallow and small/deep fields, all in the Northern Sky. It is on 
the
background of these surveys that VISTA was conceived. VISTA, originally
planned to have both an infrared and an optical camera but later designed
with only an infrared camera, is an almost
(at the time of writing this thesis) completed 4-m.~telescope dedicated
almost entirely to near-infrared surveys. For this purpose, it has been
equipped with an infrared camera array of 16 detectors, covering a 
field-of-view of 0.6~square~degrees in each single exposure, see 
Fig.~\ref{fig:vistaarray}. The detectors are
not buttable, i.e. they cannot be placed adjacent to each other, but need to 
be separated by a large fraction of the width of the
detector itself. This is because the detectors interfere with each other if
they are too near each other. Thus, to get a continuous mosaic image, the 
camera needs to
be shifted in six different positions (or paw-print, as each single 
exposure is called). A plot of the resulting exposure time coverage can
be seen in the right panel of Fig.~\ref{fig:vistafovcomp}. 

The field-of-view (FOV) of each paw-print with VISTA is unprecedented as can be 
seen in the FOV comparison in the left panel of Fig.~\ref{fig:vistafovcomp}. 
The image shows the FOVs of current detectors (HST/NICMOS, VLT/ISAAC and 
UKIRT/WFCAM) and a future detector (VLT/HAWK-I). As in the picture, the
VISTA FOV is approximately 3 times the FOV of WFCAM (only northern hemisphere)
and 346 times the FOV of ISAAC.

\begin{figure}[!pb]
\begin{center}
\includegraphics[width=0.45\textwidth,height=0.55\textwidth]{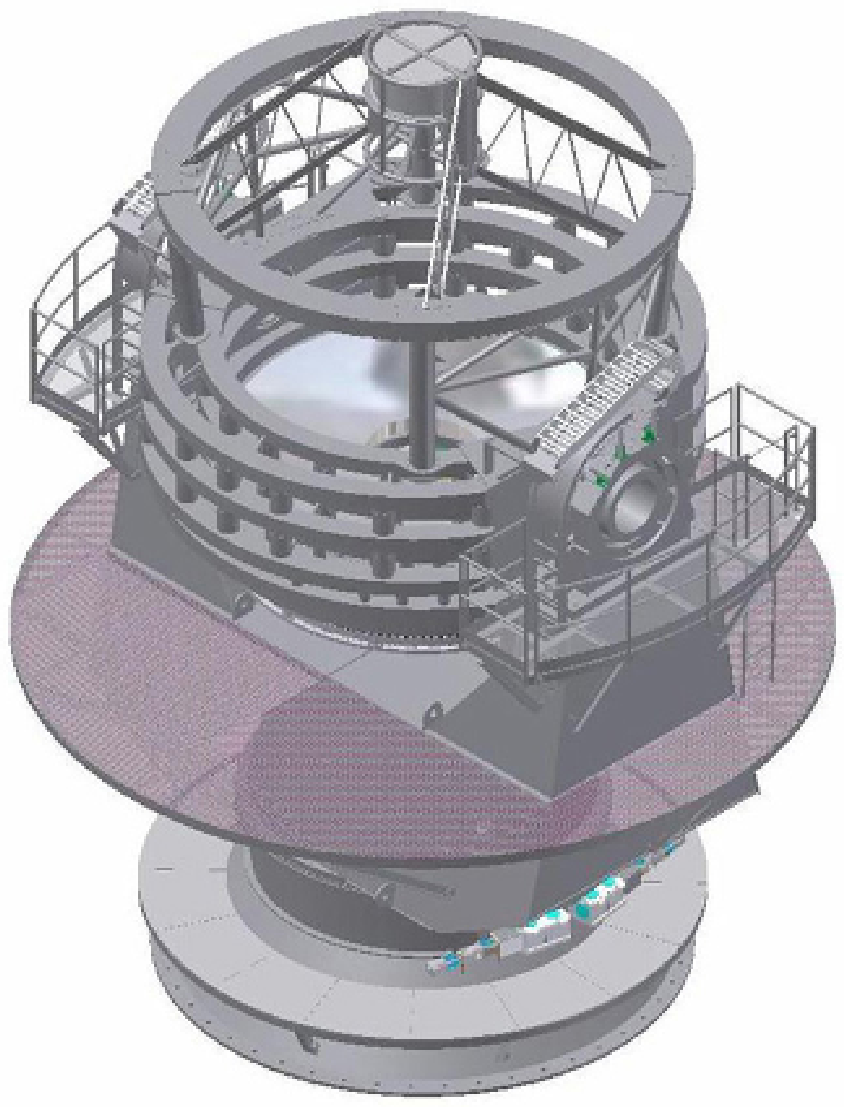}\includegraphics[width=0.45\textwidth,height=0.55\textwidth]{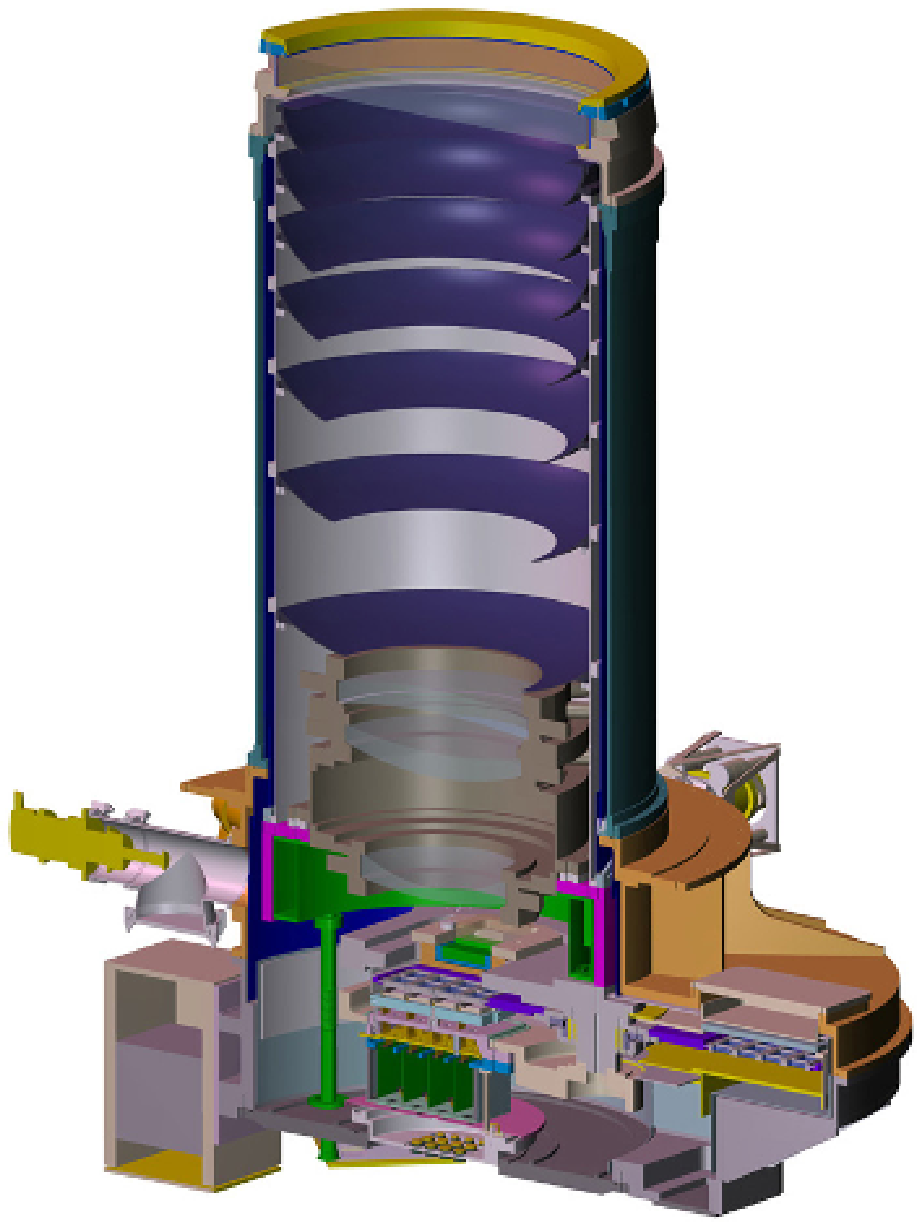}
\caption[Schematic view of VISTA IR telescope and camera]{Schematic view of the VISTA IR telescope (\emph{left}) and camera (\emph{right}). From \emph{www.vista.ac.uk}.\label{fig:vistatel}}
\end{center}
\end{figure}

A schematic lay-out of the telescope and camera
can be seen in Fig.~\ref{fig:vistatel} and the technical details of VISTA are
given in Table~\ref{tab:vistadetails}. 
\begin{table}[t]
\begin{center}
\caption[Technical details of VISTA]{Technical details of VISTA. See also 
Emerson et al. (2004) and Dalton et al. (2006). }\label{tab:vistadetails}
\vspace{0.5cm}
\begin{tabular}{|l|l|ccccccc}
\hline
M1 diameter & 3.95 m. \\
M2 diameter & 1.24 m. \\
\emph{f} ratio at instrument & 3.25 \\
Mount & Altitude-Azimuth \\
Number of detectors & 16 \\
Type of detector & Raytheon VIRGO HgCdTe 0.84-2.5 micron \\
Number of pixels per detector & $2048\times2048$ \\
Pixel size & $0.34''$/pixel \\
Read-out noise & 20.9 e$^-$ \\
Detector Quantum Efficiency & 71 \% (J), 74 \% (H), 75 \% (K$_s$) \\
Operating temperature of camera & $\sim 80$ K \\
Available filters & z, Y, J, H, K$_s$, NB1185 \\
\hline
\end{tabular}
\end{center}
\end{table}  
A point to note is that the VISTA detector array is stationary in the camera,
which has no shutter, and the filters are turned into the light path with
a filter wheel. The filter wheel has eight slots, of which one holds a blank
plate, six hold the filters mentioned in Table~\ref{tab:vistadetails} and
the final slot is empty initially. The fact that there is no shutter makes
positioning of filter sets and rotation patterns of the filter wheel crucial. 
When moving to a blue filter, or a narrow-band filter, it is undesirable to
turn the wheel so that the detector array is exposed to the light coming
through one of the redder filters, as this will ``flash'' the array with 
light and reduce the sensitivity temporarily.
VISTA is funded by 
PPARC, UK, but operated by ESO. It is placed at the ``NTT'' peak at Paranal,
Chile and will be fully operational by early 2008, according to the
plan. The majority, 75\%, of VISTAs time is dedicated to Public Surveys, 
i.e.~large scale
surveys conducted by the science community where all data, raw and reduced, 
become public. Six surveys have been approved for the initial five year 
survey period. An overview of each survey is given in 
Table~\ref{tab:vistasurveys}.
\begin{table}[!ht]
\begin{center}
\caption[Approved Public Surveys with VISTA]{Approved Public Surveys for the
first five years of operation of VISTA. See also Arnaboldi et al. 2007.}\label{tab:vistasurveys}
\vspace{0.5cm}
\begin{tabular}{|l|c|c|l|ccccc}
\hline
Name & Area (deg$^2$) & Filters & Science case \\
\hline
Ultra-VISTA & 0.9   & Y, J, H, K$_s$, NB1185 & Very high redshift universe,\\
            &       &                        & galaxy formation and evolution,\\
	    &       &                        & very deep, small area survey\\ 
VHS         & 20000 & Y, J, H, K$_s$         & Complete, shallow, southern \\
            &       &                        & hemisphere map\\
VIDEO       & 15    & z, Y, J, H, K$_s$      & AGN surveys, galaxy clusters,\\
            &       &                        & very massive galaxies\\
VVV         & 520   & z, Y, J, H, K$_s$      & Galactic bulge and plane,\\
            &       &                        & open and globular clusters,\\
	    &       &                        & variable sources\\
VIKING      & 1500  & z, Y, J, H, K$_s$      & Weak lensing and baryon \\
            &       &                        & acoustic oscillations\\
VMC         & 184   & Y, J, K$_s$            & Survey of the Magellanic system \\
\hline
\end{tabular}
\end{center}
\end{table}  

\section{The narrow-band filters}\label{sec:filters}
\begin{figure}[!ht]
\begin{center}
\includegraphics[width=0.90\textwidth,height=0.70\textwidth]{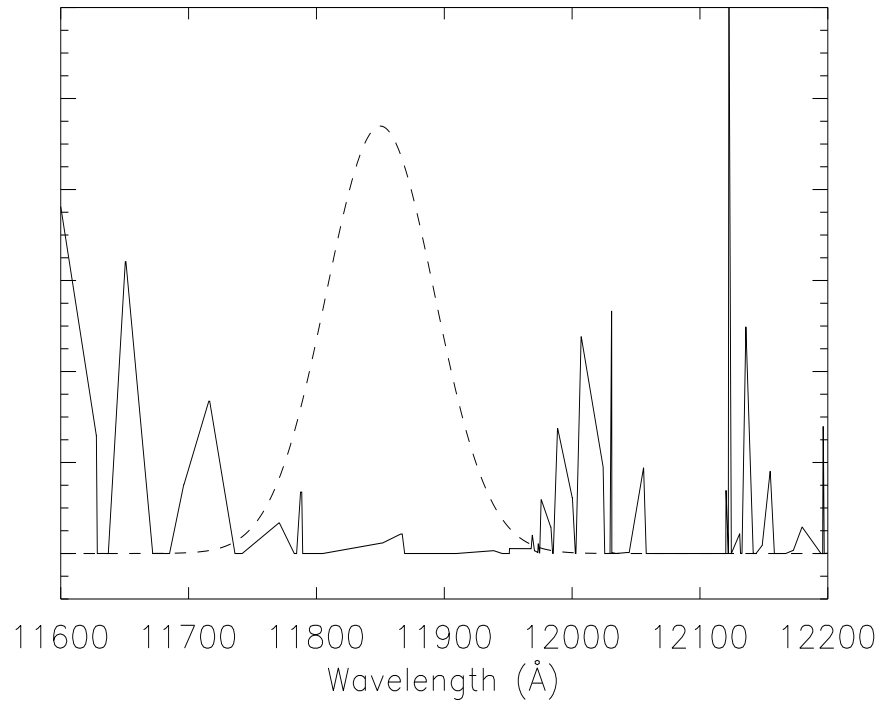}
\caption[VISTA NB filter curve]{Plot of the sky background OH-lines, with the intended filter curve
over-plotted with a dashed line in arbitrary units.\label{fig:ohwindow}}
\end{center}
\end{figure}

The width of a VISTA narrow-band filter can not be narrower
than $\Delta \lambda \approx 120$~{\AA}~due to the large shifts in wavelengths
over the field-of-view, see also sec.~\ref{sec:shiftcalc}. As the sky spectrum 
in the J band is
full of atmospheric OH-lines, there are only a small number of possible
wavelengths to place such a filter. These correspond to e.g.
$z_{\mathrm{Ly}\alpha} = $~7.73, 8.22, 8.78 etc. The first redshift, and later on 
the last, are planned to be surveyed with DAzLE (The Dark Ages z Lyman-alpha 
Explorer, Horton et al. 2004). Our first idea was thus
to place the filter at the intermediate redshift. However, it was discovered
that this window has a significant [OII]-line in the dark range, and it was
discarded as an option. The window corresponding to
$z_{\mrm{Ly}\alpha} = 8.8$ is quite wide (see Fig.~\ref{fig:ohwindow}) and it 
will only be surveyed by DAzLE at a later stage, hence we chose to place our
filter at the wavelength corresponding to $z_{\mathrm{Ly}\alpha} = 8.78$,
$\Delta z = 0.10$.

\subsection{Filter specifications}\label{sec:filterspecs}
The filter specifications were prepared in the same manner as the original
VISTA broad-band filters specifications. The camera array will experience a 
slight passband
shift over the field-of-view due to the filters operating in an f/3.25 beam,
and the outer arrays may experience a broadening of $\sim 0.5$~\% due to a
slight field-dependent effect, due to the chief ray being tilted by
$\sim 5$~degrees at the edge of the field. These effects will be especially
pronounced with the narrow-band filter, due to the small passband width. We
investigated how these effects would influence the filter specifications
thoroughly prior to sending out calls for quotations, and found that the 
effect would not be severe, see sec.~\ref{sec:shiftcalc}. A plot of such a 
study can be seen in Fig.~\ref{fig:pbshift1}.
We also anticipated the possibility of extensively
measuring the effects during commissioning of the telescope and
following that produce corrective algorithms to be used while reducing the data.
The specifications made for the filter were thus set to be:
\begin{itemize}
\item[-] Central wavelength should be 1185 nm $\pm$ 2 nm.
\item[-] Peak transmission should be at least 75\% (Goal: 85\%).
\item[-] The passband FWHM should be 10 nm $\pm$ 2 nm.
\item[-] The transmission at 1174 nm and 1195 nm should be below 10\%.
\item[-] The transmission at 1165 nm and 1210 nm should be below 1\%.
\item[-] The average transmission between 700 nm and 1140 nm should be below 0.1\%.
\item[-] The average transmission between 1250 nm and 3000 nm should be below 0.01\%.
\end{itemize}
The last point is especially important as ``red leaks'' are a severe problem 
with near-infrared narrow-band filters (Sutherland, priv. communication).
The call for quotations was sent to four firms in October 2005, and the
best bid was made by 
\emph{NDC Infrared Engineering}\footnote{http://www.ndcinfrared.com/}. 
The filters were subsequently ordered by the Dark Cosmology
Centre in January 2006 and 20 filters were delivered to the Rutherford 
Appleton Laboratory in Oxford, UK, in April 2007, see Table~\ref{tab:filtertab}.
\begin{table}[t]
\begin{center}
\caption[NB filter details delivered by NDC]{Table of delivered narrow-band filters from NDC. Values given at 
operating temperature. The filter marked with a star has been excluded from 
the selection due to a large gradient in the color on the edge of the filter.}\label{tab:filtertab}
\vspace{0.5cm}
\begin{tabular}{lccccccccc}
\hline
\hline
Filter & CWL (nm) & FWHM (nm) & Maximal T (\%) \\
\hline
911056R5\_4   &   1182  & 11.1 & 71 \\
301106R6\_1   &   1183  & 11.6 & 72 \\
011206R5\_1-1 &   1183  & 12.0 & 71 \\
021206R5\_3   &   1184  & 11.0 & 73 \\
011206R5\_1   &   1185  & 11.2 & 70 \\
260207R5\_1   &   1185  & 12.0 & 76 \\
281106R5\_3   &   1185  & 10.7 & 70 \\
230207R6\_3   &   1185  & 11.9 & 74 \\
080207R6\_1   &   1185  & 11.8 & 74 \\
230207R5\_3   &   1186  & 11.5 & 78 \\
270207R5\_4   &   1187  & 12.0 & 70 \\
260207R5\_3   &   1187  & 12.0 & 74 \\
090207R5\_4   &   1188  & 11.6 & 73 \\
230207R5\_5   &   1188  & 11.9 & 76 \\
050207R6\_1   &   1188  & 11.7 & 73 \\
090207R5\_1   &   1188  & 11.8 & 71 \\
230207R5\_2$^{\star}$   &   1188  & 11.8 & 73 \\
230207R6\_2   &   1188  & 11.8 & 76 \\
260207R5\_4   &   1188  & 11.9 & 77 \\
311006R5\_1   &   1189  & 10.5 & 76 \\
\hline
\end{tabular}
\end{center}
\end{table}

\subsection{Central wavelength and passband shift}\label{sec:shiftcalc}
The optical path of VISTA is a converging beam. This means that each pixel on 
each detector sees an annulus of light coming from the primary mirror and 
obscured by the secondary mirror. But a pixel at the edge of the 
filter/detector 
array has a tilted view of this annulus, i.e. the centre of the annulus is 
shifted into the actual light annulus. The effect is a cosine function of the 
length of a vector pointing out a light element in the annulus divided by the 
refractive index (assumed to be 1.8, Sutherland priv.~communication), 
integrated over the entire annulus. This causes the central wavelength of the 
filter to shift and the passband to increase.

\begin{figure}[!pt]
\vspace{-0.5cm}
\begin{center}
\includegraphics[width=0.45\textwidth,height=0.45\textwidth]{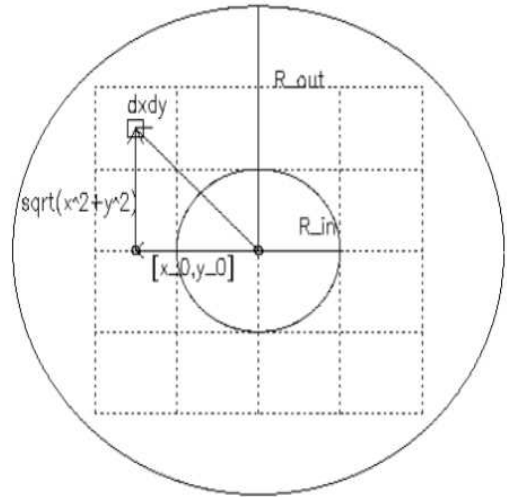}
\caption[Light annulus on VISTA camera array]{Sketch of the annulus with inner and outer radii $R_{in}$ and 
$R_{out}$ respectively. To estimate the shift in central wavelength, we need 
to integrate the incident light on a pixel, by integrating the area of the 
annulus for all $x$ and $y$. The centre of the annulus then moves with 
increasing distance from the centre of the camera array. The shift is given by 
$x_0$ and $y_0$. For an edge or corner pixel we hence need to integrate over 
the annulus, with shifted central pixels. \label{schematic}}
\end{center}
\vspace{-0.5cm}
\end{figure}

A schematic of the annulus can be seen in Fig.~\ref{schematic}.
The shifted central wavelength is given by:
\begin{equation}
\lambda_\mathrm{shifted} = \frac{\left(\int_{x=-R_{out}}^{x=R_{out}} \int_{y=-f(x)}^{y=f(x)} \mathrm{cos}(\frac{\sqrt{x^2+y^2}}{1.8}) \mathrm{dx dy}-\int_{x=-R_{in}}^{x=R_{in}} \int_{y=-f'(x)}^{y=f'(x)} \mathrm{cos}(\frac{\sqrt{x^2+y^2}}{1.8}) \mathrm{dx dy}\right)}{\pi \times (R_{out}^2-R_{in}^2)} \times \lambda_\mathrm{c}
\end{equation}
\noindent where
\begin{displaymath}
f(x) = \sqrt{R_{out}^2-x^2}
\end{displaymath}
\begin{displaymath}
f'(x) = \sqrt{R_{in}^2-x^2}
\end{displaymath}

\noindent where $R_{out}$ = 8.75 degrees, $R_{in}$ = 3.85 degrees and 
$\lambda_c$ = 1185 nm (Sutherland, priv.~communication). This is true for a 
central pixel at the very centre 
of the array. For any other pixel, a value of $x_0$ and $y_0$ has to be added 
to $x$ and $y$ in the cosine function in the top integral. The largest effect 
is achieved when $(x_0,y_0) = (5.15,5.15)$, which is at the corner of the 
field. The effect of the integrated factor is always to decrease the central 
wavelength and the effect is at its greatest a factor of 0.5 \%. The change in 
wavelength for $z_{Ly\alpha} = 8.8$ can be found in Table~\ref{tab:cwlshift}
for four positions. These positions correspond to e.g. the positions of 
detector 6 (number 1 in table), detector 5 (number 2 in table), detector
1 (number 3 in table) and detector 2 (number 4 in table). The shifts are 
symmetric in each quadrant of the array.
\begin{table}[!h]
\begin{center}
\caption[Central wavelength shift of VISTA NB filters]{Central wavelength shift
in the centre position and the middle point of four detectors according to the
text. Based on an original central wavelength of the filter of 1185 nm.}\label{tab:cwlshift}
\vspace{0.5cm}
\begin{tabular}[b]{|c|c|c|}
\hline
\multicolumn{2}{|c|}{Shifted wavelength in four points}\\
\hline
Point  & $\lambda_{\mathrm{shifted}}$ (nm) \\
\hline
Centre & 1182.46 \\
1      & 1182.34 \\
2      & 1182.01 \\
3      & 1181.43 \\
4      & 1181.76 \\
\hline
\end{tabular}
\end{center}
\end{table}

\begin{figure}[!pb]
\vspace{-0.5cm}
\begin{center}
\includegraphics[width=0.62\textwidth,height=0.42\textwidth]{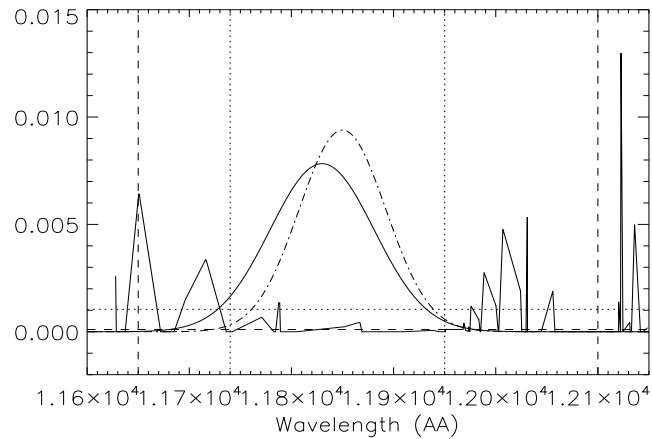}
\caption[Shift of NB transmission over the camera array]{
Plot of the sky background and the filter specifications given to the productions
company. The dot-dashed line indicates the desired filter transmission curve, 
the dotted and dashed vertical and horizontal lines are the specifications 
made to the company 
(see sec.~\ref{sec:filterspecs}). The solid line is the worst case shift at 
the corner of the
camera array. As can be seen, this effect will not significantly overlap with
the OH lines near the edge.}
\label{fig:pbshift1}
\end{center}
\vspace{-0.5cm}
\end{figure}

We assume that the filter transmission function is a Gaussian function:
\begin{equation}
T = \frac{1}{\sigma\sqrt{2\pi}} \times e^{-\frac{(\lambda - \lambda_c)^2}{2\sigma^2}}
\end{equation}  

\noindent where FWHM~$= 2\times \sqrt{2 \mathrm{ln}2}\sigma$ and 
$\lambda_c =$~central wavelength of the filter. Since the central wavelength 
shifts over each camera array, the transmission will become broadened and 
skewed towards a top-hat function. Figure~\ref{fig:pbshift1} 
show this effect on the corner array in the $z = 8.8$ OH-line free window. 
The effect is a broadening in FWHM of the order of 10~\%.

\subsection{Inspection of narrow-band filters}
In May, 2007, we visited Rutherford Appleton Laboratory (RAL) in order to 
inspect the filters. Some photographs from the inspection can be found in 
Fig.~\ref{fig:filterphoto}.

During the inspection, performed in the clean room at RAL, we carefully opened 
each package and checked each filter for the following issues:
\begin{itemize}
\item[-] Consistency in labelling on wrapping and filter.
\item[-] Overall condition of filter.
\item[-] Level of flaking.
\item[-] Existence of ``pinholes'', ``comas'', scratches or other damages.
\item[-] Discolorations.
\end{itemize}
All filters displayed varying degrees of flaking, i.e. material from the 
coating at the edges falling off and lying loose in the package. This
is very undesirable, as it may scratch the coating if it rubs against the 
surface. Two filters were mislabeled, or even unlabeled on the filter itself.
One filter showed a large discoloration along the edges, indicating that the
coating was incomplete around the edges. This filter was removed from the 
pool of usable filters. A few filters had small chips at the edges, although
small enough to not cause any complications. Many filters had small pinholes 
but they were shallow enough that they can be flat-fielded away (Dalton,
priv. communication). Overall, all
but the one with a discolored edge were accepted for use.

\subsection{Positioning of filters in the VISTA filter tray}
\begin{figure}[!pb]
\begin{center}
\vspace{-0.5cm}
\includegraphics[width=0.5\textwidth,height=0.30\textwidth]{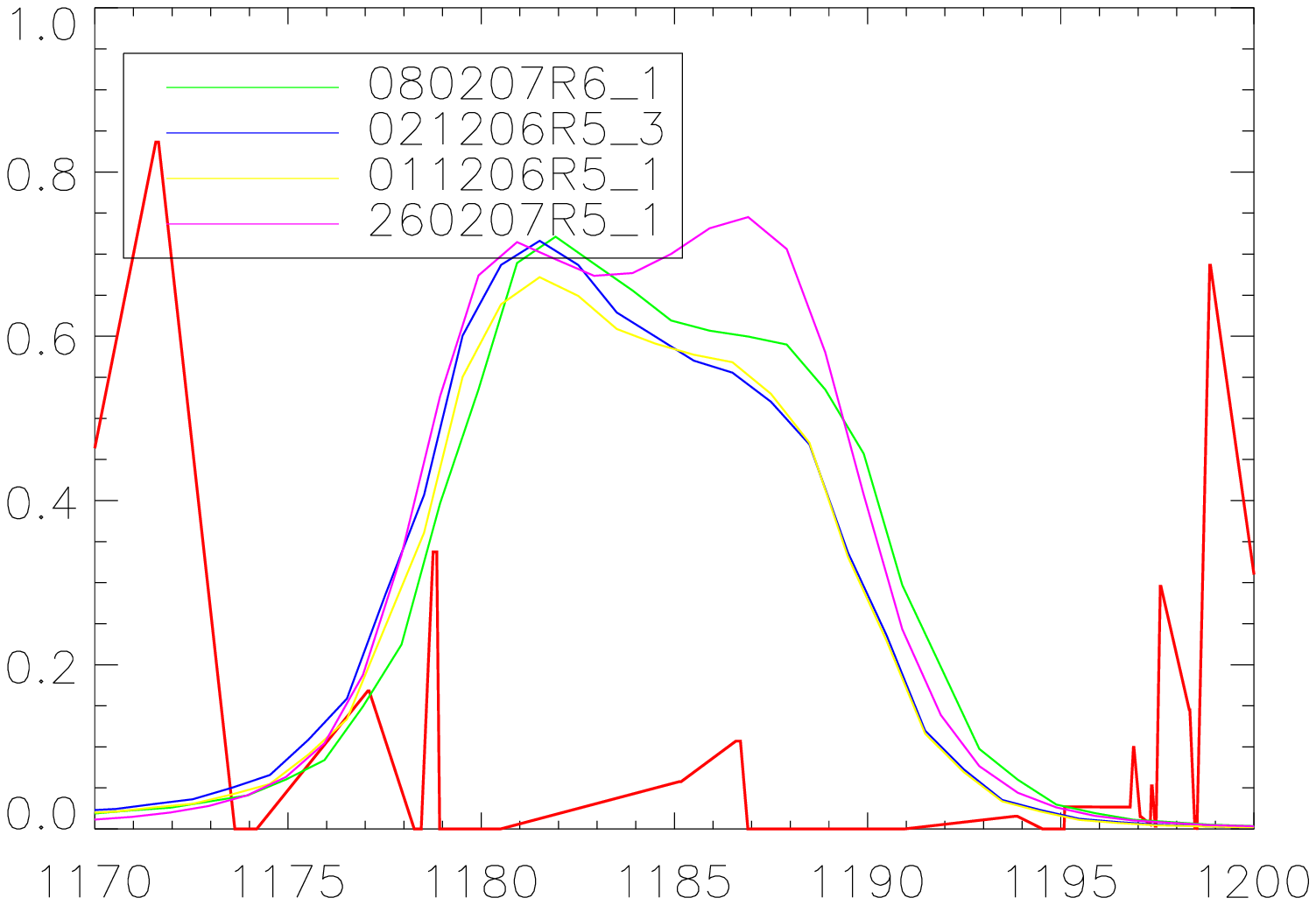}\includegraphics[width=0.5\textwidth,height=0.30\textwidth]{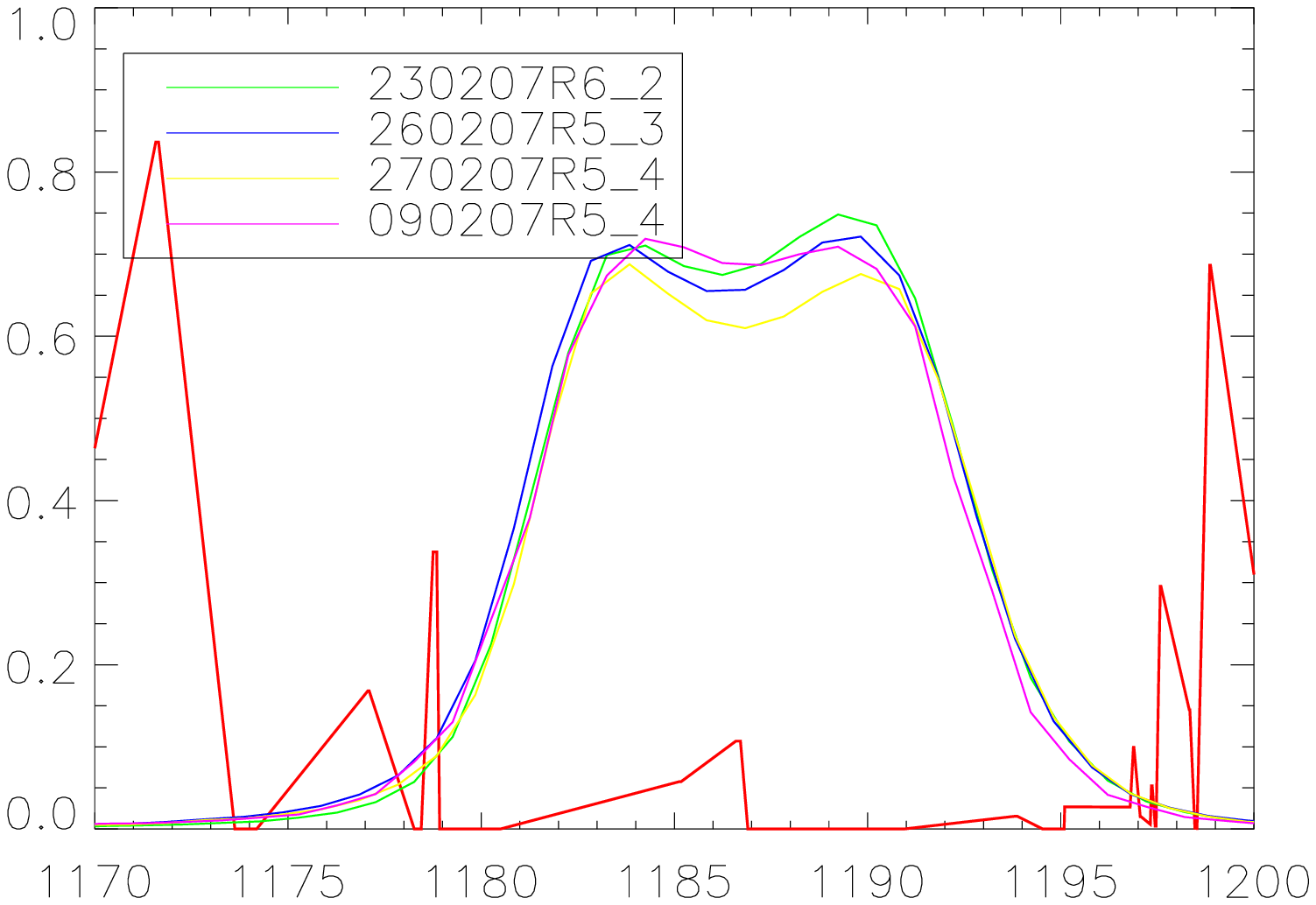}

\includegraphics[width=0.5\textwidth,height=0.30\textwidth]{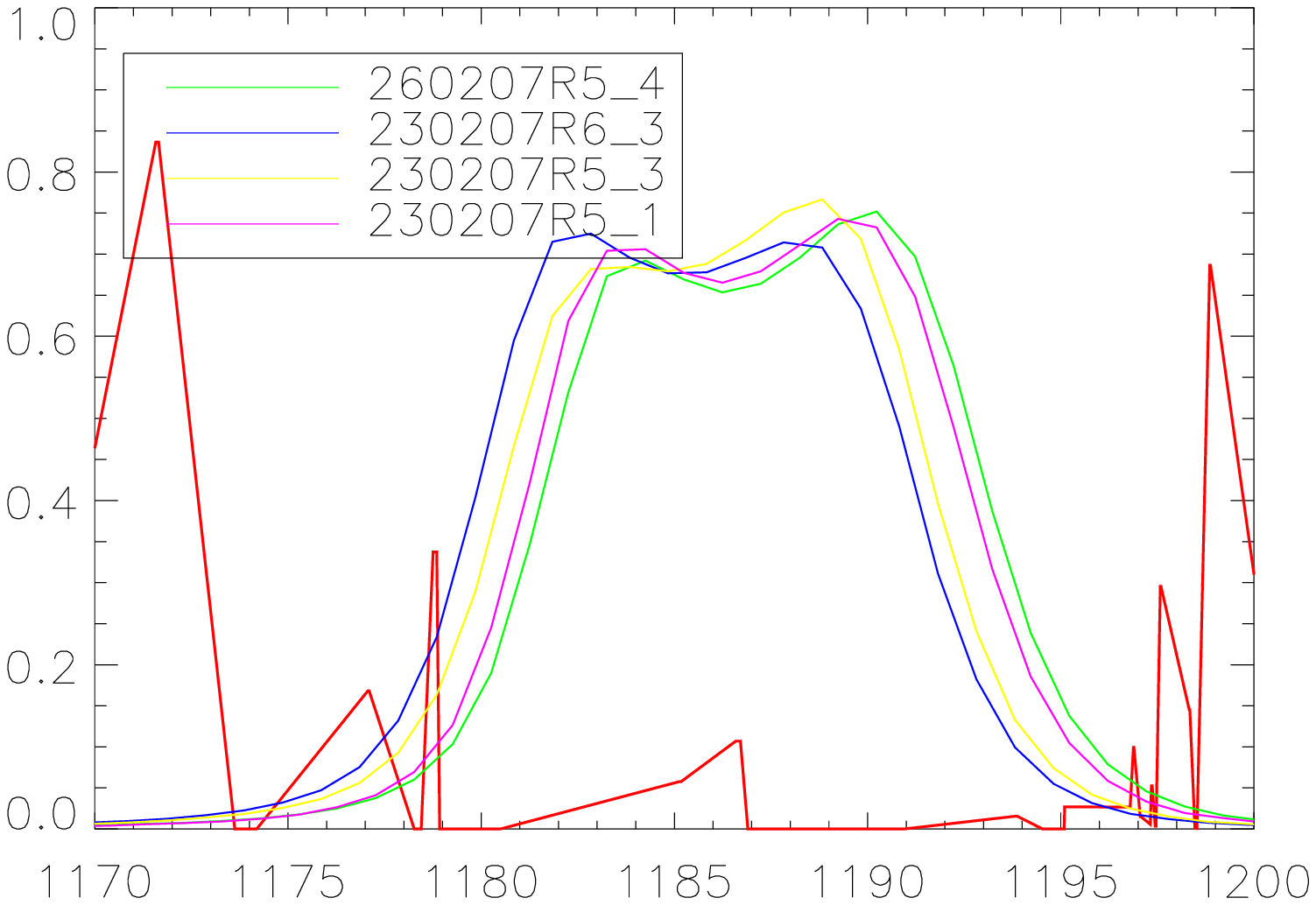}\includegraphics[width=0.5\textwidth,height=0.30\textwidth]{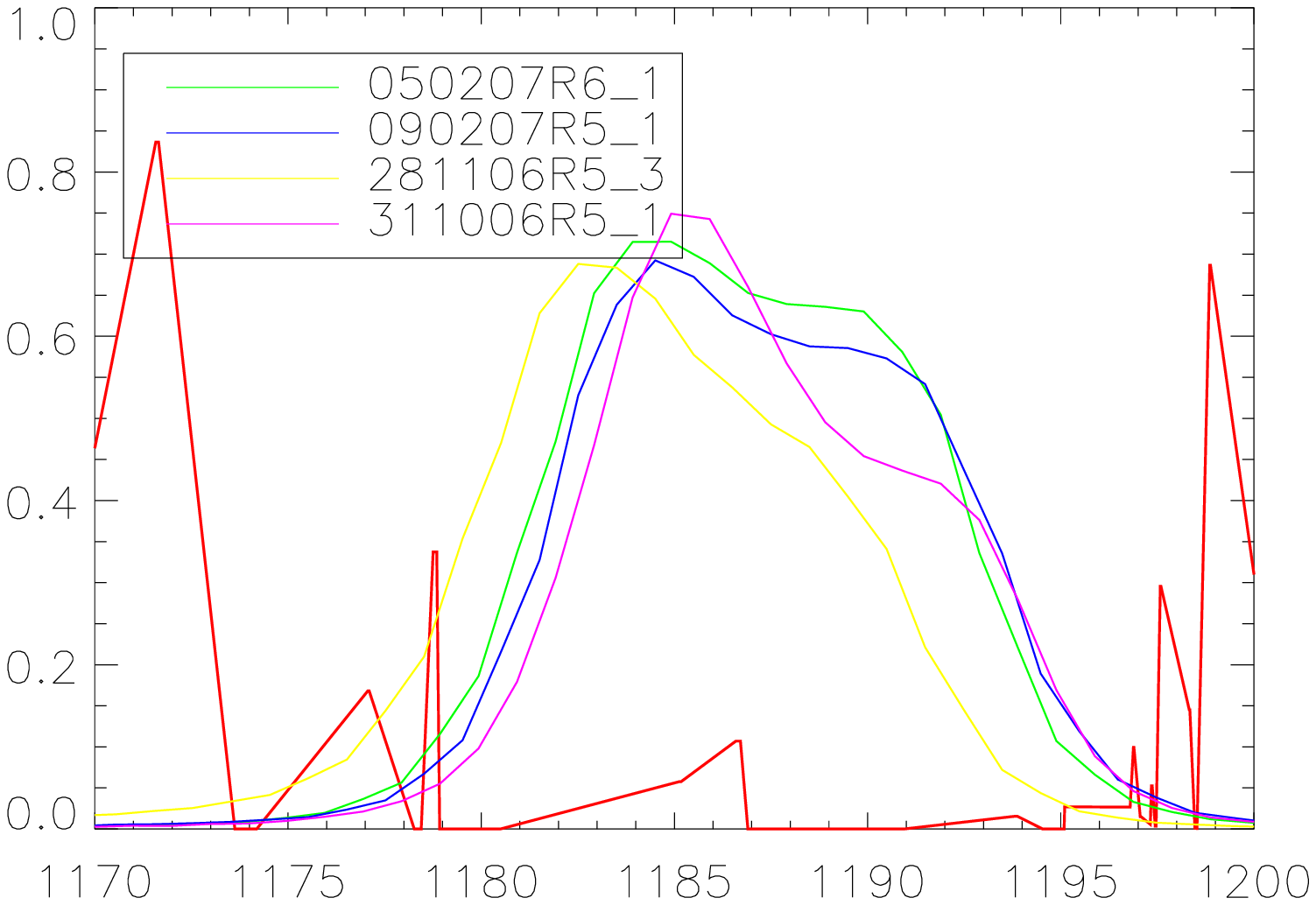}
\caption[Positioning of VISTA NB filters]{Positioning of VISTA narrow-band filters. Upper left image shows 
column 1 (detector 1-4), upper right shows column 2 (detector 5-8),
lower left shows column 3 (detector 9 - 12) and lower right shows column 4
(detector 13 - 16). Red lines display the OH sky background, differently
coloured lines display the transmission curves of each individual filter. 
Green and magenta lines represent filters to be placed at the uppermost and 
lowermost position in the tray (e.g. detector 1 and 4 in column 1) and blue 
and yellow lines show filters to be placed in the central locations (e.g.
detector 2 and 3 in column 1). Curves have been shifted to correspond to
the position they are to be placed in.\label{fig:filterpos}}
\vspace{-0.5cm}
\end{center}
\end{figure}

\begin{figure}[!pt]
\begin{center}
\vspace{-0.5cm}
\includegraphics[width=0.33\textwidth,height=0.25\textwidth]{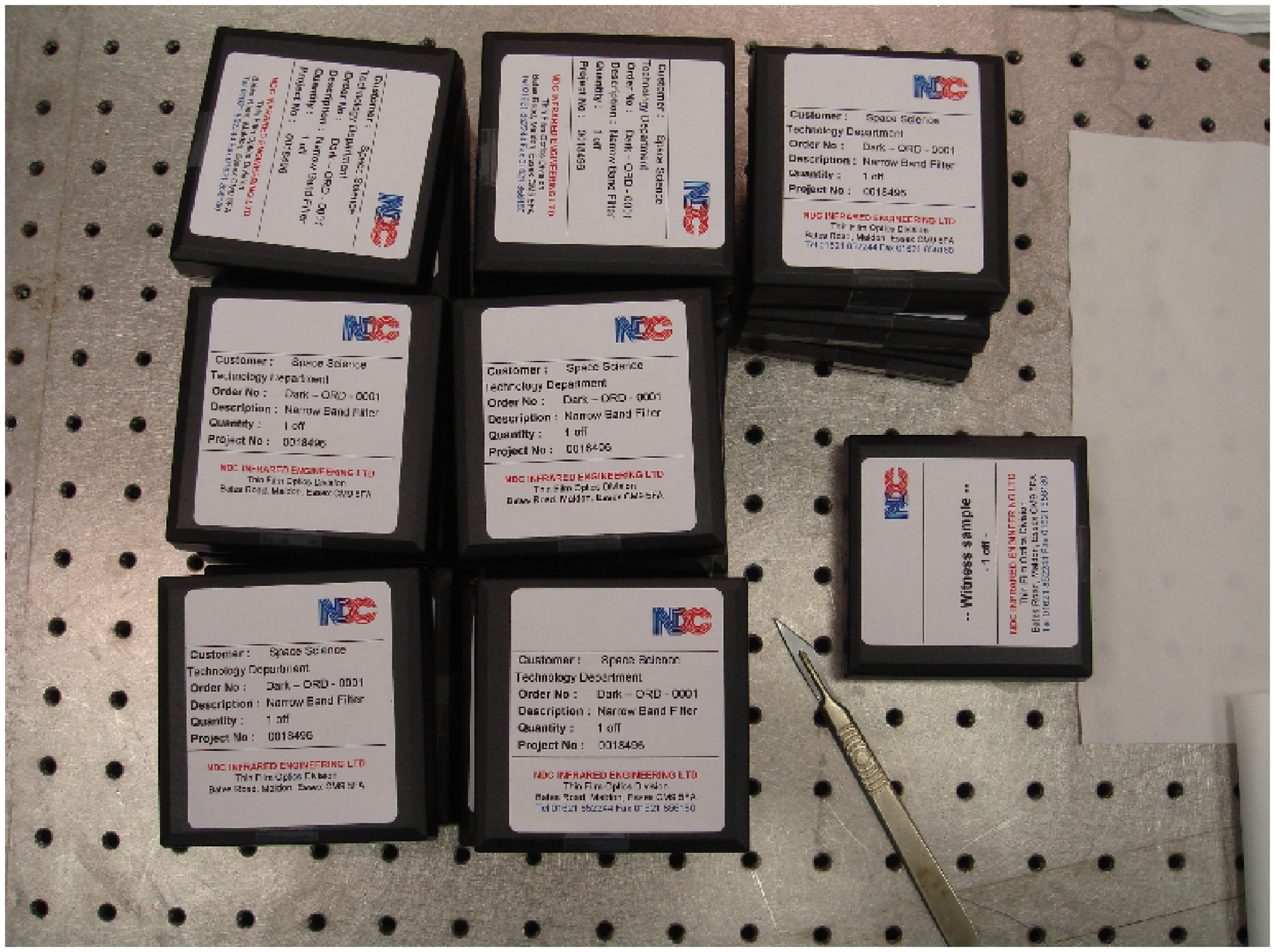}\includegraphics[width=0.33\textwidth,height=0.25\textwidth]{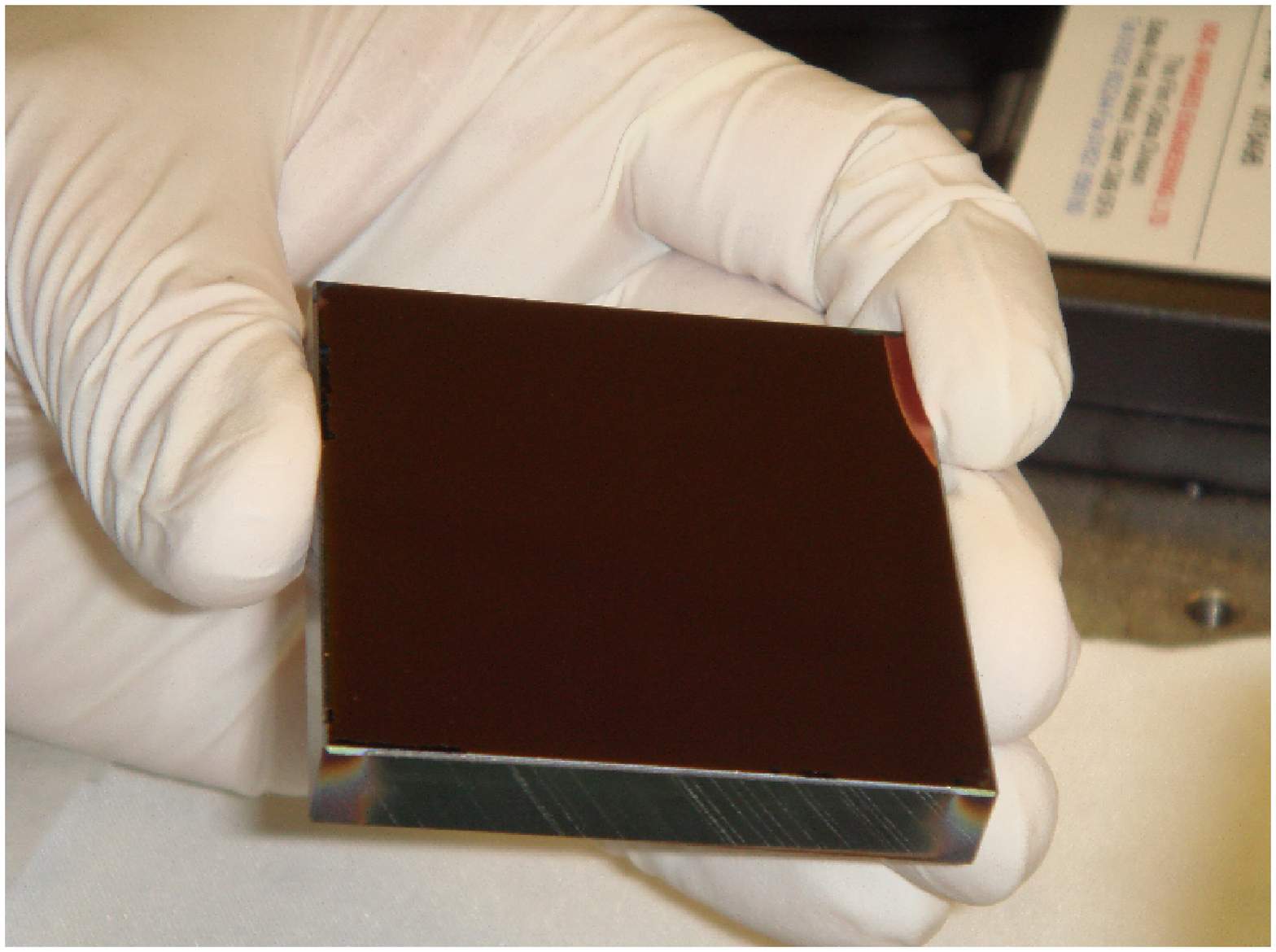}\includegraphics[width=0.33\textwidth,height=0.25\textwidth]{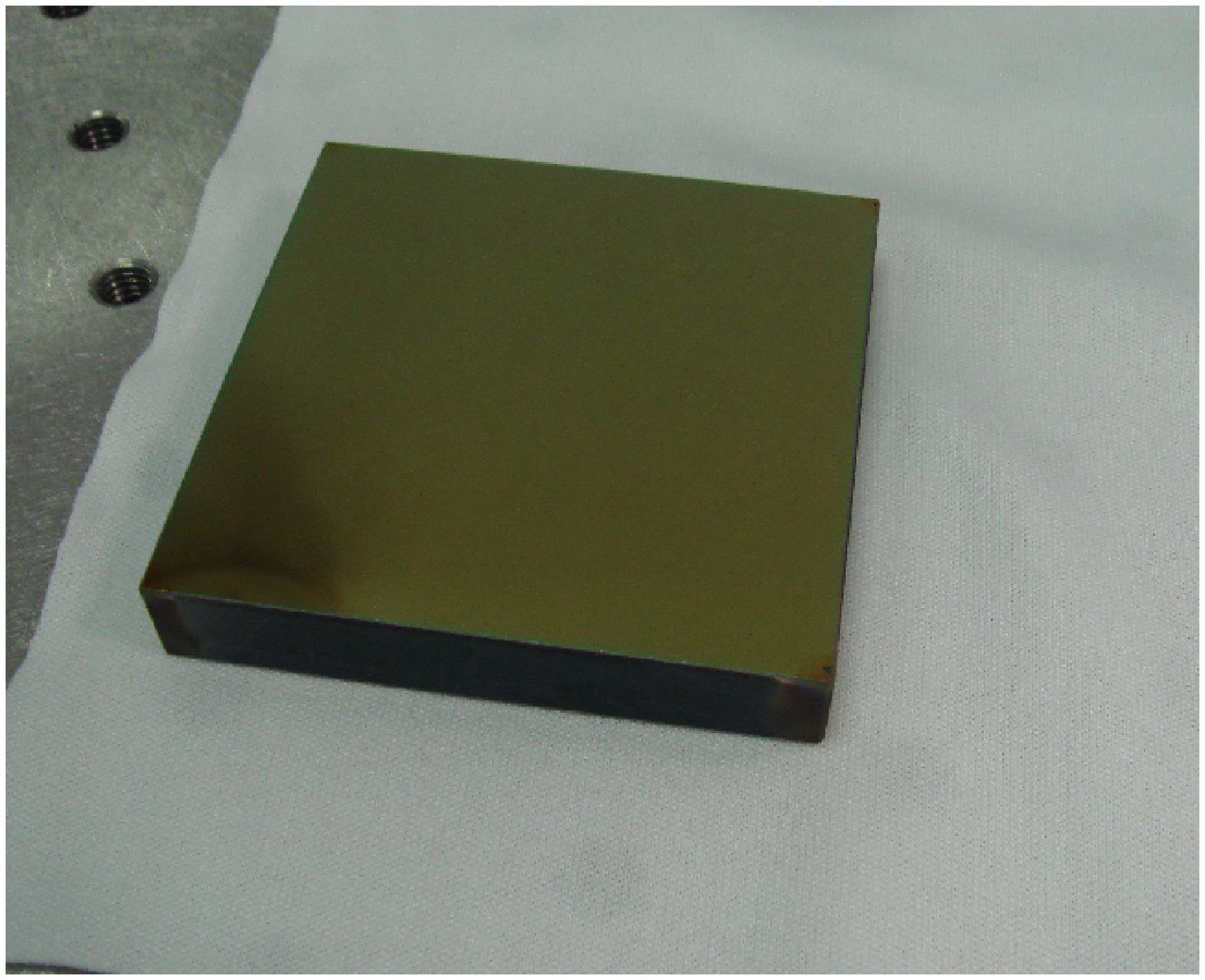}
\caption[Photographs of VISTA NB filters]{Photographs from the filter inspection. To the left are the whole stack 
of filters, including two witness pieces, delivered from \emph{NDC Infrared} to
\emph{Rutherford Appleton Laboratory}. The middle panel show the blocking side of 
the filter, in relation to the size of a finger. The right panel show the filter
transmission side of the filter. Images courtesy of Wolfram Freudling.\label{fig:filterphoto}}
\vspace{-0.5cm}
\end{center}
\end{figure}

After studying the transmission curves of the remaining, accepted 19 filters,
the three most blue filters were also excluded from further use, and kept as
spares, as they overlap significantly with sky emission lines in the blue
edge of the sky window, as seen in Fig.~\ref{fig:ohwindow}. Thus, with 16
filters left to use we attempted to ``puzzle'' these filters together so that
all detectors along each column of the detector array would have as similar
transmission curves as possible, also taking into account the shift that
occurs of the field-of-view. The result can be seen in Fig.~\ref{fig:filterpos}.

\section{ELVIS and Ultra-VISTA}
ELVIS is part of Ultra-VISTA, an ESO Public Survey that aims to do very deep
near-infrared broad- and narrow-band imaging in one pointing in the COSMOS
field (Scoville et al. 2006). The main science case for Ultra-VISTA is to 
get one of the deepest
images to date in the near-IR bands of a large area on the sky. It will focus
on detecting the first galaxies at very high redshift and to study faint
objects at intermediate redshifts. ELVIS aims specifically at searching for 
emission-line galaxies at several redshifts in order to study galaxy 
formation and evolution, at what redshift re-ionisation started as well as 
the evolution of the cosmic star formation history.

\subsection{Science goals}
As my involvement in Ultra-VISTA is almost solely focused on the ELVIS part, 
I will here only discuss the science goals of ELVIS.
There are a number of arguments why a large scale survey in the infrared for
emission line galaxies will yield new and
exciting insights into galaxy formation and star formation history. The
objective is split into three parts, one looking for very high redshift
Ly$\alpha$ emitters, and the others to observe H$\alpha$ and intermediate 
redshift emitters.

\subsubsection{Ly$\alpha$ emitters}
The quest for detecting cosmological objects is always pushing to reach further
away in redshift space, and further back in time. So far, no objects have 
been confirmed to be at a redshift larger than seven. Of the objects confirmed
to be above redshift six, the vast majority are in fact {\ly}-emitters.
To date, 22 Ly$\alpha$-emitters (Hu et al. 2002; Kodaira et al. 2003; Rhoads
et al. 2004; Kurk et al. 2004; Stern et al. 2005a; Taniguchi et al. 2005 and
Kashikawa et al. 2006), 13 quasars (Fan et al. 2006; Willott et al. 2007) and 
one GRB 
(Haislip et al. 2006) with spectroscopic
redshift above six has been published. Another two detections of
gravitationally lensed Ly$\alpha$ emitters at redshifts around $z \sim 9$ have
been suggested in Stark et al. (2007). The reason for the easy detection
of {\ly}-emitters is the relative ease with which they are found
(narrow-band imaging) and with which the redshift can be confirmed
(spectroscopic follow-up and detection of single emission line). Hence, it is
often believed that {\ly}-emitters will be the most successful tool in
discovering even higher redshift sources. Some projects have already started.
Three attempts to observe Ly$\alpha$-emitters at
redshift $z \sim 9$ have been made (Parkes, Collins \& Joseph 1994; Willis
\& Courbin 2005; Cuby et al. 2006) but the results are null-detections due to
too small observed fields and/or to shallow flux limits. The 
DAzLE project has been presented by
Horton et al. (2004). It is designed to find Ly$\alpha$-emitters at redshift
$z = 7.73$ and $z = 8.78$, using a specially made instrument to be placed in
visitor focus at the VLT. The DAzLE instrument, however, has a small field of 
view ($6.83'$x$6.83'$). As these sources will be very under-luminous due to
the distance to them, and we thus sample only the top of the {\ly} luminosity
function, the key to finding a large sample of very high redshift
galaxies is large survey area and volume. This is the main strength of the
VISTA telescope.

There are two main reasons why we wish to observe very high redshift
Ly$\alpha$ emitters;
re-ionisation and galaxy formation studies. Re-ionisation is the epoch of the
history of our Universe when its vast amounts of hydrogen gas was ionised after
being completely neutral. This happened when the first stars, and potentially
the first quasars, lit and galaxies started to form. But it is yet
unknown exactly when this event took place. Recent results from studying the
cosmic microwave background (CMB) have constrained the upper limit of the
re-ionisation redshift to be $z_{\mrm{reion}} \lesssim 10$ and observations of
galaxies
and quasars have set the lower limit to be $z_{\mrm{reion}} \gtrsim 6.5$
(e.g. Spergel et al. 2006; Malhotra \& Rhoads 2004; Fan et al. 2006). Several
authors have argued that
Ly$\alpha$-emitting high-redshift galaxies
may provide a tool to constrain the redshift of re-ionisation better
(e.g. Miralda-Escud{\'e} 1998; Miralda-Escud{\'e} \& Rees 1998; Haiman
2002; Gnedin \& Prada 2004; Haiman \& Cen 2005). These authors show that
absorption in the Gunn-Peterson trough (i.e. the absorption on the short
wavelength side of the {\ly} line due to the intergalactic medium; Gunn \&
Peterson 1965) may extend to the red side of the
Ly$\alpha$ emission line and cause damping wings. Haiman (2002) and
Haiman \& Cen (2005) suggest that this damping will change the shape of the
luminosity function of Ly$\alpha$-emitters before and after re-ionisation.
McQuinn et al. (2007) suggest that the clustering of very high redshift
Ly$\alpha$ emitters is an independent and powerful tool to diagnose the
level of re-ionisation.
With a large sample of Ly$\alpha$-emitters at
redshift $z = 8.8$, it will be possible to analyse the Ly$\alpha$ emission
line, make a comparison of luminosity functions of Ly$\alpha$-emitters at 
different redshifts, and determine the level of clustering, thus revealing when 
re-ionisation happened, or at a minimum placing constraints on the time of 
re-ionisation.

When it comes to galaxy formation and evolution, possible observables include
number density of Ly$\alpha$-emitters, SFR and $\rho_{\mathrm{SFR}}$ and
clustering effects. An interesting question to answer is how many
Ly$\alpha$-emitters existed at a specific time? Taniguchi et al. (2005) study
the number densities of several Ly$\alpha$-surveys between
$z \approx 3.4$~to~6.6 and find no evolution with comoving number density.
Another key question related to galaxy formation and evolution is that of
the shape of the star formation history. Observations at lower
redshifts indicate that Ly$\alpha$-emitters are moderately star-forming, dust
free galaxies with little or no AGN content. It is thus of interest to
investigate if the SFR was different in this class of objects at very high
redshift, what percentage of the total star
formation happened in this type of galaxy at very high redshift and what the
star formation density was at this epoch.

\subsubsection{Emission-galaxies at intermediate redshift}
At lower redshift, ELVIS will also be sensitive to emission line
galaxies with much higher surface density; H$\alpha$ at $z = 0.80$, [OIII]
at $z = 1.36$, H$\beta$ at $z = 1.43$, and [OII] at $z = 2.17$. These
redshifts span most of the time when the cosmic star formation occurred,
associated with the formation of the stars and galaxies we see in the Universe
today. The [OIII]-,  H$\beta$-, and [OII]-lines can all be used as tracers of star
formation, and [OIII] and [OII] are also frequently used as
tracers of AGN and Seyfert galaxies. The line profiles and ratios of these
emission lines give information on the kinematics of the emitting gas, and of
the properties of the AGN (Heckman et al. 1981; Boroson 2005; Gu et al.
2006). The forbidden oxygen lines are metallicity dependent, but also
affected by AGN. Nevertheless, in particular [OII] is still a
good tracer of star formation and hence we will have an interesting handle
on the star formation density at $z = 2.2$, which is complementary to
broad-band surveys targeting similar redshifts (e.g. Adelberger et al. 2004).

\subsubsection{H$\alpha$ emitters}
H$\alpha$ is a one of the best, direct tracers of the
instantaneous star
formation rate and it is particularly useful as it is relatively unaffected by
metallicity and dust extinction.
Several authors have attempted to obtain the H$\alpha$ luminosity
function at redshifts close to $z = 1$, which is believed to be the peak
of star formation in the Universe, but sample sizes are still small
(e.g. Tresse et al. 2002; Doherty et al. 2006; Hopkins
et al. 2000; Yan et al. 1999). Evidence exists for
strong evolution in the SFR from redshift zero to one, by perhaps an order
of magnitude (Hopkins 2004, and references therein). However, many
surveys are limited to the bright end of the luminosity function due
to the small sample sizes and it
is therefore not clear whether this evolution in global SFR
is a property of galaxies with well above average SFRs, or whether it also
extends to average SFR galaxies.  The H$\alpha$ luminosity of
normal galaxies is
well studied in the local universe. Comparing the full range of
H$\alpha$ luminosities at higher redshifts to those of local galaxies
could therefore provide crucial insight into the
evolution of the SFR. Another question recently posed is that of the apparent
``downsizing'' of star forming galaxies (Cowie et al. 1996; Heavens et al.
2004;
Juneau et al. 2005), claiming that more massive galaxies form earlier than
less massive
galaxies. To date, no survey has simultaneously reached
the sensitivity and the area necessary to fully
address the questions of global SFR and ``down-sizing''. The sensitivity of 
ELVIS will reach
H$\alpha$ luminosities more than three orders of magnitude below L*(H$\alpha$).
This will enable unprecedented constraints to the H$\alpha$ luminosity
function at $z = 0.8$, in particular the value of the faint-end slope
which is still a matter of debate, hence giving a direct observational
evidence as to whether the ``down-sizing'' scenario is true or merely an
observational bias.

\subsection{Survey plan}
Ultra-VISTA has three parts; a ``deep'' survey, a ``very deep'' survey and
ELVIS. All parts will observe the COSMOS field 
(R.A.~$= 10^h 00^m 28.6^s$, Dec~$= +02^{\circ} 12' 21''$ (J2000)). The VISTA
camera array consists of 16 detectors, placed in a square pattern with large
gaps in between, see Fig.~\ref{fig:vistaarray}.
The gaps can be filled in by shifting the array in a six step pattern. A 
full mosaic will then be 1.6 sq.degrees. The broad-band components of
Ultra-VISTA will observe a full mosaic and four strips respectively. The
``deep'' survey will observe a full mosaic but the ``very deep'' survey
will only shift the detectors along the y-axis, creating four strips.

For ELVIS, it was debated for some time which configuration to choose; the 
continuous mosaic or the four strips. The total integration time was 
assumed to be fixed and so the full mosaic would be more shallow and 
the configuration with four strips would go deeper. Based on the expected
number counts and clustering results (see Chapter~\ref{chapter:predictions})
it was decided that ELVIS would also observe in four strips but to a deeper
flux limit. The expected depths reached by the different Ultra-VISTA 
surveys can be seen in Table~\ref{tab:uvdepths}.
\begin{table}[!t] 
\caption[Ultra-VISTA depths]{Expected flux depths from the Ultra-VISTA survey.}\label{tab:uvdepths}
\begin{tabular}{l|cccc|c|cccc}
\hline
& \multicolumn{4}{c}{\textbf{Deep Survey}}  & \multicolumn{1}{c}{\textbf{ELVIS}} & \multicolumn{4}{c}{\textbf{Ultra Deep Survey}}\\
\hline
Filter & Y & J & H & K$_s$ & NB1185 & Y & J & H & K$_s$\\
\hline
Total exp. time (h) & 48 & 48 & 48 & 48 & 180 & 320 & 320 & 320 & 320  \\
5$\sigma$ depth, AB & 25.7 & 25.5 & 25.1 & 24.5 & 24.1 & 26.7 & 26.6 & 26.1 & 25.6 \\
\hline
\end{tabular}
\end{table}
    
The survey is planned to start early 2008, with a first release to the public
about six months later, followed by yearly releases until the survey is 
completed. If the telescope operates at peak efficiency and Ultra-VISTA
is awarded all the time in its RA range, the program will be completed in 
five years.

\subsection{Expectations}\label{subsec:vistaexpect}
The expectations, i.e.~the expected number of objects detected, in ELVIS
is divided into two parts; Ly$\alpha$-emitters, and lower redshift emitters. 

\subsubsection{Ly$\alpha$ emitters}
The attempt to predict the number of Ly$\alpha$ emitters detected with 
ELVIS sparked the project presented in Chapter~\ref{chapter:predictions}, which 
presents results from two theoretical models of very high redshift Ly$\alpha$ 
emitters; one semi-analytic model based on $\lambda$CDM and one 
phenomenological model. The semi-analytic model GALFORM has been presented
in Le Delliou et al. (2005; 2006) and follows a hierarchical evolution
of galaxies. It has been shown to well reproduce the Ly$\alpha$ luminosity 
functions at lower redshifts. The phenomenological model has been presented in
Thommes \& Meisenheimer (2005) and assumes that Ly$\alpha$ luminosity is 
proportional to star formation rate, which in turn is proportional to the
baryonic mass of the galaxy. The results of comparing the two models and
extrapolating results at higher redshifts are described in 
Chapter~\ref{chapter:predictions}, and for ELVIS the models predict the
detection of 3 - 13 Ly$\alpha$ emitters.

\subsubsection{Lower redshift emitters}
To estimate the number of H$\alpha$-emitters that may be found, we use the
H$\alpha$
luminosity function of Tresse et al. (2002) with
H$_0 = 73$, $\Omega_m = 0.3$, $\Omega_{\lambda} = 0.7$ and without
reddening correction (Doherty et al. 2006).
Using this luminosity function in a field of view of 0.9~deg$^2$, with a line
flux limit
of $3.7 \times 10^{-18}$~erg~cm$^{-2}$~s$^{-1}$ and over a redshift range $z
= 0.802 - 0.820$ we calculate that of the order 3500
H$\alpha$-emitters would be detectable.
The expected density of intermediate redshift galaxies 
will probably be smaller than the density of H$\alpha$ emitters
because of the higher distance moduli and (on average) lower equivalent widths,
even though the higher redshift increases the observed equivalent widths.
We estimate that it will be possible to detect about an order of magnitude
fewer such
objects compared to H$\alpha$ emitters, hence of the order of 350 intermediate
redshift galaxies.

\chapter{Predicting results from very high redshift Ly$\alpha$ surveys}\label{chapter:predictions}
\markboth{Chapter 5}{Very high redshift Ly$\alpha$ emitters}
This paper has been accepted for publication in 
\emph{Astronomy \& Astrophysics} on 31 August, 2007. The authors are 
Nilsson, K.K., Orsi, A., Lacey, C.G., Baugh, C.M., \& Thommes, E.

\section{Abstract}
\emph{Context} 
Many current and future surveys aim to detect the highest redshift
($z \gtrsim 7$) sources through their Lyman-{\ly} (Ly$\alpha$) emission,
using the narrow-band imaging method. However, to date the surveys have only
yielded non-detections and upper limits as no survey has reached the necessary
combination of depth and area to detect these very young star forming
galaxies.

\noindent \emph{Aims} 
We aim to calculate model luminosity functions and mock surveys of
Ly$\alpha$ emitters at $z \gtrsim 7$ based on a variety of approaches
calibrated and tested on observational data at lower redshifts.

\noindent \emph{Methods} 
We calculate model luminosity functions at different redshifts based
on three different approaches: a semi-analytical model based on CDM, a
simple phenomenological model, and an extrapolation of observed Schechter
functions at lower redshifts.
The results of the first two models are compared with observations made at
redshifts $z \sim 5.7$ and $z \sim 6.5$, and they are then extrapolated to
higher redshift.

\noindent \emph{Results} 
We present model luminosity functions for redshifts between
$z = 7 - 12.5$ and give specific number predictions for future planned
or possible narrow-band surveys for Ly$\alpha$ emitters. We also investigate
what constraints future observations will be able to place on the
Ly$\alpha$ luminosity function at very high redshift.

\noindent \emph{Conclusion} 
It should be possible to observe $z = 7 - 10$ Ly$\alpha$ emitters
with present or near-future instruments if enough observing time is
allocated. In particular, large area surveys such as ELVIS (Emission
Line galaxies with VISTA Survey) will be useful in collecting a large
sample. However, to get a large enough sample to constrain well the $z
\geq 10$ Ly$\alpha$ luminosity function, instruments further in the
future, such as an ELT, will be necessary.

\section{Introduction}
One of the most promising ways of detecting very high redshift ($z
\gtrsim 5$), star-forming galaxies is via narrow-band imaging surveys
targeting Lyman-$\alpha$ (Ly$\alpha$).  In particular,
redshifts $z \sim 5.7$~and~$6.5$ have been extensively surveyed by
several groups (e.g. Ajiki et al. 2003; Hu et al. 2004; Shimasaku et
al. 2005; Ouchi et al. 2005, 2007; Malhotra et al. 2005; Taniguchi et
al. 2005; Tapken et al. 2006; Kashikawa et al. 2006).  The current
redshift record for a spectroscopically confirmed Ly$\alpha$ emitter
(LEGO -- Ly$\alpha$ Emitting Galaxy-building Object; see
M{\o}ller \& Fynbo 2001) is $z = 6.96$ (Iye et al. 2006) although
Stark et al.  (2007) have suggested the discovery of two LEGOs at $z =
8.99$~and~$9.32$.  The reason why narrow-band surveys are restricted
to a discrete number of narrow redshift windows is the night sky
OH emission lines. According to the OH line atlas of Rousselot et
al. (2000), at Ly$\alpha$ redshifts $z_{Ly\alpha} \gtrsim 7$ ($\lambda
\gtrsim 9800$~{\AA}) there are only a few possible wavelengths where a
narrow-band filter can fit in between the OH sky lines. These
correspond to $z_{Ly\alpha} \approx 7.7$,~$8.2$,~$8.8$,~$9.4$ and
$10.1 - 10.5$. Several future surveys will target these windows in the
sky aiming to detect very high redshift galaxies.  Three narrow-band
surveys for Ly$\alpha$ at redshift $z \sim 8.8$ have already been
completed (Parkes, Collins \& Joseph 1994; Willis \& Courbin 2005;
Cuby et al. 2007) but have only yielded upper limits. Future surveys
planned for these redshifts include DaZle (Dark Ages Z
Lyman-$\alpha$ Explorer, Horton et al. 2004) and ELVIS (Emission-Line
galaxies with VISTA Survey, Nilsson et al. 2006b).
Observations of very high redshift LEGOs have been proposed as
an excellent probe of reionisation, through its effects on the
Ly$\alpha$ emission line profile (e.g. Miralda-Escud{\'e} 1998;
Miralda-Escud{\'e} \& Rees 1998; Haiman 2002; Gnedin \& Prada 2004),
the luminosity function (e.g. Haiman \& Cen 2005; Dijkstra, Wyithe \&
Haiman 2007b) and the clustering of sources (McQuinn et
al. 2007).

We here focus on Ly$\alpha$ emission from star-forming
galaxies, where the Ly$\alpha$ photons are emitted from gas which is
photo-ionised by massive young stars.  During recent years,
theoretical work on Ly$\alpha$ emitting galaxies has made significant
progress. There are three main aspects to these studies: \emph{i)}
predicting the numbers of star-forming galaxies as a function of star
formation rate and redshift, \emph{ii)} calculating the fraction of the
Ly$\alpha$ photons which escape from galaxies into the IGM and \emph{iii)}
calculating the factor by which the Ly$\alpha$ flux is attenuated by
scattering in the IGM on its way to the observer. Accurate treatments
of \emph{ii)} and \emph{iii)} are complicated because Ly$\alpha$ photons are
resonantly scattered by hydrogen atoms, with the consequences that
absorption of Ly$\alpha$ by dust in galaxies is hugely amplified,
thereby reducing the escape fraction, and that even a small neutral fraction
in the IGM can be effective at scattering Ly$\alpha$ photons out of
the line-of-sight, thus attenuating the flux. Because of these
complications, most theoretical papers have chosen to concentrate on
only one aspect, adopting simplified treatments of the other two
aspects.  Haiman \& Spaans (1999) made predictions of the number
counts of Ly$\alpha$ emitting galaxies by combining the
Press-Schechter formalism with a treatment of the inhomogeneous dust
distribution inside galaxies. Barton et al. (2004) and Furlanetto
et al. (2005) calculated the numbers of Ly$\alpha$ emitters in
cosmological hydrodynamical simulations of galaxy formation, but did
not directly calculate the radiative transfer of Ly$\alpha$
photons. Radiative transfer calculations of the escape of Ly$\alpha$
photons from galaxies include those of Zheng \& Miralda-Escud{\'e} (2002),
Ahn (2004) and Verhamme, Schaerer \& Maselli (2006) for idealised
geometries, and Tasitsiomi (2006) and Laursen \& Sommer-Larsen (2007)
for galaxies in cosmological hydrodynamical simulations. The
transmission of Ly$\alpha$ through the IGM has been investigated by
Miralda-Escud{\'e} (1998), Haiman (2002), Santos (2004) and Dijkstra, Lidz
\& Wyithe (2007a), among others. Several authors (e.g. Haiman, Spaans \&
Quataert 2000; Fardal et al. 2001; Furlanetto et al. 2005) have studied the
effect of cold accretion to describe the nature of so-called Ly$\alpha$ blobs
(Steidel et al. 2000; Matsuda et al. 2004; Nilsson et al. 2006a), see also
sec.~\ref{sec:conclusion}.

Two models in particular, dissimilar in their physical assumptions,
have been shown to be successful in reproducing the observed number
counts and luminosity functions of Ly$\alpha$ emitting galaxies at
high redshifts: firstly, the phenomenological model of Thommes \&
Meisenheimer (2005)
which assumes that Ly$\alpha$ emitters are associated with the
formation phase of galaxy spheroids, and secondly the semi-analytical
model GALFORM (Cole et al. 2000, Baugh et al. 2005), which
follows the growth of structures in a hierarchical, $\Lambda$CDM
scenario. The GALFORM predictions for Ly$\alpha$ emitters are
described in detail Le Delliou et al.  (2005, 2006) and Orsi et
al. (in prep.), who show that the model is successful in reproducing
both the luminosity functions of Ly$\alpha$ emitting galaxies in the
range $3<z<6$ and also their clustering properties.

In this paper we aim to provide model predictions to help guide
the design of
future planned or possible narrow-band surveys for very high
redshift Ly$\alpha$ emitters.  We make predictions based on three
approaches: the semi-analytical and phenomenological models
already mentioned, and an extrapolation from observations at lower
redshift.  In section~\ref{sec:models} we describe the different
models used to make the predictions, and in section~\ref{sec:lf} we
present the predicted number counts and comparisons with observed
luminosity functions at lower redshifts.  In section~\ref{future} we
make number predictions for some specific future surveys. A
brief discussion regarding what can be learned from these future
surveys is found in section~\ref{constraints}. We give our
conclusions in section~\ref{sec:conclusion}.

\vskip 5mm
Throughout this paper, we assume a cosmology with $H_0=70$
km s$^{-1}$ Mpc$^{-1}$, $\Omega _{\rm m}=0.3$ and $\Omega
_\Lambda=0.7$, apart from the mock surveys discussed in
section~\ref{constraints}, which use GALFORM models matched to the
cosmology of the Millenium Run (Springel et al. 2005), (which has
$H_0=73$
km s$^{-1}$ Mpc$^{-1}$, $\Omega _{\rm m}=0.25$ and $\Omega
_\Lambda=0.75$).

\section{Models}\label{sec:models}
We use three different approaches to predict the numbers of high
redshift ($z>7$) Ly$\alpha$ emitters. The models
are based on very disparate assumptions.  The first model is the
semi-analytical model GALFORM (Le Delliou et al. 2005, 2006), the
second is the phenomenological model of Thommes \& Meisenheimer
(2005), and the third model is based on directly extrapolating from
observational data at lower redshifts.

Both the semi-analytical and phenomenological models assume that the
fraction of Ly$\alpha$ photons escaping from galaxies is constant, and
that the IGM is transparent to Ly$\alpha$. The simple expectation
is that before reionisation, the IGM will be highly opaque to
Ly$\alpha$, and after reionisation it will be mostly transparent. However,
various effects can modify this simple behaviour; e.g. Santos (2004)
finds that the transmitted fraction could be significant even before
reionisation, while Dijkstra, Lidz \& Wyithe (2007a) argue that
attenuation could be important even after most of the IGM has been
reionised. The WMAP 3-year data on the polarisation of the microwave
background imply that reionisation occurred in the range $z \sim 8-15$
(Spergel et al. 2007), i.e. the IGM may be mostly transparent to
Ly$\alpha$ at the redshifts of most interest in this paper. In any
case, what is important for predicting fluxes of Ly$\alpha$ emitters
is the product of the escape fraction from galaxies with the
attenuation by the IGM. The two effects are in this respect
degenerate.

\subsection{Semi-analytical model}
The semi-analytical model GALFORM (Cole et al. 2000; Baugh et
al. 2005), which is based on $\Lambda$CDM, has been shown to be
successful in reproducing a range of galaxy properties at both high
and low redshift, including Ly$\alpha$ emitters in the range $z = 3 -
6$ (Le Delliou et al. 2005; 2006). A full description of GALFORM
is given in these earlier papers, so we only give a brief summary
here. GALFORM calculates the build-up of dark halos by merging, and
the assembly of the baryonic mass of galaxies through both gas cooling
in halos and galaxy mergers. It includes prescriptions for two modes
of star formation -- quiescent star formation in disks, and starbursts
triggered by galaxy mergers -- and also for feedback from supernovae
and photo-ionisation. Finally, GALFORM includes chemical evolution of
the gas and stars, and detailed stellar population synthesis to
compute the stellar continuum luminosity from each galaxy consistent
with its star formation history, IMF and metallicity (see Cole et al. 2000
for more details). The unextincted
Ly$\alpha$ luminosity of each model galaxy is then computed from the
ionising luminosity of its stellar continuum, assuming that all
ionising photons are absorbed by neutral gas in the galaxy, with case
B recombination.

The semi-analytical approach then allows us to obtain the properties
of the Ly$\alpha$ emission of galaxies and their abundances as a
function of redshift, calculating the star formation histories for the
entire galaxy population, following a hierarchical evolution of the
galaxy host haloes. In addition, when incorporated into an N-body
simulation, we also obtain spatial clustering information. This model
has been incorporated into the largest N-body simulation to date, the
Millennium Simulation (Springel et al. 2005), to predict clustering
properties of Ly$\alpha$ galaxies. These results will be presented in
a forthcoming paper (Orsi et al., in prep.).

The version of GALFORM which we use here is the one described in
Baugh et al. (2005) and Le Delliou et al. (2006), with the same values
for parameters.  The parameters in the model were chosen in
order to match a range of properties of present-day galaxies, as well
as the numbers of Lyman Break and sub-mm galaxies at $z \sim
2-3$. We assume a Kennicutt IMF for quiescent star
formation, but a top-heavy IMF for starbursts, in order to reproduce
the numbers of sub-mm galaxies. The only parameter which has been
adjusted to match observations of Ly$\alpha$ emitters is the
Ly$\alpha$ escape fraction, which is taken to have a constant value
$f_{esc}=0.02$, regardless of galaxy dust properties. Le Delliou
et al. (2006) show that the simple choice of a constant escape
fraction $f_{esc}=0.02$ predicts luminosity functions of Ly$\alpha$
emitters in remarkably good agreement with observational data at
$3<z<6$. Le Delliou et al. (2006) also compared the predicted Ly$\alpha$
equivalent widths with observational data at $3<z<5$, including some model
galaxies with rest-frame equivalent widths of several 100\AA, and found broad
consistency.  For this reason, we use the same value $f_{esc}=0.02$
for making most of our predictions at $z>7$. However, since the value
of the escape fraction at $z>7$ is {\em a priori} uncertain in the
models (e.g. it might increase with redshift if high redshift galaxies are
less dusty) we also present some predictions for other values of
$f_{esc}$.

Reionisation of the IGM affects predictions for the numbers of
Ly$\alpha$ emitters in deep surveys in two ways: \emph{i)} feedback from
photo-ionisation inhibits galaxy formation in low-mass halos and \emph{ii)}
reionisation changes the opacity of the IGM to Ly$\alpha$ photons
travelling to us from a distant galaxy, as discussed above. GALFORM
models the first effect in a simple way, approximating reionisation as
being instantaneous at redshift $z_{reion}$ (see Le Delliou et
al. 2006 for more details). We assume $z_{reion}=10$, in line with the
WMAP 3-year results (Spergel et al. 2007). As was shown in Le Delliou
et al. (2006; see their Fig.~8), as far as the feedback effect is concerned,
varying $z_{reion}$ over the  range $7 \lesssim z_{reion}
\lesssim 10$ does not have much effect on the bright end of the
Ly$\alpha$ luminosity function most relevant to current and planned
surveys. For example, varying $z_{reion}$ between 7 and 10 changes the
predicted luminosity function at $L_{Ly\alpha} >
10^{41.5}$~erg~s$^{-1}$ by less than 10\% for $z \sim 7-10$.

\subsection{Phenomenological model}
The phenomenological model of Thommes \& Meisenheimer (2005; TM05
hereafter) assumes that the Ly$\alpha$ emitters seen at high redshift
are galaxy spheroids seen during their formation phase. We summarise
the main features here, and refer the reader to TM05 for more
details. The model is normalised to give the observed mass function of
spheroids at $z=0$, which is combined with a phenomenological function
that gives the
distribution of spheroid formation events in mass and redshift. Each
galaxy is assumed to be visible as a Ly$\alpha$ emitter during an
initial starburst phase of fixed duration (and Gaussian in time),
during which the peak SFR is proportional to the baryonic mass and
inversely proportional to the halo collapse time. The effects of the
IMF and the escape fraction on the Ly$\alpha$ luminosity of a galaxy
are combined into a single constant factor (i.e. the escape fraction
is effectively assumed to be constant). With these assumptions, the
luminosity function of Ly$\alpha$ emitters can be computed as a
function of redshift. The free parameters in the model were chosen by
TM05 to match the observed number counts of Ly$\alpha$ emitters at
$3.5<z<5.7$ (analogously to the choice of $f_{esc}$ in the GALFORM
model). This model does not include any effects from
reionisation.

\subsection{Observational extrapolation}
Our third approach is to assume that the Ly$\alpha$ luminosity
function is a Schechter function at all redshifts, following
\begin{equation}
\phi(L)dL = \phi^{\star}(L/L^{\star})^{\alpha}\mathrm{exp}(-L/L^{\star})dL/L^{\star}
\end{equation}
and to derive the Schechter parameters $\alpha$, $\phi^{\star}$ and
$L^{\star}$ at high redshifts by extrapolating from the observed
values at lower redshifts.  For our extrapolation, we use fits to
observations at redshift $z \approx 3$ (van Breukelen et al. 2005;
Gronwall et al. 2007; Ouchi et al. 2007), $z = 3.7$ (Ouchi et
al. 2007), $z = 4.5$ (Dawson et al. 2007), $z \approx 5.7$ (Malhotra
\& Rhoads 2004; Shimasaku et al. 2006; Ouchi et al. 2007) and $z
\approx 6.5$ (Malhotra \& Rhoads 2004; Kashikawa et al. 2006), as found
in Table~\ref{tab:lfdata}.
\begin{table}[!t]
\begin{center} 
\caption[Parameters of observed Schechter functions]{Parameters of the fitted Schechter function in previously published
papers. References are 1) van Breukelen et al. (2005), 2) Gronwall et al.
(2007), 3) Ouchi et al. (2007), 4) Dawson et al. (2007),
5) Malhotra \& Rhoads (2004), 6) Shimasaku et al. (2006), and
7) Kashikawa et al. (2006). References $3-6$
fit for three faint end slopes
($\alpha = -1.0, -1.5$~and~$-2.0$), but here we only reproduce the results for
fits with $\alpha = -1.5$ as we fix the slope in our calculations. Malhotra
\& Rhoads (2004) do not give error bars on the fits. Dawson et al. (2007)
fix the slope to $\alpha = -1.6$.}
\vspace{0.5cm}
\begin{tabular}{lccccccccc}
\hline
\hline
Ref & Redshift & $\alpha$ & $\log{\phi^{\star} \mathrm{Mpc}^{-3}}$ & $\log{L^{\star} \mathrm{ergs/s}}$ & \\
\hline
1 & $\sim 3.2$ & $-1.6$  & $-2.92^{+0.15}_{-0.23}$ & $42.70^{+0.13}_{-0.19}$ & \\
2 &       3.1  & $-1.49^{+0.45}_{-0.54}$ & $-2.84$ & $42.46^{+0.26}_{-0.15}$ & \\
3 &       3.1  & $-1.5$  & $-3.04^{+0.10}_{-0.11}$ & $42.76^{+0.06}_{-0.06}$ & \\
3 &       3.7  & $-1.5$  & $-3.47^{+0.11}_{-0.13}$ & $43.01^{+0.07}_{-0.07}$ & \\
4 &       4.5  & $-1.6$  & $-3.77^{+0.05}_{-0.05}$ & $43.04^{+0.14}_{-0.14}$ & \\
5 &       5.7  & $-1.5$  & $-4.0$ & $43.0$ & \\
6 &       5.7  & $-1.5$  & $-3.44^{+0.20}_{-0.16}$ & $43.04^{+0.12}_{-0.14}$ & \\
3 &       5.7  & $-1.5$  & $-3.11^{+0.29}_{-0.31}$ & $42.83^{+0.16}_{-0.16}$ & \\
5 &       6.5  & $-1.5$  & $-3.3$ & $42.6$ & \\
7 &       6.5  & $-1.5$  & $-2.88^{+0.24}_{-0.26}$ & $42.60^{+0.12}_{-0.10}$ & \\
\hline
\label{tab:lfdata}
\end{tabular}
\end{center}
\end{table}
We make linear fits to $\log\phi^{\star}$ and $\log L^{\star}$ {\em
vs} $z$, and extrapolate to higher redshift.  For simplicity,
we assume a fixed faint end slope of $\alpha = -1.5$. We do not make
any corrections for any possible effects of reionisation or IGM
opacity. The extrapolated values are given in Table~\ref{lfnew}.
\begin{table}[!t]
\begin{center}
\caption[Extrapolated Schechter function parameters]{Extrapolated parameters of the observed Schechter function at higher
redshifts. The faint end slope is fixed to $\alpha = -1.5$.}
\vspace{0.5cm}
\begin{tabular}{lccccccccc}
\hline
\hline
Redshift & $\log{\phi^{\star} \mathrm{Mpc}^{-3}}$ & $\log{L^{\star} \mathrm{ergs/s}}$ & \\
\hline
7.7  & $-3.73 \pm 0.50$ & $42.88 \pm 0.24$ & \\
8.2  & $-3.80 \pm 0.50$ & $42.89 \pm 0.24$ & \\
8.8  & $-3.88 \pm 0.50$ & $42.91 \pm 0.24$ & \\
9.4  & $-3.96 \pm 0.50$ & $42.92 \pm 0.24$ & \\
12.5 & $-4.38 \pm 0.50$ & $42.99 \pm 0.24$ & \\
\hline
\label{lfnew}
\end{tabular}
\end{center}
\end{table}

\section{Luminosity functions}\label{sec:lf}
The possible Ly$\alpha$ redshifts between $z = 7$ and $z=10$ where a
narrow-band filter can be placed are $z_{Ly\alpha} = 7.7, 8.2,
8.8,$~and~$9.4$. Redshifts beyond 10 are unreachable with ground-based
instruments of the near-future. However, one possibility for $z>10$
surveys may be the James Webb Space Telescope (JWST, see
section~\ref{jwst}) and so we also make predictions for redshift
$z_{Ly\alpha} = 12.5$.

First, we compare the Ly$\alpha$ luminosity functions predicted by the
semi-analytical (GALFORM) and phenomenological (TM05) models with
current observational data at $z \sim 6$. This comparison is shown in
Fig.~\ref{fig:oldplot}, where we compare the models with the cumulative
luminosity functions measured in several published surveys at $z=5.7$
and $z=6.5$.
We can see that both models match the observational data reasonably
well, once one takes account of the observational uncertainties. The
error bars on the observational data points, omitted in the plot in
order to not confuse the points, are large, at the bright end of the
luminosity function due to small number statistics, and at the faint
end due to incompleteness in the samples. The shallow slopes at the
faint ends of the Taniguchi et al. (2005) and Ouchi et al. (2007)
luminosity functions may be due to spectroscopic incompleteness. Both
models fit the observations well.
Hence we conclude that both of these models can be used to extrapolate
to higher redshifts.

We now have three methods of extrapolating to higher redshifts,
when the direct extrapolation of the Schechter
function from lower redshifts is included.  In Fig.~\ref{newplot} we plot the
predicted luminosity functions at $z=7.7$, 8.8 and 12.5 computed by
these three methods. For other redshifts, the curves may be
interpolated. For GALFORM, we show predictions for the standard value
of the escape fraction $f_{esc}=0.02$ in the left panel, and for a
larger value $f_{esc}=0.2$ in the right panel. This illustrates the
sensitivity of the predictions to the assumed value of $f_{esc}$ at
high redshift. The predictions from the other two models are plotted
identically in both panels, since they do not explicitly include the
escape fraction as a parameter.  GALFORM predictions for the numbers
of Ly$\alpha$ emitters at $z > 7$ were also given in Le Delliou et al
(2006).
We can see that the predictions from the different methods are fairly
similar at $z = 7.7$, but gradually diverge from each other with
increasing redshift.  For the highest redshift, $z = 12.5$, the TM05
model fails in producing a prediction due to numerical problems. We
note that making predictions for $z = 12.5$ is challenging, for
several reasons. Even though only $\sim 200$~Myrs separate the ages of
the Universe between redshift 8.8~and~12.5, the Universe went through
an important transition at this time as reionisation occurred
(Spergel et al. 2007). However, we do not know exactly how and when
this happened. Also, during this epoch the structure in the dark
matter (and hence also in galaxies) was building up very rapidly. This
underlines the interest of obtaining observational constraints at
these redshifts.

The hatched regions in Fig.~\ref{newplot} show the region of
the luminosity function diagram that has been observationally excluded
at $z = 8.8$ by Willis \& Courbin (2005) and Cuby et al. (2006). The
former survey was deeper but in a smaller area, whereas the latter was
more shallow over a larger area, hence the two-step appearance of the
hatched area. From the plot, it is obvious that their non-detections
are perfectly consistent with our theoretical models, although
the GALFORM model with the non-standard escape fraction $f_{esc}=0.2$
is marginally excluded.

\section{Future surveys}\label{future}
In this  section, we discuss more specific predictions for several
planned and possible future surveys. For all calculations, we assume a
simple selection on the flux of the Ly$\alpha$ emission line, with no
additional selection on the equivalent width (i.e. we include all
galaxies with $EW_{Ly\alpha} \ge 0$). We also assume no absorption by
the neutral hydrogen in the IGM
which would reduce the measured fluxes and for GALFORM predictions
we assume an escape fraction of $f_{esc} = 0.02$.  The predictions from the
GALFORM and TM05 models for these future surveys as well as some
published surveys are summarised in Table~\ref{tab:futnum}.

\subsection{DaZle -- Dark ages \emph{z} Lyman-{\ly} Explorer}
DaZle is a visitor mode instrument placed on the VLT UT3 (Horton et
al. 2004).  The instrument is designed to use a narrow-band
differential imaging technique, i.e. observing the same field with two
very narrow filters with slightly offset central wavelength. Objects
with Ly{\ly} in one of the filters can then be selected from the
differential image of both filters. The field-of-view of DaZle is
$6.83' \times 6.83'$ and it is expected to reach a flux level of $2
\times 10^{-18}$~erg~s$^{-1}$~cm$^{-2}$ (5$\sigma$) in 10 hours of
integration in one filter. This corresponds to a luminosity limit at
redshift $z = 7.7$ of $\log{(\mathrm{L}_{Ly\alpha})} =
42.13$~erg~s$^{-1}$.  The two initial filters are centred on
$z_{Ly\alpha} = 7.68$~and~$7.73$ (with widths $\Delta z=0.006$ and
0.025 respectively) and at this redshift, the surveyed volume becomes
$1340$~Mpc$^3$ per pointing per filter pair. Thus, from
Fig.~\ref{newplot}, we can conclude that DaZle will discover $\sim
0.16 - 0.45$ candidates at $z=7.7$ with one pointing and filter pair.

\subsection{ELVIS -- Emission Line galaxies with VISTA Survey}
ELVIS\footnote{www.astro.ku.dk/$\sim$kim/ELVIS.html} is part of
Ultra-VISTA, a future ESO Public Survey with
VISTA\footnote{www.vista.ac.uk} (Visible and Infrared Survey Telescope
for Astronomy). Ultra-VISTA is planned to do very deep near-infrared
broad- and narrow-band imaging in the COSMOS field. It will observe
four strips with a total area of 0.9~deg$^2$. The narrow-band filter
is focused on the $z_{Ly\alpha} = 8.8$ sky background window with
central wavelength $\lambda_c = 1185$~nm, and redshift width $\Delta
z=0.1$.  The flux limit of the narrow-band images is expected to reach
$3.7 \times 10^{-18}$~erg~s$^{-1}$~cm$^{-2}$ (5$\sigma$) after the
full survey has been completed. Ultra-VISTA will run from early 2008
for about 5 years and all the data will be public. ELVIS is presented
further in Nilsson et al. (2006b).  ELVIS will survey several
different emission-lines (e.g. H$\alpha$ at redshift $z = 0.8$, [OIII]
at redshift $z = 1.4$ and [OII] at redshift $z = 2.2$) as well as the
Ly{\ly} line.

When the survey is complete, the final mosaic will reach a Ly{\ly}
luminosity limit of $\log{(\mathrm{L}_{Ly\alpha})} =
42.53$~erg~s$^{-1}$. The volume surveyed will be $5.41 \times
10^5$~Mpc$^3$.  From Fig.~\ref{newplot} we see that ELVIS should be
expected to detect $3 - 20$ LEGOs at $z=8.8$.

\subsection{JWST}\label{jwst}
A possibility even further into the future is to use the James Webb
Space Telescope\footnote{www.jwst.nasa.gov} (JWST). JWST is scheduled
for launch in 2013 and will have excellent capabilities within the
near- and mid-infrared regions of the spectrum. Two of the instruments
aboard JWST could be used for narrow-band surveys; NIRCam, the
near-infrared camera, and TFI, the tunable filter imager (for a review
on JWST see Gardner et al. 2006).  NIRCam will have 31 filters, of
which nine are narrow-band filters. The filter with shortest
wavelength has central wavelength $\lambda_c = 1.644$~$\mu$m (F164N;
$z_{Ly\alpha} = 12.5$, $\Delta z = 0.135$). TFI will have tunable
filters with variable central wavelength, however it is only sensitive
at wavelengths larger than $\lambda \sim 1.6$~$\mu$m. NIRCam is
expected to reach a flux limit of $\sim 1 \times
10^{-18}$~erg~s$^{-1}$~cm$^{-2}$ (5$\sigma$) in $10000$~s of exposure
time. Hence, a flux limit of $\sim 5 \times
10^{-19}$~erg~s$^{-1}$~cm$^{-2}$ (5$\sigma$,
$\log{(\mathrm{L}_{Ly\alpha}(z = 12.5))} = 42.00$~erg~s$^{-1}$) could
be reached in 10 hours, assuming that the sensitivity is proportional
to the square root of the exposure time. TFI is expected to be able to
reach a flux limit almost a factor of two deeper in the same time,
however it has a field-of-view of only half of the NIRCam (which is $2
\times 2.16' \times 2.16'$). In one NIRCam pointing at redshift $z =
12.5$, approximately $1640$~Mpc$^3$ are surveyed. Again, from
Fig.~\ref{newplot}, we can estimate that we will detect $\lesssim 0.1$
galaxies per 10-hour pointing with NIRCam. However, the number of
detections depends strongly on the escape fraction which is unknown at such
high redshifts, and thus the number of detected galaxies can be larger.

\section{Constraints on the early Universe}\label{constraints}
Of the surveys at these redshifts that have been presented in previous
articles (Horton et al. 2004; Willis \& Courbin 2005; Cuby et al. 2006; Nilsson
et al. 2006b), or are conceivable (JWST, see
section~\ref{jwst}) only ELVIS will detect a large enough sample to
start to measure the luminosity functions and the extent of
reionisation at these redshifts and to study the fraction of
PopIII stars in the population. We here discuss these issues with
respect to ELVIS.

\begin{figure}[!t]
\begin{center}
\epsfig{file=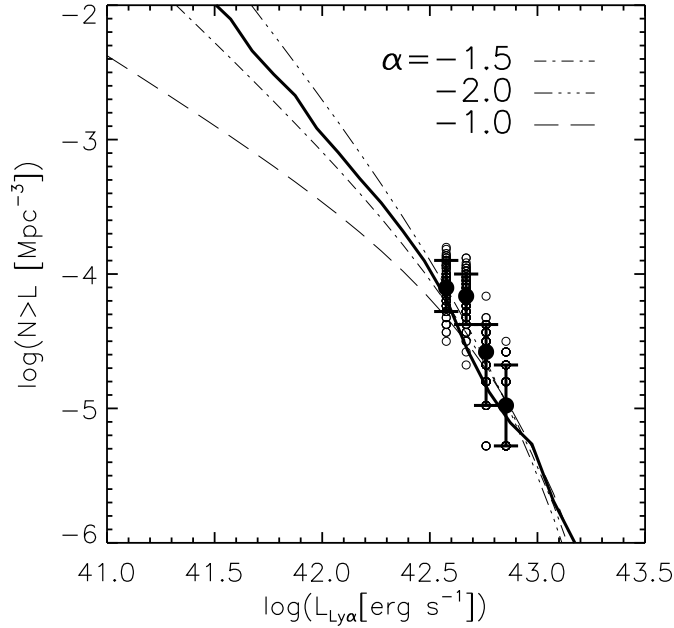,width=9.2cm}
\caption[Mock luminosity function for ELVIS]{Luminosity functions at $z=8.8$ for a set of mock ELVIS
surveys computed using GALFORM. The 112 mock surveys are identical
apart from being taken from different regions in the Millennium
simulation volume. The open circles show number counts in each mock
catalogue, in four luminosity bins. The black dot with error bars
shows the median of the mocks in each bin, with the error bars
showing the 10-90\% range. The thin lines are best fit Schechter
functions to the median points with different assumed faint end
slopes.  The thick solid line shows the ``true'' luminosity function,
as measured from model galaxies in the total Millenium simulation
volume.}
\label{cosvar}
\end{center}
\end{figure}

\begin{figure}[!t]
\begin{center}
\epsfig{file=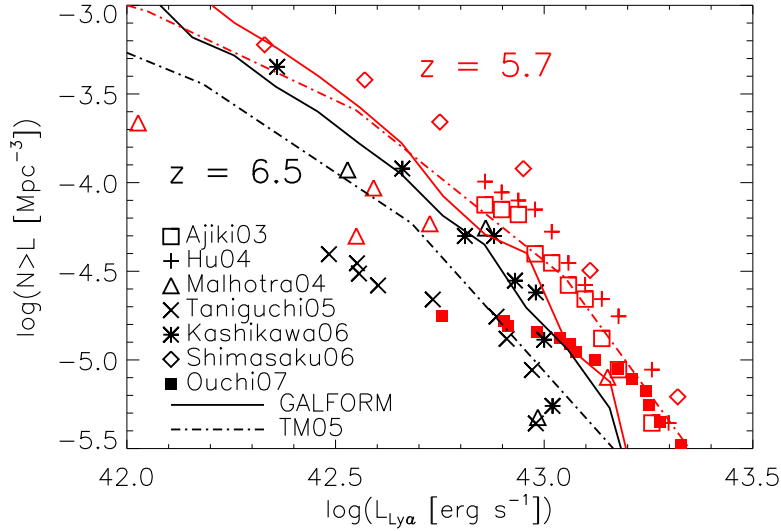,width=11.0cm} 
\caption[Plot of luminosity functions at redshifts $z = 5.7$ and
$6.5$]{Plot of luminosity functions at redshifts $z = 5.7$ and 
$6.5$. Red points and lines are at redshift $z = 5.7$, black
points/lines at redshift $z = 6.5$. Points are observations by Ajiki
et al. (2003; redshift 5.7, squares), Hu et al. (2004; redshift 5.7,
pluses), Taniguchi et al. (2005; redshift 6.5, crosses), Shimasaku et
al. (2006; redshift 5.7, diamonds), Kashikawa et al. (2006; redshift
6.5, stars), Malhotra \& Rhoads (2004; redshift 5.7 and 6.5,
triangles) and Ouchi et al. (2007; redshift 5.7, filled
squares). Solid lines show the GALFORM model (with escape fraction
$f_{esc}=0.02$), dot-dashed lines the TM05 model. Note that the Taniguchi et
al. (2005) and the Ouchi
et al. (2007) samples are the spectroscopic samples only.}
\label{fig:oldplot}
\end{center}
\end{figure}

From the semi-analytical modelling, we can make mock observations of
the ELVIS survey. The procedure to produce these catalogues is
explained in detail in Orsi et al. (in prep.), but the outline of the
process is that galaxies from GALFORM are placed in matching dark
matter haloes in the Millenium N-body simulation (Springel et
al. 2005), which is a cubical volume in a CDM universe of comoving
size 500~Mpc/$h$, thus creating a mock universe with simulated
galaxies which includes all the effects of clustering. We can then
make mock observations of this simulated Universe, including the
same limits on flux, redshift, sky area etc.~as for any real
survey, and from these observations produce mock galaxy
catalogues. From the mock catalogues, we can in turn make mock
luminosity functions of Ly$\alpha$ emitters at redshift $z = 8.8$. In
Fig.~\ref{cosvar} we plot the ``observed'' luminosity functions in the
112 mock catalogues taken from different regions of the Millenium
simulation volume. Note that for making these mock catalogues, GALFORM
was run with the same cosmological parameters as in the Millenium
simulation itself, which are slightly different from the
``concordance'' values assumed elsewhere in this paper, as described
in the Introduction (this is why the mean luminosity function for the
whole simulation volume which is plotted in Fig.~\ref{cosvar} is
slightly different from the GALFORM prediction for $z=8.8$ plotted in
Fig.~\ref{newplot}). We used escape fraction $f_{esc}=0.02$.
The figure shows that the spread in number density between the
different mock catalogues is large, almost a factor of ten in number
density in each luminosity bin. This is a consequence both of the
small numbers of galaxies in the mock surveys and of galaxy
clustering, which causes ``cosmic variance'' between different sample
volumes. The prediction from GALFORM is therefore that it will be
difficult to accurately measure the luminosity function of Ly$\alpha$
emitters at $z = 8.8$ even using the sample from the large area ELVIS survey. In
particular, there will be no useful constraint on the faint-end slope
$\alpha$. This is simply a consequence of the flux limit of narrow-band surveys,
i.e.~even if we use the median values of the luminosity function from
all the mocks, then Schechter functions with slopes in the range $-1$ to
$-2$ all give acceptable fits, as illustrated in
Fig.~\ref{cosvar}. However, if all the data are combined in one
luminosity bin, it should be possible to measure $\phi^{\star}$ assuming
values for $\alpha$ and $L^{\star}$. The possibility of including data points
from several surveys at different luminosities (e.g. also lensing surveys that
probe the faint end of the luminosity function) would also significantly
improve the results.

Two suggested methods of constraining reionisation from
observations of Ly$\alpha$ emitters, without requiring spectroscopy,
are to measure the clustering of Ly$\alpha$-sources and to compare the
Ly$\alpha$ and UV continuum luminosity functions at these redshifts
(Kashikawa et al. 2006; Dijkstra, Wyithe \& Haiman 2007b; McQuinn et
al. 2007). McQuinn et al. (2007) show that large HII bubbles may exist
during reionisation, and that these will enhance the observed
clustering of Ly$\alpha$ emitters in proportion to the fraction of
neutral hydrogen in the Universe. A sample of $\sim 50$ emitters will
be enough to constrain the level of reionisation using this effect
(McQuinn, priv. communication), almost within reach of the ELVIS
survey. A future, extended version of ELVIS would be able to place
very tight constraints on reionisation. In Kashikawa et
al. (2006) and Dijkstra, Wyithe \& Haiman (2007b) the use of the
combination of the UV and Ly$\alpha$ LFs to constrain the IGM
transmission is explored. Ly$\alpha$ emission will be much more
susceptible to IGM absorption than the continuum emission and thus the
ratio between the two LFs will give information on the level of
IGM ionisation.  However, with increasing redshift for Ly$\alpha$, the
continuum emission will be increasingly difficult to observe, and it is
unclear if this method will be feasible for surveys such as ELVIS.

It is possible that galaxies at $z = 8.8$ still have a
significant population of primordial PopIII stars. A test for the
fraction of primordial stars is the amount of HeII~1640~{\AA} emission
(Schaerer 2003; Tumlinson, Schull \& Venkatesan 2003). Depending on
models, these authors predict that the HeII~1640~{\AA} emission line
should have a flux between $1 - 10$~\% of the flux in the Ly$\alpha$
line. For ELVIS $z = 8.8$ Ly$\alpha$ emitters, the HeII~1640~{\AA}
line is redshifted to $1.61$~$\mu$m. Due to the many OH sky emission
lines in this region of the spectrum, it would be desirable to try to
observe the HeII~1640~{\AA} line from a space-based observatory such
as JWST. According to the JWST homepage, NIRSpec will achieve a
sensitivity in the medium resolution mode on an emission line at
$1.6$~$\mu$m of $\sim 7 \times 10^{-19}$~erg~s$^{-1}$~cm$^{-2}$ ($10\sigma$)
for an exposure time of $10^5$~s (30
hours). Thus, if the Ly$\alpha$ emission line has a flux of $\sim 5
\times 10^{-18}$~erg~s$^{-1}$~cm$^{-2}$, the HeII~1640~{\AA} will be
marginally detected with JWST in 30 hours of integration, depending on
the ratio of HeII~1640~{\AA} to Ly$\alpha$ flux. The NIRSpec
sensitivity increases at longer wavelengths, but the increasing
luminosity distance to galaxy candidates with HeII~1640~{\AA} emission
at longer wavelengths will most likely counteract this effect.

\section{Discussion}\label{sec:conclusion}
\begin{figure}[!pt]
\vspace{-0.5cm}
\begin{center}
\epsfig{file=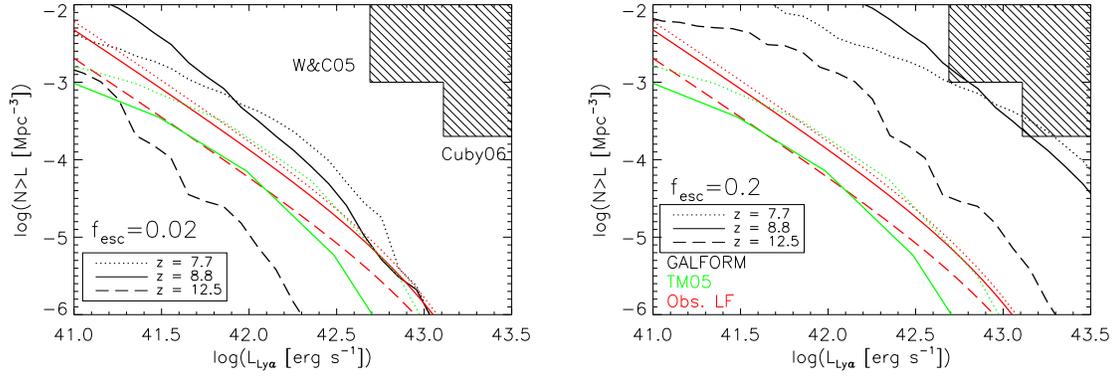,width=15cm}
\caption[Predicted Ly$\alpha$ luminosity functions at $z>7$]{Predicted Ly$\alpha$ luminosity functions at $z>7$. Red lines
are extrapolations from observed luminosity functions at lower
redshift, green lines are TM05 models and black lines are GALFORM
models. Different linestyles show different redshifts $z=7.7$, 8.8 and
12.5. No prediction is shown for the TM05 model at $z=12.5$. Hatched
area marks observational upper limits from Willis \& Courbin (2005)
and Cuby et al. (2006), both at redshift $z = 8.8$. In the left panel,
the GALFORM predictions are shown for escape fraction $f_{esc}=0.02$
(our standard value), while in the right panel, they are shown for
$f_{esc}=0.2$. The predictions from the other two methods are
identical in both panels.}
\label{newplot}
\end{center}
\vspace{-0.5cm}
\end{figure}

\begin{figure}[!pb]
\vspace{-0.3cm}
\begin{center}
\epsfig{file=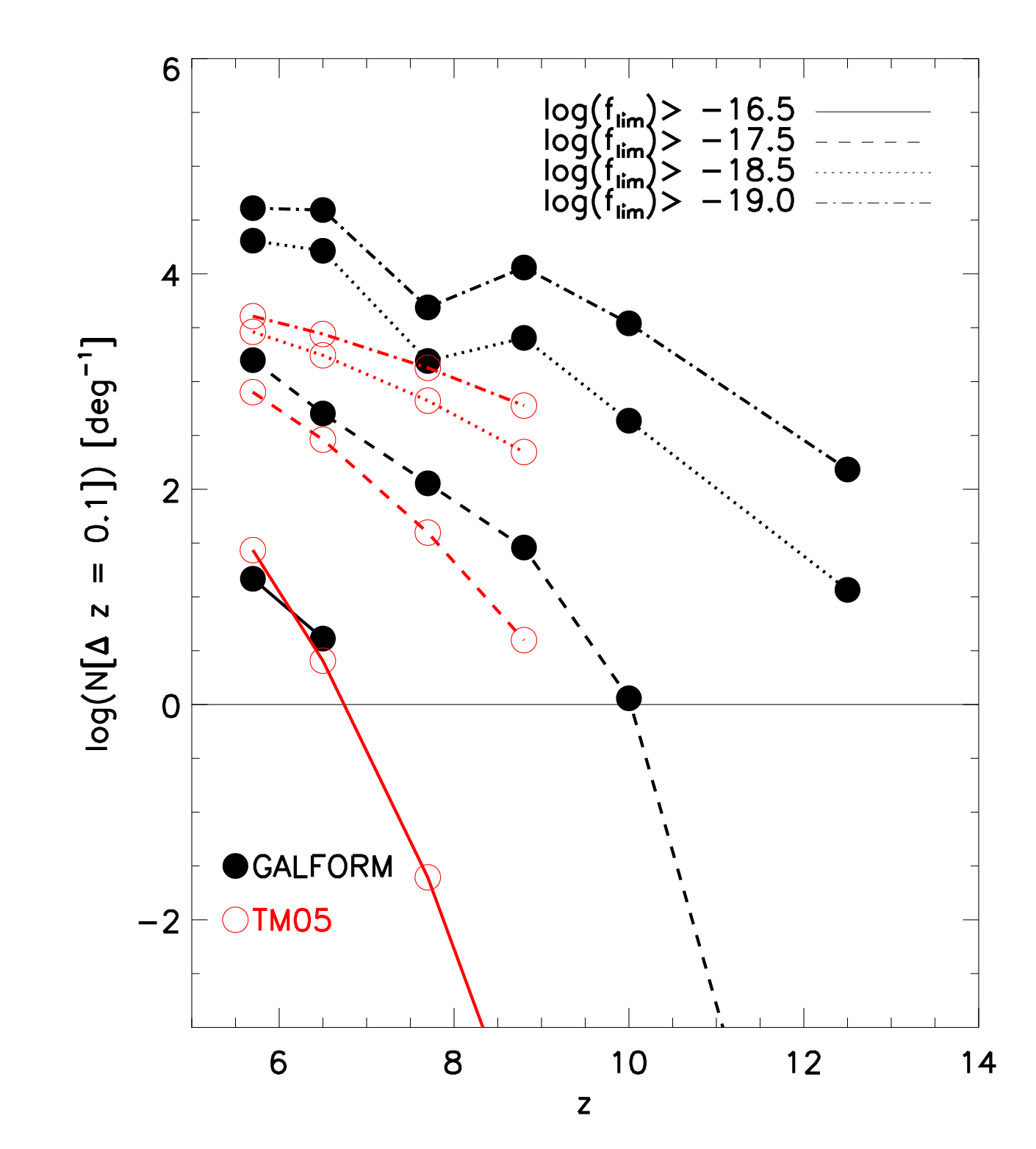,width=7.2cm}
\caption[Summary of predictions]{Summary of predictions. The plot shows the number of
Ly$\alpha$ emitting galaxies expected per square degree per redshift
interval $\Delta z = 0.1$ as a function of redshift and observed flux
limit. The predictions of GALFORM are shown in black and of the TM05
model in red. The different line styles are for different flux limits.
}
\label{ngal}
\end{center}
\vspace{-0.5cm}
\end{figure}

We summarise our predictions for number counts of Ly$\alpha$ emitters
in narrow-band surveys in Fig.~\ref{ngal}.
We also summarise the numbers of detected objects for specific current
and future surveys in Table~\ref{tab:futnum}.
\begin{table}[!t]
\begin{center}
\caption[Summary of present and future surveys]{Number of predicted/observed objects per observed field in
several present and future surveys from two theoretical models.  Data
from Subaru XMM Deep Field (SXDS) are from Ouchi et al. (2005),
Shimasaku et al. (2006) and Kashikawa et al. (2006). GALFORM
predictions are made assuming an escape fraction of $f_{esc} = 0.02$.}
\vspace{0.5cm}
\begin{tabular}{lccccccccc}
\hline
\hline
Name & $z$ & Area       & Luminosity limit & GALFORM & TM05 & Observed &  \\
     &          & (arcmin$^2$)                       & (5$\sigma$, erg~s$^{-1}$) & & & number & \\
\hline
SXDS-O       & 5.7  & 8100 & $10^{42.40}$ &  443    & 339   & 515 &  \\
SXDS-S   & 5.7  & 775  & $10^{42.40}$ &  112    & 86   & 83  &  \\
SXDS-K    & 6.5  & 918  & $10^{42.27}$ &  108    & 57   & 58   &  \\
DaZle              & 7.7  & 47   & $10^{42.13}$ &  0.45   & 0.16  & --- &  \\
ELVIS              & 8.8  & 3240 & $10^{42.50}$ &  20     & 2.8    & --- &  \\
Cuby06             & 8.8  & 31   & $10^{43.10}$ &  0.0003 & 0.0  & 0   &  \\
W\&C05             & 8.8  & 6.3  & $10^{42.25}$ &  0.015  & 0.002  & 0   &  \\
JWST               & 12.5 & 9.3  & $10^{42.00}$ &  0.018  & ---   & --- &  \\
\hline
\label{tab:futnum}
\end{tabular}
\end{center}
\end{table}
A few comments can be made on the differences in predictions between
the two models.  Firstly, as can be seen in Fig.~\ref{ngal} and also
Fig.~\ref{newplot}, the luminosity functions have steeper faint-end
slopes in the GALFORM models than in the TM05 models. Secondly, the
GALFORM and TM05 models predict similar amounts of evolution at a
given flux over the range $z=6-9$ where they can be compared.

Several factors enter into the error bars of our predictions. One
problem is the uncertainties in, and disagreement between, the
observed lower redshift luminosity functions which are used to
calibrate the theoretical models. There are many caveats in producing
Ly$\alpha$ luminosity functions, of which the selection function is
the most difficult to correct for.  The problem arises from that the
filter transmission curve is not box-shaped, but rather
gaussian. Thus, only brighter objects will be observed at the wings of
the filter, and these will be observed to have smaller than intrinsic
luminosities. Secondly, the equivalent width (EW) limit that the
survey is complete to depends on the depth of the broad-band images
used for the selection. Thirdly, if the sample is a photometric
sample, it is possible that there are lower redshift interlopers,
where the emission line is e.g. [OII], in the sample. Finally, the
samples are still so small that we have to deal with small number
statistics.  All of these problems cause the observed luminosity
function at lower redshifts to be uncertain.

Both theoretical models (semi-analytical and phenomenological) have
uncertainties resulting from how they model the galaxy formation
process, and also from the assumption that the fraction of Ly$\alpha$
photons escaping from galaxies is constant and does not change with
redshift. In addition, neither model includes attenuation of the
Ly$\alpha$ flux due to neutral hydrogen in the IGM. This attenuation
would be expected to be strong at $z>z_{reion}$, when the IGM is
neutral, and weaker at $z<z_{reion}$, when most of the IGM is
ionised. The degree of attenuation depends on a number of different
effects, as analysed in Santos (2004), and discussed in Le Delliou et
al. (2006), and is currently very uncertain. Nonetheless, this
attenuation is expected to produce observable effects on the evolution
of the Ly$\alpha$ luminosity function, if reionisation occurs within
the redshift range covered by future observations, and so estimating
the reionisation redshift and the neutral fraction after reionisation
are included in the science goals of these surveys.

It is apparent that the key to acquiring a large sample of
Ly$\alpha$-emitting galaxies at redshifts greater than 7 is both depth
and area. In a recent paper, Stark, Loeb \& Ellis (2007) suggest that
one of the most efficient means of finding very high redshift
Ly$\alpha$ emitters is through spectroscopic surveys focused on
gravitational lensing clusters. Lensing surveys could easily reach
down to a luminosity limit of $10^{40.5}$~erg~s$^{-1}$ in a few tens
of hours. However, the surveyed volumes are very small, of the order
of a hundred Mpc$^3$. For a lensed survey, the area in the source
plane is reduced by the same factor that the flux is amplified, so in
principle one gains in the total number of objects detected relative
to an unlensed survey if the luminosity function is steeper than
$N(>L) \propto L^{-1}$. In the GALFORM and TM05 models, the asymptotic
faint-end slope is shallower than this, but at higher luminosities,
the slope can be steeper. For example, GALFORM predicts that at
$z=10$, the average slope in the luminosity range
$10^{41}$--$10^{42}$~erg~s$^{-1}$ is close to $N(>L) \propto L^{-2}$
(see Fig.~8 in Le~Delliou et al. 2006), so that a lensing amplification
of 10 results in 10 times more objects being detected, with intrinsic
luminosities 10 times lower, compared to an unlensed survey with the
same area and flux limit.
Therefore lensing and narrow-band surveys are complementary to each
other as they probe different parts of the luminosity function.  With
either type of survey, reaching a significant sample of redshift $z
\sim 7 - 8$ should be possible in the next few years with
telescopes/instruments in use or soon available.

An interesting type of object found recently in narrow-band
surveys are the Ly$\alpha$ blobs, large nebulae with diameters up to
150 kpc and Ly$\alpha$ luminosities up to $10^{44}$~erg~s$^{-1}$ with
or without counterpart galaxies (e.g. Steidel et al. 2000; Matsuda et
al. 2004; Nilsson et al. 2006a). Several mechanisms have been proposed
to explain this phenomenon, including starburst galaxies and
superwinds, AGN activity or cold accretion. It is interesting to
consider if such objects would be detected in any of these surveys,
assuming they exist at these redshifts.  A typical Ly$\alpha$ blob
will have a luminosity of $\sim 10^{43}$~erg~s$^{-1}$ and a radius of,
say, 25~kpc. This will result in a surface brightness of $\sim 5
\times 10^{39}$~erg~s$^{-1}$~kpc$^{-2}$. Thus, a narrow-band survey
will have to reach a flux limit, as measured in a $2''$~radius
aperture of $\sim 1.3 \times 10^{42}$~erg~s$^{-1}$ at redshift $z =
8.8$, corresponding to $\log{L} = 42.11$. (An aperture radius of
around $2''$ is expected to be roughly optimal for signal-to-noise.)
For lower or higher redshifts, this limit is higher or lower
respectively. Thus, ELVIS will not be able to detect Ly$\alpha$ blobs
unless they are brighter and/or more compact at higher redshift than a
typical blob at lower redshift. DaZle and JWST could in principle
detect this type of object, but only if they are very abundant in the
very high redshift Universe, due to the small survey volumes of these
instruments. It is of course highly uncertain what properties such
Ly$\alpha$ blobs would have at $z \sim 7 - 9$, or their space density,
but it appears unlikely that the future surveys presented here would
detect any such objects.

To find compact Ly$\alpha$ emitters at redshifts $z \gtrsim 10$ in
significant numbers we will probably have to await instruments even
further in the future. If a future 40-m ELT (Extremely Large
Telescope) was equipped with a wide-field NIR imager and a
narrow-band filter of similar width to ELVIS, it could reach a
luminosity limit of $L \sim 10^{41.2}$~erg~s$^{-1}$ at redshift $z =
10.1$ (where a suitably large atmospheric window exists) in
approximately 20 hours. Using the GALFORM model for $z = 10$, the
number density should be N($>$L)$\approx 4 \times 10^{-3}$~Mpc$^{-3}$
at this luminosity limit. Thus, to get a sample of ten Ly$\alpha$
emitters would require imaging an area on the sky of approximately
16~square arcminutes, assuming a narrow-band filter with
redshift range $10.05 < z_{\mathrm{Ly}\alpha} < 10.15$. This could be
achieved with one pointing if the detector has a field-of-view of
6~arcmin on a side, as suggested by the ESO ELT Working
Group\footnote{http://www.eso.org/projects/e-elt/Publications/ELT\_INSWG\_FINAL\_REPORT.pdf}. It
should of course be noted that these are very tentative numbers, but
they display the possibilities of far future instruments.

\chapter{Selection methods for Ly$\alpha$ emitters}\label{chapter:fundamental}
\markboth{Chapter 6}{Selection methods for Ly$\alpha$ emitters}
This paper is in preparation. The authors will be Nilsson, K.K., M{\"o}ller, 
O., Hayes, M., M{\o}ller, P., \& Fynbo, J.P.U.

\section{Abstract}
\emph{Context} Narrow-band surveys for Ly$\alpha$ emitters is an increasingly 
popular method
to find high redshift galaxies. However, different groups have presented
different methods to find Ly$\alpha$ galaxy candidates. We here present
a method to determine the best selection criteria, combining narrow- and
broad-band observations.

\noindent \emph{Aims} To find the optimal selection sub-space in colour-colour 
space for Ly$\alpha$ emitters,
i.e. to determine the best colour selection criteria for this type of object.

\noindent \emph{Methods} We simulate a galaxy population by constructing a 
large number of galaxy
spectra and convolve with filter profiles
to get the ``observed'' fluxes. Some spectra have Ly$\alpha$ emission
superposed, which is then observed through a narrow-band filter profile.
We study the distribution in $N$ dimensional colour space and attempt to find the 
angle
in which the scatter in the normal galaxy population is the least and the
emission-line candidates are most easily and most accurately selected.

\noindent \emph{Results} We present the three most favourable combinations of galaxy 
colours to be used to select \emph{i)} Ly$\alpha$ emitters from field galaxies 
and \emph{ii)} Ly$\alpha$ emitters from [OII] emitters. We also present the 
optimal selection criteria to use with these colours.

\noindent \emph{Conclusions} All of the best combinations of colours for
selecting Ly$\alpha$ emitters from field galaxies include sampling the 
UV slope of the galaxy with two broad-band filters, one on either side of
the Ly$\alpha$ line. It is concluded that selection with two 
narrow-band/broad-band colours is superior to selection with only one 
narrow-band/broad-band colour.

\section{Introduction}
Narrow-band imaging surveys have become increasingly successful in 
discovering high and very high redshift galaxies. There are now
in total several hundreds of spectroscopically confirmed Ly$\alpha$ emitters
at redshifts $z \sim 3$ (e.g. Steidel et al. 2000; Fynbo et al. 2003; Matsuda
et al. 2005; Venemans et al. 2007; Nilsson et al. 2007), $z \sim 4.5$
(Finkelstein et al. 2007),
$z \sim 5.7$ (Malhotra et al. 2005; Shimasaku et al. 2006; Tapken et al. 2006)
and $z \sim 6.5$ (Taniguchi et al. 2005; Kashikawa et al. 2006). Most
candidates are selected using one narrow-band filter and one broad-band filter,
where objects that are bright in the narrow-band filter but comparatively
faint or non-detected in the broad-band filter are selected. In some cases,
the selection is based on determining the equivalent width (EW) of the
potential emission line. However, the calculation of the EW of an
emission-line observed through a narrow-band/broad-band filter set-up has
been shown to be uncertain (Hayes \& {\"O}stlin 2006). In 1993,
M{\o}ller \& Warren (1993) suggested for the first time
that two colours should be used for detecting emission-line galaxies, i.e.
that any object should be observed with a narrow-band filter and two
broad-band filters. This method was subsequently explored in Fynbo et al.
(2003) and Nilsson et al. (2007). However, present day and future
multi-wavelength surveys such as e.g. the GOODS
(Giavalisco et al. 2004) and COSMOS (Scoville et al. 2006) surveys warrant
a review on the selection criteria for narrow-band surveys in fields
where public, multi-wavelength data exists.

This \emph{Letter} is organised as follows; in section~\ref{sec:funmeth} we 
present the models we use to create colour plots for ``normal'' galaxies and
Ly$\alpha$ emitters as well as the method by which we find the most
efficient selection colours. In section~\ref{sec:funres} we present our results 
and in sections~\ref{sec:disc} we discuss the results and draw conclusions.

\vskip 5mm
Throughout this paper, we assume a cosmology with $H_0=72$
km s$^{-1}$ Mpc$^{-1}$ (Freedman et al. 2001), $\Omega _{\rm m}=0.3$ and
$\Omega _\Lambda=0.7$. Magnitudes are given in the AB system.

\section{Method}\label{sec:funmeth}
\subsection{Creating a mock sample}
To study which selection criteria are most successful in determining
Ly$\alpha$ emitter candidates, we create model spectra using the Starburst99
models (Leitherer et al. 1999; V{\'a}zquez \& Leitherer 2005). For the
``normal'' galaxy population,
we create $50'000$ spectra, evenly covering the parameter space given in
Table~\ref{tab:input}.
\begin{table}[t]
\begin{center}
\caption[Parameter space for field galaxy population]{Parameter space for field galaxy population. }
\vspace{0.5cm}
\begin{tabular}{@{}lcccccc} 
\hline
\hline
          & Min. value & Max. value & Number of steps & \\
\hline
Redshift (\emph{z})   &  0   &   3  & 50 &  \\
Dust (E(B-V))         &  0   &   1  & 20 &  \\
Age (Gyr)             &  0.001   &   10  & 50 (log)&  \\
\hline
\label{tab:input}
\end{tabular}
\end{center}
\end{table}
The age of each model galaxy is constrained to be less than the age of the
Universe at the redshift it is at. The mass of a galaxy is a simple
multiplication factor and as we here only work with colours, i.e. flux ratios,
the mass is irrelevant except in the cases when the mass is so
small that the flux in a certain waveband is lower than the flux detection
limit. We discuss this further in sec.~\ref{sec:disc}. All spectra are
calculated using single stellar populations (SSP), i.e. all the stars are
created instantaneously. This will cause the spectrum to be blue to the extreme
at a young age and red to the extreme when old, thus widening the distribution of 
the field galaxy
population with respect to other more extended SFHs. Particularly at late times
colours will be redder, with fewer hot stars present in the population.

For the Ly$\alpha$ emitters, we create galaxy spectra with the same range of
ages and dust amounts as for the field galaxies and redshift them to the
redshift in question. We then add Ly$\alpha$ by calculating the expected
H$\alpha$ EW from
the age of the starburst, convert this to a Ly$\alpha$ EW using case B
recombination (Brocklehurst 1971) and add such a line to the model spectra.
The Ly$\alpha$ EWs range between $0 - 240$~{\AA} (Charlot \& Fall 1993). As we 
are also interested in
distinguishing between Ly$\alpha$ emitters and interloper galaxies such
as [OII]-emitters, we also add an
[OII]-emission line to a sample of galaxy spectra with EWs ranging between
$0 - 100$~{\AA}, as [OII] is the most
common interloper in Ly$\alpha$ emission surveys. We create $50'000$
Ly$\alpha$ emitters and [OII] emitters respectively. We wish to decide the
optimal selection criteria for a variety of Ly$\alpha$ redshifts, see
Table~\ref{tab:redshifts}, that have been or will be surveyed.
\begin{table}[t] 
\begin{center}
\caption[Redshifts for Ly$\alpha$]{Redshifts used for Ly$\alpha$
emitters. The redshifts are selected to represent a wide range of redshifts
that have or will be surveyed. Representative references are given. }
\vspace{0.5cm}
\begin{tabular}{@{}llccccc}
\hline
\hline
Redshift (\emph{z}$_{Ly\alpha}$) & Reference (e.g.) & \\
\hline
2.4   &  Francis et al. 2004 &  \\
3.1   &  Nilsson et al. 2007 &  \\
4.5   &  LALA; Finkelstein et al. 2007 &  \\
5.7   &  Shimasaku et al. 2006 &  \\
6.5   &  Kashikawa et al. 2006 &  \\
7.7   &  DaZle; Horton et al. 2004 &  \\
8.8   &  ELVIS; Nilsson et al. 2006b &  \\
\hline
\label{tab:redshifts}
\end{tabular}
\end{center}
\end{table}

\begin{figure}[!t] 
\begin{center}
\epsfig{file=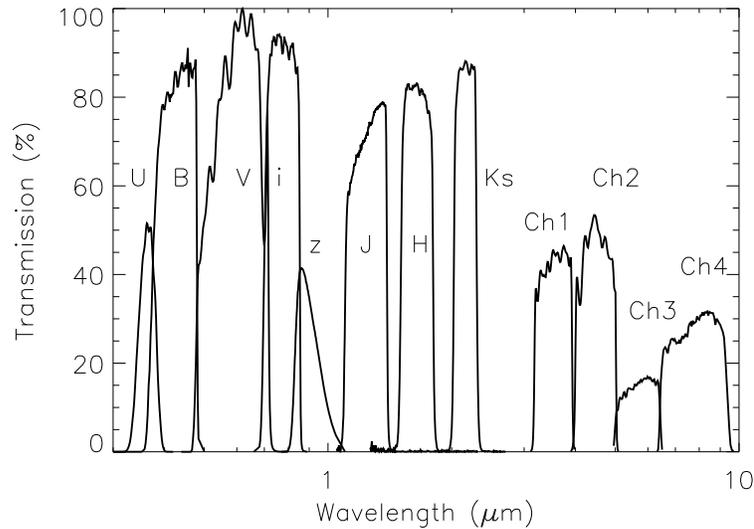,width=11.0cm}
\caption[GOODS-S data-set]{Transmission of selection filters from the GOODS-S data-set.}
\label{fig:filters}
\end{center}
\end{figure}

For each galaxy spectrum, we convolve the spectrum with a set of standard
filter transmission curves in order to calculate magnitudes and colours.
The filters used correspond, for broad-bands, to the filter set used in the
GOODS Surveys, see Table~\ref{tab:filters} and Fig.~\ref{fig:filters}, and are
very similar to standard
Johnson-Cousins filters. The narrow-band filters are defined to
be gaussian shaped, with central wavelength centred on the particular
Ly$\alpha$ redshift desired and with full-width half-maximum (FWHM) of
1\% of the central wavelength.
\begin{table}[t] 
\begin{center}
\caption[GOODS-S filters]{Filters used in this paper. These filters are used
in the convolution with the synthetic spectra in order to calculate fluxes and
colours in typical observed bands.}
\vspace{0.5cm}
\begin{tabular}{@{}lcccccc}
\hline
\hline
Filter name & Central Wavelength & FWHM & \\
\hline
$U$ (\emph{ESO 2.2-m})       &  3630 \AA    &  760 \AA     &\\
$B$ (F435W, \emph{HST})      &  4297 \AA    &  1038 \AA    &\\
$V$ (F606W, \emph{HST})      &  5907 \AA    &  2342 \AA    &\\
\emph{i} (F814W, \emph{HST}) &  7764 \AA    &  1528 \AA    &\\
$z'$ (F850LP, \emph{HST})    &  9445 \AA    &  1230 \AA    &\\
$J$ (\emph{VLT})             &  1.25 $\mu$m &  0.6 $\mu$m  &\\
$H$ (\emph{VLT})             &  1.65 $\mu$m &  0.6 $\mu$m  &\\
$K_s$ (\emph{VLT})            &  2.16 $\mu$m &  0.6 $\mu$m  &\\
$Ch1$ (\emph{Spitzer})       &  3.58 $\mu$m &  0.75 $\mu$m &\\
$Ch2$ (\emph{Spitzer})       &  4.50 $\mu$m &  1.02 $\mu$m &\\
$Ch3$ (\emph{Spitzer})       &  5.80 $\mu$m &  1.43 $\mu$m &\\
$Ch4$ (\emph{Spitzer})       &  8.00 $\mu$m &  2.91 $\mu$m &\\
\hline
\label{tab:filters}
\end{tabular}
\end{center}
\end{table}

\subsection{Optimal distinction of Ly$\alpha$ emitters}
In past work (e.g. Fynbo et al. 2003; Nilsson et al.~2007) a distinction 
between Ly$\alpha$ emitting candidates and non-emitters was performed 
by making a colour cut in two narrow- minus broad-band colours. The reasoning
was that, due to the emission in the narrow-band (NB) filter, the difference
between Ly$\alpha$ emitters
and non-emitters should be maximal. However, this was based on a
purely qualitative argument, rather than a careful analysis. In
this \emph{Letter} we want to perform a more quantitative study using the
evolutionary models developed. The data at hand are three populations of
galaxies with, in total, 78 colour combinations created from mock magnitudes
in the bands presented in Table~\ref{tab:filters} combined with a NB filter.
Thus, we wish to discriminate between three populations of galaxies in a
78 dimensional space. The goal is to find two colours in which the populations
can be plotted where the largest separation between the three populations
occur. We choose two colours as this allows a significant number of
combinations - more than one colour - but does not require a large amount of
additional data for the selection.

For each combination of colour, the catalogues of galaxies is smoothed using
a gaussian filter onto
a $100 \times 100$ grid with a step size of $0.1$ in magnitude difference. We
then optimise the selection of Ly$\alpha$ emitters by
calculating a ``merit'' number $M$ which is the ratio of distance between the
mean colour of each population over the area of the overlap between the
populations, see Eq.~\ref{eq:merit}.
\begin{equation}\label{eq:merit}
M = \frac{<C_{Ly\alpha}> - <C_{FG}>}{A_{Ly\alpha \cap FG}}
\end{equation} 
In this equation, $C$ are the colours and $A$ is the area covered by
the population in colour space. 
The best
selection colours will have the highest merit number $M$.
To calculate the optimum, we need to know the distribution of samples of both
populations. This can in principle be obtained
observationally. However, in the case here, for our evolutionary
models, we do not have that information. The sample of normal galaxies will
for example outline a region in the parameter space where outliers are extreme
types of galaxies. Within these boundaries, we do not know the actual
number distribution of galaxies in the colour space. Rather we assume that
within
the region covered by the boundaries, the distribution is uniform. 

\section{Results}\label{sec:funres}
For each redshift, all combinations in two dimensions that include the 
narrow-band magnitude in both combinations (in order to minimise the number of
observed bands
needed) are ranked according to their merit number, $M$, as described in the
previous
section. We also run the code with only [OII]-emitters and Ly$\alpha$ emitters
with the aim to find an optimal combination of colours for separating the two
types of emitters photometrically. For the emitters, we make a number of cuts
in the properties of the spectra. We set the maximum age of a Ly$\alpha$
emitter to be 40 Myrs and the maximum age of an [OII] emitter to be 100 Myrs.
The maximum E(B-V) allowed is $0.3$ for both Ly$\alpha$ and [OII] emitters.
Also, for both populations of emitters, the minimum EW of the emission line is 
taken to be 10~{\AA}.

\begin{figure}
\begin{center} 
\epsfig{file=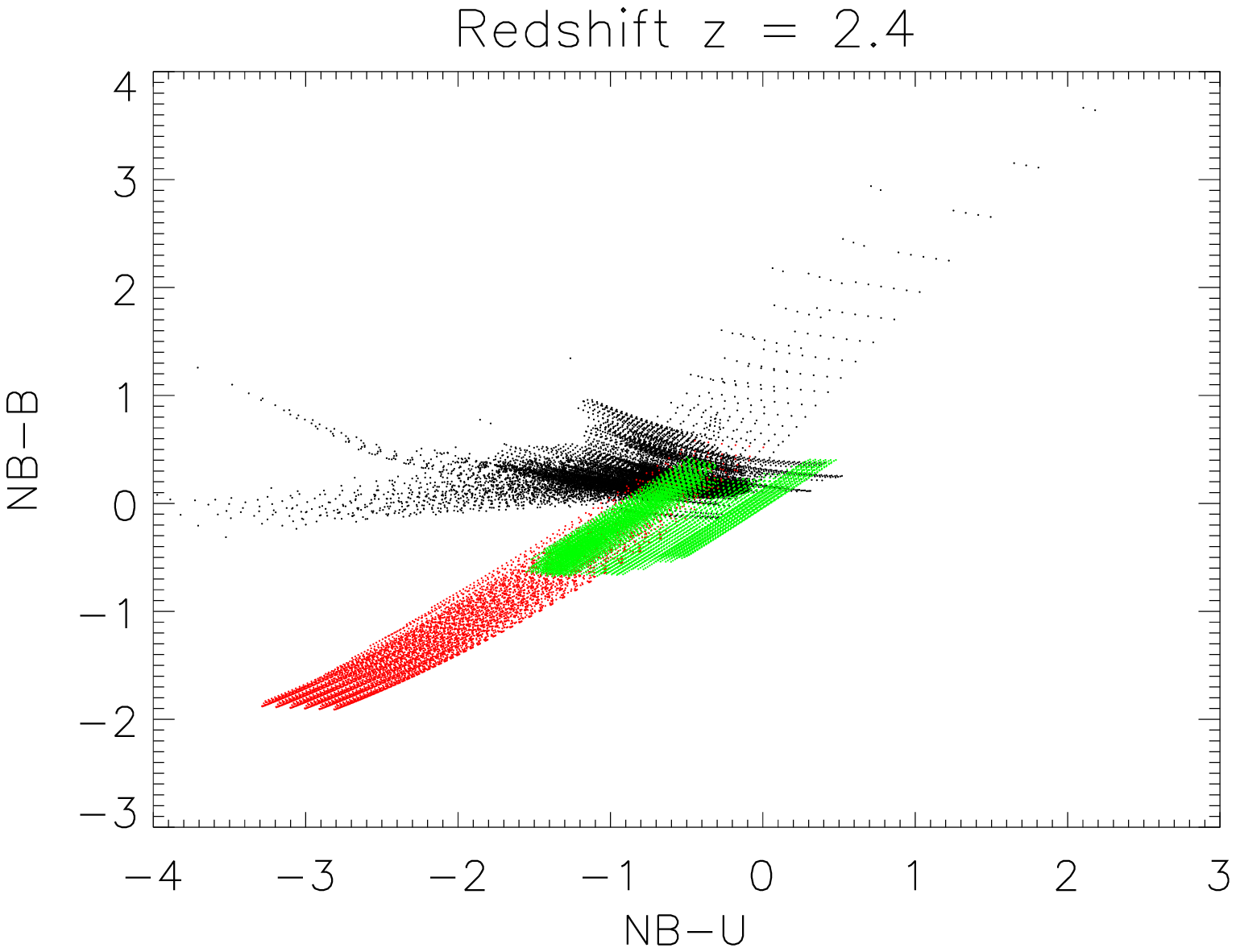,width=7.5cm}\epsfig{file=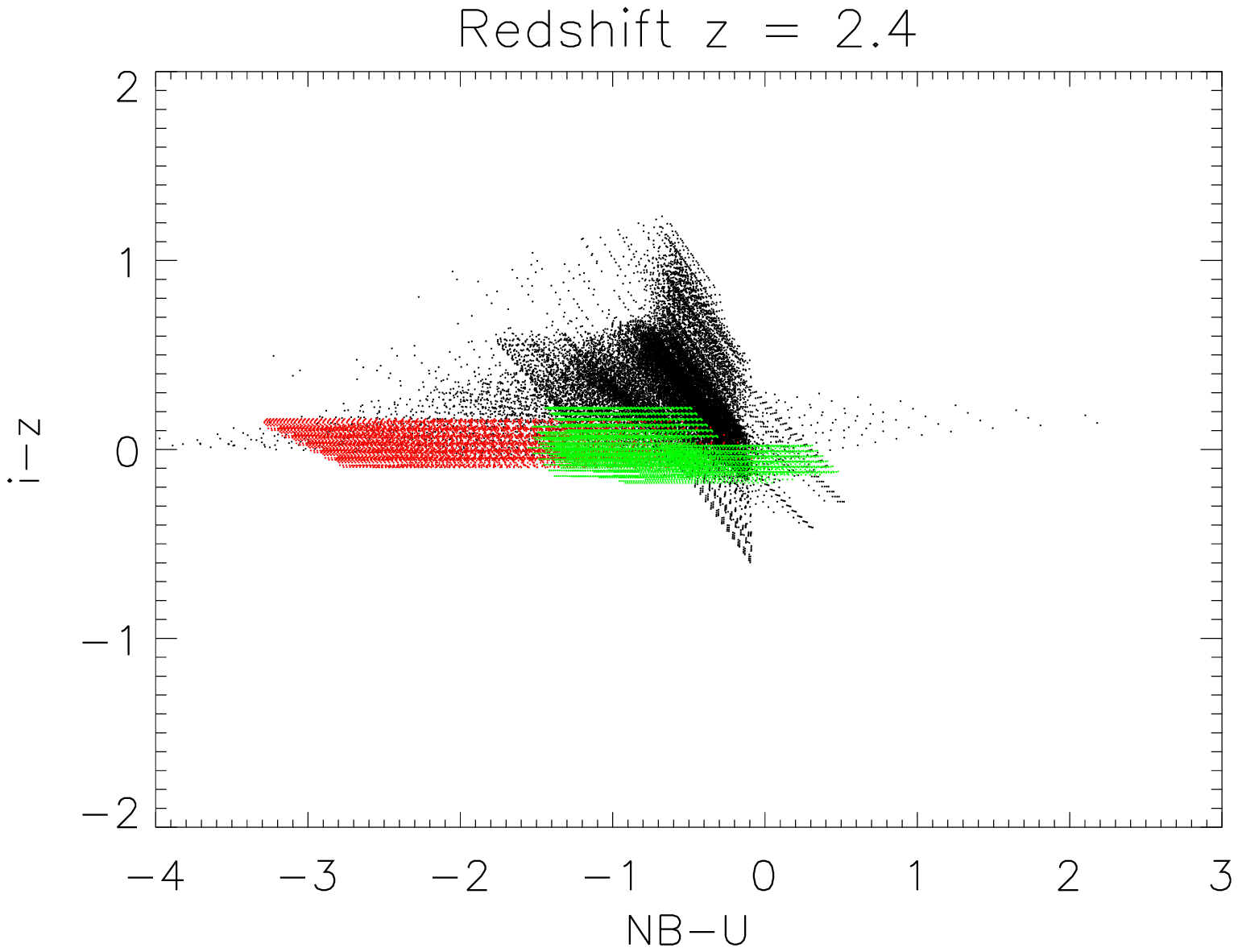,width=7.5cm}
\caption[Illustration of selection method]{Illustration of selection method for redshift $z = 2.4$. The left panel shows the optimal selection colours to distinguish
between Ly$\alpha$ emitters and field galaxies. The right panels shows the best
combination to distinguish between Ly$\alpha$ emitters and [OII] emitters.
Black points
are field galaxies, red points are Ly$\alpha$ emitters and green points are
[OII]-emitters.}
\label{fig:illustration}
\end{center}
\end{figure}

An illustration of the best
combinations for redshift $z = 2.4$ can be seen in Fig.~\ref{fig:illustration}.
The plot shows the top combinations of colours to select between Ly$\alpha$
emitters and field galaxies or [OII] emitters. As can be seen, the overlap is 
very small with the field galaxy population.
The overlap between the two emitter populations is larger.
The best three selection combinations for our selected redshifts can be
found in Table~\ref{tab:results}. In this table, the merit number has been
normalised to the most favourable value for each redshift. The merit number for the
distinction between
Ly$\alpha$ emitters and [OII]-emitters has been calculated the same way as
for the distinction with field galaxies.
\begin{table}[t]
\begin{center}
\caption[Best selection methods]{Best selection methods for Ly$\alpha$ emitters at different redshifts.
The columns give the first, second and third best selection
colours. NB denotes the narrow-band filter at the specified redshift. The
numbers in the parentheses are the relative goodness of the selection
technique normalised to the best solution. The upper part of the table give
the solutions for selecting between Ly$\alpha$ emitters and field galaxies and
the lower part between Ly$\alpha$ emitters and [OII]-emitters.}
\vspace{0.5cm}
\begin{tabular}{@{}llllllc}
\hline
\hline
$z$ & Best solution & Second solution & Third solution \\
\hline
\multicolumn{4}{c}{Ly$\alpha$ emitters vs. Field galaxies}\\
\hline
2.4 & NB-U vs. NB-B (1.00) & NB-B vs. NB-V (0.85)   & NB-B vs. NB-$i$ (0.67) \\
3.1 & NB-B vs. NB-V (1.00) & NB-V vs. NB-$i$ (0.86) & NB-B vs. NB-$i$ (0.83) \\
4.5 & NB-B vs. NB-J (1.00) & NB-B vs. NB-K$_s$ (1.00) & NB-B vs. NB-H (1.00) \\
5.7 & NB-U vs. NB-H (1.00) & NB-B vs. NB-H (0.76) & NB-B vs. NB-$Ch1$ (0.76) \\
6.5 & NB-U vs. NB-H (1.00) & NB-V vs. NB-H (0.83) & NB-V vs. NB-$Ch2$ (0.83) \\
7.7 & NB-U vs. NB-K$_s$ (1.00) & NB-B vs. NB-H (0.93) & NB-B vs. NB-K$_s$ (0.93) \\
8.8 & NB-B vs. NB-K$_s$ (1.00) & NB-V vs. NB-K$_s$ (0.55) & NB-$i$ vs. NB-$Ch3$ (0.37) \\
\hline
\multicolumn{4}{c}{Ly$\alpha$ emitters vs. [OII]-emitters}\\
\hline
2.4 & NB-U vs. $i$-$z'$ (1.00) & NB-U vs. NB-B (0.98) & NB-B vs. $i$-$z'$i (0.93) \\
3.1 & NB-$Ch1$ vs. U-B (1.00) & NB-K$_s$ vs. U-B (0.57) & NB-$Ch2$ vs. U-B (0.57) \\
4.5 & NB-K$_s$ vs. U-V (1.00) & NB-J vs. B-V (0.51) & NB-K$_s$ vs. B-V (0.51) \\
5.7 & NB-U vs. V-$Ch3$ (1.00) & NB-U vs. V-$Ch2$ (1.00) & NB-U vs. V-$Ch1$ (1.00) \\
6.5 & NB-U vs. V-$Ch1$ (1.00) & NB-U vs. V-$Ch2$ (1.00) & NB-U vs. $i$-$Ch4$ (1.00) \\
7.7 & NB-U vs. U-$Ch1$ (1.00) & NB-U vs. U-$Ch2$ (1.00) & NB-U vs. B-$Ch1$ (0.94) \\
8.8 & NB-U vs. U-$Ch2$ (1.00) & NB-B vs. U-$Ch3$ (0.96) & NB-U vs. B-$Ch2$ (0.96) \\
\hline
\label{tab:results}
\end{tabular}
\end{center}
\end{table}
For each best solution, we also want to find the actual selection criteria to
apply. Again, as the actual distribution of galaxies within colour space is
unknown, it is impossible to decide different confidence levels at which
the criteria can be applied. Instead, the only possibility is to calculate
the 100\% confidence limit, i.e.~where there is no contamination of the two
populations. For the best solutions in Table~\ref{tab:results}, these criteria
are given in Table~\ref{tab:criteria}.
\begin{table}[!t]
\begin{center}
\caption[Selection criteria]{Selection criteria. The selection criteria here are
the ``100\% confidence levels'' so that the populations are separate with the
combinations of colours corresponding to the best selection colours as 
presented in Table~\ref{tab:results}.}
\vspace{0.5cm}
\begin{tabular}{@{}llllllc}
\hline
\hline
Redshift & Selection criteria (Ly$\alpha$/FG) \\
\hline
2.4 & NB-B$ < -0.2$~$ \cap$~(NB-B)$- 0.3 \times $(NB-U)$ < 0.25$ & \\
3.1 & NB-V$ < -0.7$~$ \cup$~(NB-V)$+ 0.45 \times $(NB-B)$ < -0.7$ &  \\
4.5 & NB-J$ < -1.2$~$ \cup$~(NB-J)$+ 0.48 \times $(NB-B)$ < -1.2$ &  \\
5.7 & NB-H$ < -1.2$~$ \cup$~(NB-H)$+ 0.30 \times $(NB-U)$ < -1.2$ &  \\
6.5 & NB-H$ < -1.2$~$ \cup$~(NB-H)$+ 0.30 \times $(NB-U)$ < -1.2$ &  \\
7.7 & NB-K$_s < -1.2$~$ \cup$~(NB-K$_s$)$+ 0.22 \times $(NB-U)$ < -1.2$  \\
8.8 & NB-K$_s < -1.0$~$ \cup$~(NB-K$_s$)$+ 0.23 \times $(NB-B)$ < -1.0$ \\
\hline
 & Selection criteria (Ly$\alpha$/[OII]) \\
 \hline
2.4 & NB-U$< -1.6$ \\
3.1 & U-B$ > 0.8$~$ \cup$~(U-B)$- 0.13 \times $(NB-$Ch1$)$ > 0.8$ \\
4.5 & U-V$ > 1.2$~$ \cup$~(U-V)$- 0.33 \times $(NB-K$_s$)$ > 1.0$ \\
5.7 & NB-U$ < -2.5$~$\cup$~(V-$Ch3$)$ + 2.71 \times $(NB-U)$ < -5.9$  \\
6.5 & NB-U$ < -3.0$~$\cup$~(V-$Ch1$)$ + 1.60 \times $(NB-U)$ < -3.2$ \\
7.7 & NB-U$ < -4.0$~$\cup$~(U-$Ch1$)$ + 2.15 \times $(NB-U)$ < -4.0$  \\
8.8 & NB-U$ < -4.0$~$\cup$~(U-$Ch2$)$ + 2.15 \times $(NB-U)$ < -4.6$ \\
\hline
\label{tab:criteria}
\end{tabular}
\end{center}
\end{table}

\section{Discussion}\label{sec:disc}
The motivation of this \emph{Letter} is to study how to best select
Ly$\alpha$-emitter candidates photometrically. The necessity of such a 
review is two-fold; firstly, several surveys have or are being conducted
that offer a large set of public, multi-wavelength data in fields or
varying size on the sky. These data-sets can be used to make very accurate
photometric redshifts and should hence provide excellent opportunities to
select Ly$\alpha$ emitting candidates from narrow-band surveys with great
accuracy. Secondly, narrow-band surveys will try to reach fainter fluxes
and larger area/samples at all redshifts, and also to much higher redshifts
($z_{Ly\alpha} \gtrsim 7$). In both cases
spectroscopic follow-up to confirm the candidates can be challenging or
even beyond the capacity of spectrographs. Without spectroscopic
confirmation, it is necessary to make very accurate photometric selections
with as few interlopers as possible.

Hence, we have here reviewed the most efficient colour combinations that
can be made with a large multi-wavelength data-set for selecting Ly$\alpha$
emitters from field galaxies, as well as distinguishing between Ly$\alpha$
emitters and [OII]-emitters. For the first part of the results, it is
interesting to note that all of the best selection criteria for Ly$\alpha$
emitters/field galaxies involve one
combination with a broad-band filter on the blue side of the narrow-band
filter and one combination with a broad-band filter on the red side
of the narrow-band filter. For the best combination for all redshifts,
the filter on the red side of the Ly$\alpha$ line is also always located
on the blue side of the Balmer break in the spectrum. This result can be
understood intuitively, as two broad-band magnitudes on either side
of the Ly$\alpha$ line, but both blue-ward of the Balmer break, will
constrain the UV slope of the galaxy and thus one easily gets a measure of the
equivalent width of the emission line when this value is combined with the
narrow-band flux.

For redshift $z = 3.1$ the best selection
colour is identical to what has been used in other publications (Fynbo et al.
2003; Nilsson et al. 2007). Fynbo et al. (1999; 2002) have also used a two-colour
approach in trying to find Ly$\alpha$ emitters at redshifts $z \sim 2$. They
applied the colours NB-U and NB-$i$ with successful results. This criteria is
similar to the third best solution that we find for redshift $z = 2.4$. At other 
redshifts, the method of selecting
emitters using two filters has not previously been used. However, it is
obvious from our results that a selection with two colours including 
narrow-band/broad-band combinations is better than a selection with
only one narrow-band/broad-band colour as is most commonly used. This
is obvious since our method finds selection criteria for all redshifts 
(in Table~\ref{tab:criteria}) where both colours are included. If one
colour would be the best selection combination, the criteria would have 
been a cut parallel to either the $x-$ or $y-$axis. This is not the case. This
is also an intuitive result. More information will always yield a more accurate
result. In 
future work it would be interesting to further quantify how much better
results a two colour selection yields compared to a one colour selection. 


As for the distinction between Ly$\alpha$ emitters and [OII] emitters,
for several redshifts the best selection requires the narrow-band minus
U band colour. This is presumably because Ly$\alpha$ emitters with redshifts
above $z \sim 2$ should at all times have extremely faint emission in U, as
this filter will sample the far-UV, whereas [OII]-emitters are at lower
redshift and will thus have continuum emission in the observed U filter.

The method used here of maximising the merit number $M$ has the advantage that
it is very simple to calculate and to understand. It also involves very few
bands, in our case only three for the selection between Ly$\alpha$ emitters
and field galaxies, and so it can be applied to almost any survey. The
disadvantage is that for large multi-wavelength surveys such as GOODS or
COSMOS, it does not make use of all the data available. Better solutions
may be found if more bands are allowed. The method is not very rigorous
mathematically, nor is it a simple task to define the merit number. A more
careful analysis would take a more complex approach,
using a principal component analysis such as e.g. Fisher Linear Discriminants 
(Fisher 1936).

A caveat with the input to the method is the unknown mass distribution. In
order to
facilitate the calculations made here, all galaxies were assumed to have a
mass of $M_{\star} = 10^{11} M_{\odot}$. The catalogues were also cut at
a magnitude of mag$_{AB} = 30$ which introduces another arbitrary cut in
mass properties. Thus, we do not know the actual distribution of galaxies in
colour space, and we will miss extreme cases of colours in low or high mass
galaxies. The next step would thus also be to try to incorporate a mass function of
galaxies into the model. Yet another caveat is the lack of understanding of the
properties of Ly$\alpha$ emitters and [OII] emitters. The true age and dust
content distribution of these galaxies are not known. Previous SED fitting
of Ly$\alpha$ emitters have shown them to be in general young, low mass
and dust free (Gawiser et al. 2006; Lai et al. 2007; Finkelstein et al. 2007;
Nilsson et al. 2007). However, this may be a selection bias arising from
the selection criteria which often include high EW and blue colours.
Finally, in future work it would be desirable to try to incorporate the effects
of IGM opacity in the spectra. This effect will be most important at very high
redshifts, but also very uncertain as we have no conclusive information about how 
and when re-ionisation happened.

\chapter{Conclusions}\label{chapter:conclusions}
\markboth{Chapter 7}{Conclusions}
The technique of finding high redshift galaxies through their Ly$\alpha$
emission and narrow-band imaging, and hence the class of high redshift
galaxies called Ly$\alpha$ emitters,
is a relatively young branch of observational cosmology. Even though it was
suggested already by Partridge \& Peebles in 1967, the first successful 
observations did not occur until the beginning of the 1990's. Since the year
2000, many Ly$\alpha$ emitters have been published though, reaching in total
more than 500 spectroscopically confirmed galaxies between redshift 
$2 < z < 7$, including the highest redshift galaxy recorded to date
at $z = 6.96$ (Iye et al. 2006). The relative ease with which this type of
galaxy can be found, i.e. with moderate exposure times on medium sized to
large telescopes, promises to make this technique one of the best selection
techniques for future high and very high redshift galaxy surveys. This in
turn will enable a better understanding of the young Universe and galaxy 
formation and evolution through the study of these galaxies, and an
insight into the so-called ``Dark Ages'', before the Universe was
re-ionised.

However, although the sample of Ly$\alpha$ emitters is steadily growing, 
relatively little is known about the nature of these galaxies, or their 
relation to 
other classes of high redshift galaxies. In the first part of this thesis,
we attempt to study a sample of redshift $z = 3.15$ Ly$\alpha$ emitters found 
through narrow-band imaging in
GOODS-S, a field with a large amount of public, multi-wavelength data
(Chapter~\ref{chapter:blob} and~\ref{chapter:goodss}). The main objective
to get this narrow-band image was to use the multi-wavelength data to 
try to understand the nature (e.g.~the masses, ages, star formation rates, 
redshift distribution and spatial distribution) of these galaxies, and to 
compare their 
properties to those of other classes of high redshift galaxies. In part,
this goal has been achieved. The following main results have been presented
in this thesis.
\begin{itemize}
\item \textbf{SED fitting analysis: properties of Ly$\alpha$ emitters at $z = 3.15$}
\end{itemize}
In Chapter~\ref{chapter:goodss}, the
multi-wavelength data was used to perform an SED fitting of the spectra of the
Ly$\alpha$ emitters, thus allowing a study of their properties, including 
stellar mass, metallicity and dust content.
We found that the stacked SED of these Ly$\alpha$-emitters, which covered 
the spectra of the galaxies from the restframe far-UV to restframe 
near-infrared, was best fitted by a very low metallicity 
($Z = 0.005 \times Z_{\odot}$), had low dust extinction 
($A_V = 0.26^{+0.11}_{-0.17}$), medium ages ($0.85^{+0.13}_{-0.42}$~Gyrs)
and medium stellar masses 
(M$_\ast = 4.7^{+4.2}_{-3.2} \times 10^8$~M$_{\odot}$). These results are
in reasonably good agreement compared to results from other 
studies (Gawiser et al. 2006; Lai et al. 2007; Finkelstein et al. 2007),
see also Table~\ref{tab:sedcompare}. In one respect our results differ
from the others; our age is higher. However, the age of an SED is the
parameter which is the hardest to constrain, as can also be seen in 
Fig.~\ref{ch3:contours}. Thus, relatively little weight should be put on these
differences, as well as the actual ages themselves. 

\begin{figure}[t]
\begin{center}
\includegraphics[width=0.80\textwidth,height=0.50\textwidth]{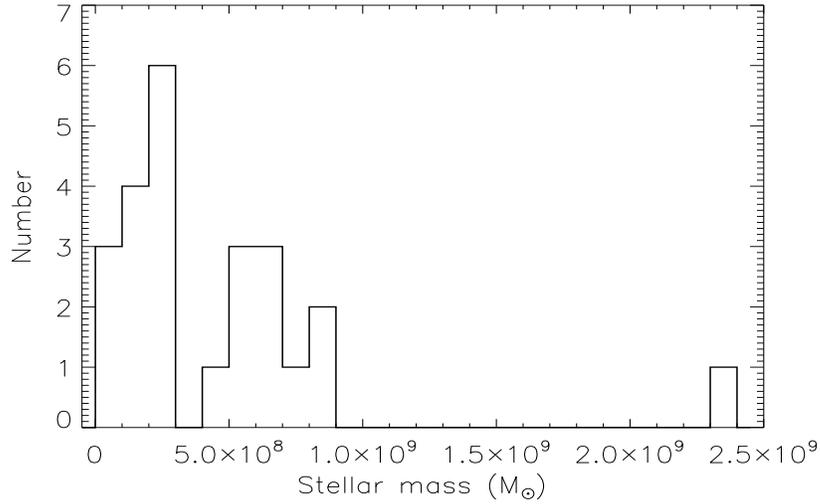}
\caption[Mass function of GOODS-S LEGOs]{
Mass function of GOODS-S LEGOs.\label{fig:massfuncLEGO}}
\end{center}
\end{figure}

To expand on the work presented in the paper in Chapter~\ref{chapter:goodss}, we 
here evaluate
the mass function of our Ly$\alpha$ emitters. The properties cited above
are for the stacked SED, hence a sort of average of all the properties of
all candidates. It is interesting to consider what the distribution of
masses for these objects look like. To do this, we take the mass
derived from the SED fitting, multiply it with the number of objects in 
the stack (this number is 23) and calculate individual masses by assuming
that the stellar mass of the galaxy is proportional to the $V$ band luminosity.
The reason for choosing the $V$ band as a mass estimator is because it is in
this band that we have the largest signal-to-noise detections of the 
candidates. This is not the best estimator, as the observed $V$ band
corresponds to the restframe UV and the galaxy light in this regime can be
heavily absorbed by dust, i.e. a more massive galaxy with a large amount of
dust would appear smaller in this calculation. However, it is the only band in
which we can do this analysis, as it is the only band with adequate 
signal-to-noise and most detections. The resulting distribution can be seen in 
the histogram plot in Fig.~\ref{fig:massfuncLEGO}.
It appears that a large fraction, more than 20\%, of the total mass of the
sample is concentrated to one object (LEGO\_GOODS-S\#14), and more than half
the sample has masses below M$_\ast < 3 \times 10^8$~M$_{\odot}$. With these
results in mind, it will be of great interest in the future to try to
improve the statistics on this kind of Ly$\alpha$ emitter mass function,
to see if this type of massive Ly$\alpha$ emitters are very rare occurrences
or if they are more abundant. Whatever the result, it can be said with
certainty that the majority of Ly$\alpha$ emitters have small masses.

We can compare the results of our SED fitting with similar results from
LBG surveys. Three such surveys have been published in
Papovich, Dickinson \& Ferguson (2001), Shapley et al. (2001) and Verma et al.
(2007). In all papers, the authors use samples of 
spectroscopically confirmed LBGs at redshift $z \sim 3$~or~$5$ respectively. 
They use Bruzual \&
Charlot (2003) models, with a Calzetti extinction law and metallicity fixed
at solar levels. They fit for star formation history, dust, age and stellar 
mass. The median results for the sample of 74 LBGs presented in Shapley et 
al.~(2001) are
$E(B-V) = 0.155$, age~$= 320$~Myrs and 
M$_{\ast} \sim 2 \times 10^{10}$~M$_{\odot}$. Papovich, Dickinson \& Ferguson 
(2001) find similar results, but with a larger dust content of up to
$A_V \sim 2$. Verma et al. (2007) find their LBGs to be less massive 
(M$_{\star} \approx 2 \times 10^{9}$~M$_{\odot}$) and 
younger, but with larger star formation rates. To compare the dust 
results, one has to take into account that the ratio between $A_V$~and~$E(B-V)$
is $R \approx 3.1$. Thus, the dust extinction in the LBGs of Shapley et al.
(2001) is approximately $A_V \approx 0.48$, in good agreement with the results
of Verma et al. (2007). This is similar, or a bit higher
than the value found in Chapter~\ref{chapter:goodss}. The ages of the LBGs
are slightly lower than those of the Ly$\alpha$ emitters, although this is 
a difficult quantity to constrain, as discussed above. Finally, a large
difference also lies in the stellar masses. LBGs at $z \sim 3$ are much more 
massive
than Ly$\alpha$ emitters at the same redshift. Another difference is the
star formation rate. Papovich, Dickinson \& Ferguson (2001), Shapley et al. 
(2001) and Verma et al. (2007) all find star formation rates
in the range of a few ten to several hundred solar masses per year, whereas
we find star formation rates less than ten solar masses per year.
Thus, there is a clear difference in the properties of LBGs and Ly$\alpha$
emitters at this redshift; LBGs are more massive, more star forming and 
possibly more dusty than Ly$\alpha$ emitters although the two types of 
galaxies have similar ages. 

\begin{itemize}
\item \textbf{LEGO\_GOODS-S\#16: A Ly$\alpha$ emitting dusty starburst galaxy?}
\end{itemize}
During the analysis of the multi-wavelength data of the candidates 
in the GOODS-S field, we discovered one candidate with a very unusual
SED profile, LEGO\_GOODS-S\#16. This candidate was undetected in the bluest
band, the HST $B$ band, and marginally detected in the $V$ band with the
greatest signal-to-noise. But as we followed the candidate in the redder
bands, its flux increased dramatically, see also Fig.~\ref{ch3:thumbs19}.
Although with a small offset between Ly$\alpha$ and counterpart centroid,
it appears as if this is a very dusty starburst galaxy with Ly$\alpha$
emission emerging from one side. We are hoping to follow this detection up
with future mm/sub-mm observations, that could hopefully give more insight
into the nature of this anomalous object.

\begin{itemize}
\item \textbf{Discovery of filamentary structure: Two parallel filaments at $z = 3.15$}
\end{itemize}
A further result from that paper was the discovery of a filamentary structure,
consisting of two, seemingly parallel filaments stretching beyond the 
length of the image. Filaments have been observed before (M{\o}ller \&
Fynbo 2003; Matsuda et al. 2005), but never two parallel filaments. It
would be of great interest to get spectroscopy of the candidates in order
to confirm the structure, and plot it in three dimensions. Ultimately, further
imaging at the ends of the filament would be interesting, to map out the
structure completely.

\begin{itemize}
\item \textbf{First detection of cold accretion: a Ly$\alpha$ blob in GOODS-S}
\end{itemize}
In Chapter~\ref{chapter:blob} we present the discovery of a Ly$\alpha$ blob.
This blob has an angular diameter of up to $\sim 100$~kpc and a Ly$\alpha$
line luminosity of almost $10^{43}$~erg~s$^{-1}$. After studying the 
complete set of deep multi-wavelength data available, we saw that this
blob, unlike all other published blobs, had no obvious optical/infrared
counter-part. It was realised, after taking all possible explanations into 
account, that this was the first observed Ly$\alpha$ blob that could
only be described by cold accretion in a satisfactory way.

\begin{itemize}
\item \textbf{ELVIS: a promising future survey for very high redshift Ly$\alpha$ emitters}
\end{itemize}
A large part of the time during my PhD has been spent on the project which
evolved into ``ELVIS''. The initial phase of the project included writing
numerous applications for funding of the filters and observing time. I was also
involved in the designing of
the filters and the initial contacts with the filter producing companies.
At later stages, I was involved in writing a Public Survey proposal 
and in the testing of the filters. After comparing the survey parameters (i.e.
area, flux limit expected, filter specifications) of different
near-infrared surveys for very high redshift Ly$\alpha$ emitters,
including ISAAC surveys, DaZle, ELVIS, JWST etc., it is clear that
ELVIS will be one of the best opportunities to observe the very high redshift
Universe. The combination of large area (0.9~deg$^2$) and relatively deep
flux limit for a near-infrared narrow-band image 
($F_{lim,5\sigma} = 3.7 \times 10^{-18}$~erg~s$^{-1}$~cm$^{-2}$) should
enable ELVIS to detect a few times ten Ly$\alpha$ emitters at $z = 8.8$. These
detections can be used to study galaxy formation and will probably
give one of the best constraints on re-ionisation at this redshift from
observed high redshift galaxies. Moreover, the survey will produce a vast 
catalogue of H$\alpha$ emitters, as well as other emission line objects at
intermediate redshift. This catalogue will be a treasure for anyone interested
in studying the star formation and metallicity history of the Universe
and galaxy evolution in general. Again, this sample will be gathered in
the COSMOS field, a field with a vast amount of public, multi-wavelength data,
including the very deep near-infrared data collected by 
Ultra-VISTA. This will enable further multi-wavelength studies of, in
particular, the lower redshift galaxy sample.

\begin{itemize}
\item \textbf{Predictions for very high redshift Ly$\alpha$ surveys}
\end{itemize}
In Chapter~\ref{chapter:predictions} we studied very high redshift Ly$\alpha$
luminosity functions. This was a follow-up work from the preparatory work
done for ELVIS.
In this chapter, we extrapolate two published theoretical models for 
Ly$\alpha$ emitters at lower redshifts to higher redshifts. We also try to
extrapolate the parameters of observed Ly$\alpha$ luminosity function fits to
Schechter functions. The result is a plot of $7 < z < 12.5$ luminosity 
functions which can be used for groups planning future surveys for 
high redshift Ly$\alpha$. Using these luminosity functions, we made specific
number predictions for several present and future narrow-band surveys.
The conclusion of the project was that detections of small samples of 
$z > 7$ galaxies are to be expected in the next five years, but to get a 
good handle on the very high redshift Universe, yet more future observatories
will be necessary.

\begin{itemize}
\item \textbf{How to best select Ly$\alpha$ emitters}
\end{itemize}
In the final chapter of this thesis, Chapter~\ref{chapter:fundamental}, we 
explore the
colour space spanned by a population of mock galaxies, including ``normal''
galaxies, Ly$\alpha$ emitters and [OII]-emitters. The aim of this work in 
progress is to 
determine how to best select Ly$\alpha$ emitters from a general population of
observed galaxies and how to distinguish them from interloper emitters such
as [OII]-emitters. The motivation was to see if it is possible to select
Ly$\alpha$ emitters in a confident way so that spectroscopy is unnecessary, using
the strength of present and future public multi-wavelength surveys such as
GOODS and COSMOS. In our method, we calculated a ``merit number'' which 
quantifies how good any combination of colours is in separating two populations 
of galaxies. This merit number is related to the projected distance between
the two populations and to the area of the overlap between the two projections.
As a result, we presented the three best selection combinations to distinguish
Ly$\alpha$ emitters from field galaxies or [OII] emitters. The conclusion of 
the project is a set of selection colours and selection criteria that may be
used by the community both to find Ly$\alpha$ emitters with observed data,
as well as when preparing observing proposals. Two interesting results
from the project were the conclusions that for selection, two colour combinations
including the narrow-band magnitude offer better results than one colour
combination and that these two combinations should include a broad-band filter
on either side of the Ly$\alpha$ line, but both blue-ward of the Balmer break.
This is work that will be refined in the future.

\section*{Summary}
To summarise, in the first part of the thesis, the emphasis was on using
the available multi-wavelength data in GOODS-S to study the nature of 
Ly$\alpha$ emitters at $z \sim 3$. This data turned out to be crucial in
several aspects of my work. It was necessary, for instance, to determine
that the Ly$\alpha$ blob that was detected was indeed the first one to be
best explained by cold accretion. This would have been impossible if the
broad-band coverage had been smaller. The X-ray and mid-IR data ruled out
an AGN nature of the blob and the optical/near-IR data excluded the 
possibility of a star-forming galaxy. Further, this data was also crucial in 
performing an accurate SED fitting for the candidates. It is true that many of 
the bands only included upper limits, but in many cases these limits
constrained the SED. Especially the infrared data was useful in the SED 
fitting, in order to constrain the masses of the objects. It was also using 
the multi-wavelength data that the
special SED of LEGO\_GOODS-S\#16 was identified as a potential Ly$\alpha$
emitting dusty starburst galaxy. If this is confirmed, LEGO\_GOODS-S\#16 will be 
the first dusty starburst galaxy detected through its Ly$\alpha$ emission.
The increase in public multi-wavelength data also prompted the work presented
in the final chapter, where we tried to find new selection criteria for
Ly$\alpha$ emitters at several different redshifts for surveys where
a large complimentary data-set is available.

In the second part of this thesis, a project to detect very high redshift
($z = 8.8$) Ly$\alpha$ emitters was presented. This is a project which has
not yielded any observational results yet, although it prompted the work 
presented in
Chapter~\ref{chapter:predictions}, where new Ly$\alpha$ luminosity functions 
are derived and discussed in the context of future narrow-band surveys. ELVIS
is a project which I am hoping to be part of for the coming years, and which
will produce very interesting results within the next five years.

The topic of Ly$\alpha$ emitters is a field of research which is growing
intensively at the moment, with increasing interest from the science community.
It is a very promising method to easily detect high redshift, low luminosity,
low mass, dust- and AGN-free star forming galaxies. It is also one of the most,
if not \emph{the} most, promising methods in detecting the highest redshift 
galaxies in the Universe. In the next chapter I write about what future
ideas, plans and hopes I have regarding this subject.

\chapter{Future ideas, plans and hopes}\label{chapter:future}
\markboth{Chapter 8}{Future ideas, plans and hopes}
The subject of exploring the Universe using Ly$\alpha$ as a tool has
only begun, but it is increasing in popularity and it is quite 
imaginable that many new and interesting revelations about galaxy
formation will come from this field in the future. A few aspects that
I see as being crucial in understanding these galaxies, and which I hope to
be able to work on are mentioned here.

\subsubsection{Understanding the stellar population in Ly$\alpha$ emitters and their role in the scheme of high redshift galaxies.}
Today, the sample of confirmed or candidate Ly$\alpha$ emitters is growing 
rapidly. The
number of spectroscopically confirmed LEGOs at redshifts between
$z = 1.8 - 9$ approaches 600 and probably
an equal amount of candidates have been presented in the literature.
However, the exact nature of these galaxies is not yet known to any great
detail. A few papers have attempted SED fitting of the continuum of
LEGOs at redshifts $\sim 3$ (Gawiser et al. 2006; Nilsson et al. 2007), 
$\sim 4.5$ (Finkelstein et al. 2007) and $\sim 5.7$ (Lai et al. 2007). 
Unfortunately, the samples on which these analysis has been done are
still very small and larger samples are needed before statistically
robust results will come. However, the future is promising. Several large
public multi-wavelength surveys are under way. A project that will make
a significant improvement in the field is the COSMOS survey at redshift
$z = 2.3$ that we have started with the WFI imager in Chile. This study will 
greatly increase the 
number of LEGOs with sufficient multi-wavelength coverage for a large 
statistical study of the continuum SED. 

Another burning issue is that of the scheme of high redshift galaxies.
There are today many different methods of finding high redshift galaxies,
and just as many ``classes'' of objects; LEGOs, LBGs, DLAs, sub-mm galaxies,
GRB host galaxies, DRGs, EROs etc. etc. How are these classes of galaxies
related? It has been suggested that LEGOs are the predecessors of LBGs
(Stark, Loeb \& Ellis 2007) or that they are part of the same population
(Giavalisco 2002), but neither suggestion has been proved. And where does 
the population of Ly$\alpha$
emitting and dusty sub-mm galaxies fit in? During which phases of galaxy
formation/evolution do galaxies emit Ly$\alpha$? These are very important
questions that I hope will be, if not explained, than at least better
understood in the next decade.

\subsubsection{Understanding how Ly$\alpha$ emitters trace Large Scale Structure and using it to derive information about the geometry of the Universe. }
LEGOs appear to be very good tracers of the Large Scale Structure (LSS) of the
Universe. They have been observed in candidate proto-clusters (Venemans et al. 
2007 and references therein) and in filamentary structures 
(M{\o}ller \& Fynbo 2001; Matsuda et al. 2005; Nilsson et al. 2007). Are LEGOs 
more prone to 
follow LSS than other classes of galaxies? If so, why? If it is true
that LEGOs are tracers of LSS, then they could be used as an independent test
on cosmological models, as described by Weidinger et al. (2002).
Further, large scale surveys such as the COSMOS survey and also 
specifically designed instruments such as the VIRUS instrument for the
HETDEX\footnote{www.as.utexas.edu/hetdex} survey may be able to make use 
of this property of LEGOs to draw conclusions about the cosmological 
parameters that describe our Universe.

\subsubsection{Understanding the evolution of Ly$\alpha$ emitters with redshift. Are they the same population at different redshifts? }
As mentioned earlier, the sample of Ly$\alpha$ emitting galaxies now
extend between redshifts from 1.8 to almost 9. This covers almost 
3 billion years in time, or almost 25\% of the age of the Universe. 
Are the galaxies that emit Ly$\alpha$ the same across this period
of time, or are they different types of objects? Are the mechanisms that
create Ly$\alpha$ the same? Yet another question is how LEGOs evolve with 
time, if they are only short-lived ``flares'', or if it is one of many 
stages in galaxy formation. It would also be interesting to compare the 
properties of high redshift LEGOs with Ly$\alpha$ emitting galaxies in the
local Universe. Can we detect Ly$\alpha$ emitters at redshift zero? Again, 
larger samples and better knowledge
of the nature of these objects at different redshifts will hopefully add
understanding within this subject.

\subsubsection{Exploring the star formation history of the Universe. }
Ly$\alpha$ is a direct tracer of the star formation rate of a galaxy,
albeit potentially affected by dust. Hence, using observations of Ly$\alpha$ 
at different redshifts will put constraints on the star formation rate and 
star formation density history of the Universe. Together with observations
from other emission-line galaxies, which are sometimes by-products of
narrow-band surveys for Ly$\alpha$, and from other means of estimating
star formation rates in galaxies, we can expect to get more and more accurate
measurements of the star formation history of the Universe.

\subsubsection{Detecting the first stars, understanding the very first stages of galaxy formation and how and when re-ionisation occurred.}
LEGOs are truly one of the best tools in detecting the first stars and the
first galaxies. As we move to higher and higher redshifts, the continuum of
galaxies will become more and more faint and reach a level that is
almost impossible to observe with the instruments of today and to some 
extent of tomorrow as well. However, it has been suggested that the 
equivalent width of very high redshift galaxy Ly$\alpha$ emission can be
very high (Schaerer 2003; Dijkstra \& Wyithe 2007) due to the Population III 
stars at 
those redshifts. It will then be possible to observe Ly$\alpha$ emission to
very high redshifts, even with todays instruments. ELVIS is one such example.
With ELVIS we will hopefully discover a large sample of redshift $z = 8.8$
galaxies and observe the first stages of galaxy formation.

Re-ionisation is also a popular topic of study at the moment. When did it happen? 
How? Two sources of information about the re-ionisation are conceivable. 
One is mapping the HII-emission (e.g. Iliev et al. 2002; Furlanetto, 
Zaldarriaga \& Hernquist 2004; McQuinn et al. 2006) in the radio regime and the
other one is Ly$\alpha$ observations. With ELVIS we will hopefully be
able to constrain the level of ionised gas at redshift $z = 8.8$, 
providing key information to help improve models of re-ionisation.

\subsubsection{ }
In short, the future for Ly$\alpha$ studies appears to be bright, with
many interesting results to look forward to. I hope I can be part of them.

\chapter*{Bibliography}
\markboth{Bibliography}{Bibliography}
\addcontentsline{toc}{chapter}{Bibliography}
\begin{flushleft}
\begin{verse}
Adelberger, K.L., Steidel, C.C., Shapley, A.E., et al., 2004, ApJ, 607, 226\\
Ahn, S.-H., 2004, ApJ, 601, L25\\
Ajiki, M., Taniguchi, Y., Fujita, S.S., et al., 2003, AJ, 126, 2091\\
Alonso-Herrero, A., P{\'e}rez-Gonz{\'a}lez, P.G., Alexander, D.M., et al., 2006, ApJ, 640, 167\\
Arnaboldi, M., Neeser, M.J., Parker, L.C., et al., 2007, \emph{ESO Messenger}, 127, 28\\
Barton, E.J., Dav{\'e}, R.,; Smith, J.-D.T., et al., 2004, ApJ, 604, L1\\
Baugh, C.M., Lacey, C.G., Frenk, C.S., et al., 2005, MNRAS, 356, 1191\\
Bertin, E., Arnouts, S. 1996, A\&AS, 117, 393\\
Bahcall, J.N., 1966, ApJ, 145, 684\\
Bahcall, N.A., \& Soneira, R.M., 1983, ApJ, 270, 20\\
Birnboim, Y. \& Dekel, A., 2003, MNRAS, 345, 349\\
Blain, A.W., Smail, I., Ivison, R.J., Kneib, J.P., \& Frayer, D.T. 2002, Phys. Rep., 369, 111\\
Bloom, J.S., Djorgovski, S.G., Kulkarni, S.R., \& Frail, D.A., 1998, ApJ, 507, L25\\
Bolzonella, M., Miralles, J.-M., Pell{\'o}, R., 2000, A\&A 363, 476\\
Boroson, T., 2005, AJ, 130, 381\\
Bower, R.G., Morris, S.L., Bacon, R. et al., 2004, MNRAS, 351, 63\\
Brocklehurst, M., 1971, MNRAS, 153, 471\\
Bruzual, G.A., \& Charlot S., 2003, MNRAS, 344, 1000\\
Bunker, A.J., Stanway, E.R., Ellis, R.S., \& McMahon, R.G., 2004, MNRAS, 355, 374\\
Calzetti, D., Kinney, A.L., \& Storchi-Bergmann, T. 1994, ApJ, 429, 582\\
Calzetti, D., Meurer, G.R., Bohlin, R.C., et al., 1997, AJ, 114, 1834\\
Calzetti, D., Armus, L., Bohlin, R.C., et al., 2000, ApJ, 533, 682\\
Cantalupo, S., Porciani, C., Lilly, S. J., \& Miniati, F., 2005, ApJ, 628, 61\\
Carilli, C.L., \& Yun, M.S., 1999, ApJ, 513, L13\\
Chandrasekhar, S., 1945, ApJ, 102, 402\\
Chapman S.C., Scott D., Windhorst R.A. et al., 2004, ApJ, 606, 85\\
Charlot, S., \& Fall, S.M, 1993, ApJ, 415, 580\\
Charlot, S., \& Fall, S.M, 2000, ApJ, 539, 718\\
Cole, S., Lacey, C.G., Baugh C.M., \& Frenk, C.S., 2000, MNRAS, 319, 168\\
Collin, S., 2001, in the lectures given at ``GH Advanced Lectures on the Starburst-AGN Connection'', INAOE, June 2000, eds. D. Kunth, I. Aretxaga, astro-ph/0101203\\
Condon, J.J., 1992, ARAA, 30, 575\\
Cowie, L.L., Songaila, A., Hu, E.M., \& Cohen, J.G., 1996, AJ, 112, 839\\
Cowie, L.L., \& Hu, E.M., 1998, AJ, 115, 1319\\
Cuby, J.-G., Hibon, P., Lidman, C., et al., 2007, A\&A, 461, 911\\
Dalton, G.B., Caldwell, M., Ward, A.K., et al., 2006, SPIE, 6269, 34\\
Davis, M., \& Wilkinson, D.T., 1974, ApJ, 192, 251\\
Dawson, S., Rhoads, J.E., Malhotra, S., et al., 2004, ApJ, 617, 707\\
Dawson, S., Rhoads, J.E., Malhotra, S., et al., 2007, submitted to ApJ, arXiv:0707.4182\\
Dickinson, M., Giavalisco, M. \& the GOODS team, 2001, in ``The Mass of Galaxies at Low and High Redshift,'' Proceedings of the ESO Workshop held in Venice, Italy, 24-26 October 2001; eds. R. Bender \& A. Renzini, p. 324, astro-ph/0204213\\
Dekel, A. \& Birnboim, Y., 2006, MNRAS, 368, 2\\
Dey A., Bian C., Soifer B.T. et al., 2005, ApJ, 629, 654\\
Dijkstra, M., Haiman, Z., \& Spaans, M., 2006a, ApJ, 649, 14\\
Dijkstra, M., Haiman, Z., \& Spaans, M., 2006b, ApJ, 649, 37\\
Dijkstra, M., Lidz, A., \& Wyithe, J.S.B., 2007a, MNRAS, 377, 1175\\
Dijkstra, M., Wyithe, S., \& Haiman, Z., 2007b, MNRAS, 379, 253\\
Dijkstra, M., \& Wyithe, S., 2007c, MNRAS, 379, 1589\\
Doherty, M., Bunker, A., Sharp, R., et al., 2004, MNRAS, 354, L7\\
Doherty, M., Bunker, A., Sharp, R., et al., 2006, MNRAS, 370, 331\\
Eisenstein, D.J., Zehavi, I., Hogg, D.W., et al., 2005, ApJ, 633, 560\\
Emerson, J.P., Sutherland, W.J., McPherson, A.M., et al., 2004, \emph{ESO Messenger}, 117, 27\\
Fabian, A.C., 1994, ARA\&A, 32, 277\\
Fan, X., Strauss, M.A., Becker, R.H., et al., 2006, AJ, 132, 117\\
Fardal, M.A., Katz, N., Gardner, J.P. et al., 2001, ApJ, 562, 605\\
Fernandez, E.R., \& Komatsu, E., 2007, submitted to MNRAS, arXiv:0706.1801\\
Finkelstein, S.L., Rhoads, J.E., Malhotra, S., Prizkal, N., \& Wang, J., 2007, ApJ, 660, 1023\\
Fisher, R. 1936, Annals of Eugenics, 7, 179-188\\
Francis, P.J., Woodgate, B.E., Warren, S.J. et al., 1996, ApJ, 457, 490\\
Francis, P.J., Williger, G.M., Collins N.R. et al., 2001, ApJ, 554, 1001\\
Francis, P.J., Palunas, P., Teplitz, H.I., Williger, G.M., \& Woodgate, B.E., 2004, ApJ, 614, 75\\
Freedman, W.L., Madore, B.F., Gibson, B.K., et al., 2001, ApJ, 553, 47\\
Fruchter, A. S., Levan, A. J., Strolger, L. et al., 2006, Nature, 441, 463\\
Fujita, S.S., Ajiki, M., Shioya, Y., et al., 2003, AJ, 125, 13\\
Furlanetto, S.R., Schaye, J., Springel, V., \& Hernquist, L., 2003, ApJ, 599, L1\\
Furlanetto, S.R., Zaldarriaga, M., \& Hernquist, L., 2004, ApJ, 613, 1\\ 
Furlanetto, S.R., Schaye, J., Springel, V., \& Hernquist, L., 2005, ApJ, 622, 7\\
Fynbo, J.P.U., M{\o}ller, P., \& Warren, S.J. 1999, MNRAS, 305, 849\\
Fynbo, J.P.U., M{\o}ller, P., \& Thomsen, B. 2001, A\&A 374, 443\\
Fynbo, J.P.U., M{\o}ller, P., Thomsen, B, et al., 2002, A\&A, 388, 425\\
Fynbo, J.P.U., Ledoux, C., M{\o}ller, P., Thomsen, B., \& Burud, I., 2003, A\&A, 407, 147\\
Fynbo, J.P.U., Gorosabel, J., Smette, A., et al., 2005, ApJ, 633, 317\\
Gallagher, J.S., Hunter, D.A., \& Bushouse, H., 1989, AJ, 97, 700\\ 
Gardner, J.P., Mather, J.C., Clampin, M., et al., 2006, Space Science Reviews, 123, 485, astro-ph/0606175\\
Gawiser, E., Van Dokkum, P.G., Gronwall, C., et al., 2006, ApJL, 642, 13\\
Giavalisco, M., Steidel, C.C., Adelberger, K.L., et al., 1998, ApJ, 503, 543\\
Giavalisco, M., 2002, ARAA, 40, 579\\
Giavalisco, M., Ferguson, H.C., Koekemoer, A.M., et al., 2004, ApJ, 600, L93\\
Gilks, W.R., Richardson, S., \& Spiegelhalter, D.J., 1995, Markov Chain Monte Carlo in Practice, Chapman \& Hall, ISBN 0412055511\\
Gnedin, N.Y. \& Prada, F., 2004, ApJ, 608, L77\\
Gorosabel, J., P{\'e}rez-Ram{\'i}rez, D., Sollerman, J., et al., 2005, A\&A, 444, 711\\
Gronwall, C., Ciardullo, R., Hickey, T., et al., 2007, accepted for publication
in ApJ, arXiv:0705.3917\\
Gu, Q., Melnick, J., Fernandes, R.C., et al., 2006, MNRAS, 366, 480\\
Gunn, J.E., \& Peterson, B.A., 1965, ApJ, 142, 1633\\
Haiman, Z., \& Spaans, M., 1999, ApJ, 518, 138\\
Haiman, Z., \& Loeb, A., 1999, ApJ, 519, 479\\
Haiman, Z., Spaans, M. \& Quataert, E., 2000, ApJL 537, L5\\
Haiman, Z. \& Rees, M.J., 2001, ApJ 556, 87\\
Haiman, Z., 2002, ApJ, 576, L1\\
Haiman, Z. \& Cen, R., 2005, ApJ, 623, 627\\
Haislip, J.B., Nysewander, M.C., Reichart, D.E., et al., 2006, Nature, 440, 181\\
Hamilton, A.J.S., 1993, ApJ, 417, 19\\
Hashimoto, Y., Oemler, A., Lin, H., \& Tucker, D.L., 1998, ApJ, 499, 589\\
Hawkins, E., Maddox, S., Cole, S., et al., 2003, MNRAS, 346, 78\\
Hayashino, T., Matsuda, Y., Tamura, H., et al., 2004, AJ, 128, 2073\\
Hayes, M., \& {\"O}stlin, G., 2006, A\&A, 460, 681\\
Heavens, A., Panter, B., Jimenez, R. \& Dunlop, J., 2004, Nature, 428, 625\\
Heckman, T.M., Miley, G.K., van Breugel, W.J.M. \& Butcher, H.R.,  1981, ApJ, 247, 403\\
Hill, G.J., Gebhardt, K., Komatsu, E., \& MacQueen, P.J., 2004, in the proceedings of the ``THE NEW COSMOLOGY: Conference on Strings and Cosmology; The Mitchell Symposium on Observational Cosmology'' conference held in Texas, US, March 2004, ed. R.E. Allen, D.V. Nanopoulos \& C.N. Pope, AIPC, 743, 224\\	
Hjorth, J., Sollerman, J., M{\o}ller, P., et al., 2003, Nature, 423, 847\\
Hjorth, J., Watson, D., Fynbo, J.P.U., et al., 2005, Nature, 437, 859\\
Hogan, C.J.,  \& Rees, M.J., 1979, MNRAS, 188, 791\\
Holland, W.S., Robson, E.I., Gear, W.K., et al., 1999, MNRAS, 303, 659\\
Horton, A., Parry, I., Bland-Hawthorne, J., et al., 2004, astro-ph/0409080\\
Hopkins, A.M., Connolly, A.J., \& Szalay, A.S., 2000, AJ, 120, 2843\\
Hopkins, A.M., 2004, ApJ, 615, 209\\
Hopkins, A.M., \& Beacom, J.F., 2006, ApJ, 651, 142\\
Hu, E.M., \& McMahon, R.G., 1996, Nature, 382, 231\\
Hu, E.M., Cowie, L.L., \& McMahon, R.G., 1998, ApJ, 502, L99\\
Hu, E.M., Cowie, L.L., McMahon, R.G., et al., 2002, ApJ, 568, L75\\
Hu, E.M., Cowie, L.L., Capak, P., et al., 2004, ApJ, 127, 563\\
Iliev, I.T., Shapiro, P.R., Ferrara, A., \& Martel, H., 2002, ApJ, 572, L123\\
Ivison, R.J., Smail, I., Barger, A.J., et al., 2000, MNRAS, 315, 209\\
Ivison, R.J., Greve, T.R., Serjeant, S. et al., 2004, ApJS, 154, 124\\
Ivison, R.J., Smail, I., Dunlop, J.S., et al., 2005, MNRAS, 364, 1025\\
Iye, M., Ota, K., Kashikawa, N. et al. 2006, Nature, 443, 186\\
Jaunsen, A.O., Andersen, M.I., Hjorth, J., et al., 2003, A\&A, 402, 125\\
Johansson, P.H., V{\"a}is{\"a}nen, P., \& Vaccari, M., 2004, A\&A, 427, 795\\
Juneau, S., Glazebrook, K., Crampton, D., et al., 2005, ApJ, 619, L135\\
Kashikawa, N., Shimasaku, K., Malkan, M.A., et al., 2006, ApJ, 648, 7\\
Keel W.C., Cohen S.H., Windhorst R.A., Waddington I., 1999, AJ, 118, 2547\\
Kennicutt, R.C., 1983, ApJ, 272, 54\\
Kennicutt, R.C., 1998a, ARAA, 36, 189\\
Kennicutt, R.C., 1998b, ApJ, 498, 541\\
Keres, D., Katz, N., Weinberg, D.H., \& Dave, R., 2005, MNRAS, 363, 2\\
Klaas, U., Haas, M., M{\"u}ller, S.A.H., et al., 2001, A\&A, 379, 823\\
Kobayashi, M.A.R., Totani, T., \& Nagashima, M., 2007, submitted to ApJ, arXiv:0705.4349\\
Kodaira, K., Taniguchi, Y., Kashikawa, N., et al., 2003, PASJ, 55, L17\\
Kova$\check{\mathrm{c}}$, K., Somerville, R.S., Rhoads, J.E., Malhotra, S., \& Wang, J.X., 2007, accepted for publication in ApJ, arXiv:0706.0893\\
Kudritzki, R.-P., M{\'e}ndez, R.H., Feldmeier, J.J., et al., 2000, ApJ, 536, 19\\ 
Kurk, J.D., Cimatti, A., Serego Alighieri, S.d., et al., 2004, A\&A, 422, L13\\
Lai, K., Huang, J.-S., Fazio, G., et al., 2007, ApJ, 655, 704\\  
Lacy, M., Storrie-Lombardi, L.J., Sajina, A., et al., 2004, ApJS, 154, 166\\
Landy, S.D., \& Szalay, A.S., 1993, ApJ, 412, 64\\
Landy, S.D., Szalay, A.S., \& Broadhurst, T.J., 1998, ApJ, 494, L133\\
Laursen, P., \& Sommer-Larsen, J., 2007, ApJ, 657, L69\\
Lawrence, A., Warren, S.J., Almaini, O., et al., 2007, MNRAS, 379, 1599\\
Leitherer, C., Schaerer, D., Goldader, J.D., et al., 1999, ApJS, 123, 3\\
Le Delliou, M., Lacey, C.G., Baugh, C.M., et al., 2005, MNRAS, 357, L11\\
Le Delliou, M., Lacey, C.G., Baugh, C.M., \& Morris, S.L., 2006, MNRAS, 365, 712\\
Loeb, A., \& Rybicki, G.B., 1999, ApJ, 524, 527\\
Lowenthal, J.D., Hogan, C.J., Green, R.F., et al., 1991, ApJ, 377, L73\\
Macchetto, F., Lipari, S., Giavalisco, M., Turnshek, D.A., \& Sparks, W.B., 1993, ApJ, 404, 511\\
Madau, P., 1995, ApJ, 441, 18\\
Madau, P., Ferguson, H.C., Dickinson, M.E., et al., 1996, MNRAS, 283, 1388\\
Maier, C., Meisenheimer, K., Thommes, E., et al., 2003, A\&A, 402, 79\\
Malhotra, S., \& Rhoads, J.E. 2002, ApJL, 565, L71\\
Malhotra, R. \& Rhoads, J., 2004, ApJ, 617, L5\\
Malhotra, S., Rhoads, J.E., Pirzkal, N., et al., 2005, ApJ, 626, 666\\
Maller, A.H. \& Bullock, J.S., 2004, MNRAS, 355, 694\\
Matsuda Y., Yamada T., Hayashino T. et al., 2004, AJ, 128, 569\\
Matsuda, Y., Yamada, T., Hayashino, T. et al., 2005, ApJ, 634, L125\\
McCarthy, P.J., 1993, ARAA, 31, 639\\
McQuinn, M., Zahn, O., Zaldarriaga, M., Hernquist, L., \& Furlanetto, S.R., 2006, ApJ, 613, 815\\
McQuinn, M., Hernquist, L., Zaldarriaga, M., \& Dutta, S., 2007, submitted to MNRAS, astro-ph/0704.2239\\
Meier, D.L., 1976, ApJ, 207, 343\\
Menzel, D.H, 1926, PASP, 38, 295\\
Miralda-Escud{\'e}, J., 1998, ApJ, 501, 15\\
Miralda-Escud{\'e}, J. \& Rees, M.J., 1998, ApJ, 497, 21\\
Monaco, P., M{\o}ller, P., Fynbo, J.P.U., et al., 2005, A\&A, 440, 799\\
Mori, M., Umemura, M., \& Ferrara, A., 2004, ApJ, 613, L97\\
Murayama, T., Taniguchi, Y., Scoville, N.Z., et al., 2007, accepted for publication in ApJ, astroph/0702458\\
M{\o}ller, P., \& Warren, S.J., 1993, A\&A, 270, 43\\
M{\o}ller P., \& Warren S.J. 1998, MNRAS 299, 661\\
M{\o}ller, P., \& Fynbo, J.U. 2001, A\&A, 372, L57\\
M{\o}ller, P., Warren, S.J., Fall, S. M., Fynbo, J.U.,\& Jakobsen, P. 2002, ApJ, 574, 51\\
Neugebauer, G., \& Leighton, R.B., 1969, Two Micron Sky Survey, a Preliminary Catalog (NASA SP-3047; Washington: GPO)\\
Nilsson, K.K., Fynbo, J.P.U., M{\o}ller, P., Sommer-Larsen, J. \& Ledoux, C., 2006a, A\&A, 452, L23\\
Nilsson, K.K., Fynbo, J.P.U., M{\o}ller, P., \& Orsi, A., 2006b, to appear in the ASP Conference proceedings of 'At the Edge of the Universe', eds. J. Afonso, H. Ferguson and R. Norris, astro-ph/0611239\\
Nilsson, K.K., M{\o}ller, P., M{\"o}ller, O., et al., 2007, A\&A, 471, 71\\
Ohyama, Y., Taniguchi, Y., Kawabata, K.S. et al., 2003, ApJ, 591, L9\\
Ouchi, M., Shimasaku, K., Furusawa, H. et al., 2003, ApJ, 582, 60\\
Ouchi, M., Shimasaku, K., Okamura, S., et al., 2004a, ApJ, 611, 660\\
Ouchi, M., Shimasaku, K., Okamura, S., et al., 2004b, ApJ, 611, 685\\
Ouchi, M., Shimasaku, K., Akyiama, M. et al., 2005, ApJ, 620, L1\\
Ouchi, M., Shimasaku, K., Akiyama, M., et al., 2007, submitted to ApJ, arXiv:0707.3161\\
Overzier, R.A., Miley, G.K., Bouwens, R.J., et al., 2006, ApJ, 637, 58\\
Palunas, P., Teplitz H.I., Francis P.J., Williger G.M., Woodgate B.E., 2004, ApJ, 602, 545\\
Papovich, C., Dickinson, M., \& Ferguson, H.C., 2001, ApJ, 559, 620\\
Parkes, I.M., Collins, C.A., \& Joseph, R.D., 1994, MNRAS, 266, 983\\
Partridge, R.B., \& Peebles, P.J.E., 1967, ApJ, 147, 868\\
Partridge, R.B., 1974, ApJ, 192, 241\\
Pascarelle, S.M., Windhorst, R.A., Driver, S.P., Ostrander, E.J., \& Keel, W.C., 1996, ApJ, 456, L21\\
Peacock, J.A., 1983, MNRAS, 202, 615\\
Peebles, P.J.E., 1974, ApJ, 189, L51\\
Petitjean, P., Pecontal, E., Valls-Gabaud, D., \& Charlot, S., 1996, Nature, 380, 411\\
Pettini, M., Shapley, A.E., Steidel, C.C., et al., 2001, ApJ, 554, 981\\ 
Portinari, L., \& Sommer-Larsen, J., 2007, MNRAS, 375, 913\\
Pritchet, C.J., \& Hartwick, F.D.A., 1989, ApJ, 320, 464\\
Pritchet, C.J., \& Hartwick, F.D.A., 1990, ApJ, 355, L11\\
Pritchet, C.J., 1994, PASP, 106, 1052\\
Pope, A., Borys, C., Scott, D., et al., 2005, MNRAS, 358, 149\\
Ranalli, P., Comastri, A., \& Setti, G., 2003, A\&A, 399, 39\\
Rhee, G.F.R.N., Webb, J.K., Katgert, P., 1989, A\&A, 217, 1\\
Rhoads, J.E., Malhotra, S., Dey, A., et al., 2000, ApJ, 545, L85\\
Rhoads, J.E., Dey, A., Malhotra, S. et al., 2003, AJ, 125, 1006\\
Rousselot, P., Lidman, C., Cuby, J.-G., Moreels, G., \& Monnet, G., 2000, A\&A, 354, 1134\\
Santos, M.R., 2004, MNRAS, 349, 1137\\
Schaerer, D., 2003, A\&A, 397, 527\\
Schaerer, D., \& Pell{\'o}, R., 2005, MNRAS, 362, 1054\\
Schechter, P., 1976, ApJ, 203, 297\\
Schmidt, M., 1965, ApJ, 141, 1295\\
Schmidt, M., 1963, Nature, 192, 1040\\
Schmitt, H.R., Calzetti, D., Armus, L., et al., 2006, ApJ, 643, 173\\
Scoville, N., Aussel, H., Brusa, M., et al., 2006, astro-ph/0612305\\
Shapley, A.E., Steidel, C.C., Adelberger, K.L., et al., 2001, ApJ, 562, 95\\
Shimasaku, K., Ouchi, M., Okamura, S., et al., 2003, ApJ, 586, L111\\
Shimasaku, K., Kashikawa, N., Doi, M., et al., 2006, PASJ, 58, 313\\
Silva, L., Granato, G.L., Bressan, A., \& Danese, L., 1998, ApJ, 509, 103\\
Silverman, J., Green, P., Barkhouse, W., et al., 2005, in the proceedings of the ``The X-ray Universe 2005'' conference held in San Lorenzo de El Escorial, Spain, September 2005, ed. A. Wilson, p. 795, astro-ph/0511552\\
Skrutskie, M.F., Cutri, R.M., Stiening, R., et al., 2006, AJ, 131, 1163\\
Smail, I., 2002, RSPTA, 360, 2697\\
Smail, I., Chapman, S.C., Blain, A.W., \& Ivison, R.J., 2004, ApJ, 616, 71\\
Sokolov, V.V., Fatkhullin, T.A., Castro-Tirado, A.J., et al., 2001, A\&A, 372, 438\\
Sommer-Larsen, J., 2005, in the proceedings of the ``Island Universes: Structure and Evolution of Disk Galaxies'' conference held in Terschelling, Netherlands, July 2005, ed. R de Jong (Springer Dordrecht), astro-ph/0512485\\
Spergel, D.N., Bean, R., Dor{\'e}, O., et al., 2007, ApJS, 170, 377\\
Springel, V., White, S.D.M., Jenkins, A., et al., 2005, Nature, 435, 629\\
Stanway, E.R., Bunker, A.J., McMahon, R.G., et al., 2004, ApJ, 607, 704\\
Stark, D.P., Ellis, R.S., Richard, J., et al., 2007, ApJ, 663, 10\\
Stark, D.P., Loeb, A., \& Ellis, R.S., 2007, submitted to ApJ, astro-ph/0701882\\
Steidel, C.C., Giavalisco, M., Pettini, M., Dickinson, M., \& Adelberger, K.L., 1996, ApJ, 462, L17\\
Steidel, C.C., Adelberger, K.L., Dickinson, M., et al., 1998, ApJ, 492, 428\\
Steidel, C.C., Adelberger, K.L., Giavalisco, M., Dickinson, M., \& Pettini, M., 1999, ApJ, 519, 1\\
Steidel, C.C., Adelberger, K.L., Shapley, A.E., et al., 2000, ApJ, 532, 170\\
Steidel, C.C., Adelberger, K.L., Shapley, A.E., et al., 2003, ApJ, 592, 728\\ 
Stern, D., Yost, S.A., Eckart, M.E., et al., 2005a, ApJ, 619, 12\\
Stern, D., Eisenhardt, P., Gorjian, V., et al., 2005b, ApJ, 631, 163\\
Taniguchi, Y., Shioya, Y. \& Kakazu, Y., 2001, ApJL 562, L15\\
Taniguchi, Y., Ajiki, M., Nagao, T. et al., 2005, PASJ, 57, 165\\
Tapken, C., Appenzeller, I., Gabasch, A., et al., 2006, A\&A, 455, 145\\
Tasitsiomi, A., 2006, ApJ, 645, 792\\
Thommes, E. \& Meisenheimer, K., 2005, A\&A, 430, 877\\
Totsuji, H., \& Kihara, T., 1969, PASJ, 21, 221\\
Tresse, L., Maddox, S.J., Le F{\`e}vre, O. \& Cuby, J.-G., 2002, MNRAS, 337, 369\\
Tumlinson, J., Shull, J.M., \& Venkatesan, A., 2003, ApJ, 584, 608\\
Urry, C.M., \& Padovani, P., 1995, PASP, 107, 803\\
van Breukelen, C., Jarvis, M. J., \& Venemans, B. P. 2005, MNRAS, 359, 895\\
V{\'a}zquez, G.A., \& Leitherer, C., 2005, ApJ, 621, 695\\
Venemans, B.P., Kurk, J.D., Miley, G.K., et al., 2002, ApJ, 569, L11\\
Venemans, B.P., R{\"o}ttgering, H J A,; Overzier, R A, et al., 2004, A\&A, 424, L17\\
Venemans, B.P., R{\"o}ttgering, H.J.A., Miley, G.K. et al., 2005, A\&A, 431, 793\\
Venemans, B.P., R{\"o}ttgering, H.J.A., Miley, G.K., et al., 2007, A\&A, 461, 823\\
Verhamme, A., Schaerer, D., \& Maselli, A., 2006, A\&A, 460, 397\\
Verma, A., Lehnert, M.D., F{\"o}rster Schreiber, N.M., Bremer, M.N., \& Douglas, L., 2007, MNRAS, 377, 1024\\
Villar-Mart{\'i}n, Sanchez, S.F., De Breuck, C., 2005, MNRAS, 359, L5\\
Willis, J.P. \& Courbin, F., 2005, MNRAS, 357, 1348\\
Wadadekar, Y., Casertano, S., \& de Mello, D., 2006, AJ, 132, 1023\\
Wang, J.X., Rhoads, J.E., Malhotra, S., et al., 2004, ApJL, 608, 21\\
Warren, S.J., \& M{\o}ller, P., 1996, A\&A, 311, 25\\
Warren, S.J., M{\o}ller, P., Fall, S.M.,\& Jakobsen, P., 2001, MNRAS, 326, 759\\
Weidinger, M.; M\o ller, P., Fynbo, J.P.U., Thomsen, B., \& Egholm, M.P., 2002, A\&A, 391, 13\\
Weidinger, M., M\o ller, P., Fynbo, J.P.U., 2004, Nature, 430, 999\\
Weidinger, M., M\o ller, P., Fynbo, J.P.U., Thomsen, B., 2005, A\&A, 436, 825\\
Willis, J.P. \& Courbin, F., 2005, MNRAS, 357, 1348\\
Willott, C.J., Delorme, P., Omont, A., et al., 2007, submitted to AJ, arXiv:0706.0914\\
Wilman, R.J., Gerssen, J., Bower, R.G. et al., 2005, Nature, 436, 14\\
Wolfe, A.M., Turnshek, D.A., Smith, H.E, \& Cohen, R.D., 1986, ApJS, 61, 249\\
Wolfe, A.M., Lanzetta, K.M., Turnshek, D.A., \& Oke, J.B., 1992, ApJ, 385, 151\\
Wolfe, A.M., Gawiser, E., \& Prochaska, J.X., 2005, ARAA, 43, 861\\
Worsley, M.A., Fabian, A.C., Bauer, F.E., et al., 2005, MNRAS, 357, 1281\\
Yan, L., McCarthy, P. J., Freudling, W., et al., 1999, ApJ, 519, L47\\
Zanstra, H., 1927, ApJ, 65, 50\\
Zheng, Z., \& Miralda-Escud{\'e}, J.  2002, ApJ 578,33\\
\end{verse}
\end{flushleft}

\backmatter
\chapter{Appendix: Co-author statements for papers presented in Chapter~\ref{chapter:blob}, \ref{chapter:goodss}, \ref{chapter:predictions} and \ref{chapter:fundamental}}
\markboth{Appendix}{Appendix}
\vspace{-0.5cm}
\section*{Paper I --- Chapter~\ref{chapter:blob}}
\begin{center}
A Lyman-$\alpha$ blob in the GOODS South field: evidence for cold accretion onto a dark matter halo

by
K.K. Nilsson, J.P.U Fynbo, P. M{\o}ller, J. Sommer-Larsen \& C. Ledoux

\emph{Astronomy \& Astrophysics}, \textbf{452}, L23-L26 (2006)

\end{center}

\noindent In this paper, the observations, image reduction and source 
extraction were 
done by J. Fynbo (JF) and P. M{\o}ller (PM). The spectroscopic observations 
were prepared by K. Nilsson (KN). The spectroscopic data was delivered bias 
subtracted,
flat-fielded and wavelength calibrated. KN did the spectrum extraction and 
stacking, sky subtraction, extraction of 1-D spectrum and flux calibration.
KN also did most of the analysis of the multi-wavelength data and the 
imaging/spectroscopic data available and the photometric redshifts. The 
calculations in the beginning of the Discussion were made by Jesper
Sommer-Larsen (JSL), who also did the simulation of cold accretion described
in the Discussion. The text in the paper was mainly written by KN with support
from JF and PM, the plots were all made by KN.

\subsection*{}
\vspace{-1.5cm}
\begin{figure}[!h] 
\begin{center}
\epsfig{file=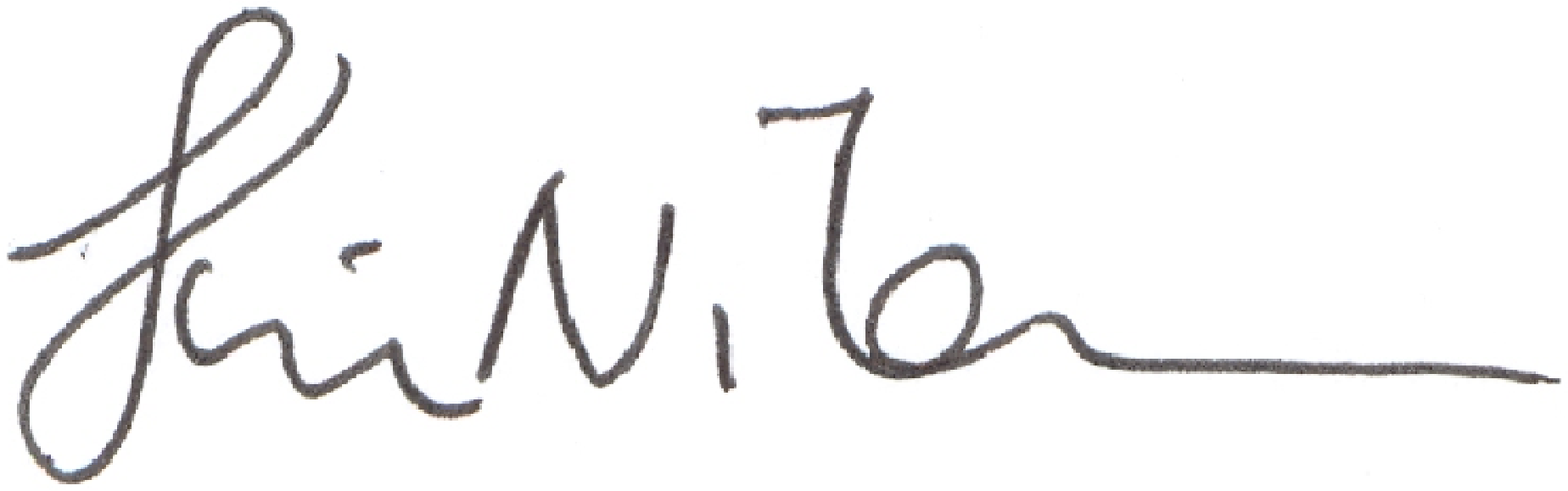,width=6.0cm}\hspace{2.0cm}\epsfig{file=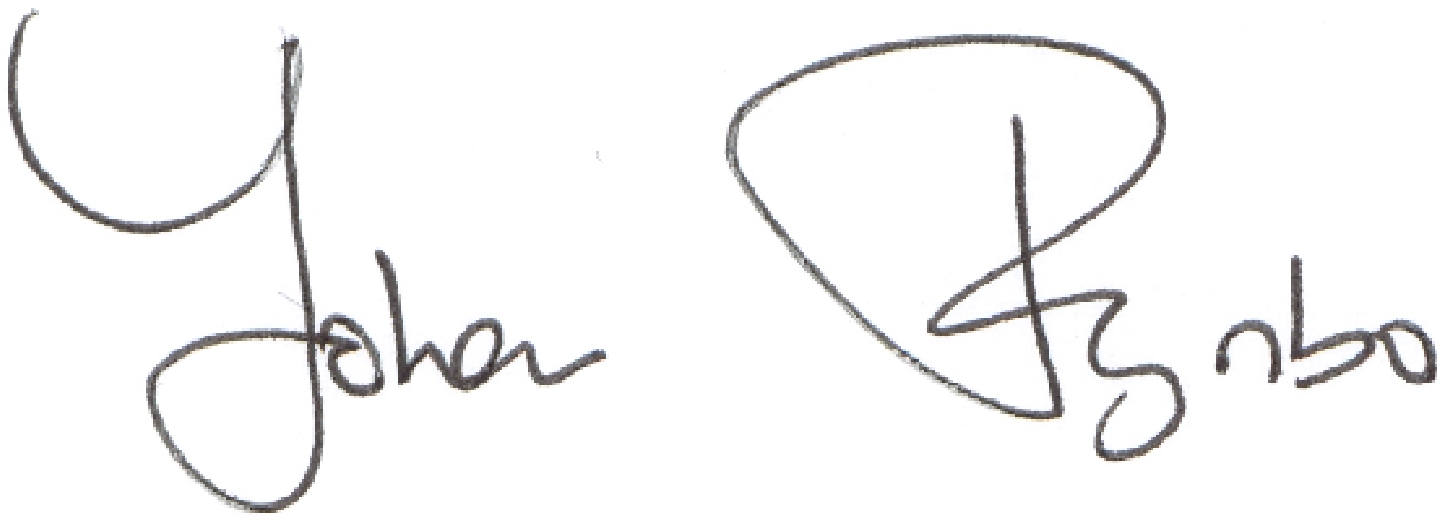,width=6.0cm}
\end{center}
\end{figure}
\vspace{-0.5cm}

\begin{table}[!h]
\begin{center}
\begin {tabular}{llllllllllllllll}
Kim Nilsson &&&&&&&&&&&&&&& Johan Fynbo \\
\end{tabular}
\end{center}
\end{table}

\newpage

%
%
%
%
%
%
%

%

%
\section*{Paper II --- Chapter~\ref{chapter:goodss}}
\begin{center}
A multi-wavelength study of $z = 3.15$ Lyman-$\alpha$ emitters in the GOODS South Field

by
K.K. Nilsson, P. M\o ller, O. M{\"o}ller, J.P.U. Fynbo, M.J Micha{\l}owski, D. Watson, C. Ledoux, P. Rosati, K. Pedersen \& L.F. Grove

\emph{Astronomy \& Astrophysics}, \textbf{471}, 71-82 (2007)

\end{center}

\noindent The results of this paper was based on the same narrow-band imaging
as in Paper I, hence the original narrow-band image data reduction was
performed by Johan Fynbo (JF) and Palle M{\o}ller (PM). The source extraction
was performed by Kim Nilsson (KN), as well as counterpart galaxy 
identification. The spectra were the same as in Paper I, i.e.~basic data
reductions were supplied from the ESO pipeline but KN performed stacking, 
sky subtraction, 1-D spectra extraction and flux calibration. All photometry
in all bands was made by KN. The discovery of the filamentary structure
and the statistical test was performed by KN. For the SED fitting, 
KN stacked the candidates and extracted photometry of the SED, and also gave
input to the SED fitting code. However, the code itself was written by
Ole M{\"o}ller (OM). KN did most of the analysis on LEGO\_GOODS-S\#16, except
for the GRASIL fits which were made by Michal Micha{\l}owski (MM). 
The comparison to LBGs was performed by KN. KN wrote most of the text,
excluding small parts of the introduction, counter-part selection (section 2.2
and 2.3) and SED fitting method section (section 5.1). KN made all plots,
excluding Fig.~8 and 12.

\subsection*{}

\begin{figure}[!h] 
\begin{center}
\epsfig{file=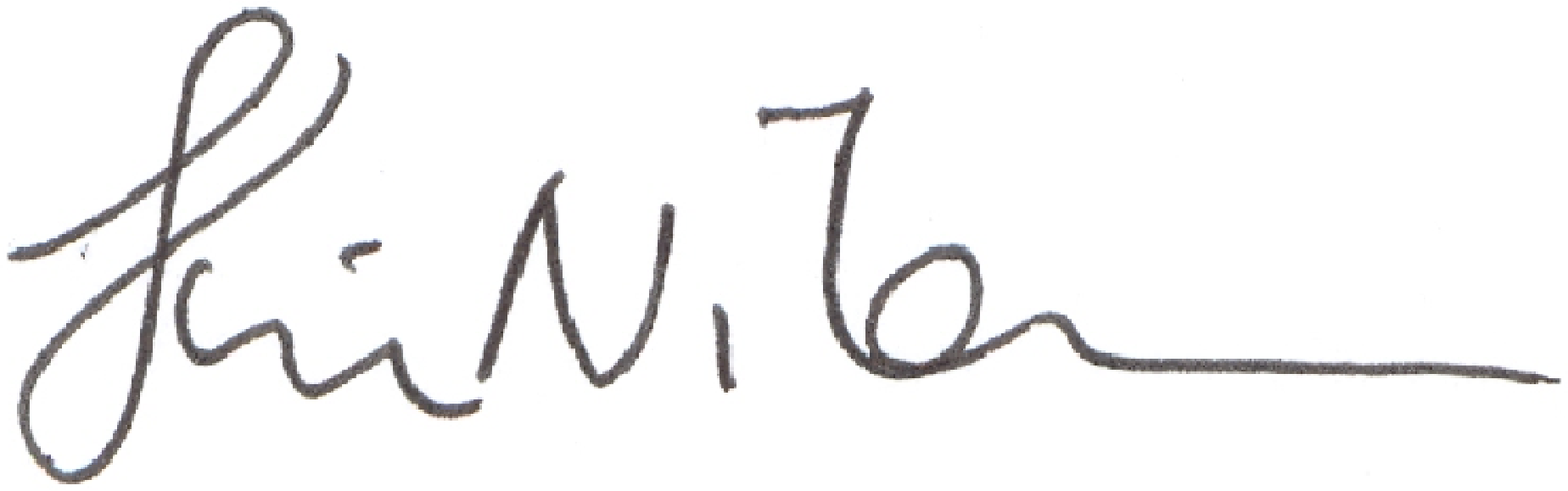,width=6.0cm}\hspace{2.0cm}\epsfig{file=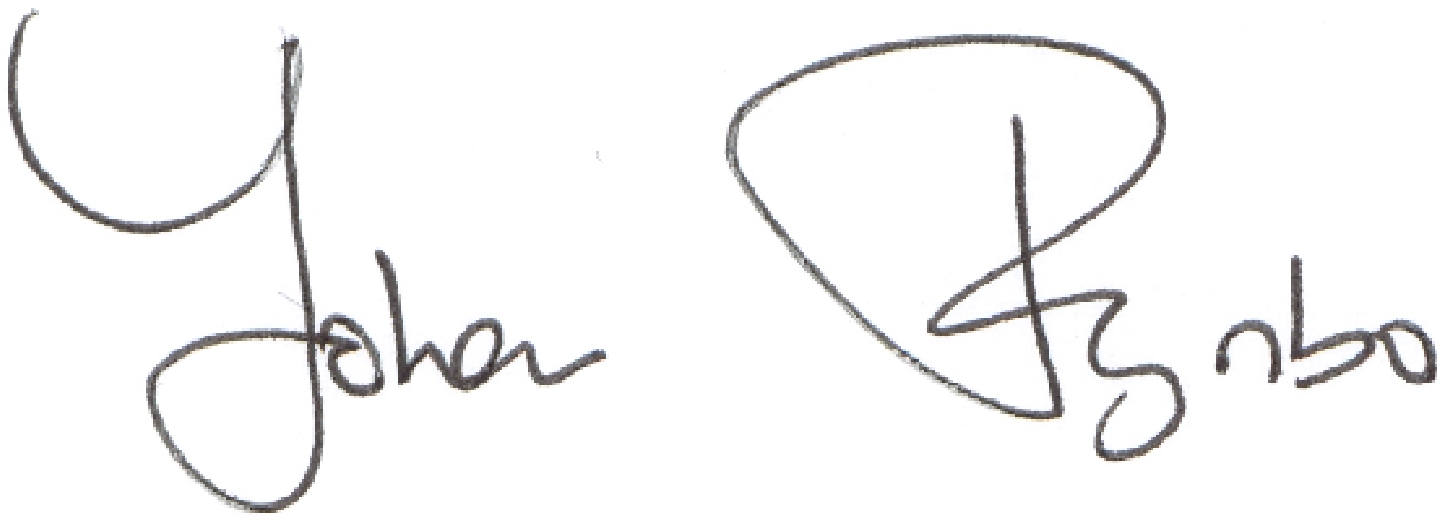,width=6.0cm}
\end{center}
\end{figure}

\begin{table}[!h]
\begin{center}
\begin {tabular}{llllllllllllllll}
Kim Nilsson &&&&&&&&&&&&&&& Johan Fynbo \\
\end{tabular}
\end{center}
\end{table}

\newpage

\section*{Paper III --- Chapter~\ref{chapter:predictions}}
\begin{center}
Narrow-band surveys for very high redshift Lyman-$\alpha$ emitters

by
K.K. Nilsson, A. Orsi, C.G. Lacey, C.M. Baugh \& E. Thommes

accepted in \emph{Astronomy \& Astrophysics}

\end{center}

\noindent The two models (the semi-analytical and the phenomenological model) 
used in this paper have been presented in earlier publications and are not
new to this work. The calculations made from the semi-analytical model were 
made by Alvaro Orsi (AO), Cedric Lacey (CL) and Carlton Baugh (CB) and the 
calculations from the Thommes \& Meisenheimer model were made by Eduard 
Thommes (ET). Kim Nilsson (KN) compared the models with lower redshift results 
and calculated the fit to the Schechter functions. The calculation for
Cosmic Variance for ELVIS were made by AO. The text of the paper was written 
mainly by KN, except for section 2.1 which was written by AO. KN made
all plots except Fig.~3 and 4.

\subsection*{}

\begin{figure}[!h] 
\begin{center}
\epsfig{file=APPENDIX/kim2.ps,width=6.0cm}\hspace{2.0cm}\epsfig{file=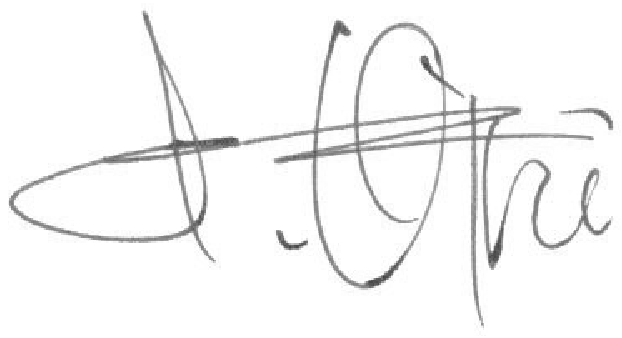,width=5.0cm}
\end{center}
\end{figure}

\begin{table}[!h]
\begin{center}
\begin {tabular}{llllllllllllllll}
Kim Nilsson &&&&&&&&&&&&&&& Alvaro Orsi \\
\end{tabular}
\end{center}
\end{table}

\newpage

\section*{Paper IV --- Chapter~\ref{chapter:fundamental}}
\begin{center}
Selection methods for Lyman-$\alpha$ emitters

by
K.K. Nilsson, O. M{\"o}ller, M. Hayes, P. M{\o}ller \& J.P.U. Fynbo

in prep.

\end{center}

\noindent In this paper, the mock sample of galaxies was created by Matthew
Hayes (MH) and the code to calculate the merit number of each combination
of selection colours was written by Ole M{\"o}ller, all with input from
Kim Nilsson (KN). KN wrote the text and made the plots. The selection criteria
in Table~\ref{tab:criteria} were also retrieved by KN.

\subsection*{}

\begin{figure}[!h] 
\begin{center}
\epsfig{file=APPENDIX/kim2.ps,width=6.0cm}\hspace{2.0cm}\epsfig{file=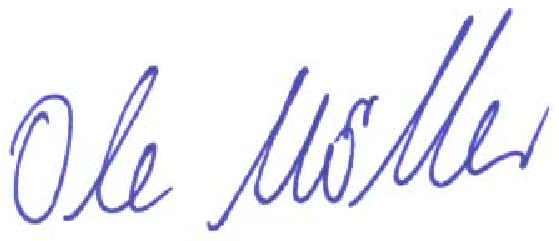,width=5.0cm}
\end{center}
\end{figure}

\begin{table}[!h]
\begin{center}
\begin {tabular}{llllllllllllllll}
Kim Nilsson &&&&&&&&&&&&&&& Ole M{\"o}ller \\
\end{tabular}
\end{center}
\end{table}

\end{document}